\documentclass[a4paper]{article} %

\usepackage{overpic}
\usepackage{graphicx, latexsym, verbatim}
\usepackage{graphics}
\usepackage{amssymb, amsmath}
\usepackage{color}
\usepackage{multirow}
\usepackage{epstopdf}
\usepackage{mdframed}
\definecolor{darkblue}{rgb}{0,0,0.55}
\definecolor{comfrey}{rgb}{0.85,0.85,0.85}
\usepackage{rotating}
\usepackage{pgfplots,tikz}
\usetikzlibrary{arrows}
\usetikzlibrary{patterns}
\usetikzlibrary{shapes.geometric}
\usepackage{pgffor}
\newcommand{\ud}{\mathrm{d}}
\newcommand{\mr}{\mathbf{r}}

\newcommand{\mG}{\mathbf{G}}

\newcommand{\mE}{\mathbf{E}}
\newcommand{\mJ}{\mathbf{J}}
\newcommand{\mH}{\mathbf{H}}
\newcommand{\mB}{\mathbf{B}}
\newcommand{\mD}{\underline{\underline{\mathbf{D}}}}
\newcommand{\mS}{\mathbf{S}}
\newcommand{\mDb}{\underline{\underline{\mathbf{D}}}^\ddagger}
\newcommand{\mA}{\mathbf{A}}

\newcommand{\mF}{\underline{\mathbf{F}}}
\newcommand{\mjt}{\tilde{\mathbf{j}}}
\newcommand{\mft}{\tilde{\mathbf{f}}}
\newcommand{\mfb}{\tilde{\mathbf{f}}^\ddagger}
\newcommand{\mgt}{\tilde{\mathbf{g}}}
\newcommand{\mgb}{\tilde{\mathbf{g}}^\ddagger}
\newcommand{\mFt}{\tilde{\underline{\mathbf{F}}}}
\newcommand{\mFb}{\tilde{\underline{\mathbf{F}}}^\ddagger}
\newcommand{\tlo}{\tilde{\omega}}
\newcommand{\tlk}{\tilde{k}}
\newcommand{\mmG}{\underline{\underline{\mathbf{G}}}}
\newcommand{\mmW}{\underline{\underline{\mathbf{W}}}}
\newcommand{\mmRho}{\underline{\underline{\boldsymbol{\rho}}}}
\newcommand{\mmChi}{\underline{\underline{\boldsymbol{\chi}}}}
\newcommand{\mmDelta}{\underline{\underline{\boldsymbol{\Delta}}}}
\newcommand{\ft}{\tilde{\text{f}}}
\newcommand{\gt}{\tilde{\text{g}}}
\newcommand{\f}{\text{f}}

\newcommand{\me}{\mathbf{e}}
\newcommand{\ma}{\mathbf{a}}

\newcommand{\hz}{\mathbf{\hat{z}}}

\newcommand{\hy}{\mathbf{\hat{y}}}

\definecolor{myBlue}{rgb}{0.0430,0.5156,0.7773}
\definecolor{light-gray}{gray}{0.90}

\newcommand{\reminder}[1]{{\color{purple}}}
\newcommand{\question}[1]{{\color{brown}#1}}

\newcommand{\takeout}[1]{{}}

\newcommand{\fran}[1]{{\color{orange}#1}}

\newcommand{\change}[1]{{#1}}

\begin{document}

\title{Modeling electromagnetic resonators using quasinormal modes}

\author{Philip Tr\o st Kristensen,$^{1*}$ Kathrin Herrmann,$^{1}$  Francesco Intravaia,$^{1}$\\ and Kurt Busch$^{1,2}$ }

\date{\small{%
    $^1$Institut f\"ur Physik, Humboldt-Universit\"at zu Berlin, Newtonstra{\ss}e 15, 12489 Berlin, Germany\\[1ex]%
    $^2$Max-Born Institut, Max-Born-Stra{\ss}e 2a, 12489 Berlin, Germany\\[1ex]
	$^*$Corresponding author: philipk@physik.hu-berlin.de\\[2ex]%
}
    \today
}
\maketitle

\begin{abstract}
We present a bi-orthogonal approach for modeling the response of localized electromagnetic resonators using quasinormal modes, which represent the natural, dissipative eigenmodes of the system with complex frequencies. For many problems of interest in optics and nanophotonics, the quasinormal modes constitute a powerful modeling tool, and the bi-orthogonal approach provides a coherent, precise, and accessible derivation of the associated theory, enabling an illustrative connection between different modeling approaches that exist in the literature.
\end{abstract}

\tableofcontents

\section{Introduction}
Electromagnetic resonators are omnipresent in science and engineering and come in diverse sizes and shapes, ranging from microwave resonators, via cavities for gas and semiconductor lasers, to optical micro cavities and plasmonic nano resonators. Common to all of them is the fact, that the resonances --- i.e. the discrete set of special frequencies that show up as peaks in scattering spectra --- all have associated  electromagnetic field distributions, which are often referred to as ``modes'' of the resonators. In all \change{realistic physical resonators}, moreover, there is a certain degree of \change{dissipation} of energy to the environment. This \change{effect} is responsible for a broadening of the peaks in the spectra and is typically quantified in terms of the so-called quality factor $Q$ --- the higher the $Q$ factor, the longer it takes before an initial excitation of the resonator has \change{dissipated away --- either through radiation to the environment or absorption in the material}. %
\change{From the panoply of different physical systems exhibiting electromagnetic resonances,} %
it is not obvious, that a common mathematical framework exists, which can be used to precisely capture their physical properties in terms of the resonant and possibly dissipative modes. Nevertheless, such a framework does exist, and it is the goal of this Tutorial to present it in some detail along with a number of relevant applications. \change{Indeed, from} a modeling perspective, it is an interesting fact, that the \change{dissipative} %
modes of electromagnetic resonators can be calculated as solutions to a specific eigenvalue problem, namely the sourceless Maxwell wave equation subject to a radiation condition to allow only solutions propagating away from the resonator as large distances. These solutions, which have complex resonance frequencies, are known in the literature as resonant states~\cite{Garcia-Calderon_NP_A265_443_1976,Calderon_2010}, morphology dependent resonances~\cite{Hill_1988, Johnson_JOSAA_10_343_1993}, or quasinormal modes (QNMs)~\cite{Ching_1996,Ching_RevModPhys_70_1545_1998}.

\subsubsection*{A brief comment on nomenclature}
\change{In the history of physics modeling}, the first use of solutions to the wave equation having complex energies is commonly attributed to Gamow~\cite{Gamow_Z_Phys_51_204_1928}, \change{and Zel'dovich appears to be the first to have introduced a method for normalization of such solutions~\cite{Zeldovich_SPJ_12_542_1961}. Both Gamow and Zel'dovich were concerned with problems occurring in quantum mechanics, where the solutions are typically referred to as ``resonant states'' or ``resonance states'', yet the wave nature of the Schr{\"o}dinger equation implies that mathematical approaches developed in this context are useful also in electromagnetism \cite{Weinstein_1969}.} %
``Morphology dependent resonances'' is the name originally given to the resonances observed in various optical micro particles, and which clearly depend on the shape and structure of the particles. For spheres, in particular, the resonances can be immediately related to poles in the Mie expansion coefficients~\cite{van_de_Hulst_1981} and the associated dissipative modes of the spheres~\cite{Fuchs_JOSA_58_319_1968}. By comparing to the Schr{\"o}dinger equation, one can appreciate that the electromagnetic problem of the sphere is very similar to the problem of an electron in a spherical potential well, but with a radially dependent potential which vanishes at large distances~\cite{Johnson_JOSAA_10_343_1993}. The terminology ``quasinormal modes'' appears to have been first used in the context of decaying perturbative solutions to the Einstein equations close to a black hole~\cite{Vishveshwara_Nature_227_936_1970,Frolov_1998}. The similarity between the linearized Einstein equations and Maxwell's equations~\cite{Campbell_AJP_44_356_1976}, as well as the open nature of the universe around a black hole, means that the mathematics are similar to that of electromagnetic resonators in free space. Indeed, some results originally derived for optical resonators have been subsequently applied also to gravitational systems~\cite{Ching_PRL_74_4588_1995}. We shall use the term ``quasinormal mode (QNM)'' to emphasize the fact, that in many ways they represent a precise generalization of the well-known modes of \change{closed cavities (with infinite $Q$ factor)} to the case of \change{general dissipative} resonators, in which the local electromagnetic energy may be lost in the form of radiation to the environment and possibly absorption in the material.

\subsubsection*{Motivation for QNM models}
In many expositions of resonator optics, the \change{dissipation} of the cavity mode is treated in a perturbative manner, where the cavity mode is first calculated for the closed cavity \change{with no absorption} and subsequently coupled to the environment via mostly phenomenological coupling constants. %
Such models, for example, have been extremely successful in modeling optical cavities of high $Q$ factors. Also when the evolution of electromagnetic resonators has led to smaller and smaller devices (in particular in the broad research area known as nanophotonics) the associated modeling has largely been based on ideas rooted in the perturbative approch. %
The strive for smaller resonators (often quantified by the effective mode volume $V_\text{eff}$) has largely been driven by the associated increase in the attainable electromagnetic field strength, even at the expense of lower $Q$ factors, %
as long as the ratio $Q/V_\text{eff}$ remains large; a lower $Q$ factor can even be beneficial, as it relaxes the fabrication tolerances or allows larger bandwidth operation. The limit of low $Q$ factors, however, is exactly where the perturbative closed-cavity resonator models  break down, and where %
a model of the cavity modes in terms of QNMs %
may be particularly advantageous. \change{In the case of metallic nano particles, for example}, the QNMs describe the localized surface plasmon polaritons that are supported by the nano structures.%

A QNM framework is attractive from a conceptual as well as a computational point of view, since in the general case there may be no practical way of calculating or even defining a closed cavity\change{. This is the case, for example, in the second introductory example in Section~\ref{Sec:plasmonic_dimer_of_gold_spheres}.} %
Even in cases where one can make such a definition, the coupling to the environment may lead to substantial frequency shifts and field distortions, which cannot be accurately predicted based on closed-cavity modes. %
\change{Moreoever, typical} descriptions based on QNMs are no more complicated than the perturbative closed-cavity models --- the main difference is that a QNM framework usually provides an explicit and \change{precise} way of calculating the various coupling parameters entering the model. \change{In cases} %
where a single QNM dominates the response, \change{for example}, the result of a QNM model of the Purcell factor can be immediately written in exactly the same form of the original formula due to Purcell~\cite{Purcell_PhysRev_69_681_1946}, but with a slightly modified expression for the effective mode volume~\cite{Kristensen_OL_37_1649_2012}; this definition of a mode volume for a leaky resonator can even be extended to plasmonic systems by properly accounting for \change{absorption and dispersion in} %
the material~\cite{Sauvan_PRL_110_237401_2013}. Finally, In systems described by several QNMs, there will generally be a phase difference between the complex QNM amplitudes and so a formal framework based on QNMs may be very useful.

\subsection{An overview of the existing literature}
\label{Section:existing_literature}
Below we attempt to provide an overview of the existing literature on QNMs. %
As already mentioned in the introduction, the generality of the wave equation means that the usefulness of a framework based on dissipative modes extends through many branches of physics. \change{In keeping with the scope of this Tutorial, however,} %
we shall  limit the overview to electromagnetism.

\subsubsection{Theoretical developments}
Due to the relatively large computational costs of calculating QNMs for general structures, an impressive body of work has been developed for problems in one dimension as well as the analytically tractable cases of cylinders and spheres in two and three dimensions, where the QNMs can be calculated relatively easily. Perturbation theory for QNMs was presented by Lai \emph{et al.} in Ref. \cite{Lai_PRA_41_5187_1990}, and, in a series of papers, Leung \emph{et al.} have treated completeness~\cite{Leung_PRA_49_3057_1994, Leung_JOSAB_13_805_1996, Leung_JPhysA_30_2139_1997}, perturbation~\cite{Leung_PRA_49_3068_1994, Leung_JPhysA_30_2153_1997} and dispersive materials~\cite{Leung_PRA_49_3982_1994}. Lee \emph{et al.} discussed the QNM completeness and electromagnetic Green tensor expansions~\cite{Lee_JOSAB_16_1409_1999} as well as perturbation theory~\cite{Lee_JOSAB_16_1418_1999} with applications to dielectric spheres; the case of degenerate perturbation theory was later discussed in Ref.~\cite{Ng_JOSAB_19_154_2002}. In Ref.~\cite{Muljarov_EPL_92_50010_2010} both one and three-dimensional problems  were treated by Muljarov \emph{et al.} using direct expansions in a subspace of QNMs --- a method generally referred to as resonant state expansion~\cite{Lind_PRC_47_1903_1993} --- to calculate the effects of material changes; this method was later extended to the case of dispersive media in Ref.~\cite{Muljarov_PRB_93_075417_2016}. The one-dimensional problem was also investigated by Settimi in Refs.~\cite{Settimi_PRE_68_026614_2003, Settimi_JOSAB_26_876_2009}, and Doost \emph{et al.} have treated slabs~\cite{Doost_PRA_85_023835_2012}, cylinders~\cite{Doost_PRA_87_043827_2013} and spheres~\cite{Doost_PRA_90_013834_2014} by a QNM expansion of the Green tensor. For modeling using a bi-orthogonal basis in one dimension, see also Refs.~\cite{Leung_PRE_57_6101_1998} and \cite{Li_Commun_Theor_Phys_51_1017_2009}. Recently, the completeness of QNM expansions in spheres made from dispersive materials was investigated by Mansuripur \emph{et al.}~\cite{Mansuripur_PRA_96_013846_2017}. Armitage \emph{et al.} used a QNM expansion of the Green tensor to model planar waveguides with oblique incidence of light in Ref.~\cite{Armitage_PRA_89_053832_2014}, and Lobanov \emph{et al.} recently suggested the use of a resonant state expansion based on the QNMs of a sphere in combination with the Dyson equation to calculate the scattering properties of general resonators~\cite{Lobanov_PRA_98_033820_2018}. %

For treating QNMs in general structures, a variety of numerical methods have been employed and are still under active development for both QNM calculation and normalization. These include the use of volume~\cite{Kristensen_OL_37_1649_2012, deLasson_JOSAB_30_1996_2013} or surface~\cite{Wiersig_JOA_5_53_2002, Makitalo_PRB_89_165429_2014, Alpeggiani_SR_6_34772_2016} integral equation formulations, as well as the Fourier Modal Method (FMM) --- also known as rigorous coupled wave analysis --- for periodic structures \cite{Weiss_PRL_116_237401_2016, Weiss_PRB_96_045129_2017}, or for single resonators by use of so-called Perfectly Matched Layers (PMLs)~\cite{Sauvan_PRL_110_237401_2013,Sauvan_PRA_89_043825_2014}. In two-dimensional coupled cavity-waveguide structures, the QNMs have been calculated by a Dirichlet-to-Neumann technique in Ref.~\cite{Hu_OE_16_17383_2008}, by FMM~\cite{deLasson_JOSA_A_31_2142_2014}, or by Finite Element (FEM) calculations with a nonlocal boundary condition in Ref.~\cite{Kristensen_OL_39_6359_2014}. The latter work also discussed normalization of the QNMs via the theory of divergent series. %
R{\"o}mer \emph{et al.}~\cite{Romer_JOSAB_25_31_2008} used QNMs obtained as the solutions to the wave equation in FEM calculation with PMLs to study spontaneous emission from emitters in photonic crystal cavities. To this end, the QNMs of interest were normalized by a volume integration which extended through the PML, a technique which was later \change{used also} by Sauvan~\emph{et al.} in combination with FMM calculations~\cite{Sauvan_PRL_110_237401_2013,Sauvan_PRA_89_043825_2014}. A FEM formulation leads essentially to one \change{numerical eigenmode} per degree of freedom in the problem. The vast majority of \change{these eigenmodes}, however, are connected with the PMLs and at first sight do not appear to be useful for modeling. Nevertheless, as shown by Vial \emph{et al.} for two-dimensional open systems~\cite{Vial_PRA_89_023829_2014}, the full set of modes containing the dominant QNMs and the auxiliary PML modes provide a useful basis for expansion of the solutions to the wave equation. These ideas were subsequently further developed by Yan \emph{et al.}~\cite{Yan_PRB_97_205422_2018} in three dimensions. Muljarov \emph{et al.} have contributed to the discussion of usefulness of various normalization methods in Ref.~\cite{Muljarov_PRB_94_235438_2016}; see also Refs.~\cite{Kristensen_PRA_92_053810_2015, Muljarov_PRA_96_017801_2017, Kristensen_PRA_96_017802_2017}. 

Numerical eigenmode calculations are typically expensive. As an alternative, therefore, direct calculations of the electromagnetic response at complex frequencies may be advantageous, as was pointed out by Bai \emph{et al.}~\cite{Bai_OE_21_27371_2013} and Perrin~\cite{Perrin_OE_24_27137_2016}. Such approaches are similar in spirit to the Riesz projection approaches~\cite{Zschiedrich_PRA_98_043806_2018, Binkowski_arXiv_1811_11624v2_2018}, which, in turn, are closely related to the Green tensor expansion usually employed in resonant state expansion literature. Although the QNMs are defined in the frequency domain, they have been successfully calculated also with time domain methods --- notably the finite difference time domain (FDTD) method --- in combination with PMLs and a Fourier transformation~\cite{Kristensen_OL_37_1649_2012, Ge_NJP_16_113048_2014, Ge_OL_39_4235_2014}. %

An inherent exponential divergence of the QNMs at large distances from the resonator means that they cannot be directly used to describe fields far from the resonators. To handle these limits, the use of a Dyson equation approach based on QNMs has been suggested~\cite{Ge_NJP_16_113048_2014, Dezfouli_PRB_97_115302_2018}. The far field problem has also been treated by Abdelrahman \emph{et al.} in Ref. \cite{Abdelrahman_OSAC_1_340_2018}, and in the context of coupled mode theory (CMT) in Ref.~\cite{Kristensen_JLT_35_4247_2017}. The problem is closely connected to that of scattering calculations, which have been treated in a number of ways in Refs.~\cite{Alpeggiani_PRX_7_021035_2017, Yan_PRB_97_205422_2018, Weiss_PRB_98_085433_2018, Lobanov_PRA_98_033820_2018, Unger_PRL_121_246802_2018}. %
Similar to the divergence at large distances, the dramatic increase of the electromagnetic feedback close to metal surfaces cannot be captured by a single QNM. In such cases, the response can be conveniently handled by an additional quasi-static contribution to the Green tensor~\cite{Ge_NJP_16_113048_2014}. %

\subsubsection{Practical applications}
Even with some fundamental questions still unsolved, %
there has been a recent bloom in the use of QNMs for practical modeling tasks in nanophotonics. Many of these applications can be seen as refinements of well-established calculation methods for resonant structures, where now the QNMs provide explicit and precise definitions of parameters that would normally be inferred by fitting to calculation data or measurements. The close connection between the QNMs and the resonances in scattering matrices of general structures has been clarified~\cite{Lobanov_PRA_98_033820_2018, Alpeggiani_PRX_7_021035_2017, Weiss_PRB_98_085433_2018}, and QNMs have been used as inputs to laser models~\cite{Andreasen_AOP_3_88_2011,Cartar_PRA_96_023859_2017} and for the derivation of the so-called temporal CMT %
~\cite{Kristensen_JLT_35_4247_2017}, including applications to switching in nonlinear materials~\cite{Kristensen_APL_102_041107_2013}. Yang~\emph{et al.} used QNM perturbation theory for sensing applications~\cite{Yang_NL_15_3439_2015}, and QNMs of resonators modeled with a nonlocal material response were presented in Ref.~\cite{Dezfouli_Optica_4_1503_2017}. The QNMs in coupled cavity-waveguide systems were used for perturbation theory and Purcell factor calculations~\cite{Kristensen_OL_39_6359_2014} as well as broadband local density-of-states (LDOS) calculations in two dimensional systems~\cite{deLasson_OL_40_5790_2015}. Malhotra~\emph{et al.} subsequently applied the theory in three dimensions to describe on-chip single photon emitters \cite{Malhotra_OE_24_13574_2016}. Also, QNM descriptions have been employed for LDOS calculations in hybrid plasmonic photonic structures~\cite{Dezfouli_PRA_95_013846_2017} and used for theoretically predicting~\cite{Ge_J_Opt_18_054002_2016} and interpreting~\cite{Hoerl_ACS_Phot_2_1429_2015} electron energy loss spectroscopy results. %

Classical laser cavity models can be said to implicitly rely on a QNM picture, namely that of a cavity with a multitude of resonant modes\change{, each of which
repeats itself after a full round trip}. Partial transmittance at the cavity end facets means that part of the electromagnetic energy escapes, leading to complex resonance frequencies. By compensating the \change{radiative energy loss} %
through a gain medium, the resonance frequency can be shifted towards the real axis, which leads to the characteristic line width narrowing at the onset of lasing~\cite{Andreasen_AOP_3_88_2011,Tureci_PRA_74_043822_2006}. In the limit of single energy quanta, %
microscopic semi-classical theories of light-matter interaction can be set up in a number of ways, often based on the Green tensor. In these limits, a QNM expansion immediately leads to physically appealing models for modified spontaneous emission calculations. %
As an alternative to methods related to the Green tensor, the QNMs have been used as input to single mode master equations~\cite{Alpeggiani_SR_6_34772_2016, Ge_PRB_92_205420_2015, Mork_Cavity_Photonics_2015, Dezfouli_arXiv_1805_10153v2_2018}. %
The problem of full quantum models based on QNMs %
has been discussed by Ho~\emph{et al.}~\cite{Ho_PRE_58_2965_1998}, Dutra~\emph{et al.}~\cite{Dutra_PRA_62_063805_2000}, Severini \emph{et al.}~\cite{Severini_PRE_70_056614_2004}, and recently by Franke~\emph{et al.}~\cite{Franke_PRL_122_213901_2019}.

To round off this overview, we remark that the QNMs of localized electromagnetic resonators are related to the modes of optical fibers and general waveguides, where absorption or radiation will also lead to inherently dissipative modes, and one can conveniently perform projections by use of suitably defined adjoint modes as we do in the derivations to follow. %
In addition, as \change{mentioned} above, QNMs have been \change{studied} in areas of physics other than electromagnetic resonators, notably in \change{quantum mechanics~\cite{Moiseyev_2011} and in general relativity~\cite{Konoplya_RevModPhys_83_793_2011}, but also in acoustics~\cite{Kergomard_JASA_119_1356_2006}. For a number of reviews on QNMs of electromagnetic resonators,} see Refs.~\cite{Ching_RevModPhys_70_1545_1998, Kristensen_ACS_Phot_1_2_2014, Lalanne_LPR_12_1700113_2018}.

\subsection{Scope and structure}
In this Tutorial, we present a variant of the %
well established expansion and projection method using %
bi-orthogonal modes~\cite{Cole_1968} to model light scattering by localized electromagnetic resonators in terms of QNMs. %
At first sight, this approach, which is independent of the particular calculation technique used to obtain the QNMs, suffers from %
the fact, that the QNMs in two and three dimensions obey the asymptotic requirement of a radiation condition instead of a boundary condition. For this reason, the traditional approach~\cite{Cole_1968} is not directly applicable in dimensions higher than one. Nevertheless, as we shall see, it is possible to extend the theory to higher dimensions by ideas originally developed in the literature on resonant states. For definiteness, we limit most of the derivations and illustrations %
to resonators embedded in homogeneous materials, where the scattered electromagnetic fields obey the Silver-M{\"uller} radiation condition. This does not include all technologically relevant geometries. Indeed, the electromagnetic field in resonators on top of substrates~\cite{Weiss_PRL_116_237401_2016, Weiss_PRB_96_045129_2017, Yan_PRB_97_205422_2018}
or coupled to optical waveguides~\cite{Hu_OE_16_17383_2008, deLasson_JOSA_A_31_2142_2014, Kristensen_OL_39_6359_2014, deLasson_OL_40_5790_2015, Malhotra_OE_24_13574_2016}
obey different radiation conditions and are not immediately covered by the theory laid out in this article. %
Such cases must be treated either by modifications of the theory using suitable radiation conditions, or by theoretical approaches specific to the calculation method of the QNMs. References~\cite{Sauvan_PRL_110_237401_2013, Yan_PRB_97_205422_2018}, for example, describe modal techniques in which the calculation domain truncation by use of PMLs are an integral part of the modeling approach, in particular of the normalization procedure. Similarly, the normalization scheme by use of divergent series in Refs.~\cite{Kristensen_OL_39_6359_2014, deLasson_OL_40_5790_2015, Malhotra_OE_24_13574_2016} arises naturally from the waveguide radiation condition of the QNMs in coupled cavity-waveguide systems. These and other interesting cases are beyond the scope of this Tutorial, and we refer instead to the literature, cf. Section~\ref{Section:existing_literature}.

In Section \ref{Sec:Introductory_example}, we present two introductory examples of practical QNM modeling applications; transmission through a dielectric barrier in one dimension, and Purcell factor calculations for a plasmonic dimer of nano spheres in three dimensions. In the course of the Tutorial, we shall repeatedly return to these example structures to exemplify the various calculations. The one-dimensional example is sufficiently simple that most of the calculations can be handled analytically, and we much encourage the interested reader to repeat them. As a supplement to this Tutorial, we provide a number of Matlab files implementing some of the one-dimensional examples~\cite{dielectricBarrier_arXiv} as well as the code necessary for calculating QNMs of three-dimensional resonators~\cite{plasmonicDimer_arXiv} using the freeware code MNPBEM~\cite{Hohenester_CPC_183_370_2012, Waxenegger_CPC_193_138_2015, Hohenester_CPC_22_209_2018}.

The remainder of this Tutorial is organized as follows. In Section~\ref{Sec:QNM_calculation_methods}, we discuss various methods currently in use for practical QNM calculations. Section~\ref{Sec:Theory} lays out the basic elements of the theory for general, three-dimensional resonators in homogeneous environments. We define the QNMs as the solutions to the sourceless wave equation subject to the Silver-M{\"u}ller radiation condition and show how this requirement naturally leads to the definition of adjoint QNMs and a projection operator for projection of certain solutions to the wave equation onto the QNMs. Using the projection operator, we construct formal expansions of general electromagnetic fields and the electromagnetic Green tensor in terms of QNMs. In addition, we discuss how the QNMs are directly related to the residues of the Green tensor, as typically exploited in the literature on resonant states, and we show how such an approach leads naturally to a convenient alternative normalization procedure for the QNMs. In Section~\ref{Sec:Completeness}, we discuss the question of convergence of the formal expansions in terms of QNMs which, for %
QNM expansions, can be assessed by the limiting behavior of the Green tensor. We discuss, how an %
investigation of the Green tensor can be used to define a region of convergence for the QNM expansions, and how one can subsequently exploit this knowledge to extend the region of convergence. Section~\ref{Sec:Applications} is devoted to a number of practical applications of QNM modeling. Section~\ref{Sec:Scattering_calculations} presents the CMT and scattering calculations, Section~\ref{Sec:Coupled_resonators} is concerned with hybridization and discusses how one can expand the QNMs of coupled resonators in terms of the QNMs of the individual resonators, Section~\ref{Sec:Perturbation_theory} presents various applications of perturbation theory, and Section~\ref{Sec:Purcell_factor_calculations} discusses the use of QNMs for Purcell factor calculations. Finally, Section~\ref{Sec:Conclusions} holds the conclusions.

\subsection{Introductory examples}
\label{Sec:Introductory_example}
To illustrate the usefulness of the QNMs in modeling responses of electromagnetic resonators, we consider two examples --- a one-dimensional example of a dielectric block, and a three-dimensional example of a dimer made from two gold nano spheres. We shall repeatedly come back to these introductory examples throughout the article, to illustrate the explicit application of the various results. %

\subsubsection{Dielectric barrier in one dimension}
\label{Sec:1Dresonator} %
\begin{figure}[htb!]
\centering %
\begin{overpic}[width=9.4cm]{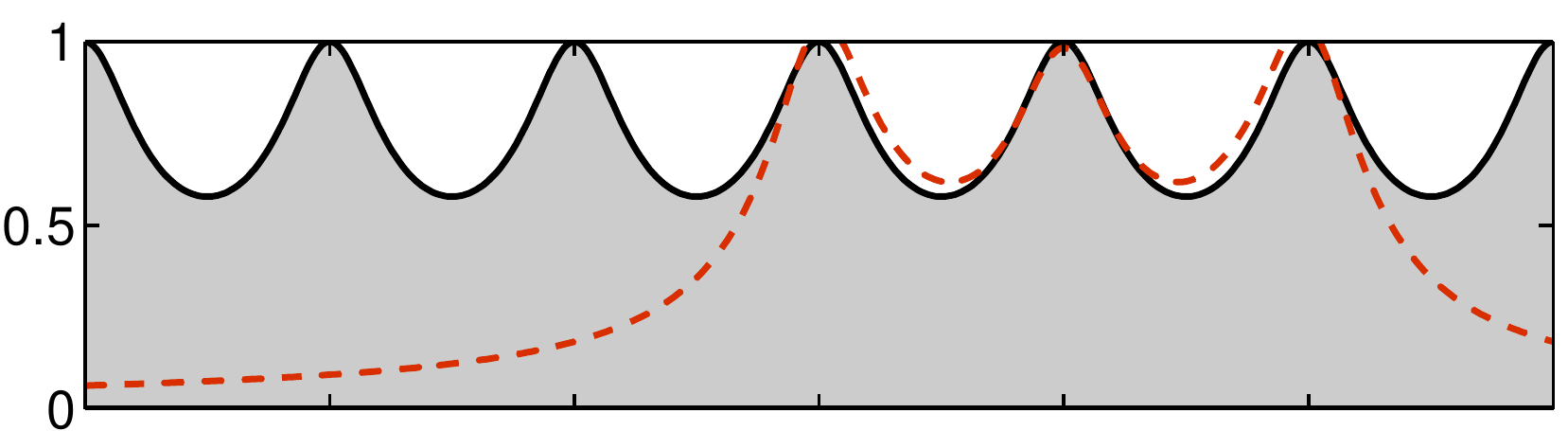}
\put(-6,1.5){\begin{sideways}Transmission\end{sideways}}
\end{overpic}\\
\begin{overpic}[width=9.65cm]{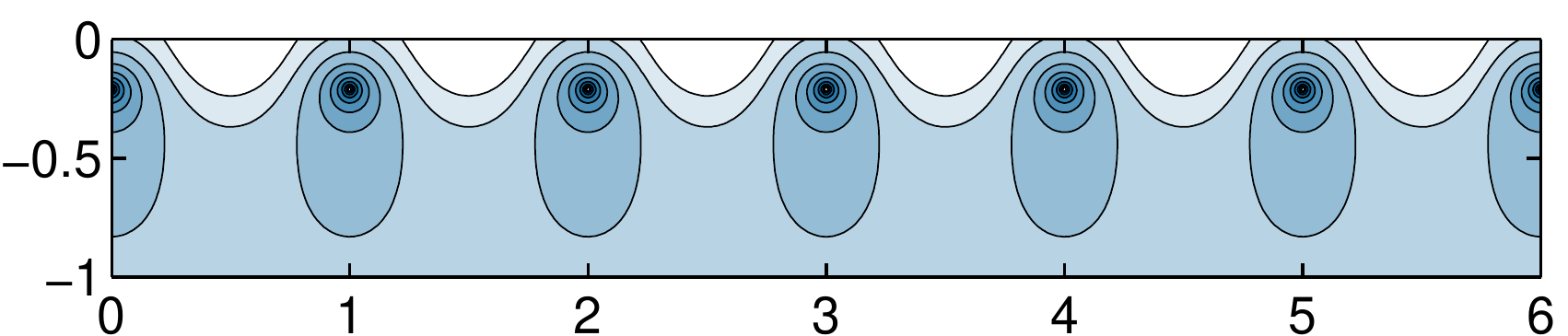}
\put(-10,3){\begin{sideways}Frequency\end{sideways}}
\put(-5,7){\begin{sideways}$\omega_\text{I}L/\text{c}$\end{sideways}}
\put(38,-5){Frequency, $\omega_\text{R}L/\text{c}$}
\end{overpic}
\\[8mm]
\begin{overpic}[width=9.65cm]{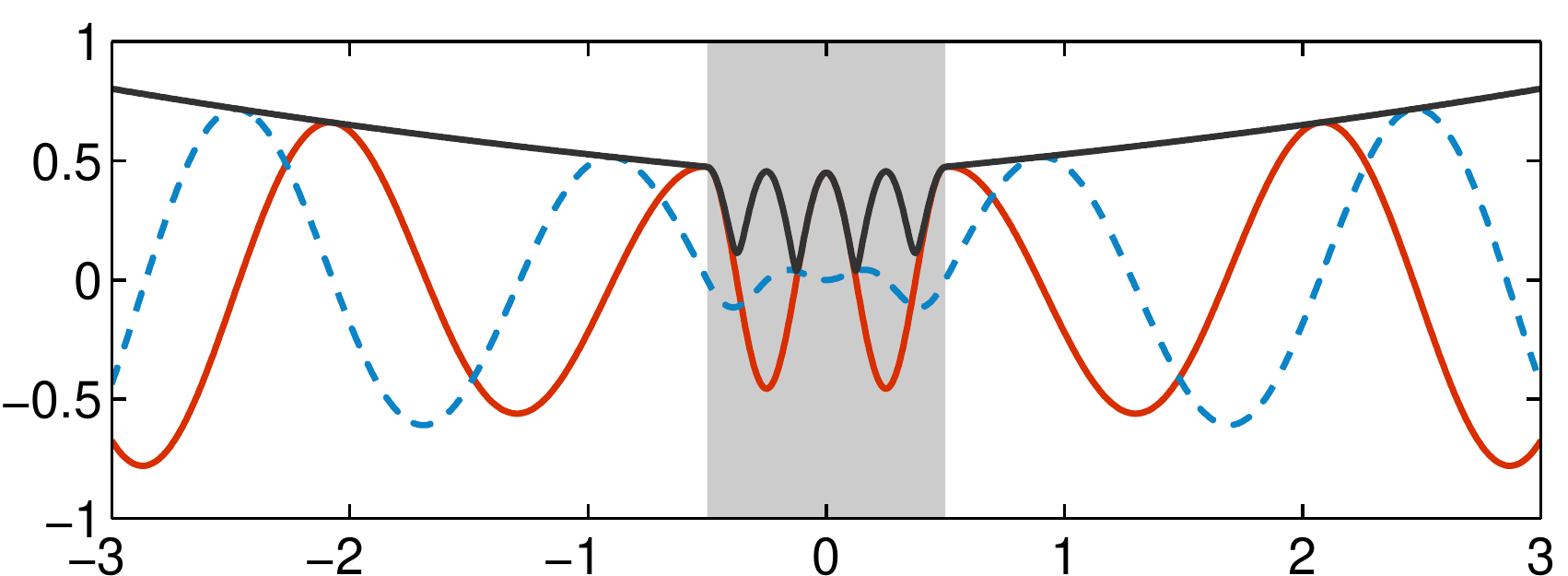}
\put(-10,3){\begin{sideways}Electric field QNM\end{sideways}}
\put(-5,14){\begin{sideways}$\ft_4/\sqrt{L}$\end{sideways}}
\put(42,-6){Position, $x/L$}
\end{overpic}
\\[5mm]
\caption{\label{Fig:transmission_spectrum_dielectricBarrier_nR_pi}Top: Transmission through the dielectric barrier as a function of frequency for the case of $n_\text{R}=\pi$. Dashed red curve shows the approximate transmission calculated using three QNMs. Center: Complex spectrum, in which the discrete QNM frequencies show up as dark spots in the lower half of the complex plane. The logarithmic color coding indicates the magnitude of the left-hand side of \change{Eq.~(\ref{Eq:1D_errorMap}).} %
Bottom: Mode profile of the electric field QNM with $\text{Re}\{\tlo\}=4\text{c}/L$. Red solid and blue dashed curves show the real and imaginary parts of the field, respectively, and the black curve shows the absolute value, which increases (exponentially) as a function of distance from the resonator. Gray shading indicates the extent of the dielectric barrier.} 
\end{figure}
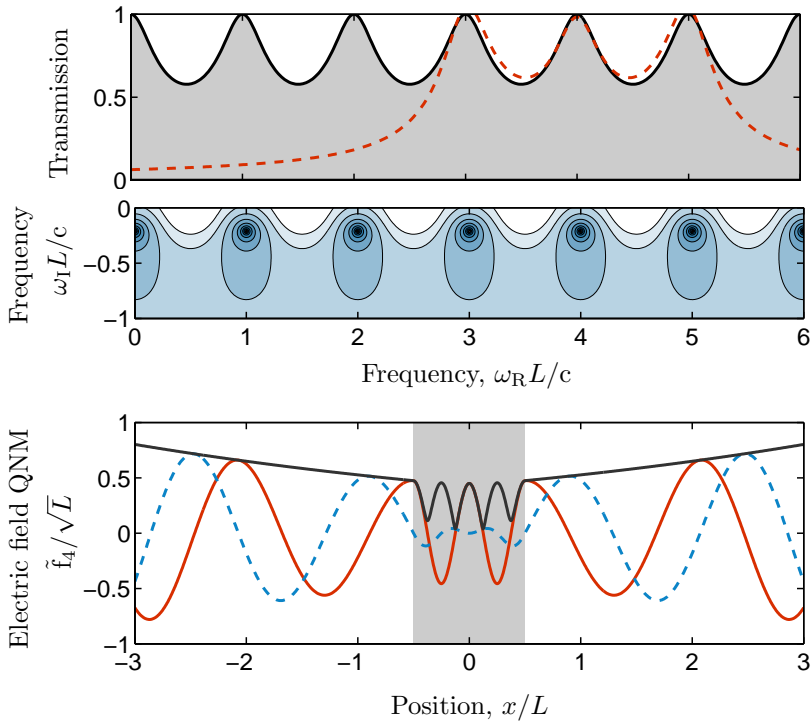

We consider first the classical example of a dielectric barrier with constant refractive index $n_\text{R}=\pi$ in a background with refractive index $n_\text{B}=1$. For definiteness, we consider propagation along the $x$ axis, and we take the electric field to point in the $y$ direction; the barrier has a width of $L$ and is centered on the origin. The top panel in Fig.~\ref{Fig:transmission_spectrum_dielectricBarrier_nR_pi} shows the transmission through the barrier as a function of frequency.
It features a number of distinct peaks with unity transmission; these are the well-known Fabry-Perot resonances, each of which can be directly associated with a QNM. The center panel shows a part of the complex frequency spectrum below the real axis, in which the QNM frequencies show up as dark spots. For this particular choice of refractive index, the real part of the QNM frequencies are evidently spaced equally by $L/\text{c}$, and all QNM frequencies have the same imaginary part. %
The bottom panel of Fig.~\ref{Fig:transmission_spectrum_dielectricBarrier_nR_pi} shows an example of a QNM field in the vicinity of the resonator. As one might expect, the QNMs represent standing waves inside the resonator, as would be the case also for a closed cavity made from perfect reflectors at each side. Contrary to the closed cavity case, the field is non zero outside, where it propagates away from the resonator at both sides. Notably, the field magnitude clearly increases in the direction away from the resonator, which is a general feature of QNMs and  %
necessitates the use of a normalization which is different from that of the closed cavity case. Once normalized, however, the QNMs can be used to calculate quantities of typical interest in the description of electromagnetic resonators. As an example, the dashed red curve in the top panel shows the approximate transmission as calculated using the three QNMs with $3\,\text{c}/L\leq\text{Re}\{\tlo\}\leq\,5\,\text{c}/L$. The approximation is best in the center of the frequency interval, where the relative error is less than one percent, and by systematically increasing the number of QNMs the error can be made arbitrarily small, as we shall see in Section~\ref{Sec:Scattering_calculations_for_the_dielectric_barrier}.

\subsubsection{Plasmonic dimer of gold spheres}
\label{Sec:plasmonic_dimer_of_gold_spheres}
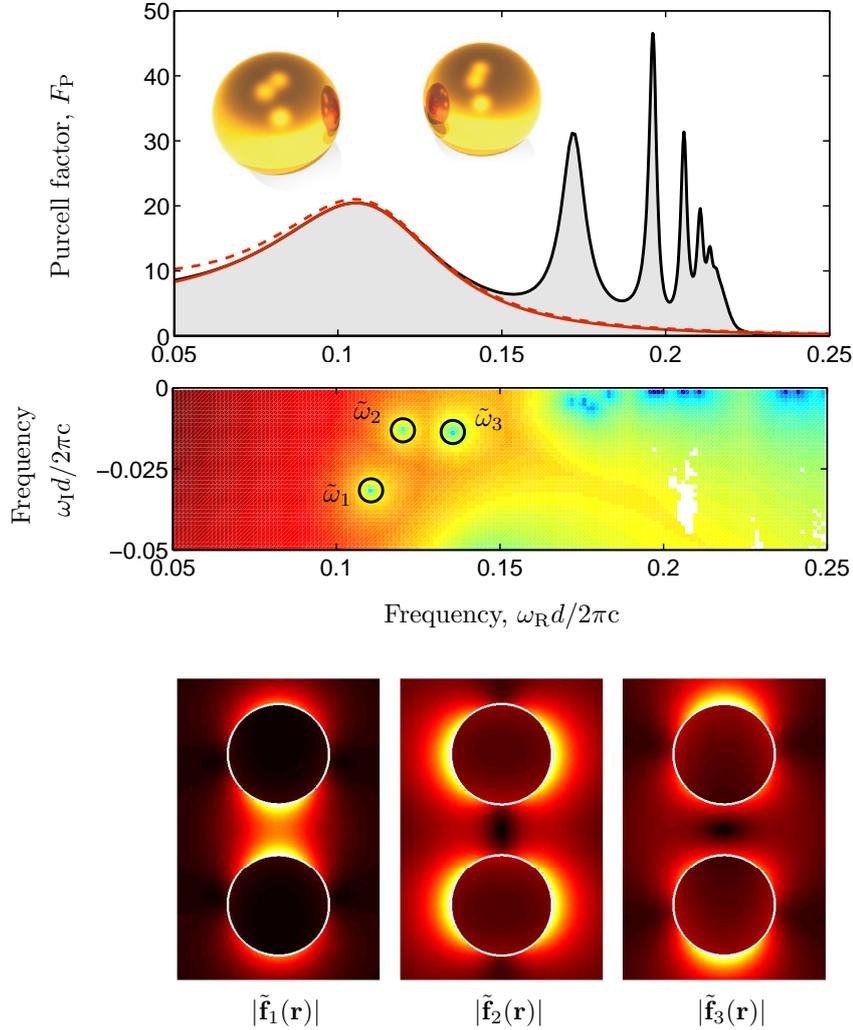
\begin{figure}[htb!]
\centering %
\quad\;\;
\begin{overpic}[width=9.4cm]{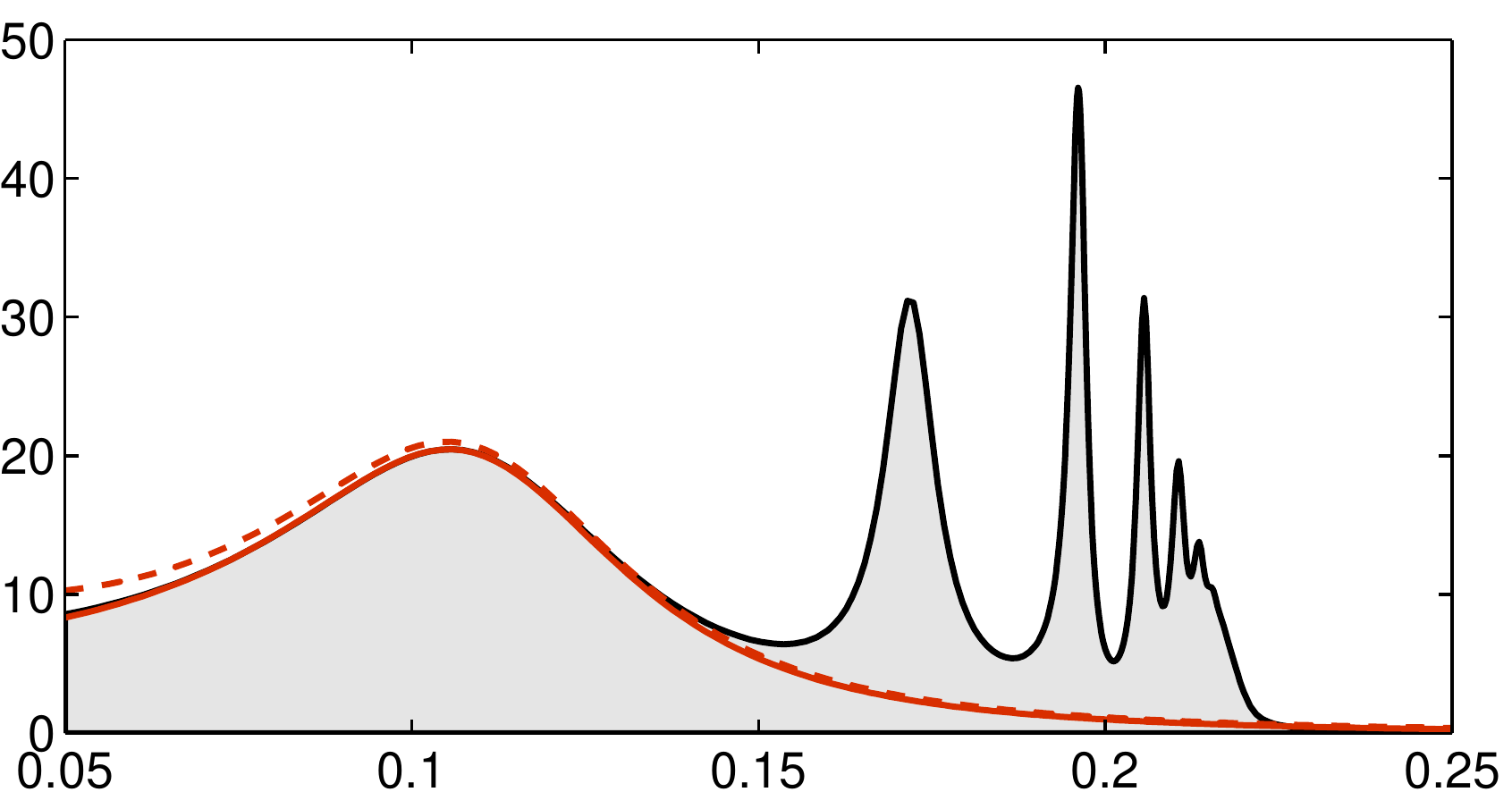}
\put(-13,12){\begin{sideways}Purcell factor, $F_\text{P}$\end{sideways}}
\put(9,24){\includegraphics[width=4.5cm]{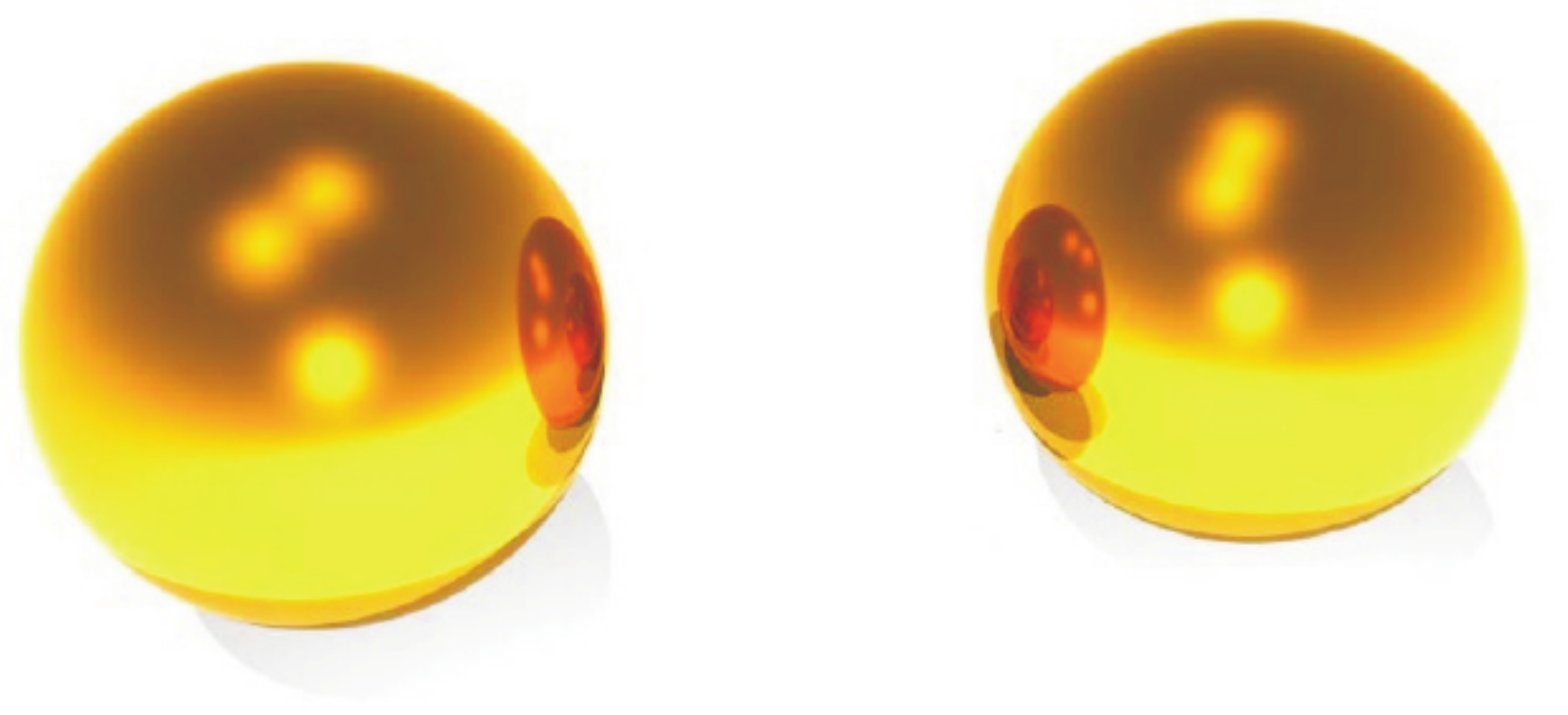}}
\end{overpic}\\[1mm]
\begin{overpic}[width=10cm]{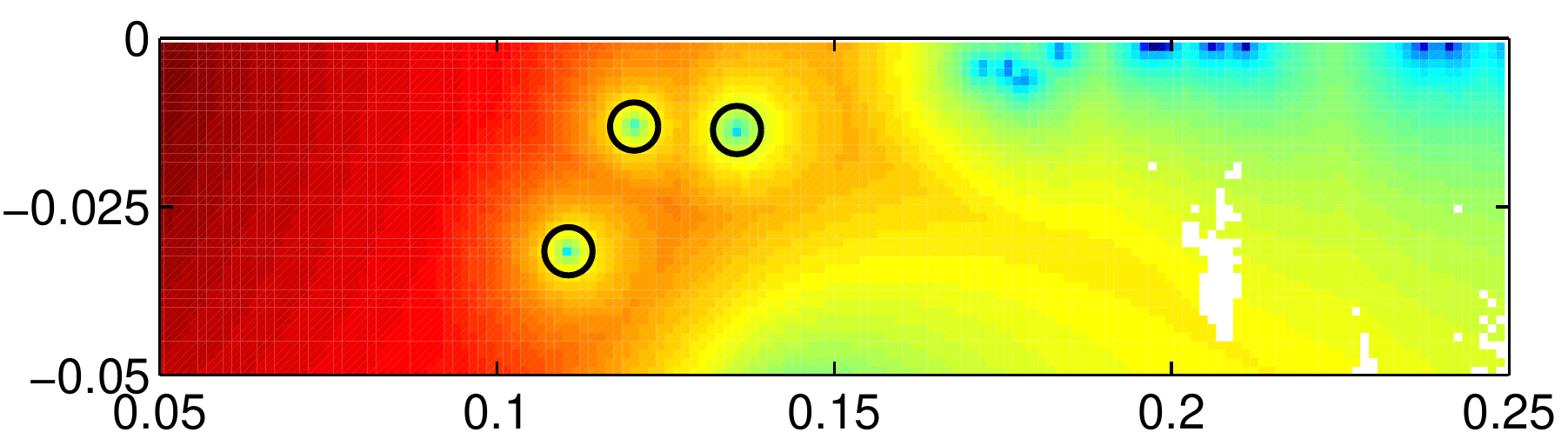}
\put(30,10){$\tlo_1$}
\put(34,21){$\tlo_2$}
\put(50,20){$\tlo_3$}
\put(-11,7){\begin{sideways}Frequency\end{sideways}}
\put(-6,8){\begin{sideways}$\omega_\text{I}d/2\pi\text{c}$\end{sideways}}
\put(38,-6){Frequency, $\omega_\text{R}d/2\pi\text{c}$}
\end{overpic}\\[10mm]
\hspace{.7cm}\begin{overpic}[width=2.75cm]{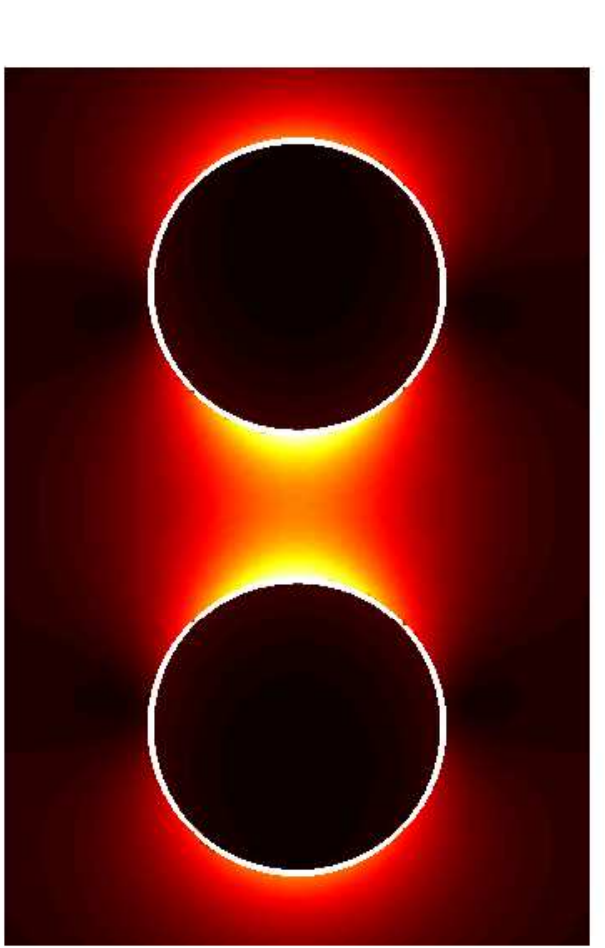}
\put(23,-12){$|\mft_1(\mr)|$}
\end{overpic}
\;\begin{overpic}[width=2.75cm]{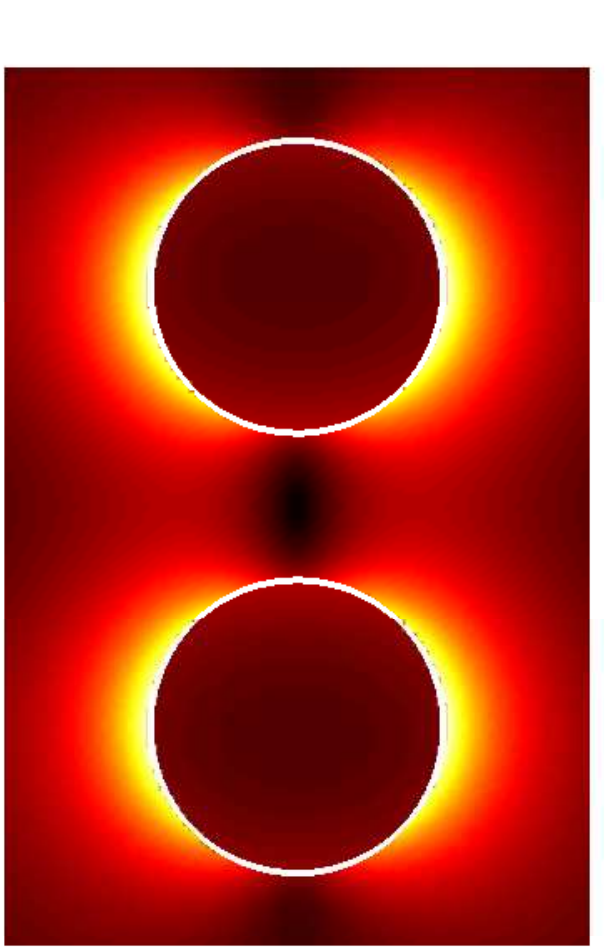}
\put(23,-12){$|\mft_2(\mr)|$}
\end{overpic}
\;\begin{overpic}[width=2.75cm]{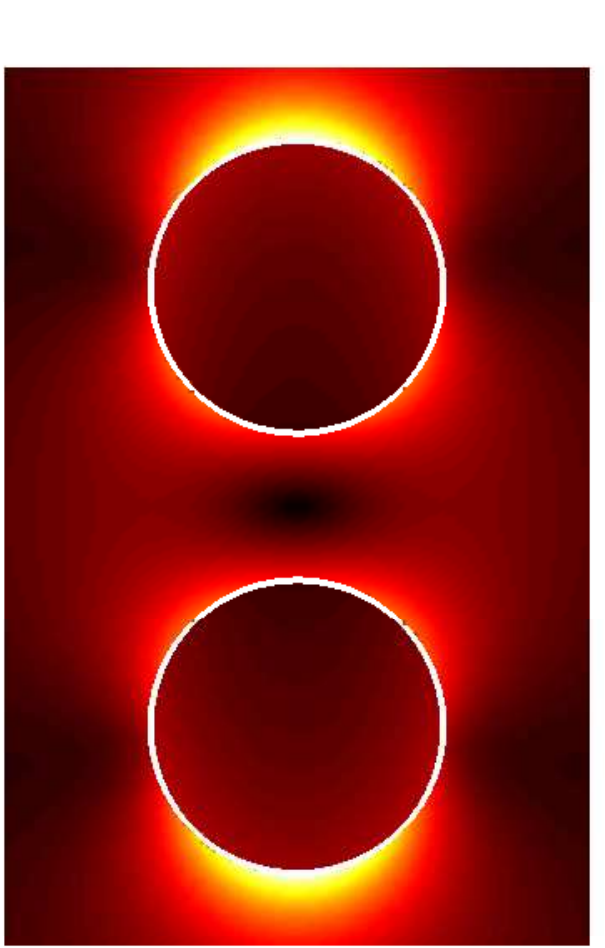}
\put(23,-12){$|\mft_3(\mr)|$}
\end{overpic}\\[5mm]
\caption{\label{Fig:Purcell_avs_Lmax_6_rdip_0p5_0p25_0_tloRange_0p05_to_0p25}Top: Purcell factor of a dimer made from gold nano spheres (see inset) %
as a function of frequency. Red curves shows the approximate responses calculated using only the QNMs $\mFt_1(\mr)$ (dashed) or $\mFt_1(\mr)$ and $\mFt_1^*(\mr)$ (solid). Center: Complex QNM spectrum, in which the QNM resonance frequencies show up as dark blue spots. The logarithmic color coding indicates the smallest magnitude eigenvalue of the operator  $\mathbf{M}_\mG(\omega)-1$, cf. Eq.~(\ref{Eq:LippmannSchwinger_discretized}). The lowest order eigenvalues of interest are indicated with black circles and numbered. Bottom: Field profiles showing the absolute values of the three QNM fields corresponding to the frequencies indicated in the spectrum above. Only $\mft_1(\mr)$ contributes to the Purcell factor in the gap center.} \end{figure}

As a second example, we consider a dimer made from two gold spheres of radius $R=50$\,nm embedded in air and separated by a distance $d=50$\,nm. %
The electromagnetic response of the gold is modeled with a Drude model of the form
\begin{align}
\epsilon_\text{r}(\omega) = 1 - \frac{\omega_\text{p}^2}{\omega^2+\text{i}\gamma\omega},
\end{align}
where we use $\hbar\omega_\text{p}=7.9$\,eV and $\hbar\gamma=0.06$\,eV for the plasma frequency and damping rate\change{, respectively}. The top panel in Fig.~\ref{Fig:Purcell_avs_Lmax_6_rdip_0p5_0p25_0_tloRange_0p05_to_0p25} shows the Purcell factor --- the spontaneous emission enhancement relative to the homogeneous background medium --- as a function of frequency of a dipole emitter oriented along the dimer axis and located in the gap center. %
The panel below shows a part of the complex frequency spectrum below the real axis, in which the QNM frequencies show up as dark blue spots. A clear correlation is visible between the peaks in the spectrum of Purcell factors and some of the QNM resonance frequencies. In particular, the real part of $\tlo_1$ is (almost) aligned to the first peak in the spectrum. The associated QNM $\mFt_1(\mr)$ is the so-called dipole mode of the resonator, with a $Q$ factor of only $Q=1.75$; the mode profile of its electric component $\mft_1(\mr)$ is depicted in the first figure below the spectrum. From the mode profiles of the modes $\mft_2(\mr)$ and $\mft_3(\mr)$, we can immediately see that they vanish in the gap center, which is why they do not contribute to the Purcell factor at this location (they will contribute at other locations). In fact, the Purcell factor in the gap center is almost entirely due to the response of the dipole mode, as seen from the red dashed curve showing the result of a QNM model using only this mode. The relative difference between the curves at the peak maximum is on the order of a few percent, which is typical of --- or even slightly worse than --- the level of accuracy found in many practical applications of QNM theory using one or very few QNMs. %
The level of accuracy can be dramatically increased by including also the mode $\mFt^*_1(\mr)$ in the model, as seen from the solid red curve; this can be done at no additional computational cost. Also, the bandwidth of the QNM approximation can be increased to accurately cover the higher lying peaks in the spectrum (not shown in Fig.~\ref{Fig:Purcell_avs_Lmax_6_rdip_0p5_0p25_0_tloRange_0p05_to_0p25}). These facts are discussed in detail in Section~\ref{Sec:Purcell_factor_of_plasmonic_dimer}. %

\section{QNM calculation methods}
\label{Sec:QNM_calculation_methods}
Although the \change{dissipative} nature of \change{physical} electromagnetic resonators is easy to accept, the introduction of associated dissipative modes of the electromagnetic field appears to be less intuitive. Indeed, scientists will routinely refer to the $Q$ factor of a resonator, but rarely to the $Q$ factor of the dissipative mode of the electromagnetic field associated with the resonator. One conceptual difficulty, it appears, is that the word ``mode'' is sometimes reserved to the solutions of Hermitean eigenvalue problems, and any mode associated with dissipation would fall outside this category. It is conceptually very fruitful, however, to broaden the scope of the word ``mode'' to mean simply the eigenfunctions of a given differential equation problem. In electromagnetism, for example, this would be the eigenfunction \change{of} the Maxwell \change{curl equations.} %
A central point in the theory of differential equations, \change{however,} is that a differential equation in itself does not specify a well-defined problem; only by introducing a set of boundary or radiation conditions do we obtain a well-posed mathematical problem, the solutions of which we can then refer to as ``modes''. From this point of view, the modes of a closed cavity belong to a different class of modes than the modes of a leaky cavity, because they fulfill different boundary conditions. Mathematically, therefore, the conceptual as well as computational difficulties associated with QNMs can be seen as a consequence of the fact that the QNMs obey a radiation condition instead of a boundary condition, and radiation conditions are comparably difficult to handle numerically, even if important progress has been made over the past decades. %

In practical QNM calculations, one typically does not enforce the radiation condition %
explicitly, but rather imposes the correct radiation behavior by one of several different strategies. One option is to directly look for solutions to the wave equation in terms of analytic continuation of functions obeying the correct radiation condition at real frequencies. %
For spheres in free space, in particular, this provides an analytically tractable approach for calculating the QNMs in terms of spherical wave functions~\cite{Lai_PRA_41_5187_1990, Muljarov_EPL_92_50010_2010}, and we use this approach also for the case of the dielectric barrier in one dimension below. %
A numerical variant of this approach is QNM calculations via a Fourier Modal Method framework~\cite{Sauvan_PRL_110_237401_2013, Weiss_PRL_116_237401_2016, Weiss_PRB_96_045129_2017,Sauvan_PRA_89_043825_2014,deLasson_JOSA_A_31_2142_2014}. Another option is to calculate the QNMs via an integral equation formulation~\cite{Kristensen_OL_37_1649_2012, deLasson_JOSAB_30_1996_2013, Makitalo_PRB_89_165429_2014, Alpeggiani_SR_6_34772_2016}. In such approaches, the radiation condition is built in via analytical continuation of the Green tensor, which manifestly fulfills the correct radiation condition. %
A third option, which is likely the most widely adopted, relies on PML type truncations of the calculation domain to eliminate as much as possible reflections at the domain boundaries and effectively emulate radiation in free space~\cite{Kristensen_OL_37_1649_2012, Sauvan_PRL_110_237401_2013, Bai_OE_21_27371_2013}. This calculation method has traditionally been the workhorse for many optical cavity calculations, %
but the lack of an explicit correspondence with the radiation condition meant that many properties of the modes were not broadly appreciated - in particular the divergent nature of the QNMs at large distances. The explicit correspondence between QNM calculations with an integral equation formulation and with PML truncations was illustrated in Ref.~\cite{Kristensen_OL_37_1649_2012}, and %
Maes \emph{et al.}~\cite{Maes_OE_21_6794_2013}, de Lasson \emph{et al.}~\cite{deLasson_OE_26_11366_2018} and Lalanne \emph{et al.}~\cite{Lalanne_JOSAA_36_6862019} recently compared a number of different calculation methods commonly in use in the literature. In addition to direct numerical solutions, numerous approximate methods exist, which may be used to device simple analytical descriptions of the QNM fields as well as insight to the physical mechanisms responsible for the partial trapping of the electromagnetic field. Models of linear defect cavities have been presented in Ref.~\cite{Lalanne_LPR_2_514_2008} and plasmonic nanorods and resonators are treated in Refs.~\cite{Taminiau_NL_11_1020_2011} and~\cite{Yang_JOpt_18_035003_2016}, respectively. For resonators with high $Q$ factors, such as micro toroids, for example, one can also obtain approximations to the QNMs by calculating the modes of the resonator embedded in a closed cavity and subsequently estimate the radiative \change{energy loss} by integration of the field on the resonator surface~\cite{Oxborrow_arXiv_0607156_2006}. %

\change{In} Sections~\ref{Sec:QNMs_of_dielectric_barrier} and \ref{Sec:QNMs_of_plasmonic_dimer} below, we present details of the calculations for the QNMs of the two example material systems \change{from Section~\ref{Sec:Introductory_example}}. %

\subsection{QNMs of the dielectric barrier}
\label{Sec:QNMs_of_dielectric_barrier}
A general electromagnetic field in a one dimension system with piecewise homogeneous materials can be expanded in forward and backwards propagating plane waves. \change{For the dielectric barrier, we find} %
that the electric field solutions to the wave equation %
with the requirement of purely outgoing waves in the regions left and right of the resonator can be written in the form
\begin{align}
\ft_m(x)=&\begin{cases}
    A_m \text{e}^{-\text{i}n_\text{B}\tlk_m x}& x<-L/2\\[2mm]
    \text{e}^{\text{i}n_\text{R}\tlk_mx}+\text{e}^{-\text{i}n_\text{R}\tlk_mx + \text{i}m\pi}& -L/2<x<L/2\\[2mm]
    A_m \text{e}^{\text{i}n_\text{B}\tlk_mx + \text{i}m\pi} & L/2<x
   \end{cases}
\label{Eq:QNM_dielectric_barrier_eField}
\end{align}
where $\tlk_m=\tlo_m/\text{c}$, and %
\begin{align}
A_m = \left(\text{e}^{-\text{i}(n_\text{R}+n_\text{B})\tlk_mL/2} + \text{e}^{\text{i}(n_\text{R}-n_\text{B})\tlk_mL/2 + \text{i}m\pi}\right),
\end{align}
see Appendix~\ref{App:QNMs_of_dielectric_slab} for details. The field inside the barrier has a cosine or a sine dependence, and the parity of $m$ is reflected in the parity of the QNMs with respect to the point $x=0$; odd and even values of $m$ correspond, respectively, to odd and even electric field functions. The fields outside the barrier at $|x|>L/2$ manifestly respects the radiation condition.  %
Demanding continuity of $\ft_m(x)$ and the associated magnetic field $\gt_m(x)$ at $x=\pm L/2$, it follows that the QNM frequencies appear as solutions to the equation
\begin{align}
\tan\big(n_\text{R}\tlo L/\text{c}\big) +2\text{i}\frac{n_\text{R}n_\text{B}}{n_\text{R}^2+n_\text{B}^2}=0,
\label{Eq:1D_errorMap}
\end{align}
and evaluating the absolute value of the left hand side, we can map out the complex frequency spectrum, as shown in Fig.~\ref{Fig:transmission_spectrum_dielectricBarrier_nR_pi}. The solutions to Eq.~(\ref{Eq:1D_errorMap}) can be written explicitly as
\begin{align}
\tlo_mL/\text{c} = \frac{2\pi m + \text{i} \ln\big[ (n_\text{R}-n_\text{B})^2/(n_\text{R}+n_\text{B})^2\big]}{2n_\text{R}},
\label{Eq:QNM_freqs_dielectric_barrier}
\end{align}
where $m\in\mathbb{Z}$ counts the position along the real axis. The frequencies in Eq.~(\ref{Eq:QNM_freqs_dielectric_barrier}) all have negative imaginary parts. From the functional form of the QNMs, we can then immediately appreciate that the QNMs diverge (exponentially) in the regions $x<-L/2$ and $x>L/2$, cf. Eq.~(\ref{Eq:QNM_dielectric_barrier_eField}). %

As an alternative to the analytical approach taken here, we could have solved the problem numerically with algebraic boundary conditions connecting the electric and magnetic QNM fields,
\begin{align}
\sqrt{\frac{\epsilon_0}{\mu_0}}n_\text{B}\ft_m(\pm L/2) \mp \gt_m(\pm L/2) = 0.
\label{Eq:QNM_boundary_conditoion_1D}
\end{align}
This is always possible in one dimension, where the Silver-M{\"u}ller radiation condition turn into a regular boundary condition, which can be applied at any distance from the resonator.  %

\change{
\subsubsection*{Supplementary code} 
As a supplement to this Tutorial we provide Matlab code to conveniently calculate and plot the QNMs of the dielectric barrier~\cite{dielectricBarrier_arXiv}, and we encourage interested readers to investigate the variation in the QNM resonance frequencies and field distributions as a function of the mode index $m$ or the permittivity difference. 
}

\subsubsection{Limit of vanishing permittivity difference}
It is instructive to consider the limit of vanishing permittivity difference between the resonator and the surrounding medium, for which one might expect to recover the plane wave solutions of the wave equation in homogeneous media. From Eq.~(\ref{Eq:QNM_freqs_dielectric_barrier}) it follows immediately that the spacing of the real parts of the complex resonance frequencies tend to $\Delta\omega=\pi/n_\text{B}$, whereas the imaginary parts diverge along paths parallel to the negative imaginary axis. The logarithmic divergence, however, is extremely slow. For a 10\% difference in refractive index, setting $n_\text{R}=1.1$ and $n_\text{B}=1$, we find $\gamma_mL/\text{c}\approx2.8$. Reducing the refractive index difference by a factor ten by setting $n_\text{R}=1.01$ and $n_\text{B}=1$, we find $\gamma_mL/\text{c}\approx5.3$.

As the refractive index contrast between the dielectric barrier and the surrounding medium vanishes, the QNMs change in a continuous fashion and become more and more leaky as the imaginary part of the resonance frequency tends to minus infinity (albeit extremely slowly). In this limit, the real-frequency electromagnetic response due to a given QNM blends with that of the other QNMs, in a subtle way mimicking the continuous spectrum of the free one-dimensional problem. The individual QNMs, however, retain the form in Eq.~(\ref{Eq:QNM_dielectric_barrier_eField}), which is manifestly different from the plane waves because the QNMs obey different boundary conditions than the plane waves. Also, whereas the plane waves can be used as a basis along the entire real line, the QNM expansions converge to the correct solution only inside or close to the resonator, as we shall discuss further in Section~\ref{Sec:Completeness}.

\subsection{QNMs of the plasmonic dimer}
\label{Sec:QNMs_of_plasmonic_dimer}
Even though the plasmonic dimer consists of only two spheres, the geometry is too complicated for any analytically tractable approach, and we must resort to numerical calculations. All results in Fig.~\ref{Fig:Purcell_avs_Lmax_6_rdip_0p5_0p25_0_tloRange_0p05_to_0p25} were calculated with the volume integral equation (VIE) method described in Ref.~\cite{deLasson_JOSAB_30_1996_2013}. This method is specialized to the problem of collections of spherical scatterers, for which it provides relatively precise results. To illustrate the generality of the results, and to comment on the \change{issues of convergence and consistency}, we also carry out nominally identical calculations with the use of the free code MNPBEM by Hohenester~\cite{Hohenester_CPC_183_370_2012,Waxenegger_CPC_193_138_2015,Hohenester_CPC_22_209_2018}, which is an implementation of a boundary element method (BEM) formulation by de Abajo and Howie~\cite{deAbajo_PRB_65_115418_2002}. For the QNM calculations using MNPBEM, we follow the approach in Ref.~\cite{Alpeggiani_SR_6_34772_2016}\change{, and we provide supplementary code to enable the reader to repeat several of the calculations~\cite{plasmonicDimer_arXiv}.} Both integral equation methods benefit from the fact that they are defined by use of the electromagnetic Green tensor. Consequently, they manifestly respect the correct radiation condition, and the numerical errors are therefore expected to derive primarily from the discretization.

\subsubsection{Calculations using spherical wave functions}
\label{Sec:Calculations_using_VIE}
\begin{figure}[htb]
\centering %
\begin{overpic}[width=8cm]{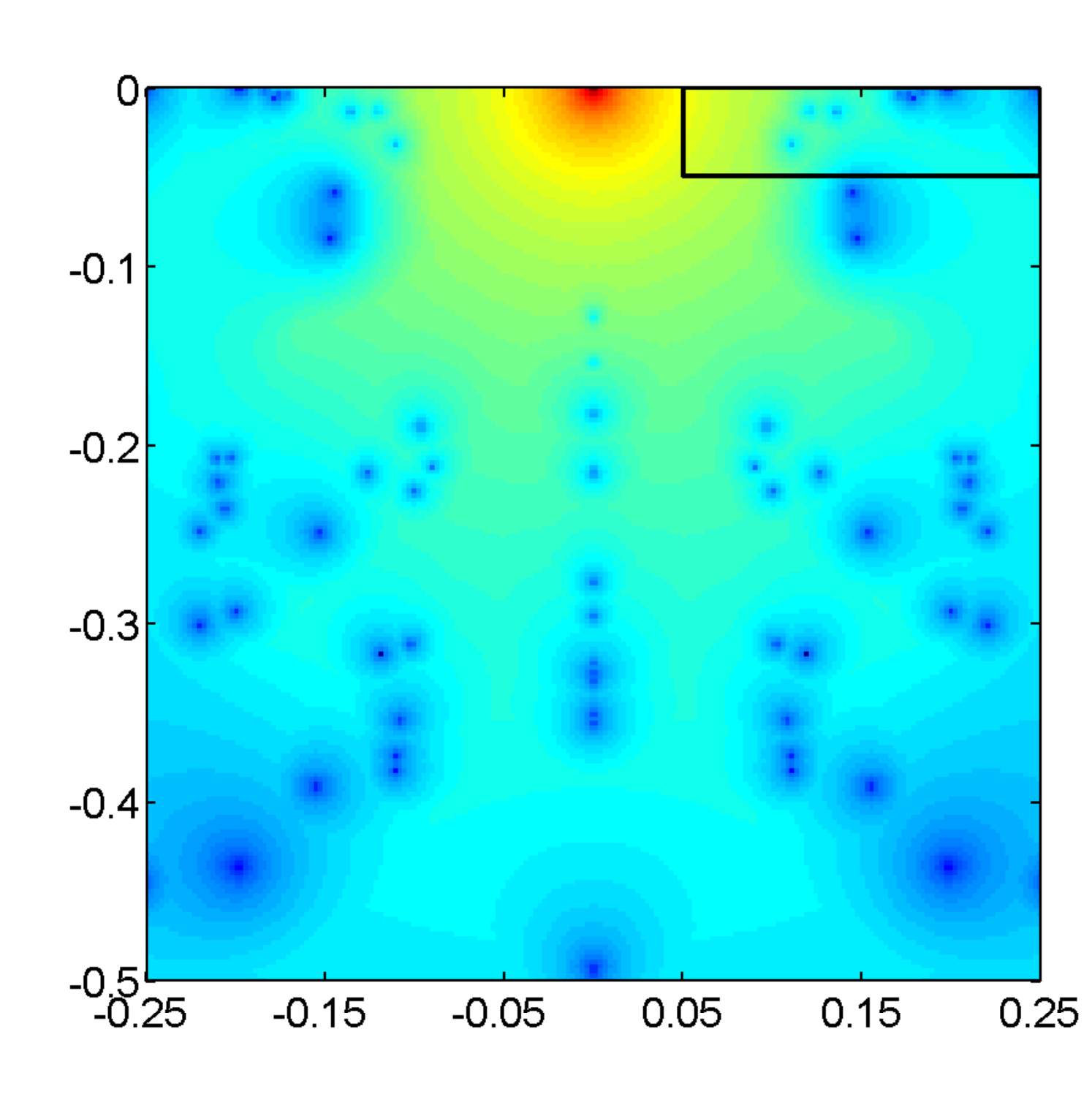}
\put(-8,32){\begin{sideways}Frequency, $\omega_\text{I}d/2\pi\text{c}$\end{sideways}}
\put(35,-8){Frequency, $\omega_\text{R}d/2\pi\text{c}$}
\end{overpic}\\[5mm]
\caption{\label{Fig:errorMap_wideRange}Complex QNM spectrum for the dimer of gold nano spheres, showing, at each position $\omega=\omega_\text{R}+\text{i}\omega_\text{I}$ in the lower half of the complex plane, the logarithm of the smallest magnitude eigenvalue of the operator  $\mathbf{M}_\mG(\omega)-1$, cf. Eq.~(\ref{Eq:LippmannSchwinger_discretized}). The QNM frequencies show up as distinct dark blue spots indicating the minima in the modal landscape. The black frame in the upper right corner indicates the extent of the frequency map displayed in Fig.~\ref{Fig:Purcell_avs_Lmax_6_rdip_0p5_0p25_0_tloRange_0p05_to_0p25}. Evidently, the whole spectrum extends far below this region.}
\end{figure}

In the VIE formulation of Ref.~\cite{deLasson_JOSAB_30_1996_2013}, a general electric field QNM is expanded in spherical wave functions inside each sphere $i$ as
\begin{align}
\mft_m(\mr_i) = \sum_{\alpha, l,m}e_{i\alpha lm} j_l(n_i\tlk_mr_i)Y_l^m(\theta_i,\phi_i)\me_\alpha,
\end{align}
where $e_{i\alpha lm}$ are the expansion coefficients to be determined, $j_l(r)$ denotes the spherical Bessel function of order $l$, \change{and $n_i$ and $r_i$ are, respectively, the refractive index and local radial coordinate in sphere $i$} . $Y_l^m$ denotes the spherical harmonics of order $l,m$, and $\me_\alpha$ is a unit vector in the direction $\alpha$. The index $l$ takes integer values in the range $0$ to $l_{max}$, and $m$ takes integer values in the range $-l$ to $l$. After discretization, the electric field QNMs appear as
solutions to a %
generalized eigenvalue equation of the form
\begin{align}
\ma = \mathbf{M}_\mG(\omega)\ma,
\label{Eq:LippmannSchwinger_discretized}
\end{align}
and the QNM frequencies can thus be found as the points in the complex plane at which the eigenvalues of the operator $\mathbf{M}_\mG(\omega)-1$ vanish. With this approach, one can scan the complex plane and create a frequency landscape where the QNM frequencies are located at the bottom of the valleys as shown in Fig.~\ref{Fig:errorMap_wideRange}.
The spectrum is clearly mirror symmetric with respect to the imaginary axis, which is a general feature of the QNM spectrum\change{, and we shall return to it in Section~\ref{Sec:differential_equation_problem_and_QNMs}}.  %
The possible degeneracy of the QNMs is fully captured by Eq.~(\ref{Eq:LippmannSchwinger_discretized}), but is not immediately clear from Fig.~\ref{Fig:errorMap_wideRange}. %

\begin{figure}[htb!]
\centering %
\begin{overpic}[width=9.4cm]{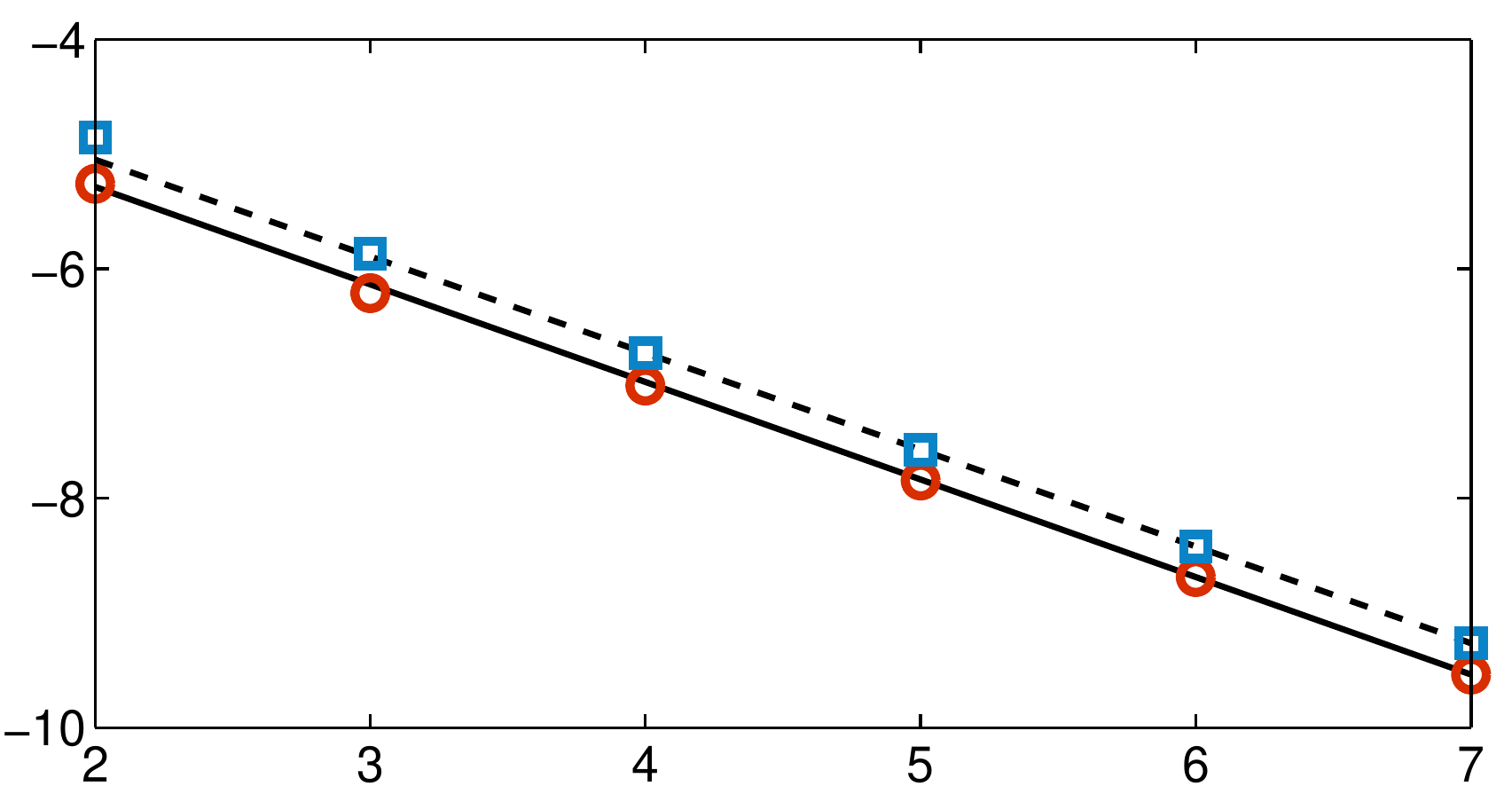}
\put(-5,3){\begin{sideways}Difference, $\log_{10}\{D_{1+}d/2\pi\text{c}\}$\end{sideways}}
\put(33,-6){Cut-off parameter, $l_\text{max}$} %
\end{overpic}
\\[5mm]
\caption{\label{Fig:Dp1_vs_Lmax_real_and_imag}Convergence analysis for the calculation of $\tlo_1$ using spherical wave functions showing, as a function of $l_\text{max}$, the difference $D_{1+}(l_\text{max})$ between results obtained using $l_\text{max}$ and $l_\text{max}+1$. Red circles and blue squares correspond to real and imaginary parts of the difference, respectively. Solid and dashed lines show the corresponding fits to the last two data points of each set.}
\end{figure}

The accuracy is governed by the cut-off parameter $l_\text{max}$. In practice, therefore, the complex resonance frequency can be thought of as a (discrete) function of $l_\text{max}$, and to estimate the accuracy of a given calculation, we consider the difference between results obtained using $l_\text{max}$ and $l_\text{max}+1$,
\begin{align}
D_{+1}(l_\text{max}) = \tlo(l_\text{max})-\tlo(l_\text{max}+1).
\end{align}
For the case of the complex QNM resonance frequency $\tlo_1$, Fig.~\ref{Fig:Dp1_vs_Lmax_real_and_imag} shows the logarithm of $D_{+1}$ as a function of $l_\text{max}$. %
To a very good approximation, the points corresponding to large values of $l_\text{max}$ fall on a straight line, indicating an exponential convergence of the form
\begin{align}
\tlo(l_\text{max}) = \tlo_1 + \mathcal{E}k^{-l_\text{max}}.
\end{align}
With this assumption, we can use fits to the data points in Fig.~\ref{Fig:Dp1_vs_Lmax_real_and_imag} to estimate the true value of $\tlo_1$ to an accuracy of approximately one part in a billion,
\begin{align}
\tlo_1d/2\pi\text{c} \approx 0.11057832294(5) -0.03161631327(9)\text{i},
\label{Eq:plasmonic_dimer_d1_resonance_freq}
\end{align}
see Appendix~\ref{App:Practical_convergence_studies} for details. Such an extreme accuracy is seldom necessary in practical calculations. Indeed, any uncertainty in the parameters defining the material system will likely lead to a much larger uncertainty in the result. Nevertheless, a convincing convergence plot as in Fig.~\ref{Fig:Dp1_vs_Lmax_real_and_imag} gives us confidence that the numerical method is properly implemented and that the calculation method is generally sound. In addition, as we shall see below, it can serve as a useful reference in cases where one does not have such a nice convergence behavior.

\subsubsection{Calculations using a boundary element method}
In the BEM approach of Refs.~\cite{Hohenester_CPC_183_370_2012, Waxenegger_CPC_193_138_2015, Hohenester_CPC_22_209_2018, deAbajo_PRB_65_115418_2002}, the electric and magnetic vector potentials $\phi(\mr)$ and $\mA(\mr)$ are cast in terms of surface charges and currents, $\sigma(\mr)$ and $\mathbf{h}(\mr)$, inside and outside the material boundaries. The material boundary itself is then discretized in surface elements, and we can formally write the corresponding discretization of the surface charges and currents as
\begin{align}
\sigma(\mr) = \sum_j\sigma_jp_j(\mr)\qquad \mathbf{h}(\mr) = \sum_j\mathbf{h}_jp_j(\mr),
\end{align}
where $\sigma_j$ and $\mathbf{h}_j$ are the expansion coefficients to be determined, and $p_j(\mr)$ is the so-called pulse basis function, which is unity for positions in element $j$ and vanishes everywhere else. After discretization, the QNMs appear once again as the solutions to a generalized eigenvalue problem~\cite{Alpeggiani_SR_6_34772_2016},
\begin{align}
\mathbf{\Sigma}(\omega)\mathbf{x}=0,
\label{Eq:BEM_discretized}
\end{align}
where now the expansion coefficients can be calculated by post processing of the eigenvector $\mathbf{x}$~\cite{Alpeggiani_SR_6_34772_2016}.

\begin{figure}[htb!]
\centering %
\begin{overpic}[width=9.4cm]{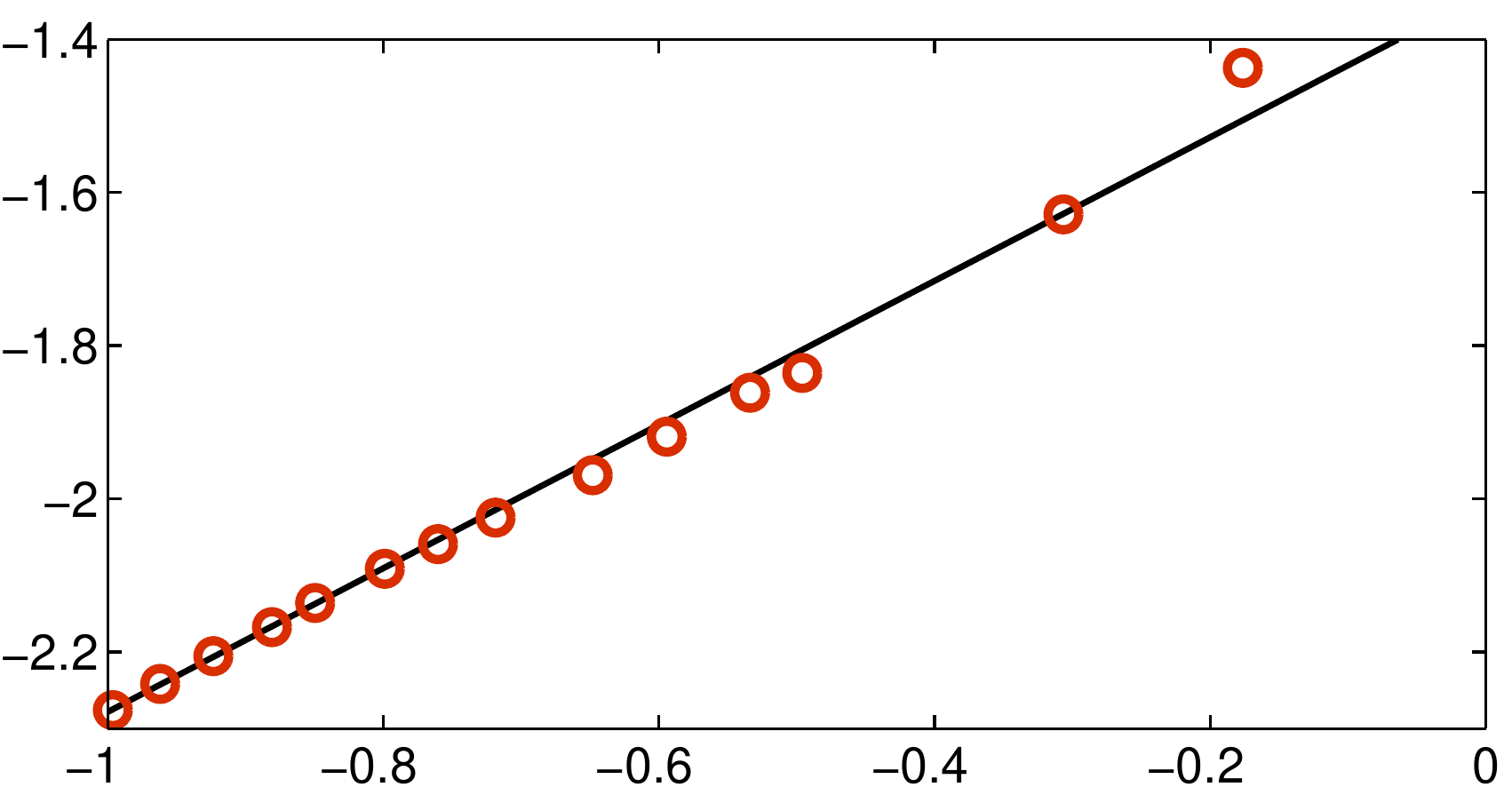}
\put(-5,3){\begin{sideways}Relative error, $\log_{10}\{\mathcal{E}_\text{BEM}\}$\end{sideways}}
\put(19.5,-6){Average triangle side length, $\log_{10}\{h/d\}$} %
\put(62,29){$y=0.94\,x - 1.34$} %
\put(10,25){\includegraphics[width=3.25cm]{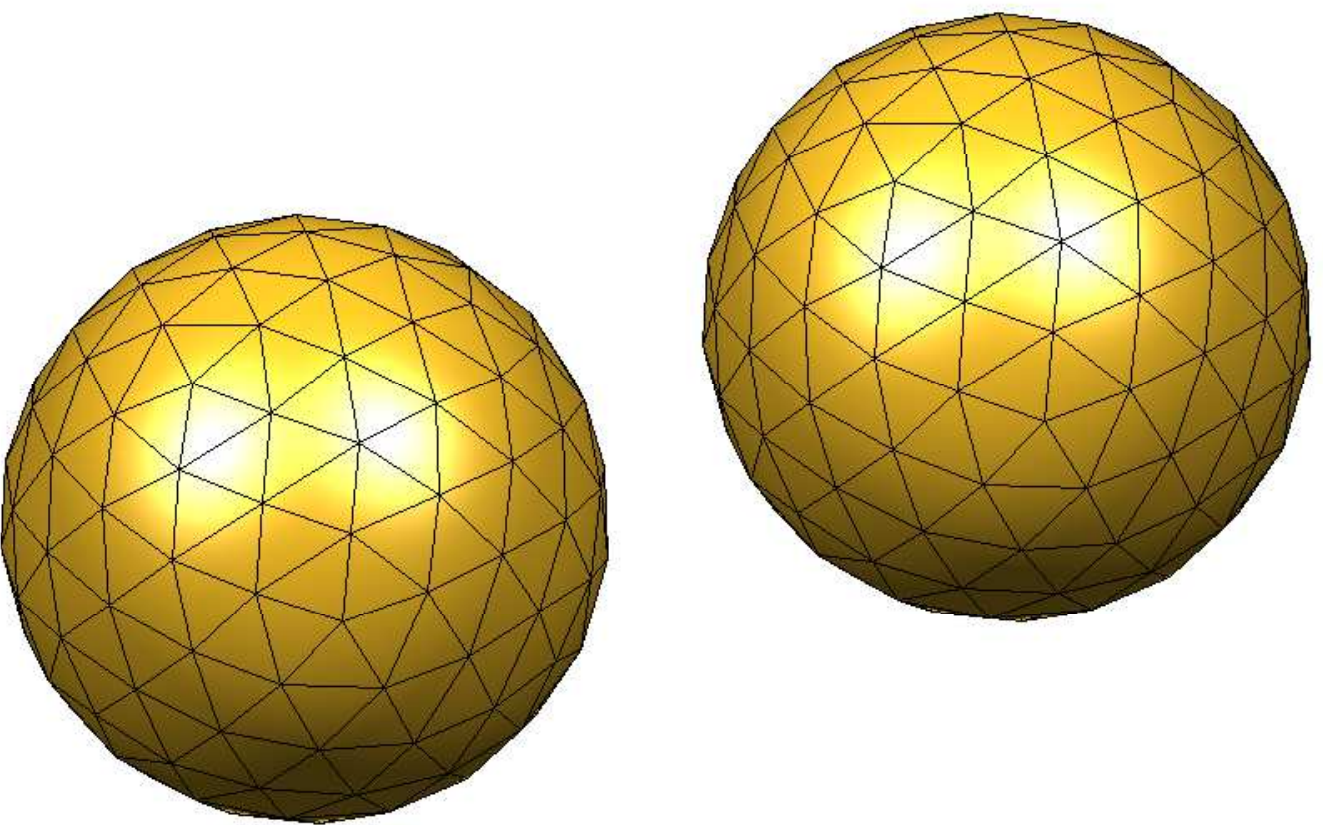}}
\end{overpic}
\\[5mm]
\caption{\label{Fig:MNPBEM_rel_err_vs_hs}Convergence analysis for the BEM calculation of $\tlo_1$ showing the relative error $\mathcal{E}_\text{BEM}=|\tlo_\text{BEM}-\tlo_\text{ref}|/|\tlo_\text{ref}|$ as a function of average side length $h$ of the triangular mesh. The reference value $\tlo_{\text{ref}}$ is given in Eq.~(\ref{Eq:plasmonic_dimer_d1_resonance_freq}), while $\tlo_\text{BEM}$ is the value obtained with the BEM approach. Circles indicate the calculated values, and the black line shows the corresponding fit to the last four data points. Inset shows an example of a relatively coarse mesh with $\log_{10}\{h/d\}=-0.5$.}
\end{figure}

For a given calculation mesh approximating the resonator geometry, we can set up Eq.~(\ref{Eq:BEM_discretized}) to map out the complex frequency plane as in Fig.~\ref{Fig:errorMap_wideRange} and find the QNMs at the positions where the eigenvalues of the operator $\mathbf{\Sigma}(\omega)$ vanish. The discretization introduces numerical errors via the discrete approximation to the surfaces charges and currents and via the approximation of the spherical surface with piecewise flat triangles. As in the case of the VIE method, we can investigate the change in resonance frequency as a function of the fineness of the mesh to assess the convergence properties of the BEM method. Because of the high estimated accuracy of the VIE method, we can use the result in Eq.~(\ref{Eq:plasmonic_dimer_d1_resonance_freq}) as a reference value for calculating the error. %
Figure.~\ref{Fig:MNPBEM_rel_err_vs_hs} shows the relative error as a function of average side length in a double logarithmic plot.
The calculated relative errors in Fig.~\ref{Fig:MNPBEM_rel_err_vs_hs} are consistent with a polynomial convergence in which the calculated value depends on the triangle side length $h$ as
\begin{align}
\tlo_\text{BEM}(h) = \tlo_\text{ref} + \mathcal{E}_0 h^\alpha,
\label{Eq:tlo_BEM_error_model}
\end{align}
\change{where $\mathcal{E}_0$ and $\alpha$ are initially unknown parameters characterizing the functional behavior of the error, see  Appendix~\ref{App:Practical_convergence_studies}. Defining} %
$y=\log_\text{10}\{|\tlo_\text{BEM}-\tlo_\text{ref}|/|\tlo_\text{ref}|\}$ and $x=\log_{10}\{h/d\}$, we can rewrite Eq.~(\ref{Eq:tlo_BEM_error_model}) in the form $y=\alpha x+\beta$, and comparing to the fit in the inset of Fig.~\ref{Fig:MNPBEM_rel_err_vs_hs}, we can immediately appreciate that the important order of the polynomial convergence is $\alpha\approx1$. %

\change{
\subsubsection*{Supplementary code} 
As a supplement to this Tutorial we provide a number of Matlab files which enable the calculation of QNMs using MNPBEM~\cite{plasmonicDimer_arXiv}. In particular, we provide the files necessary to reproduce the Purcell factor spectrum and the mode profile(s) in Fig.~\ref{Fig:Purcell_avs_Lmax_6_rdip_0p5_0p25_0_tloRange_0p05_to_0p25} as well as the convergence analysis in Fig.~\ref{Fig:MNPBEM_rel_err_vs_hs}. Even though the examples focus on the plasmonic dimer of gold nanospheres, they may also serve as a convenient template for analysis of QNMs in other geometries.}

\section{Theoretical framework}
\label{Sec:Theory}
In this Section, we lay out the basic elements of QNM modeling theory for general, three-dimensional resonators in homogeneous environments and discuss how the resulting expressions relate to various \change{formulations} in the literature.

The main emphasis is on a bi-orthogonal approach, in which the differential equation and the radiation condition is used in a systematic way to define an adjoint problem and an associated projection operation. In Section~\ref{Sec:differential_equation_problem_and_QNMs}, we start by precisely defining the differential equation problem of interest, namely the Maxwell curl equations subject to the Silver-M{\"u}ller radiation condition; the QNMs are defined as solutions to this problem in the absence of sources. Section~\ref{Sec:adjoint_QNMs_and_normalization} shows how one can use the differential equation problem and the radiation condition as inspiration to define the adjoint QNMs and a useful projection operator with the interesting property that the projection of one QNM onto another is zero, and the projection of a QNM onto itself provides a well-known normalization formula for QNMs. The projection operator can be used to expand certain electromagnetic fields, such as the electromagnetic Green tensor, in the vicinity of electromagnetic resonators, as we discuss in Section~\ref{Sec:Formal_expansions}. A conceptually different complementary approach to QNM modeling is offered by the Green tensor approach presented in Section \ref{Sec:QNMs_as_residues}, which is inherently very generally applicable because the Green tensor by definition obeys the correct radiation condition, whatever that may be. In practical calculations along such a route, one can use the normalization procedure in Section \ref{Sec:Alternative_normalization_scheme}, provided one can calculate the Green tensor at complex frequencies.

\subsection{The differential equation problem and QNMs}
\label{Sec:differential_equation_problem_and_QNMs}
Assuming a time dependence of the form $\exp\{-\text{i}\omega t\}$, we shall be interested in electric and magnetic fields, $\mE(\mr,\omega)$ and $\mH(\mr,\omega)$, which solve the %
Maxwell equations with electric and magnetic source currents, $\mathbf{J}_\text{s}(\mr,\omega)$ and $\mathbf{M}_\text{s}(\mr,\omega)$ %
and no free charges. %
In particular, we require $\nabla\cdot\mE(\mr,\omega)=\nabla\cdot\mH(\mr,\omega)=0$ and
focus on the curl equations for the six component electromagnetic field vector $\mF(\mr,\omega)=[\mE(\mr,\omega),\mH(\mr,\omega)]^\text{T}$ in the form
\begin{align}
\mD\,\mF(\mr,\omega) = -\text{i}\omega\mF(\mr,\omega) + \mmW^{-1}\underline{\mJ}(\mr,\omega),
\label{Eq:MaxwellEquations_matrixForm}
\end{align}
where
\begin{align}
\mD=\begin{bmatrix}
0& [\epsilon_0\epsilon_\text{r}(\mr)]^{-1}\nabla\times \\
-[\mu_0\mu_\text{r}(\mr)]^{-1}\nabla\times&0
\end{bmatrix},
\end{align}
$\mmW=\text{diag}\{\epsilon_0\epsilon_\text{r}(\mr), \mu_0\mu_\text{r}(\mr)\}$, $\underline{\mJ}(\omega)=[\mJ_\text{s}(\mr,\omega),\mathbf{M}_\text{s}(\mr,\omega)]^\text{T}$, $\epsilon_0$ and $\mu_0$ denote the permittivity and permeability of free space, and $\epsilon_\text{r}(\mr)$ and $\mu_\text{r}(\mr)$ denote the relative permittivity and permeability, which we assume, initially, to be
dispersionless and \change{absorptionless}, i.e., $\epsilon_\text{r}(\mr)$ and $\mu_\text{r}(\mr)$ are real; %
the case of \change{dispersive and absorptive} materials is treated in Section \ref{Sec:DispersiveMaterials}\change{, and in general we shall limit the discussion to passive materials without gain}. %
At sufficiently large distances, moreover, we shall assume that $\epsilon_\text{r}(\mr)$ and $\mu_\text{r}(\mr)$ take on the constant values $\epsilon_\text{B}=n_\text{B}^2$ and $\mu_\text{B}=1$. In addition to the curl equations, we require the electromagnetic fields to %
obey the Silver-M{\"u}ller radiation condition~\cite{Silver_1949,Martin_MultipleScattering}, in the form
\begin{align}
\mathbf{\hat{r}}\times\mE(\mr,\omega) &\rightarrow \sqrt{\frac{\mu_0}{\epsilon_0\epsilon_\text{B}}}\mH(\mr,\omega) ,\quad r\rightarrow \infty \label{Eq:SM_1}\\
\mathbf{\hat{r}}\times\mH(\mr,\omega) &\rightarrow -\sqrt{\frac{\epsilon_0\epsilon_\text{B}}{\mu_0}}\mE(\mr,\omega) ,\quad r\rightarrow \infty \label{Eq:SM_2},
\end{align}
where $\mathbf{\hat{r}}$ is the unit vector in the direction $\mathbf{r}$  and, for real frequencies, the limit is to be understood in the sense $A\rightarrow B$ if $A-B\rightarrow0$. The Silver-M{\"u}ller radiation condition derives from the requirement that there be no sources of radiation in the far field~\cite{Silver_1949} and, therefore, admits only fields that, at larges distances, travel away from the resonator. Scattering calculations, notably, often involve an incoming field, such as a plane wave, traveling from far away and hitting the scattering geometry; the incoming field does not obey Eqs.~(\ref{Eq:SM_1}) and (\ref{Eq:SM_2}), but the resulting scattered field does. Combining the second curl equation with Eq.~(\ref{Eq:SM_2}), we can write the Silver-M{\"u}ller radiation condition in the form
\begin{align}
\mathbf{\hat{r}}\times \nabla\times\mE(\mr,\omega) &\rightarrow -\text{i}n_\text{B}\frac{\omega}{\text{c}}\mE(\mr,\omega) ,\quad r\rightarrow \infty,
\label{Eq:SM_nabla_form}
\end{align}
from which it follows that $\mathbf{\hat{r}}\cdot\mE(\mr,\omega)$ vanishes at large distances.

In general, at positions sufficiently far away, the solutions to Eqs.~(\ref{Eq:MaxwellEquations_matrixForm}), %
(\ref{Eq:SM_1}) and (\ref{Eq:SM_2}) can be written in terms of outgoing waves with exponential factors of the form
\begin{align}
\mF(\mr,\omega) \propto \text{e}^{\text{i} k_\text{B}r}.
\label{Eq:mF_limiting_exponential_form}
\end{align}
where $k_\text{B}=n_\text{B}\omega /\text{c}$. In one dimension, this fact was used explicitly in the ansatz for the QNMs of the dielectric barrier in Section~\ref{Sec:QNMs_of_dielectric_barrier}. The other possible solution to the wave equation of the form $\exp\{-\text{i}k_\text{B} |x|\}$ does not satisfy Eq.~(\ref{Eq:SM_nabla_form}). Similarly, the QNMs of the plasmonic dimer in Section~\ref{Sec:QNMs_of_plasmonic_dimer} can be expanded at positions outside the spheres in terms of outwards propagating spherical wavefunctions of the form %
\begin{align}
h_l(k_\text{B} r)\mathbf{P}_{lm}(\mathbf{\hat{r}}),
\label{Eq:outgoing_spherical_wavefunction}
\end{align}
where $h_l(z)$ is the spherical Hankel function of the first kind of order $l$, and $\mathbf{P}_{lm}(\mathbf{\hat{r}})$ is any one of the vector spherical harmonics of order $(l,m$)~\cite{Barrera_EJP_6_287_1985}. This expansion is an integral part of the VIE formulation in Ref.~\cite{deLasson_JOSAB_30_1996_2013} or the BEM approach of Refs.~\cite{Hohenester_CPC_183_370_2012, Waxenegger_CPC_193_138_2015, Hohenester_CPC_22_209_2018, deAbajo_PRB_65_115418_2002}. The spherical Hankel functions of the first kind, in general, can be written in the form %
$%
h_n(z) = R_n(z)\exp\{\text{i}z\},
$ %
where $R_n(z)$ is a rational function of polynomials~\cite{Abromowitz_1964}, see also Appendix~\ref{Sec:Regions_of_completeness_analytical}. Again, the other possible solutions to the wave equation in terms of spherical Hankel functions of the second kind do not satisfy Eq.~(\ref{Eq:SM_nabla_form}).

In the absence of sources, Eqs.~(\ref{Eq:MaxwellEquations_matrixForm}), %
(\ref{Eq:SM_1}) and (\ref{Eq:SM_2}) have no solutions at real frequencies (except possibly the case $\omega=0$, which leads to non-trivial solutions only for longitudinal fields). Nevertheless, it is possible to find solutions at complex frequencies, which behave as outgoing waves and obey %
Eqs.~(\ref{Eq:SM_1}) and (\ref{Eq:SM_2}) in the weaker sense
$A\rightarrow B$ if $A/B\rightarrow 1$. These solutions can be expanded in outgoing spherical wave functions of the form in Eq.~(\ref{Eq:outgoing_spherical_wavefunction}), whose analytical continuations onto the real frequency axis obey %
Eqs.~(\ref{Eq:SM_1}) and (\ref{Eq:SM_2}) in the stronger sense \change{$A\rightarrow B$ if $A-B\rightarrow0$}. %
These solutions are what we refer to as the QNMs. To explicitly distinguish them from general fields, we use the notation $\mFt_n(\mr)=[\mft_n(\mr),\mgt_n(\mr)]^\text{T}$ for the six component QNMs, in which case %
we can write the defining equation for the QNMs in compact form as
\begin{align}
\mD \,\mFt_n(\mr) = -\text{i}\tlo_n\mFt_n(\mr).
\label{Eq:Maxwell_QNM_spinorForm}
\end{align}
Since $\mD$ is real, it is clear from Eq.~(\ref{Eq:Maxwell_QNM_spinorForm}) that if $\mFt_n(\mr)$ is a QNM with resonance frequency $\tlo_n$, then $\mFt_n^*(\mr)$ is another QNM with resonance frequency $-\tlo_n^*$.

From the chosen time dependence of the form $\exp\{-\text{i}\omega t\}$ \change{and the assumption of no gain in the materials}, %
we can infer that the imaginary part of the complex QNM resonance frequency $\tlo_n=\omega_n-\text{i}\gamma_n$ must be negative ($\gamma_n>0$) in order to correspond to an %
overall temporal decay of the local electromagnetic field as energy radiates away from the resonator or is absorbed in the material. This is consistent with the example calculations in Sections~\ref{Sec:QNMs_of_dielectric_barrier} and \ref{Sec:QNMs_of_plasmonic_dimer}. Indeed, as we shall see, the QNM resonance frequencies can be associated with the poles of the electromagnetic Green tensor, which must all be located in the lower half of the complex plane in order to ensure causality.  %
Also, as we shall see, from the definition of the $Q$ factor of a resonance as the ratio of the angular resonance frequency to the full-width at half-maximum (FWHM) of the spectral response, we can calculate the $Q$ factor pertaining to a single QNM resonance frequency as
\begin{align}
Q_n = \frac{\omega_n}{2\gamma_n}.
\label{Eq:Q_n_def}
\end{align}

\begin{figure}[htb!]
\centering %
\begin{overpic}[width=9.4cm]{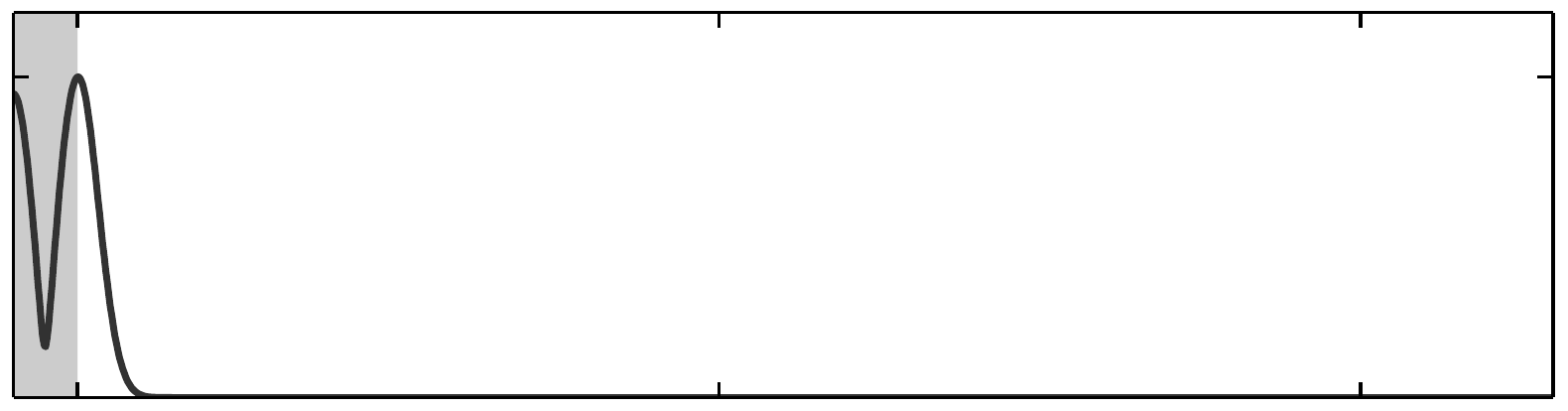}
\put(52,19){Time $t=0$} %
\end{overpic}
\begin{overpic}[width=9.4cm]{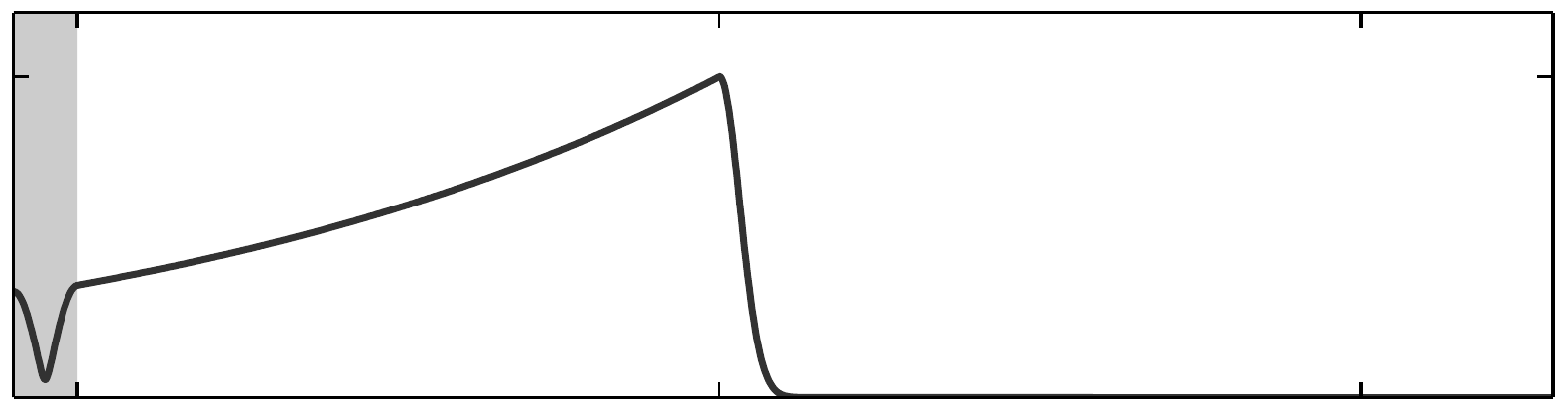}
\put(52,19){Time $t=5L/\text{c}$} %
\end{overpic}
\begin{overpic}[width=9.4cm]{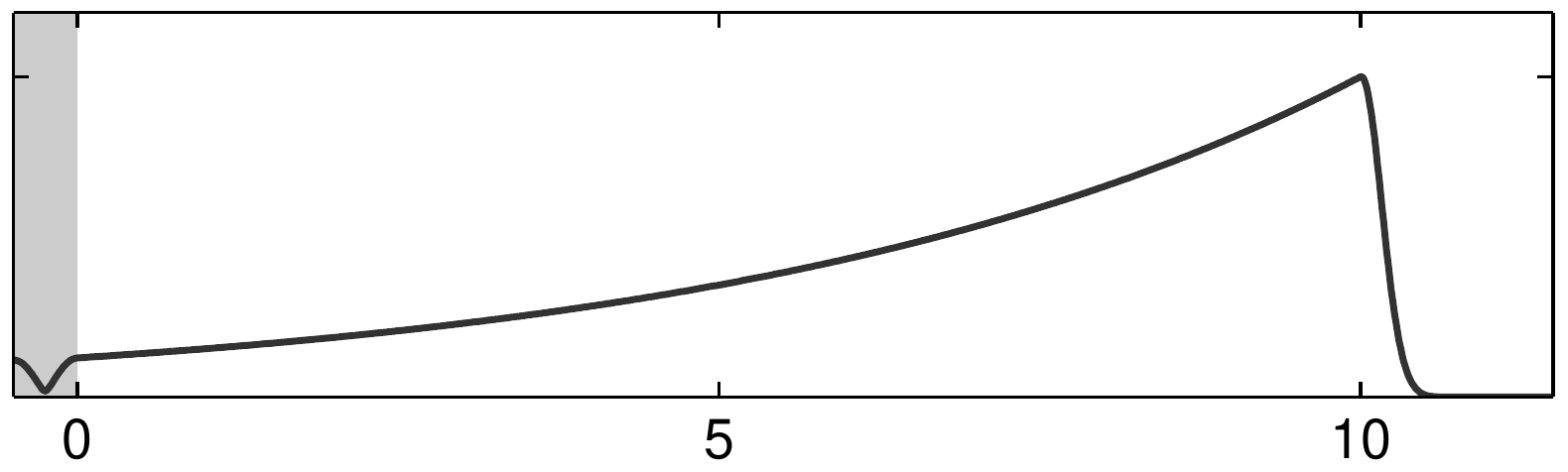}
\put(52,23){Time $t=10L/\text{c}$} %
\put(-8,18){\begin{sideways}Magnitude of electric field\end{sideways}}
\put(27,-6){Distance from resonator $x/L$} %
\end{overpic}
\\[5mm]
\caption{\label{Fig:time_domain_decay_example}\change{Sketch of the exponential decay of a field in a resonator due to radiation. At $t=0$, the field is approximately confined to the resonator, illustrated by gray shading. At all times, the field is continuous across the resonator boundary, and as the field inside the resonator decays in time, the field outside propagates away (at the speed of light). In this way, the exponential temporal decay of the field inside the resonator is mapped inversely onto the spatial distribution outside, and the field at the wavefront becomes exponentially large compared to the field in the resonator.}} 
\end{figure}

The temporal decay of the electromagnetic fields associated with each of the QNMs has an interesting effect on the spatial variation of the QNMs. From Eq.~(\ref{Eq:mF_limiting_exponential_form}) it is clear, that if the \change{frequency} %
is complex with a negative imaginary part, \change{then} $\mFt_n(\mr)$ must increase exponentially at sufficiently large distances outside the resonators. %
This increase is visible, for example, in the QNM of the dielectric barrier in Fig.~\ref{Fig:transmission_spectrum_dielectricBarrier_nR_pi}. The exponential increase of the QNMs is a real effect, %
in the sense that if one excites a resonator close to the resonance frequency of a QNM, the electromagnetic field in the resonator will subsequently decay exponentially in time as energy leaks to the environment. From the exponential nature of the decay, the  field just outside the resonator is directly proportional to the  field in the resonator; this can be seen clearly in the CMT calculations in Section~\ref{Sec:Scattering_calculations}, for example. Therefore, the exponential temporal decay for a time-harmonic field dependence is correspondingly mapped onto \change{what appears to be} an exponential growth of the spatial dependence of the field propagating away. %
As time goes on, the field in the resonator decays \change{to zero}, and the field outside the resonator propagates further away. In this way, the field at the wavefront becomes exponentially large \change{relative to the field in the resonator~\cite{Weinstein_1969}, as illustrated in Fig.~\ref{Fig:time_domain_decay_example}}.

For practical problems of interest in nanophotonics, the calculations can often, if not always, be carried out in regions close to the resonators, wherefore the divergence of the field does not lead to unphysical results. Indeed, consistency of QNM approximations is restricted to regions close to the resonator, as we discuss in Section~\ref{Sec:Regions_of_completeness}. As a result, the exponential growth of the fields outside the resonator should under no circumstances be interpreted in such a way that the coupling strength of, say, an electric dipole to the \change{field in the} resonator increases with increasing separation. \change{On a related note, we remark that there is no immediate connection between the magnitude of the individual QNM fields and the electromagnetic energy density. In particular, this means that the exponential growth of the QNM fields outside the resonators cannot in any way lead to the conclusion that the fields carry infinite energy.}

The complex resonance frequencies, and the associated divergent nature of the QNMs, are consequences of the fact, that they appear as solutions to a non-Hermitian eigenvalue problem. For this reason, they cannot immediately be used for modeling by the well-known mathematical framework developed for Hermitian eigenvalue problems --- in particular, they cannot be normalized by the norm that is often used for such problems. Instead, a slightly more general approach based on the adjoint eigenvalue problem is needed, as we discuss in the next Section.

\subsection{Adjoint QNMs and normalization} %
\label{Sec:adjoint_QNMs_and_normalization}
As in usual approaches to field expansions, we now seek to define an appropriate operator for projection of a given field onto the QNMs. As a starting point, we consider the familiar inner product $\langle\mF_A(\mr)|\mF_B(\mr)\rangle$ between two arbitrary fields $\mF_A(\mr)=[\mE_A(\mr),\mH_A(\mr)]^\text{T}$ and $\mF_B(\mr)=[\mE_B(\mr),\mH_B(\mr)]^\text{T}$ defined with a weight function $\mmW=\text{diag}\{\epsilon_0\epsilon_\text{r}(\mr), \mu_0\mu_\text{r}(\mr)\}$ as
\begin{align}
\langle\mF_A(\mr)|\mF_B(\mr)\rangle &= \frac{1}{2\epsilon_0}\int_V\mF_A^\text{T}
\mmW\,
\mF_B(\mr)\ud V\\
&=\frac{1}{2\epsilon_0}\int_V\mE_A(\mr)\epsilon_0\epsilon_\text{r}(\mr)\mE_B(\mr) + \mH_A(\mr)\mu_0\mu_\text{r}(\mr)\mH_B(\mr) \ud V,
\label{Eq:innerProduct_startingPoint}
\end{align}
where the integral is over a volume $V$ enclosing the electromagnetic scattering geometry. At real frequencies, and for $\mF_A(\mr)=\mF_B(\mr)$ this inner product is proportional to the electromagnetic field energy in the volume $V$. %
With the definition in Eq.~(\ref{Eq:innerProduct_startingPoint}), we consider the special case of the inner product between an as yet undetermined field $\mFb_m(\mr)=[\mfb_m(\mr),\mgb_m(\mr)]^\text{T}$ and the field $\mD\mF(\mr,\omega)$,%
\begin{align}
\langle\mFb_m(\mr)|\mD\,\mF(\mr,\omega)\rangle = \frac{1}{2\epsilon_0}\int_V\mfb_m(\mr)\cdot[\nabla\times\mH(\mr,\omega)] - \mgb_m(\mr)\cdot[\nabla\times\mE(\mr,\omega)]\ud V.
\end{align}
Using the relation $\nabla\cdot[\mA\times\mB] = \mB\cdot[\nabla\times\mA] - \mA\cdot[\nabla\times\mB]$ and the divergence theorem, one can rewrite the expression as
\begin{align}
\langle\mFb_m(\mr)|\mD\,\mF(\mr,\omega)\rangle = - &\langle\mD\,\mFb_m(\mr)|\mF(\mr,\omega)\rangle+ I_{\partial_V}\big(\mFb_m(\mr),\mF(\mr,\omega)\big),
\label{Eq:innerProd_adjointDerivation}
\end{align}
where
\begin{align}
I_{\partial V}\big(\mF_A(\mr),\mF_B(\mr)\big)&=\frac{1}{2\epsilon_0}\int_{\partial V} [\mH_B(\mr)\times\mE_A(\mr) - \mE_B(\mr)\times\mH_A(\mr)]\cdot\mathbf{\hat{n}}\, \ud A.
\label{Eq:I_dV}
\end{align}
where $\mathbf{\hat{n}}$ is the unit vector describing the orientation of the infinitesimal surface element.
It is clear from Eq.~(\ref{Eq:innerProd_adjointDerivation}), that the operator $\mD$ is not self adjoint under the choice of inner product in Eq.~(\ref{Eq:innerProduct_startingPoint}). Nevertheless, as will be shown below, we can use Eq.~(\ref{Eq:innerProd_adjointDerivation}) as inspiration to determine both the adjoint QNMs and a useful projection operator. %

To determine the adjoint QNMs, one must determine both the associated differential operator and the radiation condition~\cite{Cole_1968}. From Eq.~(\ref{Eq:innerProd_adjointDerivation}), it follows that one should define the adjoint operator as $\mDb = -\mD$. Moreover, in the limit of large volume (and real frequencies), the fields $\mE(\mr,\omega)$ and $\mH(\mr,\omega)$ fulfill the Silver-M{\"u}ller radiation condition on the boundary $\partial V$. In this limit, therefore, one can rewrite the surface integral as
\begin{align}
I_{\partial V} = \frac{1}{2\epsilon_0}\int_{\partial V} \mE(\mr,\omega)\cdot\left[-\sqrt{\frac{\epsilon_0}{\mu_0}}\mfb_m(\mr) + \mathbf{\hat n}\times\mgb_m(\mr)\right]\, \ud A,
\end{align}
from which it follows, that (at real frequencies) the integrand vanishes if one defines %
the adjoint QNM field radiation condition as
\begin{align}
\mathbf{\hat{r}}\times\mfb_m(\mr)&\rightarrow-\sqrt{\frac{\mu_0}{\epsilon_0}}\mgb_m(\mr),\quad r\rightarrow \infty \label{Eq:adjoint_SM_1}\\
\mathbf{\hat{r}}\times\mgb_m(\mr)&\rightarrow \sqrt{\frac{\epsilon_0}{\mu_0}}\mfb_m(\mr),\quad r\rightarrow \infty. \label{Eq:adjoint_SM_2}
\end{align}
As compared to Eqs.~(\ref{Eq:SM_1}) and (\ref{Eq:SM_2}), the signs have flipped, so that the analytical continuation of the Poynting vector $\tilde{\mS}_m^\ddagger(\mr) = \mfb_m(\mr)\times\mgb_m(\mr)$ points inwards. Writing out the matrix equation, one can verify, that if $\mFt_n(\mr)=[\mft_n(\mr),\mgt_n(\mr)]^\text{T}$ is a solution to Eq.~(\ref{Eq:Maxwell_QNM_spinorForm}) obeying the Silver-M{\"u}ller condition in Eqs.~(\ref{Eq:SM_1}) and (\ref{Eq:SM_2}), then $\mFb_m(\mr)=[\mft_m(\mr),-\mgt_m(\mr)]^\text{T}$ is a solution to the equation
\begin{align}
\mDb\mFb_m(\mr) = -\text{i}\tlo_m\mFb_m(\mr),
\label{Eq:adjointQNMsDiffEq}
\end{align}
obeying the adjoint radiation condition formulated in Eqs.~(\ref{Eq:adjoint_SM_1}) and (\ref{Eq:adjoint_SM_2}).
From Eqs.~(\ref{Eq:innerProd_adjointDerivation}) and (\ref{Eq:adjointQNMsDiffEq}) it now follows that if $\mF(\mr,\omega)$ is a solution to Eq.~(\ref{Eq:MaxwellEquations_matrixForm}), then
\begin{align}
0 &= \langle\mDb\mFb_m(\mr)|\mF(\mr,\omega)\rangle-\langle\mFb_m(\mr)|\mD\,\mF(\mr,\omega)\rangle + I_{\partial_V} \nonumber\\[2mm]
&= -\text{i}(\tlo_m-\omega)\langle\mFb_m(\mr)|\mF(\mr,\omega)\rangle + I_{\partial_V} \label{Eq:orthogonality_adjointDerivation_step_1}
\\[2mm]
&= -\text{i}(\tlo_m-\omega)\langle\langle\mFb_m(\mr)|\mF(\mr,\omega)\rangle\rangle%
,\label{Eq:orthogonality_adjointDerivation}
\end{align}
where
\begin{align}
\langle\langle\mFb_m(\mr)|\mF(\mr,\omega)\rangle\rangle &= \langle\mFb_m(\mr)|\mF(\mr,\omega)\rangle + \frac{\text{i}}{\tlo_m-\omega}I_{\partial_V}\big(\mFb_m(\mr),\mF(\mr,\omega)\big).
\label{Eq:QNM_innerprod_generalized}
\end{align}
The adjoint QNMs were chosen so as to make the surface integral vanish in the limit of large volume (and real frequencies). At finite distances (and complex frequencies), the surface integral is non-zero in general, and we must keep it. %
For the special case of $\omega=\tlo_m$, it follows from Eq.~(\ref{Eq:orthogonality_adjointDerivation_step_1}) that the surface integral vanishes identically, %
but so does the denominator in the second term of Eq.~(\ref{Eq:QNM_innerprod_generalized}). %
To investigate this limit, we %
thus expand the electromagnetic field $\mF(\mr,\omega)$ to first order as
\begin{align}
\mF(\mr,\omega) \approx \mFt_m(\mr) + (\omega-\tlo_m)\partial_\omega\mF(\mr,\omega)\vert_{\omega=\tlo_m}.
\end{align}
and use the central enabling insight, as already pointed out in Ref.~\cite{Muljarov_EPL_92_50010_2010}, that outside the resonator, the functional form of the electromagnetic fields of interest is of the general form $\mF(\mr,\omega)=\mF(\tlk_m\mr)$, where $\tlk_m=\tlo_m/\text{c}$. This justifies the relation
\begin{align}
\omega\partial_\omega\mF(\mr,\omega) = r\partial_r\mF(\mr,\omega),
\label{Eq:Muljarov_insight}
\end{align}
where $\partial_\omega$ and $\partial_r$ denote partial derivatives with respect to $\omega$ and $r$, respectively. Inserting in Eq.~(\ref{Eq:QNM_innerprod_generalized}) and taking the limit $\omega\rightarrow\tlo_m$, %
we \change{can introduce a QNM normalization, by defining} %
\begin{align}
\langle\langle\mFt_m(\mr)|\mFt_m(\mr)\rangle\rangle &= \frac{1}{2\epsilon_0}\int_V\epsilon_0\epsilon_\text{r}(\mr)\mft_m(\mr)\cdot\mft_m(\mr) - \mu_0 \mgt_m(\mr)\cdot\mgt_m(\mr)\,\ud V \nonumber \\
&+\frac{\text{i}}{2\epsilon_0\tlo_m}\int_{\partial V} \left[[r\partial_r\mft_m(\mr)]\times\mgt_m(\mr) - \mft_m(\mr)\times[r\partial_r\mgt_m(\mr)]\right]\cdot\mathbf{\hat n}\; \ud A,
\label{Eq:MuljarovNorm_E_H_form}
\end{align}
as found also in Ref.~\cite{Muljarov_OL_43_1978_2018} via a different approach. \change{Note, that in general Eq.~(\ref{Eq:MuljarovNorm_E_H_form}) will produce a complex number, and the QNM normalization then proceeds by scaling the QNMs by the square root of this number.} %
To connect to other formulations of the \change{QNM normalization} that exist in the literature, we note that the expression in Eq.~(\ref{Eq:QNM_innerprod_generalized}) %
is identical to %
Eq.~(B2) in Ref.~\cite{Kristensen_PRA_96_017802_2017}. Therefore, one can use the same procedure to rewrite the expression and arrive at the exact expression for the %
normalization due to Muljarov \emph{et al.} in Ref.~\cite{Muljarov_EPL_92_50010_2010}, where \change{it} is expressed in terms of the electric field QNMs only. %
If one does not keep the surface integral $I_{\partial V}$ in Eq.~(\ref{Eq:I_dV}), or the second term in Eq.~(\ref{Eq:MuljarovNorm_E_H_form}), %
the resulting normalization is identical to the formula introduced by Sauvan \emph{et al.}~\cite{Sauvan_PRL_110_237401_2013}, which, in turn, is intimately related to the normalization formula due to Lai \emph{et al.}~\cite{Lai_PRA_41_5187_1990}; in both cases, one must in principle regularize the resulting integral, in which case the results of the three normalization procedures are identical, as demonstrated in Ref.~\cite{Kristensen_PRA_92_053810_2015}; see also \cite{Muljarov_PRA_96_017801_2017, Kristensen_PRA_96_017802_2017}. We shall hereafter implicitly assume all QNMs to be normalized in the above sense. It follows from the discussion above, that we can substitute $\mFt_n(\mr)$ for $\mF(\mr,\omega)$ in Eq.~(\ref{Eq:QNM_innerprod_generalized}) and take the limit $\omega\rightarrow\tlo_n$ to find that, due to Eq. (\ref{Eq:orthogonality_adjointDerivation}) the projection vanishes, unless $m=n$. %
It is tempting, then, to sum this up as %
\begin{align}
\langle\langle\mFb_m(\mr)|\mFt_n(\mr)\rangle\rangle = \delta_{mn}.
\label{Eq:projection_operator}
\end{align}
This relation, although valid as written, is slightly deceptive, in that it might suggest that one can directly insert any sum of QNMs in Eq.~(\ref{Eq:QNM_innerprod_generalized}) to get the projection onto the QNM of interest. Such an approach will not work in general, however, because of the frequency dependence of the second term in Eq.~(\ref{Eq:QNM_innerprod_generalized}). In practice, therefore, we find meaningful expressions only when working with solutions to the wave equation %
(which is not the case for general superpositions of QNMs). 

\change{For} the special class of electromagnetic fields $\mF(\mr,\omega)$, which solve Eqs.~(\ref{Eq:MaxwellEquations_matrixForm}), (\ref{Eq:SM_1}) and (\ref{Eq:SM_2}), %
 the operation $\langle\langle\mFb_m(\mr)|\mF(\mr,\omega)\rangle\rangle$ %
does provide the projection %
onto the QNM $\mFt_m(\mr)$, as we show in  Section~\ref{Sec:Formal_expansions}. %
The necessity of the fields $\mF(\mr,\omega)$ to obey the Silver-M{\"u}ller radiation condition lies in the fact that only if this is true does the surface integral %
vanishes in the limit $\omega\rightarrow\tlo_m$. In particular, Eq.~(\ref{Eq:orthogonality_adjointDerivation}) is valid for any $\mF(\mr,\omega)$, which solves Eq.~(\ref{Eq:MaxwellEquations_matrixForm}), but only if $\mF(\mr,\omega)$ behaves as the proper analytical continuation of the QNM is it true that $\mF(\mr,\omega)\rightarrow\mFt_m(\mr)$ as $\omega\rightarrow\tlo_m$.
Moreover, as shown in Appendix \ref{App:integrationAppendix}, the sum of the two integrals in Eq. (\ref{Eq:QNM_innerprod_generalized}) %
is independent of size and shape of the volume $V$, as long as it contains the electromagnetic resonator. If the integral is extended to infinity using a complex coordinate transformation, for example by the use of PMLs, the surface integral vanishes, and the projection reduces to the exact expression for the projection due to Sauvan \emph{et al.}~\cite{Sauvan_PRL_110_237401_2013}.

In one dimension, the radiation condition turns into an algebraic boundary condition, and the second term in Eq.~(\ref{Eq:QNM_innerprod_generalized}) vanishes identically. The first term then defines a projection of an arbitrary sum of QNMs onto $\mFt_m(\mr)$, as suggested from Eq.~(\ref{Eq:projection_operator}).

\subsubsection{Degenerate modes}
\label{Sec:Degenerate_modes}
For degenerate QNMs with $\tlo_m=\tlo_n$, but $\mFt_m(\mr)\neq\mFt_n(\mr)$, it is possible that the modes are not immediately orthogonal under the operation in Eq.~(\ref{Eq:QNM_innerprod_generalized}). %
In such cases, as long as the individual QNMs are normalizable, one can always define a new set of QNMs via the Gram-Schmidt orthogonalization process. Moreover, %
in certain cases of degenerate QNMs, Eq.~(\ref{Eq:MuljarovNorm_E_H_form}) may not be immediately useful for normalization. In the case of a spherical resonator, for example, the QNMs can be chosen to have an azimuthal dependence of the form $\exp\{\pm\text{i}\varphi\}$ for which the integral in Eq.~(\ref{Eq:MuljarovNorm_E_H_form}) vanishes. \change{Any small perturbation of the circular shape, however,} %
will break the degeneracy and result in QNMs with azimuthal dependencies of the form  $\exp\{\text{i}\varphi\}\pm\exp\{-\text{i}\varphi\}$. In this case, therefore, one can define the QNMs by these linear combinations, for which Eq.~(\ref{Eq:QNM_innerprod_generalized}) is immediately applicable~\cite{Kristensen_PRA_92_053810_2015}. In the particular case of spheres, for which the QNMs can always be written as products of the form $\mF_m(\mr) = \mathbf{R}_m(r)\mathbf{P}_{lm}(\theta,\varphi)$, where $\mathbf{P}_{lm}(\theta,\varphi)$ is any of the vector spherical harmonics of order $(l,m)$, one can also define the adjoint modes via complex conjugation of the angular dependence only, as was done in Refs.~\cite{Lee_JOSAB_16_1409_1999, Lee_JOSAB_16_1418_1999}. Throughout, however, we shall use exclusively the definition $\mFb_m(\mr)=[\mft_m(\mr),-\mgt_m(\mr)]^\text{T}$ and assume that any problems arising from degeneracies can be handled by defining the QNMs as a suitable linear combinations, as described above.

\subsubsection{Dispersive \change{and absorptive} materials}
\label{Sec:DispersiveMaterials}
The bi-orthogonal framework can be immediately extended to the technologically interesting case of resonators made from dispersive \change{and absorptive} materials by use of auxiliary fields governing the material response~\cite{Yan_PRB_97_205422_2018, Raman_PRL_104_087401_2010}. In the case of a Drude material response, for example, for which
\begin{align}
\epsilon_\text{D}(\mr,\omega) = 1-\frac{\omega_\text{p}^2(\mr)}{\omega^2+\text{i}\omega\gamma(\mr)},
\label{Eq:Drude_model}
\end{align}
we introduce an additional field $\mJ(\mr,\omega)$ describing the current density governed by the equation
\begin{align}
-\text{i}\omega\mJ(\mr,\omega) = \epsilon_0\omega_\text{p}^2(\mr)\mE(\mr,\omega) -\gamma(\mr)\mJ(\mr,\omega),
\label{Eq:DrudeCurrent_governingEq}
\end{align}
where $\omega_\text{p}(\mr)$ and $\gamma(\mr)$ denote the plasma frequency and the damping coefficient of the material, respectively. As described in detail in Appendix~\ref{App:dispersiveMaterials}, one can follow an approach completely analogous to the one used for the dispersionless case above, and %
define generalized QNM fields $\mFt_m(\mr) = [\mft_m(\mr),\mgt_m(\mr),\mjt_m(\mr)]^\text{T}$ as the solutions to an equation of the same form as Eq.~(\ref{Eq:Maxwell_QNM_spinorForm}) along with a corresponding projection operator %
\begin{align}
\langle\langle\mFb_m(\mr)|\mF(\mr,\omega)\rangle\rangle &= \frac{1}{2\epsilon_0}\int_V\epsilon_0\eta_m(\mr,\omega)\mft_m(\mr)\cdot\mE(\mr,\omega)- \mu_0 \mgt_m(\mr)\cdot\mH(\mr,\omega)\,\ud V \nonumber \\
&\quad+\frac{\text{i}}{2\epsilon_0(\tlo_m-\omega)}\int_{\partial V} [\mE(\mr,\omega)\times\mgt_m(\mr) - \mft_m(\mr)\times\mH(\mr,\omega)]\cdot \mathbf{\hat n}\, \ud A.
\label{Eq:QNM_innerprod_generalized_Drude}
\end{align}
where now
\begin{align}
\eta_m(\mr,\omega) = 1-\frac{\omega_\text{p}^2(\mr)}{(\gamma(\mr)-\text{i}\tlo_m)(\gamma(\mr)-\text{i}\omega)}.
\label{Eq:eta_m_def}
\end{align}
In the limit $\omega\rightarrow\tlo_n$, we find, that the normalization \change{can be written as in Eq.~(\ref{Eq:MuljarovNorm_E_H_form}) with the substitution $\epsilon_\text{r}(\mr)\rightarrow\eta(\mr,\tlo_m)$, %
where}
\begin{align}
\eta(\mr,\omega) &= 1-\frac{\omega_\text{p}^2(\mr)}{(\gamma(\mr)-\text{i}\omega)^2}  =\partial_\omega\big(\omega\epsilon_\text{D}(\mr,\omega)\big),
\end{align}
as found also in Refs.~\cite{Sauvan_PRL_110_237401_2013,Yan_PRB_97_205422_2018}. %
Upon rewriting the expression in terms of the electric fields only~\cite{Kristensen_PRA_96_017802_2017,Ge_NJP_16_113048_2014}, the resulting expression is identical to \change{the formulation with a slightly different weight function $\sigma(\mr,\omega)=[2\omega]^{-1}\partial_\omega[\omega^2\epsilon_\text{r}(\mr,\omega)]$ used for dispersive materials in %
Refs.~\cite{Leung_PRA_49_3982_1994, Muljarov_PRB_94_235438_2016}.} %

\subsubsection{Adjoint QNMs and normalization for the dielectric barrier}
\label{Sec:Adjoint_QNMs_and_normalization_for_the_dielectric_barrier}
In one dimension, the radiation condition turns into an algebraic boundary condition, cf. Eq.~(\ref{Eq:QNM_boundary_conditoion_1D}). \change{Therefore, if we choose the integration volume for the normalization of the QNM of the dielectric barrier to be $|x|\le L/2$}, the second term in Eq.~(\ref{Eq:QNM_innerprod_generalized}) vanishes identically\change{, provided} we choose the adjoint QNMs obey the boundary conditions
\begin{align}
\sqrt{\frac{\epsilon_0}{\mu_0}}n_\text{B}\ft_m(\pm L/2) \pm \gt_m(\pm L/2) = 0. %
\end{align}
As in the general case, we can now see, that if $\ft_m(x)$ and $\gt_m(x)$ are electric and magnetic field QNMs, %
then $\ft_m^\ddagger(x)=\ft_m(x)$ and $\gt_m^\ddagger(x)=-\gt_m(x)$ are the associated adjoint electric and magnetic field QNMs. In the absence of the second term in Eq.~(\ref{Eq:QNM_innerprod_generalized}), we can immediately take the limit $\omega\rightarrow\tlo_m$. %
In this way, we can write the integral for the QNM normalization as
\begin{align}
\langle\langle\mFb_m(\mr)|\mFt_m(\mr)\rangle\rangle &= \frac{1}{2}\int_{-L/2}^{L/2}n_\text{R}^2\ft^{\,2}_m(x) - \frac{\mu_0}{\epsilon_0} \gt^{\,2}_m(x)\,\ud x
\label{Eq:dielectric_barrier_normalization}
\end{align}
\change{Note, that even though we chose the integration to be $|x|\le L/2$ for convenience, the value is independent of this choice, as long as the integration boundaries are beyond the extent of the barrier}. In Eq.~(\ref{Eq:dielectric_barrier_normalization}), this property is a result of the fact that the integrand vanishes identically outside the resonator. Inserting the explicit form of $\ft_m(x)$ and $\gt_m(x)$ %
the integral simplifies substantially, and we find
\begin{align}
\langle\langle\mFb_m(\mr)|\mFt_m(\mr)\rangle\rangle &= (-1)^m 2n_\text{R}^2L. %
\label{Eq:dielectric_barrier_normalization_Efield_only}
\end{align}

\subsection{Formal expansions in terms of QNMs}
\label{Sec:Formal_expansions}
We now first take the point of view that a given field can be expanded in QNMs within the volume $V$ enclosing the resonator, in which case the projection operator arises naturally to provide the expansion coefficients. Subsequently, we shall worry about the validity of the expansion, i.e. the convergence and consistency of the formal expansion.

\subsubsection{Formal expansion of a general electromagnetic field}
\label{Sec:Formal_expansion_of_general_field}
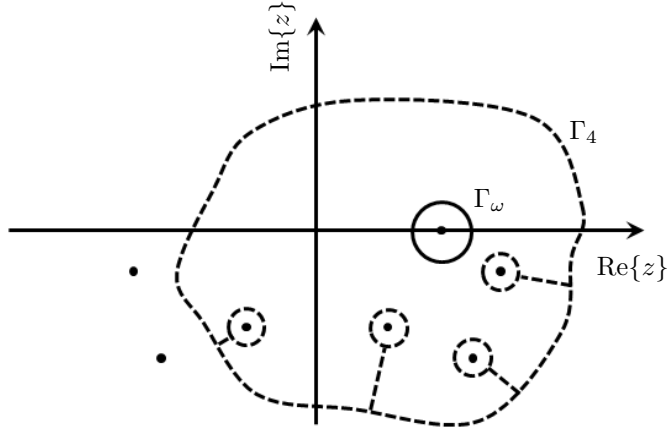
\begin{figure}[htb!]
\centering %
\begin{overpic}[width=9cm]{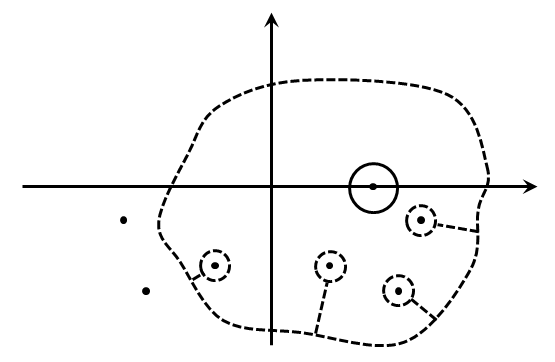}
\put(90,25){$\text{Re}\{z\}$}
\put(42,55){\begin{sideways}$\text{Im}\{z\}$\end{sideways}}
\put(72,35){$\Gamma_\omega$}
\put(86,45){$\Gamma_4$}
\end{overpic}
\caption{\label{Fig:Riesz_outline}Integration contours of the Riesz projection technique. An initial integration contour $\Gamma_\omega$ is enlarged into a contour $\Gamma_N$ while in the process excluding all poles $\tlo_N$ by integration around them as illustrated for the case of $N=4$. } \end{figure}

We consider a general electromagnetic field $\mF(\mr,\omega)$, which solves Eqs.~(\ref{Eq:MaxwellEquations_matrixForm}), (\ref{Eq:SM_1}) and (\ref{Eq:SM_2}). Multiplying from the left with $\mFb_m(\mr)\mmW$ and integrating over a volume $V$ containing the resonator and all sources, we find that
\begin{align}
\langle\langle\mFb_m(\mr)|\mF(\mr,\omega)\rangle\rangle = \frac{\text{i}}{\tlo_m-\omega}\langle\mFb_m|\mmW^{-1}\underline{\mJ}(\mr,\omega)\rangle.
\label{Eq:projection_of_general_field_w_source}
\end{align}
From the peaks in Fig.~\ref{Fig:Purcell_avs_Lmax_6_rdip_0p5_0p25_0_tloRange_0p05_to_0p25}, we know that the QNMs appear as poles in the scattering matrix. We shall therefore assume, that the electromagnetic fields of interest can be represented by the series
\begin{align}
\mF(\mr,\omega) = \sum_n\frac{a_n(\omega)}{\omega-\tlo_n}\mFt_n(\mr),
\label{Eq:mF_pole_expansion_form}
\end{align}
where $a_n(\omega)$ are analytic functions representing the unknown expansion coefficients to be determined. To see that the operator $\langle\langle\mFb_m(\mr)|\mF(\mr,\omega)\rangle\rangle$ acts to project the solution onto the QNM $\mFt_m(\mr)$, we use an approach similar to the Riesz projection technique~\cite{Calderon_2010, Lin_PRA_98_043806_2018} by first rewriting it using the Cauchy integral theorem as
\begin{align}
\langle\langle\mFb_m(\mr)|\mF(\mr,\omega)\rangle\rangle = \frac{1}{2\pi\text{i}}\oint_{\Gamma_\omega}  \frac{\langle\langle\mFb_m(\mr)|\mF(\mr,z)\rangle\rangle}{(z-\omega)}\ud z,
\label{Eq:general_F_Riesch_projection_form}
\end{align}
where the integral is taken along a closed counterclockwise oriented curve $\Gamma_\omega$ around the point $z=\omega$ and sufficiently small that no other poles are encircled, as illustrated in Fig.~\ref{Fig:Riesz_outline}. We next express $\mF(\mr,z)$ in the numerator of Eq.~(\ref{Eq:general_F_Riesch_projection_form}) using Eq.~(\ref{Eq:mF_pole_expansion_form}), but explicitly evaluate $a_n(z)$ at $z=\omega$, which is warranted since for analytical functions $\f(\omega)$ and $g(\omega)$, we have
\begin{align}
f(\omega)g(\omega) = \frac{1}{2\pi\text{i}}\oint_{\Gamma_\omega}\frac{f(z)g(z)}{z-\omega}\ud z=\frac{1}{2\pi\text{i}}\oint_{\Gamma_\omega}\frac{f(\omega)g(z)}{z-\omega}\ud z=\frac{1}{2\pi\text{i}}\oint_{\Gamma_\omega}\frac{f(z)g(\omega)}{z-\omega}\ud z.
\end{align}
With this procedure, the left hand side of Eq.~(\ref{Eq:projection_of_general_field_w_source}) can be rewritten as
\begin{align}
\langle\langle\mFb_m(\mr)|\mF(\mr,\omega)\rangle\rangle &= \oint_{\Gamma_\omega}\frac{1}{2\pi\text{i}}\sum_n\frac{a_n(\omega)}{(z-\tlo_n)(z-\omega)}\nonumber \\
&\quad\times \left\{\langle\mFb_m(\mr)|\mFt_n(\mr)\rangle + \frac{\text{i}}{\tlo_m-z}I_{\partial_V}\big(\mFb_m(\mr),\mFt_n(\mr)\big)\right\}\ud z,
\end{align}
and by expanding the integration contour to infinity --- while in the process excluding all poles $\tlo_n$ --- the projection is rewritten as a sum over the residues at $z=\tlo_n$ as
\begin{align}
\langle\langle\mFb_m(\mr)|\mF(\mr,\omega)\rangle\rangle &= \sum_n\frac{a_n(\omega)}{\omega-\tlo_n}\langle\langle\mFb_m(\mr)|\mFt_n(\mr)\rangle\rangle.
\end{align}
The additional contribution from the integral along the outer contour $\Gamma_N$ vanishes because of the functional form of the integrand. Using the orthogonality in Eq.~(\ref{Eq:projection_operator}), we conclude, that %
\begin{align}
\langle\langle\mFb_m(\mr)|\mF(\mr,\omega)\rangle\rangle &= \frac{a_m(\omega)}{\omega-\tlo_m}.
\label{Eq:projection_with_a}
\end{align}

Finally, returning to Eq.~(\ref{Eq:projection_of_general_field_w_source}), we find that if the electromagnetic field can be expanded as in Eq.~(\ref{Eq:mF_pole_expansion_form}), then %
the expansion coefficients are given as
\begin{align}
a_m(\omega) = -\frac{\text{i}}{2\epsilon_0}\int\mft_m(\mr)\cdot\mJ_\text{s}(\mr,\omega)-\mgt_m(\mr)\cdot\mathbf{M}_\text{s}(\mr,\omega)\ud V.%
\label{Eq:a_m_general_field}
\end{align}

\change{From the expansion in Eq.~(\ref{Eq:mF_pole_expansion_form}) or the projection in Eq.~(\ref{Eq:projection_with_a}), we can now appreciate, that for a constant source, the electromagnetic response due to each of the QNMs is a lorentzian centered on $\omega=\tlo_m$ and with a FWHM of $2\gamma_m$, thus justifying the definition in Eq.~(\ref{Eq:Q_n_def}). Note, however, that in general the contributions to the spectrum from different QNMs interfere, and depending on the particular resonator and the excitation conditions, this may lead to substantial deviations from lorentzian line shapes in practice.}

It is instructive to consider the problem of the source-free electromagnetic field. Setting $\underline{\mJ}(\mr,\omega)=0$, it follows from Eq.~(\ref{Eq:a_m_general_field}) that there are no non-trivial solutions to the source-free wave equation of the form in Eq.~(\ref{Eq:mF_pole_expansion_form}) with $a_n(\omega)\neq0$. Instead, the solutions are of the form $\mF(\mr,\omega)=\mFt_n(\mr)$, i.e. they are QNMs. In this case, the analytical continuation of the QNMs in the form of Eq.~(\ref{Eq:a_m_general_field}) implies that we must have \change{$a_n(\omega)=(\omega-\tlo_n)(1+\mathcal{O}(\omega))$}.

\subsubsection{Formal expansion of the Green tensor}
As a special case, the QNMs can be used for expansion of the electromagnetic Green tensor. Because of the transverse nature of the QNMs, the expansions are limited to the transverse part of the Green tensor, as further discussed in Section \ref{Sec:Convergence_of_G}. %
We define the matrix Green tensor %
as the solution to the equation
\begin{align}
\mD\,\mmG(\mr,\mr',\omega) +\text{i}\omega\mmG(\mr,\mr',\omega) = \frac{\delta(\mr-\mr')}{\text{i}\omega\mu_0} \mmW^{-1},
\label{Eq:Green_def}
\end{align}
with the additional requirement that each column of $\mmG(\mr,\mr',\omega)$ fulfills the Silver-M{\"u}ller radiation condition in Eqs.~(\ref{Eq:SM_1}) and (\ref{Eq:SM_2}). With this definition, the electromagnetic field can be calculated in the usual way via integration as
\begin{align}
\mF(\mr,\omega) = \text{i}\omega\mu_0 \int \mmG(\mr,\mr',\omega)\cdot\underline{\mathbf{J}}(\mr',\omega) \ud V',
\label{Eq:F_from_G_and_sources}
\end{align}
where the integral is over all space. %
To proceed, we assume that the matrix Green tensor can be expanded as
\begin{align}
\mmG(\mr,\mr',\omega) = \sum_n\mFt_n(\mr)\frac{\underline{\boldsymbol{\alpha}}^\text{T}_n(\mr')}{\omega-\tlo_n},
\label{Eq:G_QNM_proto_expansion}
\end{align}
and follow identical steps as in Section~\ref{Sec:Formal_expansion_of_general_field} to find, that if the matrix Green tensor can be expanded as in Eq.~(\ref{Eq:G_QNM_proto_expansion}), then
\begin{align}
\underline{\boldsymbol{\alpha}}^\text{T}_n(\mr',\omega) = -\frac{\text{c}^2}{2\omega}[\mFb_n(\mr')]^\text{T}. 
\label{Eq:alpha_in_G_QNM_expansion}
\end{align}
Inserting in Eq.~(\ref{Eq:G_QNM_proto_expansion}), the %
matrix Green tensor expansion then takes the form
\begin{align}
\mmG(\mr,\mr',\omega) = %
\frac{\text{c}^2}{2\omega}\sum_n\frac{\mFt_n(\mr)[\mFb_n(\mr')]^\text{T}}{\tlo_n-\omega},
\label{Eq:MG_QNM_expansion}
\end{align}
and it follows immediately, that the electric field Green tensor expansion is
\begin{align}
\mG^\text{EE}(\mr,\mr',\omega) = \frac{\text{c}^2}{2\omega}\sum_n\frac{\mft_n(\mr) \mft_n(\mr')}{\tlo_n-\omega},
\label{Eq:GE_QNM_expansion}
\end{align}
as also found in Refs.~\cite{Sauvan_PRA_89_043825_2014,Doost_PRA_90_013834_2014}. %
An identical expression was found for the one-dimensional problem in Ref.~\cite{Muljarov_EPL_92_50010_2010}; the two dimensional problem of scattering from a cylinder %
was discussed in Ref.~\cite{Doost_PRA_87_043827_2013}, where it was pointed out that it includes a branch cut contribution to the Green tensor in addition to a term identical to Eq.~(\ref{Eq:GE_QNM_expansion}). 

\change{We note, that because of the symmetric property of the QNM resonance frequency spectrum, as illustrated in Fig.~\ref{Fig:errorMap_wideRange}, it follows immediately from Eq.~(\ref{Eq:GE_QNM_expansion}) that the Green tensor satisfies the so-called crossing relation
\begin{align}
\mG^\text{EE}(\mr,\mr',-\omega^*) = \left[\mG^\text{EE}(\mr,\mr',\omega)\right]^*,
\label{Eq:crossing_relation}
\end{align}
which can be associated with the reality of the electromagnetic response in the time domain.}

\subsubsection{Green function expansion for the dielectric resonator}
\begin{figure}[htb!]
\centering %
\begin{overpic}[width=9.4cm]{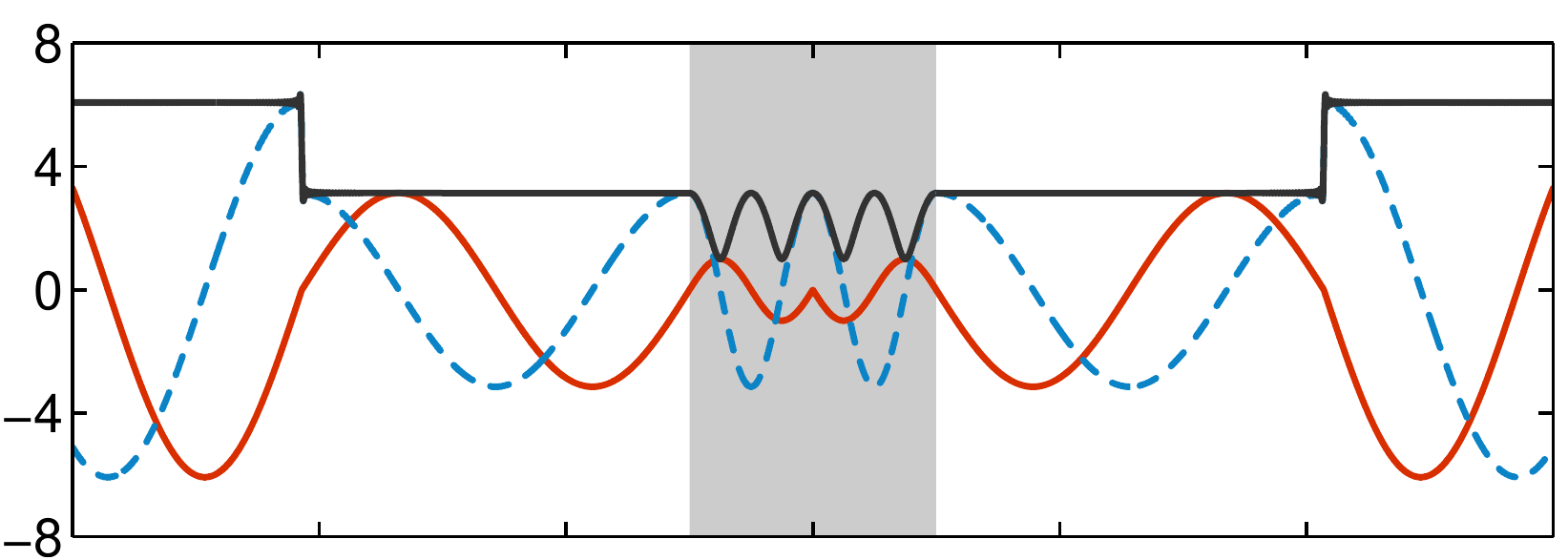}
\put(-11,6){\begin{sideways}Green function\end{sideways}}
\put(-5,6){\begin{sideways}$G_{500}(x',\omega_0)/G_0$\end{sideways}} %
\put(42,-6){Position, $x/L$}
\end{overpic}\;\\[2mm]
\begin{overpic}[width=9.5cm]{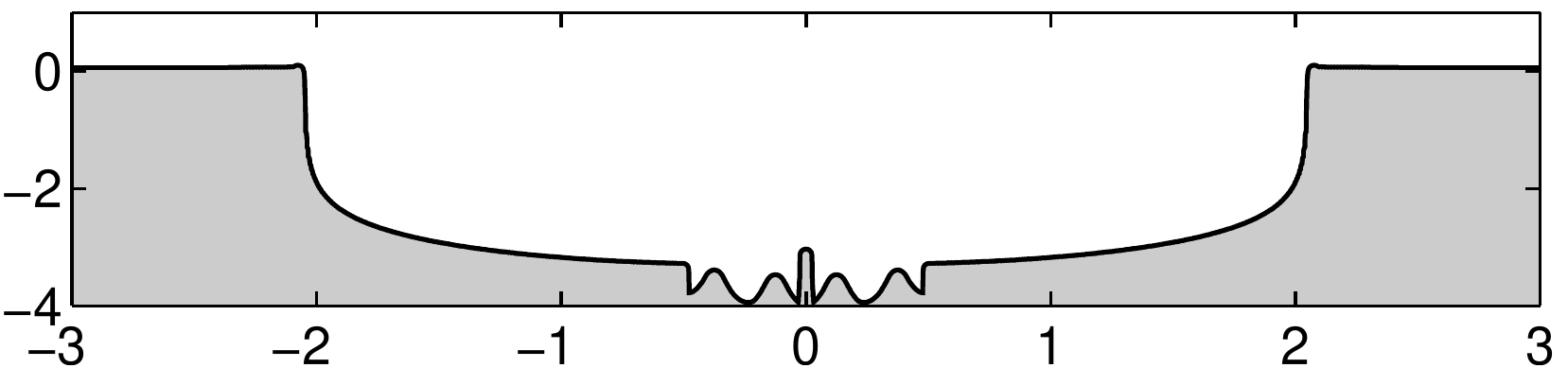}
\put(-11,4.5){\begin{sideways}Rel. Error\end{sideways}}
\put(-5,4){\begin{sideways}$\log_{10}\{\mathcal{E}_{500}\}$\end{sideways}} %
\put(42,-6){Position, $x/L$}
\end{overpic}
\\[5mm]
\caption{\label{Fig:QNM_G_wR_4_dielectricBarrier_nR_pi_N_500}
\change{Illustration of the QNM approximation to the Green function for the dielectric barrier in one dimension from Section~\ref{Sec:1Dresonator}.}
Top: QNM Green function approximation $G_{500}(x,x',\omega)$ as a function of $x$, with $x'=0$, $\omega_0=4L/\text{c}$, and in units of $G_0=\text{c}/2n_\text{R}\omega$. Red solid and blue dashed curves show the real and imaginary parts, respectively, and the black curve shows the absolute value. The gray shading indicates the extent of the dielectric barrier. Bottom: Relative error %
$\mathcal{E}_{500}(x,x',\omega_0)$. %
} 
\end{figure}
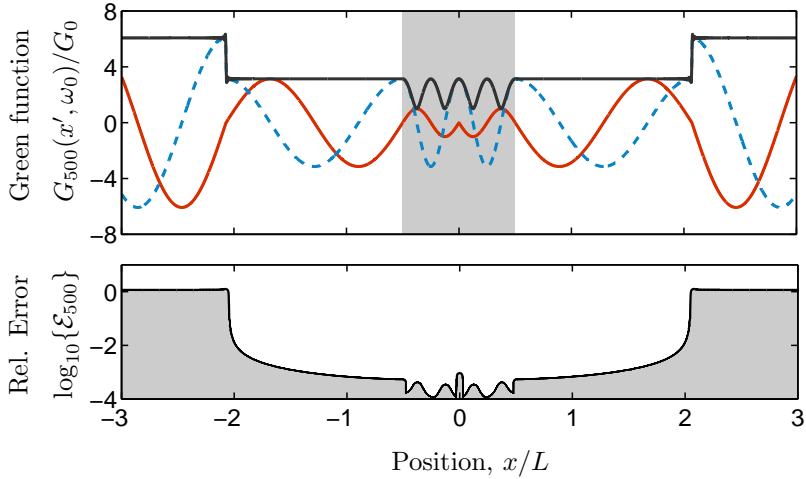

As a first test of the Green function expansion in Eq.~(\ref{Eq:GE_QNM_expansion}), %
we consider now the QNM expansion for a one-dimensional dielectric resonator system of the form
\begin{align}
G_N(x,x',\omega) = \frac{\text{c}^2}{2\omega}\sum_{n=-N}^N\frac{\ft_n(x)\ft_n(x')}{\tlo_n-\omega},
\label{Eq:G_tot_N}
\end{align}
where the index $n$ is chosen to count the real part of the QNM frequencies, cf. Fig.~\ref{Fig:transmission_spectrum_dielectricBarrier_nR_pi}. The QNMs are known analytically, cf. Section~\ref{Sec:QNMs_of_dielectric_barrier}, so one can easily sum the series to get the approximation to arbitrary accuracy, assuming the series converges. %
Figure~\ref{Fig:QNM_G_wR_4_dielectricBarrier_nR_pi_N_500} shows the approximation with $N=500$ along with the relative error
\begin{align}
\mathcal{E}_N(x,x'\omega)=\frac{|G_N(x,x',\omega)-G_\text{ref}(x,x',\omega)|}{|G_\text{ref}(x,x'\omega)|}
\label{Eq:G_N_rel_error}
\end{align}%
for the case of $x'=0$ in the center of the barrier, and where $G_\text{ref}(x,x',\omega)$ is the analytically known reference for the electric field Green function\change{, see Appendix \ref{App:Green_function_for_the_dielectric_slab}.} 
Judging from the relative error, the QNM approximation appears to break down for $|x|\gtrsim2L$, which is also clearly visible as an abrupt jump in the calculated values, with barely visible Gibbs oscillations due to high number of QNMs used. At distances further away from the resonator, the approximate Green tensor diverges in a series of steps (first of which is visible in Fig.~\ref{Fig:QNM_G_wR_4_dielectricBarrier_nR_pi_N_500}), each step of width $\pi$ and with exponentially increasing height (not shown).

\begin{figure}[htb!]
\centering %
\begin{overpic}[width=9.4cm]{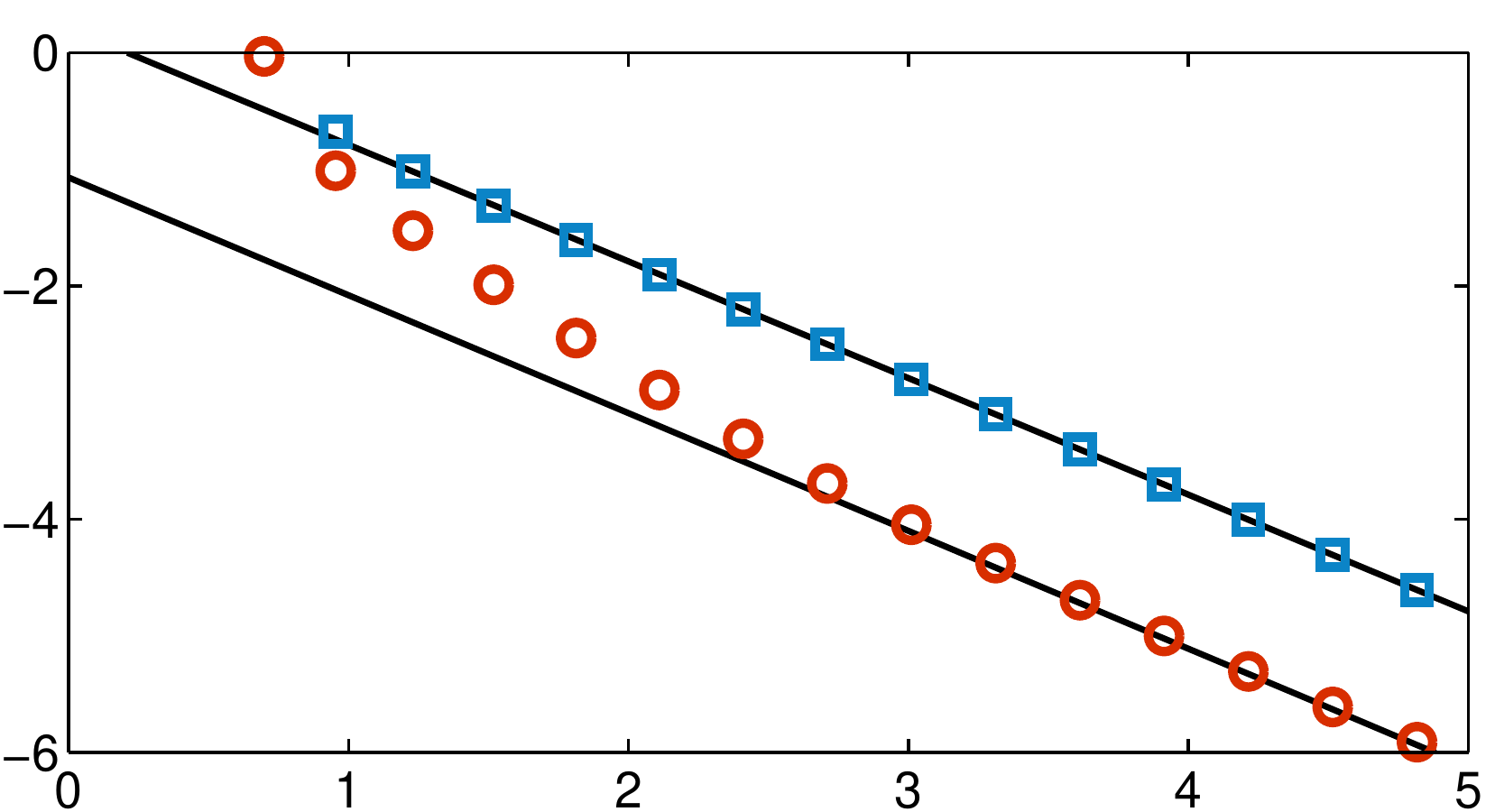}
\put(-13,16){\begin{sideways}Relative error\end{sideways}}
\put(-6,8){\begin{sideways}$\log_{10}\{\mathcal{I}_N\}$ or $\log_{10}\{\mathcal{E}_N\}$\end{sideways}}
\put(24,-6){Number of QNMs, $\log_{10}\{2N+1\}$} %
\put(64,31){$y=-1.00\,x +0.21$} %
\put(44,8){$y=-1.01\,x - 1.07$} %
\end{overpic}
\\[5mm]
\caption{\label{Fig:G_k0_4_dielectricBarrier_nR_pi_log10_relErr_vs_log10_2Np1_xp_0}Convergence analysis for the QNM Green function expansion inside the dielectric barrier. Red circles show the relative integrated error $\mathcal{I}_N(0,\omega_0)$ \change{with fixed $x'=0$} in the region $-L/4<x<L/4$ around the center of the barrier, and blue squares show the relative error $\mathcal{E}_N(0,0,\omega_0)$ at $x=x'=0$ in the center of the barrier. \change{For both data sets, $\omega_0=4L/\text{c}$, and} black lines show linear fits to the last three data points.}
\end{figure}
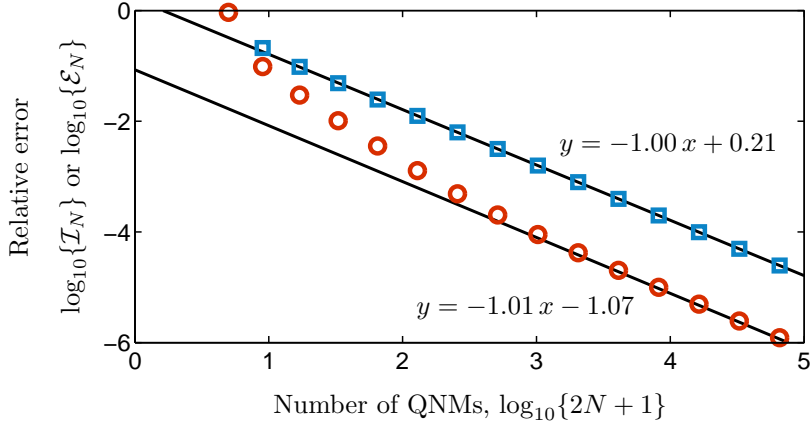

As a first glimpse at the convergence properties of the QNM Green function approximation  in Eq.~(\ref{Eq:G_tot_N}), Fig.~\ref{Fig:G_k0_4_dielectricBarrier_nR_pi_log10_relErr_vs_log10_2Np1_xp_0} shows a double logarithmic plot of the relative error in Eq.~(\ref{Eq:G_N_rel_error}), as well as the integrated error
\begin{align}
\mathcal{I}_N(x',\omega) = \frac{\int|G_N(x,x'\omega)-G_\text{ref}(x,x'\omega)| \ud x }{\int|G_\text{ref}(x,x'\omega)|\ud x},
\end{align}
where the integrals are taken in the region $-L/4<x<L/4$ around the center of the barrier. \change{The width of the integrals were chosen not to be the full width of the barrier in order that we may conveniently use the same measure of the relative error for the scattering calculations in Section~\ref{Sec:Scattering_calculations_for_the_dielectric_barrier}, where the convergence of the boundary at $x=-L/2$ turns out to problematic.} Both measures of the error appear to vanish in a first order polynomial fashion, as evidenced by the fit to the last three data points. Although this is a relatively slow convergence, this is our first indication that the formal QNM Green function approximation does indeed tend to the correct reference in the limit of large $N$.

\change{
\subsubsection*{Supplementary code}
With the supplementary code~\cite{dielectricBarrier_arXiv} we provide the files necessary to reproduce Fig.~\ref{Fig:G_k0_4_dielectricBarrier_nR_pi_log10_relErr_vs_log10_2Np1_xp_0}, and we encourage interested readers to investigate the convergence of the QNM Green function approximation when changing the number of terms in Eq.~(\ref{Eq:G_tot_N}) or study the non-trivial divergence of the QNM Green function approximation as $x\rightarrow\infty$.
}

\subsection{QNMs as the residues of the Green tensor}
\label{Sec:QNMs_as_residues}
To lay the foundation for the convergence analysis in Section~\ref{Sec:Completeness}, and as a complementary QNM modeling approach to that taken in Section~\ref{Sec:Formal_expansions}, it is illustrative to see how the QNMs appear as the residues of the Green tensor. %
The starting point, is that the Green tensor has a number of poles, all of which are located in the lower half of the complex plane. Close to a pole $\Omega_n$, we assume that the Green tensor can in general be approximated as
\begin{align}
\mmG(\mr,\mr',\omega)\approx \frac{\mmRho_n(\mr,\mr')}{\omega-\Omega_n} + \mmChi(\mr,\mr',\omega),
\label{Eq:Green_general_rho_chi_form}
\end{align}
where $(\mD+\text{i}\omega)\mmChi(\mr,\mr',\omega)$ is bounded at $\omega=\Omega_n$. %
Treating initially the case $\mr\neq\mr'$, we insert this expression in  %
Eq.~(\ref{Eq:Green_def}), multiply by $\omega-\Omega_n$, and take the limit $\omega\rightarrow\Omega_n$ to see that
\begin{align}
\mD\,\mmRho_n(\mr,\mr') + \text{i}\Omega_n\mmRho_n(\mr,\mr')  &= 0 \label{Eq_residual_rho_def_v1}.
\end{align}
It follows from Eq.~(\ref{Eq_residual_rho_def_v1}) and the inherited radiation condition, that each column of $\mmRho_n(\mr,\mr')$ solves the defining equations for the QNMs in Eqs.~(\ref{Eq:SM_1}),  (\ref{Eq:SM_2}), and (\ref{Eq:Maxwell_QNM_spinorForm}). %
Therefore, the complex QNM resonance frequencies are the poles of the Green tensor, $\Omega_n=\tlo_n$, and $\mmRho_n(\mr,\mr')$ is proportional to the QNM $\mFt_n(\mr)$, as was tacitly assumed in Eq.~(\ref{Eq:G_QNM_proto_expansion}).
In the vicinity of $\omega\approx\tlo_n$, we write this as %
\begin{align}
\mmRho_n(\mr,\mr) \approx \mF(\mr,\omega)\underline{\boldsymbol{\beta}}^\text{T}_n(\mr').
\label{Eq:mmRho_propto_QNM}
\end{align}
Relaxing now the condition $\mr\neq\mr'$, we use Eqs.~(\ref{Eq:Green_general_rho_chi_form}) and (\ref{Eq:mmRho_propto_QNM}) in %
Eq.~(\ref{Eq:Green_def}), multiply from the left with $\mFb_n(\mr)\mmW$ and integrate across the volume $V$ to find that
\begin{align}
\text{i}\langle\langle\mFb_n(\mr)|\mF(\mr,\omega)\rangle\rangle\underline{\boldsymbol{\beta}}^\text{T}_n(\mr') +(\omega-\tlo_n)\langle\mFb_n(\mr)|(\mD+\text{i}\omega)\mmChi(\mr,\mr',\omega)\rangle = -\text{i}\frac{\text{c}^2}{2\omega}\mFb_n(\mr).
\end{align}
Last, by taking the limit $\omega\rightarrow\tlo_n$, we find, that %
the residues of the matrix Green tensor can be expressed as
\begin{align}
\text{Res}\left\{\mmG(\mr,\mr',\omega),\omega=\tlo_m\right\} = -\frac{\text{c}^2}{2\tlo_m}\mFt_m(\mr)[\mFb_m(\mr')]^\text{T},
\label{Eq:QNMs_as_residues}
\end{align}
which is consistent with the formal expansion in Eq.~(\ref{Eq:MG_QNM_expansion}), but leads to a slightly different expansion of the form
\begin{align}
\mmG(\mr,\mr',\omega) = \frac{\text{c}^2}{2}\sum_m\frac{\mFt_m(\mr)[\mFb_m(\mr')]^\text{T}}{\tlo_m(\tlo_m-\omega)},
\label{Eq:MG_QNM_expansion_v2}
\end{align}
as was also found for QNMs in spherical resonators in Ref.~\cite{Lee_JOSAB_16_1409_1999}. The connection between Eqs.~(\ref{Eq:GE_QNM_expansion}) and (\ref{Eq:MG_QNM_expansion_v2}) was also discussed in Ref.~\cite{Muljarov_PRB_93_075417_2016}. For the one-dimensional problem, the expansion of the Green tensor may include an extra term in addition to the sum in Eq.~(\ref{Eq:MG_QNM_expansion_v2}), as pointed out in Ref.~\cite{Muljarov_EPL_92_50010_2010}; see also discussion in Section \ref{Sec:Alternative_expression_for_the_Green_tensor}.

\reminder{[If the application of the projection with the surface term seems a little odd, one can also multiply from the left with $[\mFb_m(\mr)]^\text{T}$ in the defining equation for $\mmG(\mr,\mr',\omega)$ and apply just the volume integral in Eq.~(\ref{Eq:innerProduct_startingPoint}). Next, one can integrate by parts to move the operator onto the adjoint QNM, which will give exactly the norm $\langle\langle\mFb_m(\mr)|\mft_m(\mr)\rangle\rangle$ multiplied onto $\boldsymbol{\beta}^\text{T}_n(\mr')$. This, however, only works if one assumes the Green tensor is dominated by the pole term - i.e. one ignores $\chi(\mr,\mr',\omega)$. One can do that, for example, by changing $\chi(\mr,\mr',\omega)$ in Eq.~(\ref{Eq:Green_general_rho_chi_form}) to $\chi(\mr,\mr',\omega)\Theta(|\omega-\tlo_m|-\Delta)$, where $\Theta(\omega)$ is the Heaviside step function (or a sufficiently smooth variant thereof. In this way, we ensure that any terms relating to $\chi(\mr,\mr',\omega)$ vanish in the projection.]
}

\reminder{
\subsubsection*{Alternative approach - under construction}
[I'm not sure if the above discussion is sufficiently rigorous. In particular, the neglect of $\mmChi(\mr,\mr',\omega)$ in the limit $\omega\rightarrow\tlo_n$ seems a bit uncontrolled. Below is an attempt at a somewhat more formal approach following Garc{\'i}a Calder{\'o}n~\cite{Calderon_2010}, but his approach will not immediately work, because it's tied up to the boundary conditions of the fields.]

Starting from Eq.~(\ref{Eq:Green_general_rho_chi_form}), inserting in the definition of the Green tensor in Eq.~(\ref{Eq:Green_def}) and adding and subtracting the term $\text{i}\Omega_n\mmRho_n(\mr,\mr')/(\omega-\Omega_n)$, we %
find %
\begin{align}
\Big[\mD+\text{i}\Omega_n\Big]\frac{\mmRho_n(\mr,\mr')}{\omega-\Omega_n} + \Big[\mD+\text{i}\omega\Big]\mmChi(\mr,\mr',\omega)+ \text{i}\mmRho(\mr,\mr') = \frac{\delta(\mr-\mr')}{\text{i}\omega\mu_0}
\begin{bmatrix}
[\epsilon_0\epsilon_\text{r}(\mr)]^{-1} & 0 \\
0 & [\mu_0]^{-1}
\end{bmatrix}.
\end{align}
Treating the two powers of $\omega-\Omega_n$ independently, we find the two equations
\begin{align}
\Big[\mD+\text{i}\Omega_n\Big]\mmRho_n(\mr,\mr') &= 0 \label{Eq_residual_rho_def}\\
\Big[\mD+\text{i}\omega\Big]\mmChi(\mr,\mr',\omega)+ \text{i}\mmRho_n(\mr,\mr') &= \frac{\delta(\mr-\mr')}{\text{i}\omega\mu_0}\begin{bmatrix}
[\epsilon_0\epsilon_\text{r}(\mr)]^{-1} & 0 \\
0 & [\mu_0]^{-1}
\end{bmatrix}.\label{Eq_residual_Chi_rho_Eq}
\end{align}
It %
follows from Eq.~(\ref{Eq_residual_rho_def}) and the inherited radiation condition, that %
$\mmRho_n(\mr,\mr')$ solves the defining equations for the QNMs in Eqs.~(\ref{Eq:SM_1}),  (\ref{Eq:SM_2}), and (\ref{Eq:Maxwell_QNM_spinorForm}). %
Therefore, the complex QNM resonance frequencies are the poles of the Green tensor, $\Omega_n=\tlo_n$, and $\mmRho_n(\mr,\mr')$ is proportional to the QNM $\mFt_n(\mr)$, as was tacitly assumed in Eq.~(\ref{Eq:G_QNM_proto_expansion}). In the vicinity of $\omega\approx\tlo_n$, we write this as %
\begin{align}
\mmRho_n(\mr,\mr) \approx \mFt_n(\mr)\boldsymbol{\beta}^\text{T}_n(\mr'),
\label{Eq:mmRho_propto_QNM}
\end{align}
and take the limit $\omega\rightarrow\tlo_n$ in the end. %

[So far so good. The next step is to use Eq.~(\ref{Eq_residual_Chi_rho_Eq}) in connection with Eq.~(\ref{Eq:mmRho_propto_QNM}) and the defining equations for the QNMs to calculate an explicit expression for $\boldsymbol{\beta}^\text{T}_n(\mr')$.In my attempt below, I would like to use the projection operator in Eq.~(\ref{Eq:QNM_innerprod_generalized}), but this will not work unless we can show that the projection of $[\mD + \text{i}\omega]\mmChi'(\mr,\mr',\omega)$ vanishes in the limit $\omega\rightarrow\tlo_m$.

Inserting this expression in Eq.~(\ref{Eq_residual_Chi_rho_Eq}), we can now apply the projection operator in Eq.~(\ref{Eq:QNM_innerprod_generalized}) to both sides of the equation. For the first term, we first rewrite $\mmChi(\mr,\mr',\omega)$ as
\begin{align}
\mmChi(\mr,\mr',\omega) =\mmChi'(\mr,\mr',\omega) + \alpha_\chi(\mr',\omega)\mFt_n(\mr),
\end{align}
where $\alpha_\chi(\mr',\omega)$ %
is the possible non-zero projection of $\mmChi(\mr,\mr',\omega)$ onto the QNM $\mFt_n(\mr)$. [Even if this is not a projection in the usual sense, the

From Eq.~(\ref{Eq:QNM_innerprod_generalized}), it follows that \begin{align}
\langle\langle\mFb_n(\mr)|\mmChi'(\mr,\mr',\omega)\rangle\rangle\rightarrow0,\; \text{for } \omega\rightarrow\tlo_m.
\end{align}
}

\subsubsection{Alternative scheme for evaluation of the normalization integral}
\label{Sec:Alternative_normalization_scheme}

From a numerical point of view, especially with integral equation techniques, the use of a volume integral for normalization is not ideal, since one must necessarily evaluate the field at numerous positions within the integration domain. Bai \emph{et al.}~\cite{Bai_OE_21_27371_2013} suggested a combined calculation and normalization approach for QNMs, which is useful when one has access to the Green tensor at complex frequencies. In a slightly different formulation, but exploiting the same idea, we can derive a variant of this normalization \change{evaluation} by noting that if
Eq.~(\ref{Eq:QNMs_as_residues}) holds, then the inverse of the \change{normalization integral} can be \change{inferred from the residue of any of the components of $\mmG(\mr,\mr',\omega)$ at $\omega=\tlo_m$. In particular, for $\alpha,\beta\in\{x,y,z\}$, one can use the $\alpha,\beta$ component of the electric field Green tensor along with the $\alpha$ and $\beta$ components of the electric field QNM at the positions $\mr$ and $\mr'$, respectively, to calculate the inverse \change{normalization integral} as} %
\begin{align}
\langle\langle\mFt_m(\mr)|\mFt_m(\mr)\rangle\rangle^{-1} =  \frac{\text{i}}{\pi}\frac{\tlo_m/\text{c}}{\ft_{m\alpha}(\mr)\ft_{m\beta}(\mr')}\oint_{\tlo_m}G^\text{EE}_{\alpha\beta}(\mr,\mr',z)\,\ud z,
\label{Eq:inverse_norm_from_contour_integral}
\end{align}
where the integral is on a closed curve around the point $z=\tlo_m$ in the complex frequency plane, and $\mr$ and $\mr'$ are any two positions where the electric field QNM \change{components $\mft_{m\alpha}(\mr)$ and $\mft_{m\beta}(\mr')$ do} not vanish. In practical calculations, one can work only with the scattered part of the Green tensor, since the background Green tensor has no poles at non-zero frequencies.

A particularly interesting property of Eq.~(\ref{Eq:inverse_norm_from_contour_integral}), is that it can be handled effectively by the trapezoidal rule because of the periodic nature of the integral~\cite{Trefethen_SIAM_review_56_385_2014}. In practice, we can always find a circle of radius $R$ centered on the QNM resonance frequency $\tlo_m$ and surrounding only this one pole. Parameterizing the curve in terms of the angle $\theta$ as $z(\theta)=\tlo_m+R\exp\{\text{i}\theta\}$, we write the integral as %
\begin{align}
\text{i}R\int_0^{2\pi}G^\text{EE}_{\alpha\beta}(\mr,\mr',z)\text{e}^{\text{i}\theta}\ud\theta,
\end{align}
which is of the same form as Eq.~(2.1) in Ref.~\cite{Trefethen_SIAM_review_56_385_2014}. As a consequence, the corresponding trapezoidal rule approximation of order $N$,
\begin{align}
\langle\langle\mFt_m(\mr)|\mFt_m(\mr)\rangle\rangle^{-1}_N = -\frac{2R}{N}\frac{\tlo_m/\text{c}}{\ft_{m\alpha}(\mr)\ft_{m\beta}(\mr')}\sum_{k=1}^NG^\text{EE}_{\alpha\beta}\left(\mr,\mr',z(\theta_k)\right)\text{e}^{\text{i}\theta_k},
\label{Eq_inverse_norm_integral_trapz_approx}
\end{align}
with $\theta_k=2\pi k/N$, converges exponentially fast as a function of $N$~\cite{Trefethen_SIAM_review_56_385_2014}. %

\subsubsection{Alternative normalization for the dielectric barrier}
\begin{figure}[htb!]
\centering %
\begin{overpic}[width=9.4cm]{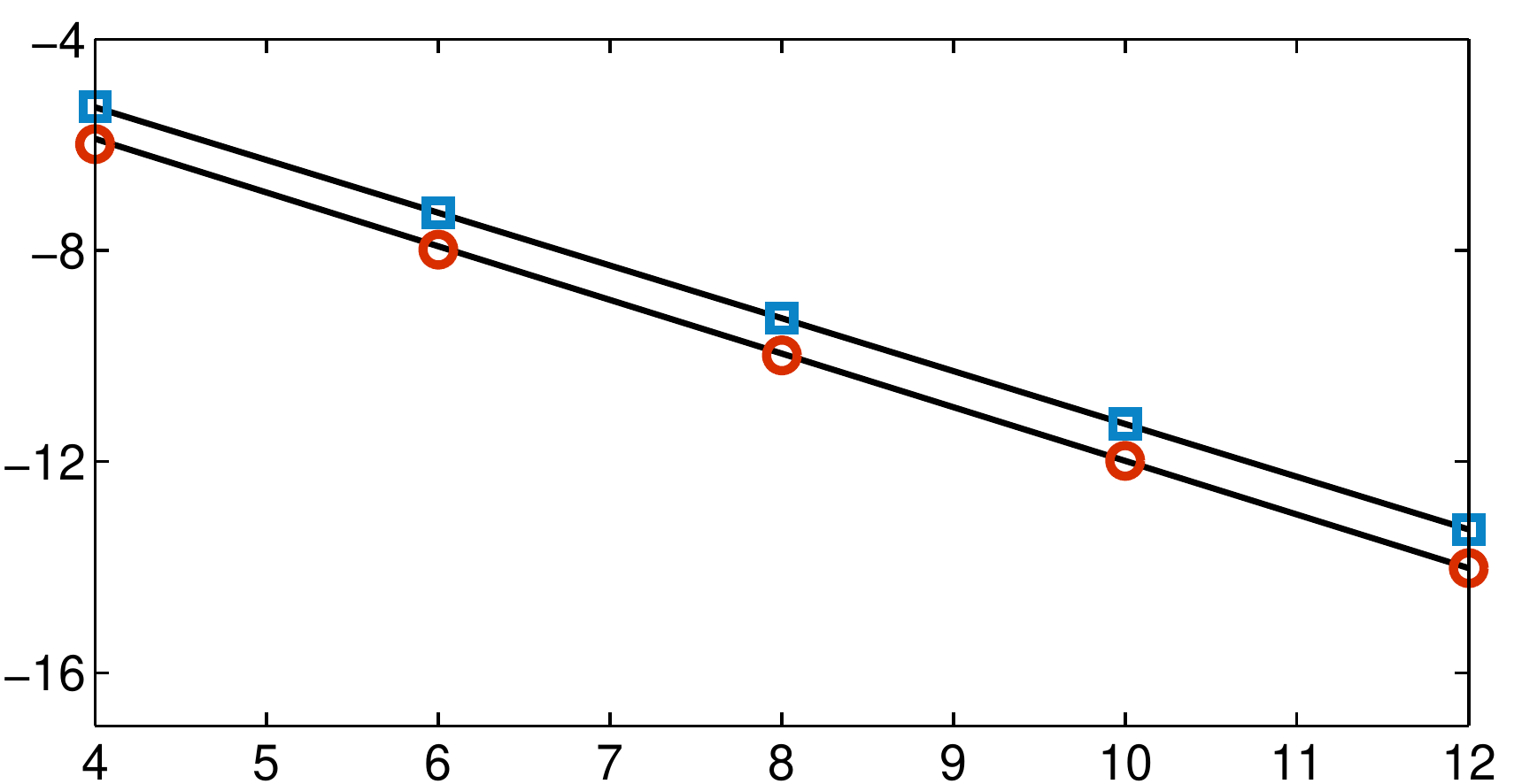}
\put(-5,6){\begin{sideways}$\log_{10}\{D_{+2}L\}$ or $\log_{10}\{\mathcal{E}\}$\end{sideways}}
\put(36,-6){Number of points, $N$} %
\end{overpic}
\\[5mm]
\caption{\label{Fig:barrier_1D_invNorm_convergence}Convergence analysis for the numerical calculation of the inverse \change{normalization integral} for the QNM of the dielectric barrier with $m=4$. The figure shows the relative error $\mathcal{E}(N)=|\langle\langle\mFt_4(x)|\mFt_4(x)\rangle\rangle^{-1}_N-\langle\langle\mFt_4(x)|\mFt_4(x)\rangle\rangle^{-1}_\text{ref}|/|\langle\langle\mFt_4(x)|\mFt_4(x)\rangle\rangle^{-1}_\text{ref}|$ (red circles) and the difference $D_{+2}(N)$ (blue squares) between results obtained using $N$ and $N+2$ points in the trapezoidal sum. Solid and lines show the corresponding fits to the last two data points of each set.}
\end{figure}

Using the analytical expression for the Green tensor in Appendix \ref{App:Green_function_for_the_dielectric_slab}, we can evaluate the right hand side of the (inverse) \change{normalization integral} in Eq.~(\ref{Eq:inverse_norm_from_contour_integral}) by explicitly calculating the residue of the Green tensor at $z=\tlo_m$. \change{Using the eigenfunctions in Eq.~(\ref{Eq:QNM_dielectric_barrier_eField}),} it takes the value
\begin{align}
\langle\langle\mFt_m(x)|\mFt_m(x)\rangle\rangle^{-1}_\text{ref} = \frac{(-1)^m}{2n_\text{R}^2L},
\end{align}
which is equivalent to the result of the direct calculation in Eq.~(\ref{Eq:dielectric_barrier_normalization_Efield_only}). %
To \change{illustrate} %
the practical numerical calculations, we can also use the analytical expression for the Green tensor to set up the sum in Eq.~(\ref{Eq_inverse_norm_integral_trapz_approx}) and evaluate it for various $N$ and thereby test the convergence properties. Figure~\ref{Fig:barrier_1D_invNorm_convergence} shows the logarithm of the relative error using Eq.~(\ref{Eq_inverse_norm_integral_trapz_approx}) as a function of $N$. To a good approximation, the points fall on a straight line indicating that the convergence is %
exponential as expected. Points corresponding to odd values of $N$ fall on a \change{parallel} straight line (not shown). To illustrate the usefulness of the approach in the general case, where one does not have access to a reference calculation, we consider also the difference between results obtained using $N$ and $N+2$ points in the trapezoidal approximation,
\begin{align}
D_{+2}(N) =  |\langle\langle\mFt_4(x)|\mFt_4(x)\rangle\rangle^{-1}_{N+2}-\langle\langle\mFt_4(x)|\mFt_4(x)\rangle\rangle^{-1}_N|.
\end{align}
The logarithm of $D_{+2}(N)$ is also shown in Fig.~\ref{Fig:barrier_1D_invNorm_convergence}, and the points fall on a straight line with a slope similar to that of the data for the relative error. \change{Using} %
the analysis in appendix~\ref{App:Practical_convergence_studies}, we can estimate the value of the inverse \change{normalization integral} to be
\begin{align}
\langle\langle\mFt_4(x)|\mFt_4(x)\rangle\rangle^{-1}L \approx 0.050660591821169,
\end{align}
with an estimated error less than $5\times10^{-16}$. Comparing directly to the reference value, one can verify that the actual error is less than the estimated error.%

\subsubsection{Normalization for the plasmonic dimer}
\label{Sec:normalization_of_plasmonic_dimer}
We have no reference calculations of the \change{normalization integral} for the QNMs of the plasmonic dimer. Moreover, we have no analytical expressions for the QNMs or the Green tensor, so both will have to be calculated numerically, and any numerical error will ultimately limit the accuracy of the calculated inverse norm.

\begin{figure}[htb!]
\centering %
\begin{overpic}[width=9.4cm]{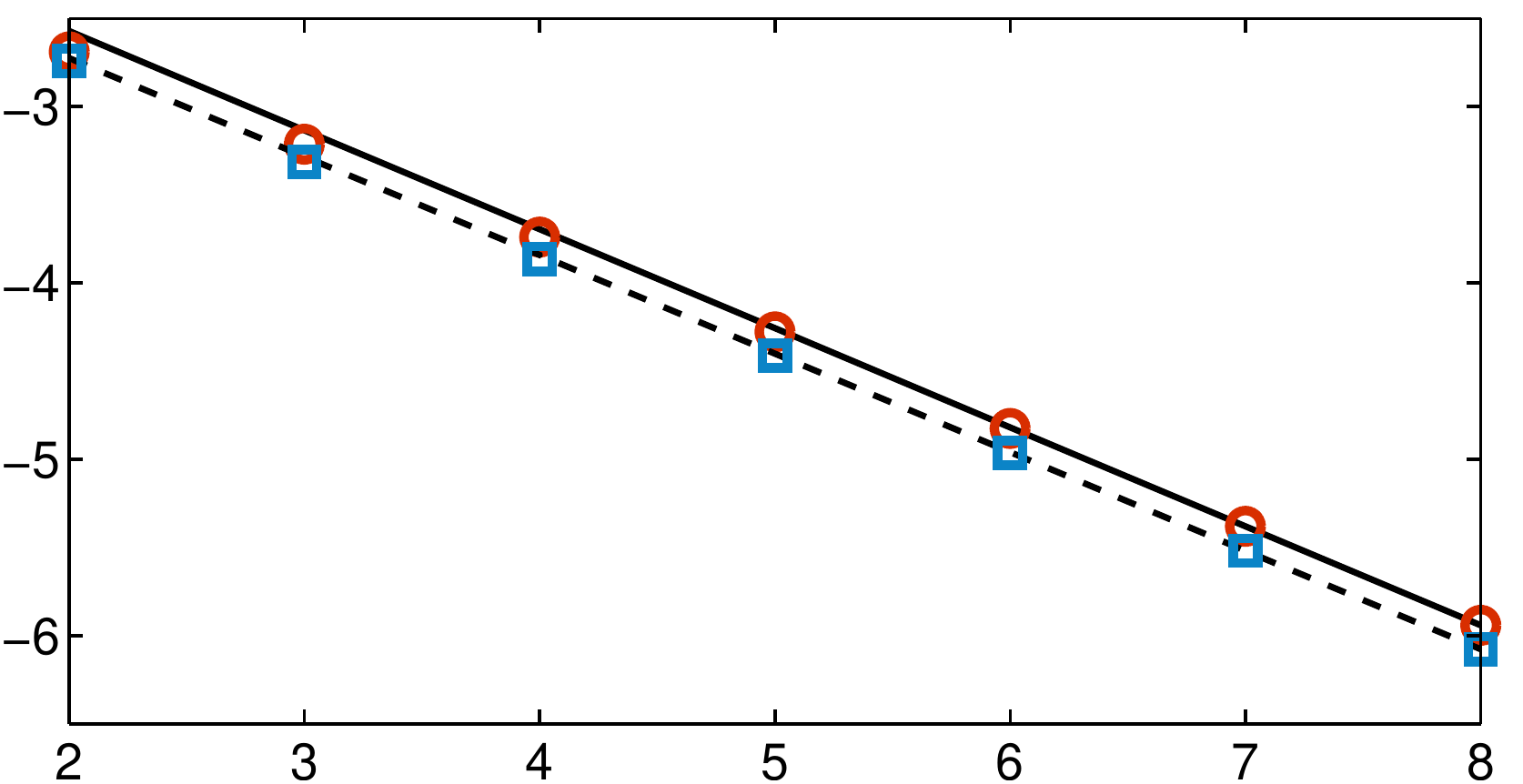}
\put(-6,7){\begin{sideways}Difference, $\log_{10}\{D_{+1}L\}$\end{sideways}}
\put(36,-6){Cut-off parameter, $l_\text{max}$} %
\end{overpic}
\\[5mm]
\caption{\label{Fig:invNorm_trapz_Dp1_vs_Lmax_real_and_imag}Convergence analysis for the calculation of the inverse \change{normalization integral} in the case of the plasmonic dimer. The figure shows, as a function of $l_\text{max}$, the difference $D_{1+}(l_\text{max})$ between results obtained using $l_\text{max}$ and $l_\text{max}+1$. Red circles and blue squares correspond to real and imaginary parts of the difference, respectively. Solid and dashed lines show the corresponding fits to the last two data points of each set.}
\end{figure}

The convergence properties of the trapezoidal rule approximation to the inverse \change{normalization integral} in Eq.~(\ref{Eq_inverse_norm_integral_trapz_approx}) is almost independent of the accuracy of the integrand. %
Therefore, we can perform calculations with a relatively low accuracy of the integrand using $l_\text{max}=2$ to find the number of elements $N$ necessary for a given accuracy of the integral. Subsequently, using this value of $N$, we then vary $l_\text{max}$ to estimate the true value of the inverse \change{normalization integral} as done also for the complex QNM frequency in Section~\ref{Sec:QNMs_of_plasmonic_dimer}. To this end, we define
\begin{align}
D_{+1}(N) =  |\langle\langle\mFt_1(\mr)|\mFt_1(\mr)\rangle\rangle^{-1}_{N+1}-\langle\langle\mFt_1(\mr)|\mFt_4(\mr)\rangle\rangle^{-1}_N|,
\end{align}
and Fig.~\ref{Fig:invNorm_trapz_Dp1_vs_Lmax_real_and_imag} shows the logarithm of $D_{+1}$ as a function of $l_\text{max}$ for constant $N=42$ which is sufficiently high to make the integration limited by the machine accuracy.
The points clearly fall approximately on straight lines, as expected from the exponential convergence of the QNMs. Choosing a scaling in which the electric field QNMs are unity in the center of the gap between the two spheres, we use the approach in Appendix~\ref{Sec:Convergence_and_consistency} to find the value \begin{align}
\langle\langle\mFt_1(\mr)|\mFt_1(\mr)\rangle\rangle^{-1}d^3\approx0.2101658(4) - 0.0527264(3)\text{i},
\label{Eq:invNorm_by_d3_for_dimer_F1}
\end{align}
with an estimated (absolute) error less than $5\times10^{-7}$.%

\section{Convergence and consistency} %
\label{Sec:Completeness}
To answer the question of convergence and consistency of the formal expansions, we follow a line of argumentation similar to that of Refs.~\cite{Leung_PRA_49_3057_1994, Leung_JOSAB_13_805_1996, Lee_JOSAB_16_1409_1999} for optical cavities and Ref.~\cite{Calderon_2010} for solutions to the Schr{\"o}dinger equation. We reserve the word ``convergence" to the property of certain series, that they tend to a finite value as the number of terms are increased. The more restrictive requirement of ``consistency" is reserved for the cases where the finite limit is also the correct function value that one hopes to approximate. Appendix~\ref{Sec:Convergence_and_consistency} discusses these properties in the context of general numerical calculations, and within the present framework, the use of $N$ QNMs can be understood simply as a specific choice of basis function set, with the parameter $h=1/N$ controlling the accuracy.

In practice, of course, we are mostly interested in the cases where the QNM expansions are consistent. As we shall see, this is not always the case, because there \change{may be} additional contributions that stem from branch cuts of the analytical continuation of the Green tensor. Moreover, as we have already seen in Fig.~\ref{Fig:QNM_G_wR_4_dielectricBarrier_nR_pi_N_500}, the QNM expansion may be convergent at some points without being consistent. Even though the solution to the general question of convergence and consistency \change{appears to be out of reach at this point,} %
we can make some useful progress by carefully analyzing the structure of the Green tensor as in Refs.~\cite{Calderon_2010,Leung_PRA_49_3057_1994, Leung_JOSAB_13_805_1996, Lee_JOSAB_16_1409_1999} and combine this insight with numerical investigations. %
In this way, we introduce a so-called region of convergence, and we further exploit the results of the analysis to device a prescription to formally extend this region.

For the one-dimensional resonator example, we can analyze the region of convergence  exactly and show that the QNM expansion of the Green function, and other expansions that can be derived from the Green function, in general converges to the correct value only for positions $x$ and $x'$ strictly within the resonator. Nevertheless, for positions $x'$ within the resonator, the region of convergence for $x$ extends beyond the boundary, in agreement with Fig.~\ref{Fig:QNM_G_wR_4_dielectricBarrier_nR_pi_N_500}. %
For the three-dimensional example, we us a numerical estimate of the region of convergence and illustrate the usefulness of the expansion method by an example using full three-dimensional calculations of a plasmonic dimer of gold nano spheres.

\subsection{Convergence of the QNM Green tensor expansion}
\label{Sec:Convergence_of_G}
The starting point is the fact that for any $\omega>0$, we can write the product of $\omega$ and the matrix Green tensor as
\begin{align}
\omega\,\mmG(\mr,\mr',\omega) = -\frac{\text{i}}{2\pi}\oint_{\Gamma_\omega}z\,\frac{\mmG(\mr,\mr',z)}{z-\omega} \ud z,
\label{Eq:omega_time_G_by_contour}
\end{align}
where the contour $\Gamma_\omega$ is a circle around $z=\omega$ with a sufficiently small radius that $z\,\mmG(\mr,\mr',z)$ has no poles inside. If, from this starting configuration, and assuming $\mG(\mr,\mr',\omega)$ has not branch cuts, we increase the radius of the contour, and if we exclude any poles encountered in the process, we find that
\begin{align}
\omega\,\mmG(\mr,\mr',\omega) &= -\frac{\text{i}}{2\pi}\oint_{\Gamma_0}z\,\frac{\mmG(\mr,\mr',z)}{z-\omega} \ud z -\frac{\text{i}}{2\pi}\oint_{\Gamma_N}z\,\frac{\mmG(\mr,\mr',z)}{z-\omega} \ud z \nonumber \\
&\qquad - \sum_{n=1}^N\text{Res}\left\{z\,\frac{\mmG(\mr,\mr',z)}{z-\omega},z=\tlo_n\right\},
\end{align}
where now the contour $\Gamma_0$ encircles the special point $z=0$ in the clockwise direction, and the contour $\Gamma_N$ encircles the origin and exactly $N$ additional poles of the matrix Green tensor in the counter clockwise direction, cf. Fig.~\ref{Fig:Riesz_outline}.

Assuming, for now, that $z\,\mmG(\mr,\mr',z)$ has no pole at $z=0$, we can use Eq.~(\ref{Eq:QNMs_as_residues}), to find that
\begin{align}
\mmG(\mr,\mr',\omega) &= -\frac{\text{i}}{2\pi\omega}\oint_{\Gamma_N}z\,\frac{\mmG(\mr,\mr',z)}{z-\omega} \ud z + \frac{\text{c}^2}{2\omega}\sum_{n=1}^N\frac{\mFt_m(\mr)[\mFb_m(\mr')]^\text{T}}{\tlo_m-\omega},
\end{align}
from which, upon comparing to Eq.~(\ref{Eq:MG_QNM_expansion}), we conclude that the formal expansion of the matrix Green tensor converges to the actual matrix Green tensor if the integral along the contour $\Gamma_N$ vanishes in the limit $N\rightarrow\infty$. %
Clearly, a sufficient condition for the integral along the contour $\Gamma_N$ to vanish, is for the product $z\,\mmG(\mr,\mr',z)$ to vanish \change{sufficiently fast} in the limit $|z|\rightarrow\infty$; this will be the condition we explore in the next Section.

\reminder{In Ref.~\cite{Ching_PRL_74_4588_1995}, it is pointed out, that sometimes there is a problem at $|\omega|\rightarrow\infty$ (on the real axis)}

\subsubsection{Alternative expression for the Green tensor}
\label{Sec:Alternative_expression_for_the_Green_tensor}
If, instead of Eq.~(\ref{Eq:omega_time_G_by_contour}), we start with the expression
\begin{align}
\mmG(\mr,\mr',\omega) = -\frac{\text{i}}{2\pi}\oint_{\Gamma_\omega}\frac{\mmG(\mr,\mr',z)}{z-\omega} \ud z,
\label{Eq:G_by_contour}
\end{align}
we can follow identical steps as above, to find that
\begin{align}
\mmG(\mr,\mr',\omega) &= -\frac{\text{i}}{2\pi}\oint_{\Gamma_0}\frac{\mmG(\mr,\mr',z)}{z-\omega} \ud z -\frac{\text{i}}{2\pi}\oint_{\Gamma_N}\frac{\mmG(\mr,\mr',z)}{z-\omega} \ud z \nonumber \\
&\qquad - \sum_{n=1}^N\text{Res}\left\{\frac{\mmG(\mr,\mr',z)}{z-\omega},z=\tlo_n\right\}.
\end{align}
If we now assume (as is sometimes the case) that $\mmG(\mr,\mr',\omega)$ has a simple pole at $\omega=0$, we can use Eq.~(\ref{Eq:QNMs_as_residues}) along with the same arguments as above to conclude, that if $\mmG(\mr,\mr',z)$ vanishes \change{sufficiently fast} in the limit $|z|\rightarrow\infty$, then the matrix Green tensor can be expressed as
\begin{align}
\mmG(\mr,\mr',\omega) &= \frac{\text{c}^2}{2}\sum_{m}\frac{\mFt_m(\mr)[\mFb_m(\mr')]^\text{T}}{\tlo_m(\tlo_m-\omega)} + \frac{1}{\omega} \text{Res}\left\{\mmG(\mr,\mr',z),z=0\right\},
\label{Eq:mmG_with_zero_pole}
\end{align}
\change{where, comparing to Eq.~(\ref{Eq:MG_QNM_expansion}), the sum over the QNMs has a slightly different weight function.}

\subsubsection{Electric field Green function in one dimension}
In one dimension, the electric field Green function in general is the sum of the background Green tensor
\begin{align}
G_\text{B}(x,x',\omega) = \text{i}\frac{\text{c}}{2\omega}\text{e}^{i\omega|x-x'|/\text{c}}
\label{Eq:G_B_1D}
\end{align}
and a scattered Green function. In cases where the scattered Green function has no pole at $\omega=0$, %
the residue of the total Green function in this point is therefore $a_0=\text{i}\text{c}/2$. By direct application of Eq.~(\ref{Eq:mmG_with_zero_pole}), we find
\begin{align}
G(x,x',\omega) = \frac{\text{c}^2}{2\omega}\sum_m\frac{\ft_m(x)\ft_m(x')}{\tlo_m-\omega} = \frac{\text{c}^2}{2}\sum_m\frac{\ft_m(x)\ft_m(x')}{\tlo_m(\tlo_m-\omega)} + \text{i}\frac{\text{c}}{2\omega},
\label{Eq:Green_function_1_D_two_different_forms}
\end{align}
as also derived using a different argumentation in Ref.~\cite{Muljarov_EPL_92_50010_2010}; note, that there is a sign difference between the definition of this Green tensor and the one in Ref.~\cite{Muljarov_EPL_92_50010_2010}.

It is illustrative to consider what happens if one substitutes the scattered part of the Green function, $G_\text{scat}(x,x',\omega)$ for $\mmG(\mr,\mr',\omega)$ in Eq.~(\ref{Eq:G_by_contour}). In this case, one can carry through the same arguments to find, that if $G_\text{scat}(x,x',z)$ vanishes \change{sufficiently fast} in the limit $z\rightarrow\infty$, then
\begin{align}
G_\text{scat}(x,x',\omega) =  \frac{\text{c}^2}{2}\sum_m\frac{\ft_m(x)\ft_m(x')}{\tlo_m(\tlo_m-\omega)},
\end{align}
which would then, by comparing with Eq.~(\ref{Eq:Green_function_1_D_two_different_forms}), lead to the wrong conclusion that $G_\text{B}(x,x',\omega)=\text{i}\text{c}/2\omega$. %
The resolution of this puzzle is the fact that $G_\text{scat}(x,x',z)$ does not vanish in the limit $z\rightarrow\infty$. %
Instead, the vanishing of the Green function in the limit $|z|\rightarrow\infty$ is governed by an interesting interplay between the two parts of the Green function.

\subsubsection{Electric field Green tensor in three dimensions}
\label{Sec:Electric_field_Green_tensor_in_three_dimensions}
In three dimensions, the electric field Green tensor can be split in a transverse (divergence free) part and a longitudinal (curl free) part as
\begin{align}
\mG^\text{EE}(\mr,\mr',\omega) = \mG^\text{EE}_\perp(\mr,\mr',\omega) + \mG^\text{EE}_{\|}(\mr,\mr',\omega).
\end{align}
Since for a local background material the scattered part is purely transverse~\cite{Buhmann_2012}, the longitudinal part in general equals the longitudinal part of the background Green tensor,
\begin{align}
\mG^\text{EE}_{\|}(\mr,\mr',\omega) = -\frac{\text{c}^2}{3\omega^2}\delta(\mathbf{R}) - \frac{\text{c}^2}{4\pi\omega^2R^3}\left[\mathbf{I}-3\frac{\mathbf{R}\mathbf{R}}{R^2} \right].
\label{Eq:G_longitudinal}
\end{align}
\change{Considering the cases, where %
the scattered part of the Green tensor has no poles at $\omega=0$, we have then the relation} %
\begin{align}
\mG^\text{EE}(\mr,\mr',\omega) = \frac{\text{c}^2}{2\omega}\sum_{m}\frac{\mft_m(\mr)[\mft_m(\mr')]^\text{T}}{\tlo_m-\omega} + \mG^\text{EE}_{\|}(\mr,\mr',\omega).
\label{Eq:G_total_QNM_Expansion_plus_longitudinal}
\end{align}
From Eq.~(\ref{Eq:G_longitudinal}), it follows immediately, that $\mG^\text{EE}_{\|}(\mr,\mr',z)\rightarrow 0$ for $|z|\rightarrow\infty$, so a necessary condition for the equality to hold is $\mG^\text{EE}_\perp(\mr,\mr',z)\rightarrow0$ for $z\rightarrow\infty$. It is clear, then, that the QNMs lead to an expansion for the transverse part of the field, only. %
To obtain the full Green tensor, one must add the longitudinal part as in Eq.~(\ref{Eq:G_total_QNM_Expansion_plus_longitudinal}).

\change{In the cases, where} the transverse part of the Green tensor has no pole at $\omega=0$ we can find a QNM expansion for it by substituting $\mG^\text{EE}_\perp(\mr,\mr',z)$ for $\mmG(\mr,\mr',z)$ in either Eq.~(\ref{Eq:omega_time_G_by_contour}) or (\ref{Eq:G_by_contour}) to find, that if the integral along the contour $\Gamma_N$ vanishes, the transverse part of the electric field Green tensor can be written in terms of the QNMs as
\begin{align}
\mG^\text{EE}_\perp(\mr,\mr',\omega) = \frac{\text{c}^2}{2}\sum_{m}\frac{\mft_m(\mr)[\mft_m(\mr')]^\text{T}}{\tlo_m(\tlo_m-\omega)}= \frac{\text{c}^2}{2\omega}\sum_{m}\frac{\mft_m(\mr)[\mft_m(\mr')]^\text{T}}{\tlo_m-\omega}.
\label{Eq:GE_transverse_two_versions}
\end{align}
The fact that the QNM expansion will provide only the transverse part of the Green tensor was explicitly pointed out in Refs.~\cite{Lee_JOSAB_16_1409_1999, Lee_JOSAB_16_1418_1999}, as well as Ref.~\cite{Doost_PRA_90_013834_2014} which discusses the need to include also the longitudinal modes. The second equality was also discussed in Ref.~\cite{Muljarov_PRB_93_075417_2016}. %

\subsection{Convergence of the general field}
\label{Sec:Convergence_of_field}
Any solution to Eqs.~(\ref{Eq:MaxwellEquations_matrixForm}), (\ref{Eq:SM_1}) and (\ref{Eq:SM_2}) can be written in terms of the Green tensor as in Eq.~(\ref{Eq:F_from_G_and_sources}). Starting from this equation, and rewriting the function $\omega\mmG(\mr,\mr',\omega)$ as in Eq.~(\ref{Eq:omega_time_G_by_contour}), we
find
\begin{align}
\mF(\mr,\omega) = \frac{\mu_0}{2\pi}\int \oint \frac{\mmG(\mr,\mr',z)}{z-\omega}\cdot\underline{\mJ}(\mr',\omega)\ud z \ud V',
\end{align}
from which we can follow the same line of argumentation as in Section~\ref{Sec:Convergence_of_G} to rewrite the Green tensor as a sum of pole terms pertaining to the QNMs. Assuming that $z\mmG(\mr,\mr',z)$ has no pole at $z=0$ and vanishes in the limit $|z|\rightarrow\infty$, we can express the solution as
\begin{align}
\mF(\mr,\omega) = -\text{i}\mu_0 \sum_{n=1}^N \int \text{Res}\left\{z\,\frac{\mmG(\mr,\mr',z)}{z-\omega},z=\tlo_n\right\}\cdot\underline{\mJ}(\mr',\omega) \ud V',
\end{align}
and by use of Eq.~(\ref{Eq:QNMs_as_residues}), we can write this in the exact form of Eq.~(\ref{Eq:mF_pole_expansion_form}) with $a_n(\omega)$ given by Eq.~(\ref{Eq:a_m_general_field}).

In this way, the expansion of the general field inherits the convergence properties of the Green tensor, and we conclude, that the formal expansion in Eq.~(\ref{Eq:mF_pole_expansion_form}) is convergent if and only if $z\mmG(\mr,\mr',z)\rightarrow0$ for $|z|\rightarrow\infty$.

\subsection{Regions of convergence}
\label{Sec:Regions_of_completeness}
From the discussion in Sections~\ref{Sec:Convergence_of_G} and \ref{Sec:Convergence_of_field}, it is clear that a sufficient condition for convergence of the QNM expansion is for the matrix Green tensor $\mG(\mr,\mr',z)$ to vanish in the limit $|z|\rightarrow\infty$ along all directions in the complex frequency plane. It follows, that for any pair of positions $\mr$ and $\mr'$, one can in principle trace out the borders of a region of convergence by investigating this limit. In practical calculations, the behavior of the Green tensor is almost always exponential, when the frequency is varied along curves parallel to the negative imaginary axis. Indeed, the background Green function is proportional to $\exp\{\text{i}z|\mr-\mr'|/\text{c}\}$ and thus tends to zero exponentially %
as $|z|\rightarrow\infty$ in the upper half of the complex plane. The scattered part of the Green tensor represents the scattering of the electromagnetic field due to the presence of material in an otherwise homogeneous background. %
From a physical point of view, therefore, it is clear that also the scattered part of the Green tensor will vanish as $|z|\rightarrow\infty$ in the upper part of the complex plane, and we shall focus only on the frequencies in the lower half, where the behavior is non-trivial.

For problems in one dimension~\cite{Leung_PRA_49_3057_1994}, and for the case of spherically symmetric resonators~\cite{Leung_JOSAB_13_805_1996}, Leung \emph{et al.} have used the WKB approximation to assess the limiting behavior of the Green tensor. The conclusion is, that the QNMs are complete %
when both positions $\mr$ and $\mr'$ are inside a boundary set by the outermost spatial discontinuity in the function (or even the derivative of the function) that describe the resonator material. These conclusions are fully consistent with very convincing direct evaluations of the associated sum closure relations~\cite{Leung_PRA_49_3057_1994,Leung_JOSAB_13_805_1996} and higher order advanced perturbation theory~\cite{Leung_PRA_49_3068_1994, Leung_JOSAB_13_805_1996, Lee_JOSAB_16_1418_1999} and resonant state expansions~\cite{Muljarov_EPL_92_50010_2010, Doost_PRA_90_013834_2014, Muljarov_PRB_93_075417_2016}.

With certain numerical calculation methods, it is possible to investigate this limit numerically by directly calculating the Green tensor at various positions in the complex plane along the negative imaginary axis. While such an analysis cannot constitute a regular proof of \change{consistency}, it can provide a convincing picture of the region of convergence for general resonators. In particular, it can be used in cases where there is no clear definition of the resonator boundary, such as in the case of the plasmonic dimer in Section \ref{Sec:Introductory_example}. As we shall see, for certain choices of $\mr'$, the region of convergence for $\mr$ may extend beyond the volume of the resonator material. What we have found consistently, however, is that %
the QNM expansion of the Green tensor seems to be convergent only when at least one of the observation points is within the resonator material. %

\subsubsection{Formally extending the region of convergence}
\label{Sec:Formally_extending_ROC}
\begin{figure}[htb!]
\centering %
\begin{tikzpicture}
\begin{axis}[
xmin = -4, xmax = 4,
ymin = 0,  ymax = 6,
width=10.2cm,
axis on top,
height=0.3\textwidth,
xlabel={Position},
ylabel={Permittivity},%
ytick={1,4},
yticklabels={$\epsilon_\text{B}$,$\epsilon_\text{R}$},
xtick=\empty,
]
\draw[thick] (axis cs: -4,1) -- (axis cs: -.5,1) -- (axis cs: -.5,4)
-- (axis cs: .5,4) -- (axis cs: .5,1) -- (axis cs: 2.5,1);
\draw[thick] (axis cs: 3,1) -- (axis cs: 4,1);
\draw[thick,dashed] (axis cs: 2.5,1) -- (axis cs: 2.5,2)
-- (axis cs: 3,2) -- (axis cs: 3,1);
\node at (axis cs: 2.5, 1.5) [left]{$\Delta\epsilon$};
\node at (axis cs: .5, 1) [below]{$x_A$};
\node at (axis cs: 3, 1) [below]{$x_B$};
\end{axis}
\end{tikzpicture}\quad\quad\;%
\caption{\label{Fig:extending_ROC_for_dielectric_barrier} Sketch of the material distribution for a dielectric barrier with permittivity $\epsilon_\text{R}$ for $-x_A<x<x_A$ and a small additional permittivity perturbation $\Delta\epsilon$. For any $\Delta\epsilon>0$, the region of convergence extends to the outermost boundary of the perturbation at $x=x_B$. }
\end{figure}
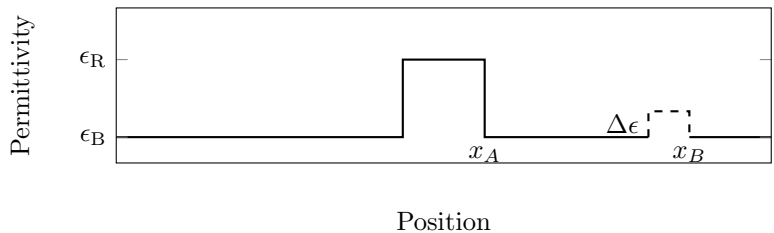

In cases where the region of convergence is bounded by the outermost discontinuity in the permittivity distribution, at a distance $x_A$ from the origin, say, one can immediately device a new geometry with an arbitrarily small, but discontinuous, perturbation $\Delta\epsilon$ of the permittivity distribution at a distance of $x_B>x_A$, as illustrated in Fig.~\ref{Fig:extending_ROC_for_dielectric_barrier}. By identical arguments as those leading to the conclusion that the QNMs are complete for $|x|<x_A$ in the original geometry, one can appreciate that the QNMs are complete for $-x_A<x<x_B$ in the new geometry, cf. Fig.~\ref{Fig:extending_ROC_for_dielectric_barrier}.

For a given QNM approximation with a fixed number of terms $N$, one can choose a $\Delta\epsilon>0$ sufficiently small so that the changes to the QNM approximation are arbitrarily small. It would seem then, that the same QNM approximation, which was convergent only for $|x|<x_A$ in the original geometry, is now convergent also for $x_A<x<x_B$. The resolution to this puzzle is the fact that even if the difference between the two QNM approximations, each with $N$ terms, can be made arbitrarily small, the difference will not remain small if more terms are added. Indeed, only one of them will %
tend to the correct value of the field as the number of terms is increased. This observation, therefore, is mostly of formal relevance. In practical calculations, for example, one is often interested in single- or few-QNM approximations at positions close to the resonator, yet possibly outside the formal region of convergence. Even if direct comparison to reference calculations show a convincing approximation, the truncation of a series which formally does not converge to the right value is arguably problematic. By extending the region of convergence as suggested above, one can immediately remove this concern by formally considering a different material system, the response of which can be chosen to be arbitrarily close to that of the original system over the bandwidth of interest. %

\subsubsection{Region of convergence for the dielectric barrier}
\label{Sec:RoC_dielectric_barrier}
\begin{figure}[htb!]
\centering %
\begin{overpic}[width=9.4cm]{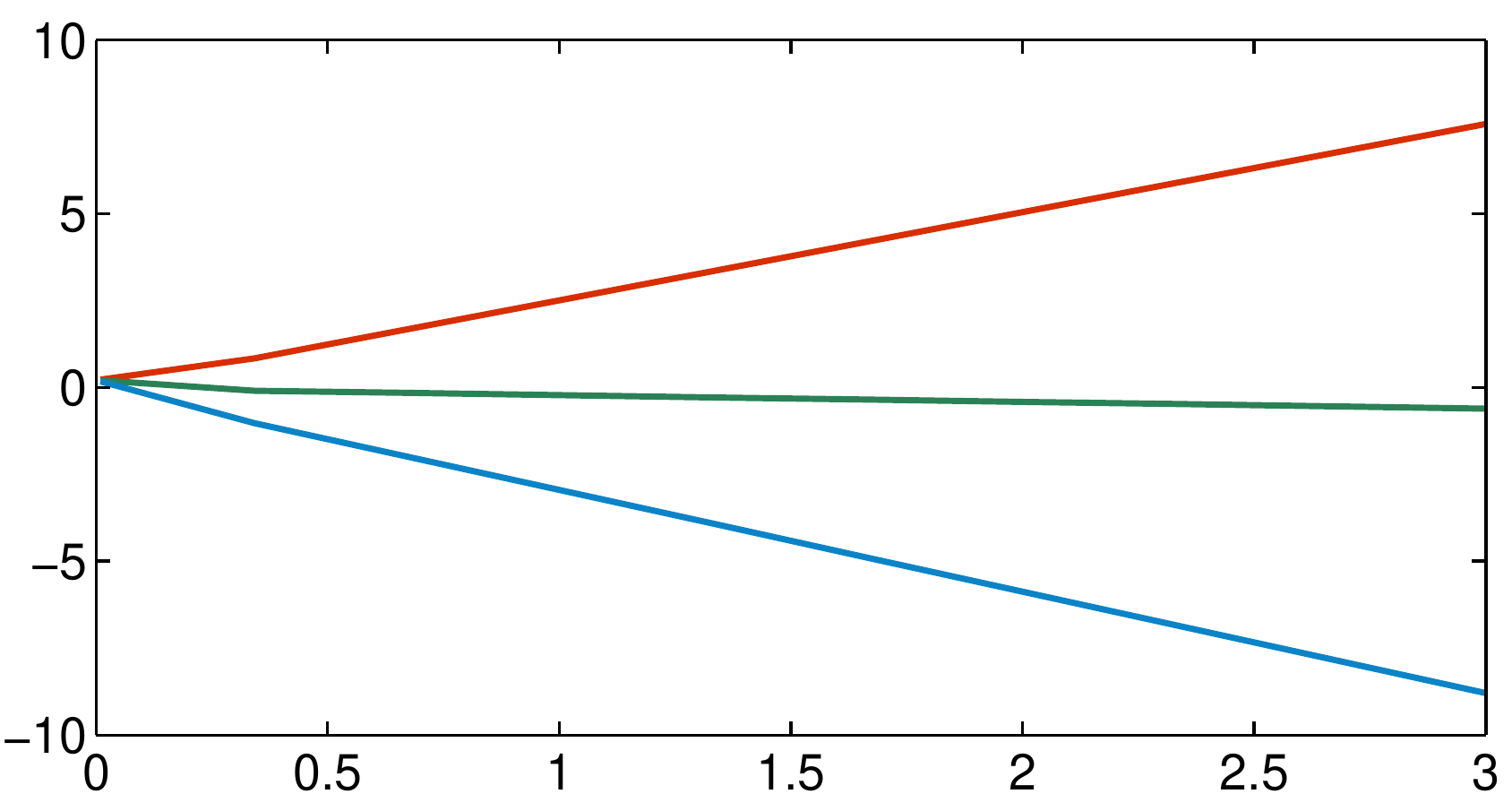}
\put(-12,5){\begin{sideways}Green function magnitude
\end{sideways}}
\put(-6,2){\begin{sideways}$\log_{10}\{|G(x,x',-\text{i}\xi)|/G_0(-\text{i}\xi)\}$
\end{sideways}}
\put(30,-5){Imaginary frequency $-\xi L/2\pi\text{c}$}
\end{overpic}
\\[5mm]
\begin{tikzpicture}
\begin{axis}[
xmin = -4, xmax = 4,
ymin = 0,  ymax = 6,
width=10.2cm,
axis on top,
height=0.3\textwidth,
xlabel={Position $x/L$},
ylabel={Permittivity $\epsilon_\text{R}$},
ytick={1,4},
xtick={-4,-3, ..., 4},
]
\fill[pattern=north east lines, pattern color=red] (axis cs: -pi/2-1/2,0) rectangle (axis cs: pi/2+1/2,6);
\fill[pattern=north west lines, pattern color=blue] (axis cs: -1/2-3*pi/4,0) rectangle (axis cs: 1/2+pi/4,6);
\draw[thick] (axis cs: -4,1) -- (axis cs: -.5,1) -- (axis cs: -.5,4)
-- (axis cs: .5,4) -- (axis cs: .5,1) -- (axis cs: 4,1);
\draw[thick, color=blue] (axis cs: .25, 0) -- (axis cs: .25,6);
\draw[thick, color=red] (axis cs: 0, 0) -- (axis cs: 0,6);
\end{axis}
\end{tikzpicture}\quad\quad\;%
\caption{\label{Fig:abs_G_1D_rel_vs_gamma_nR_pi_a_0p5_x0_0_x_m3_m2_m1}Top: Magnitude of the electric field Green function, in units of $G_0(-\text{i}\xi) = \text{c}/2\xi$, for $\omega=-\text{i}\xi$ plotted as a function of $\xi$. The position $x'=0$ is held fixed at the center of the resonator of length $L$, and $x$ is set equal to $x=-3L$ (red), $x=-2L$ (green) and $x=-L$ (blue). Bottom: Geometry of the dielectric block in one dimension showing the permittivity as a function of position (black curve). Red and blue shading indicates the region of convergence for $x'=0$ (red) and $x'=L/4$ (blue), as indicated by the vertical colored lines.} \end{figure}
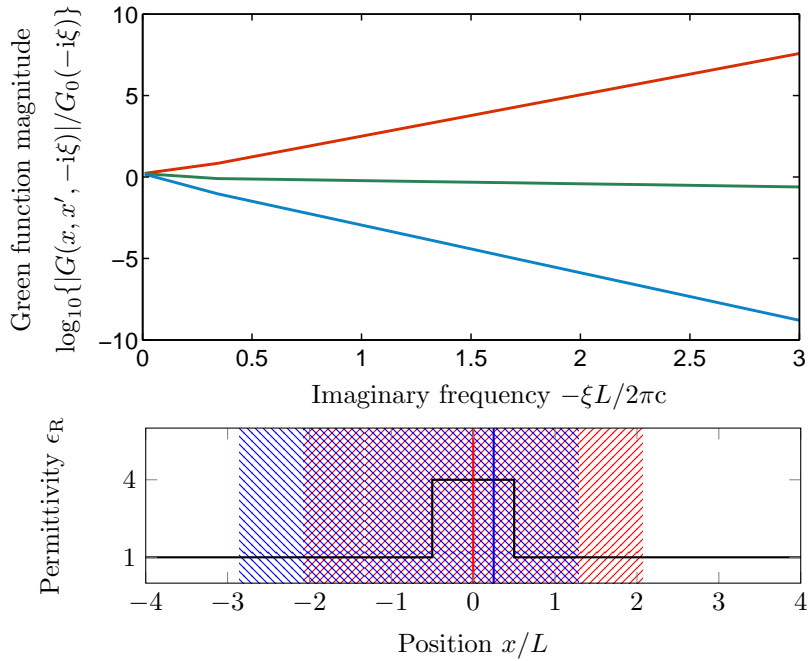

The Green function for the dielectric slab can be written in closed form, as detailed in Appendix~\ref{App:QNMs_of_dielectric_slab}, and one can therefore analytically investigate the behavior of $G(x,x',z)$ as $|z|\rightarrow\infty$. \change{In the general case, however, it is difficult to investigate the limit analytically. To illustrate the viability of a numerical approach, %
we use a one-dimensional version of the integral equation method in Refs.~\cite{deLasson_JOSAB_30_1996_2013, Kristensen_JOSAB_27_228_2010} to investigate the limiting behavior of the electric field Green tensor along the negative imaginary frequency axis, $\omega = -\text{i}\xi$}, for increasing values of $\xi$.  Figure~\ref{Fig:abs_G_1D_rel_vs_gamma_nR_pi_a_0p5_x0_0_x_m3_m2_m1} shows the magnitude of the Green function for fixed $x'=0$ and three different values of $x$ to the left of the resonator. %
From the logarithmic scale, it is evident that the magnitude of the Green tensor varies exponentially on the curve oriented downwards in the complex plane. Moreover, for $x$ sufficiently close to the resonator, there is a qualitative change in behavior so that instead of growing exponentially, the magnitude falls off exponentially. %

The change in behavior occurs at positions slightly to the left of $x=-2L$. From a detailed analysis of the Green tensor, we find that it vanishes in the limit $|z|\rightarrow\infty$ if
\begin{align}
-L/2-\pi(L/2+x')<x<L/2+\pi(L/2-x'),
\label{Eq:dielectric_barrier_region_of_convergence_nR_pi}
\end{align}
see Appendix~\ref{App:QNMs_of_dielectric_slab} for details. %
Interestingly, the region of convergence is not always symmetric around $x'$. Moving $x'$ closer to the barrier edge results in the edge of the region of convergence also moving closer to the edge (but from the opposite side), cf. Eq.~(\ref{Eq:dielectric_barrier_region_of_convergence_nR_pi}). In the general case, the Green tensor vanishes in the limit $|z|\rightarrow\infty$ only for $x$ and $x'$ both strictly within the resonator region, which is consistent with the results in Ref.~\cite{Leung_PRA_49_3057_1994}. For the case of $x'=0$, the region of convergence is symmetric and extends to $x=\pm(\pi+1)L/2\approx2.07L$, which are exactly the locations of the abrupt jumps in the QNM Green tensor approximation in Fig.~\ref{Fig:QNM_G_wR_4_dielectricBarrier_nR_pi_N_500}. 

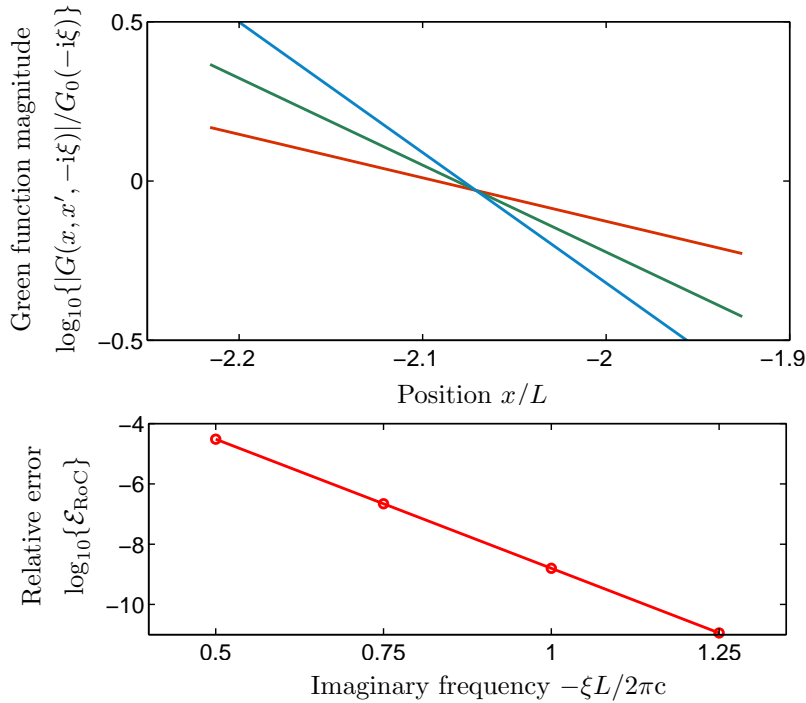
\begin{figure}[htb!]
\centering %
\begin{overpic}[width=9.4cm]{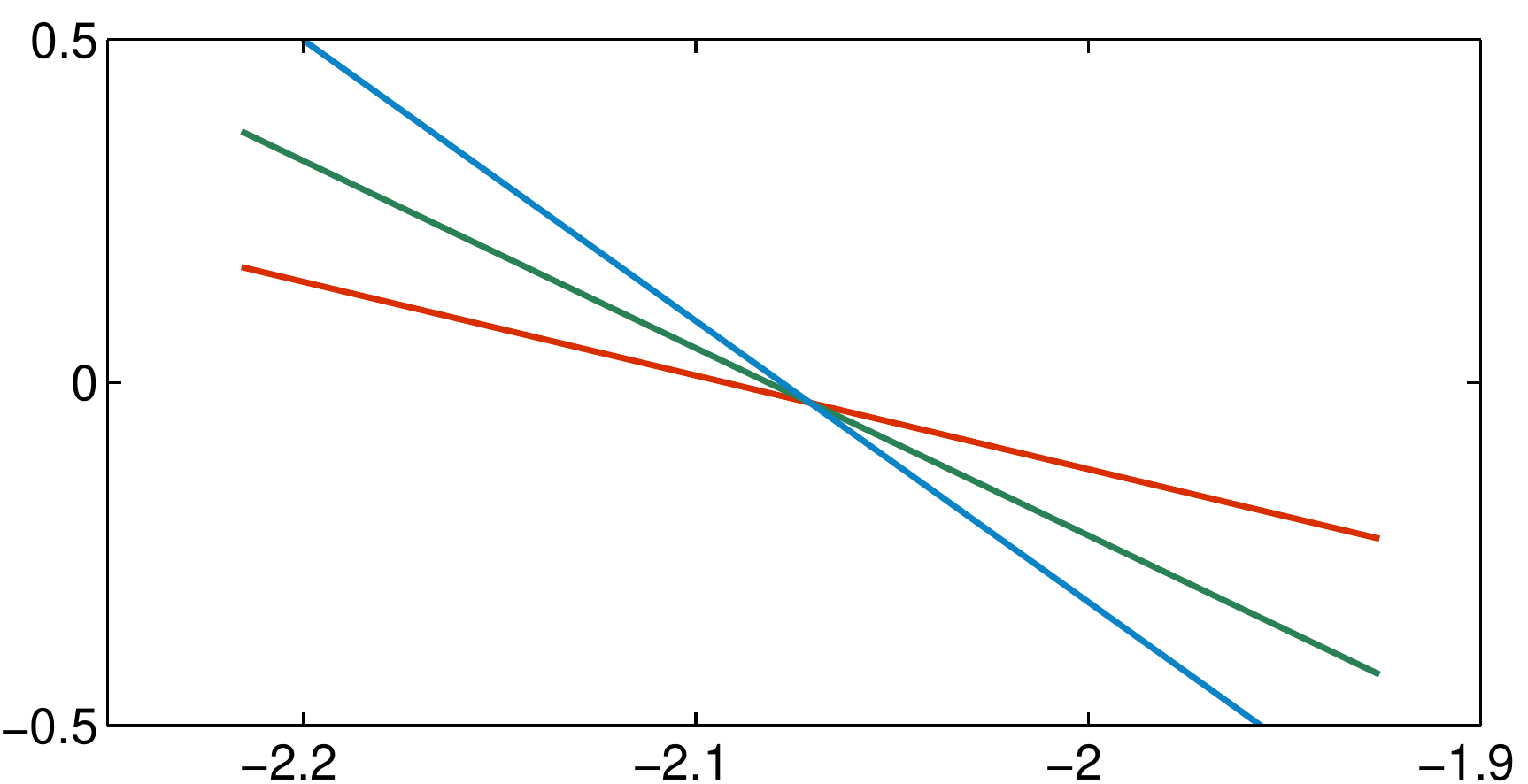}
\put(-12,5){\begin{sideways}Green function magnitude
\end{sideways}}
\put(-6,2){\begin{sideways}$\log_{10}\{|G(x,x',-\text{i}\xi)|/G_0(-\text{i}\xi)\}$
\end{sideways}}
\put(42,-5){Position $x/L$}
\end{overpic}
\\[5mm]
\begin{overpic}[width=9.05cm]{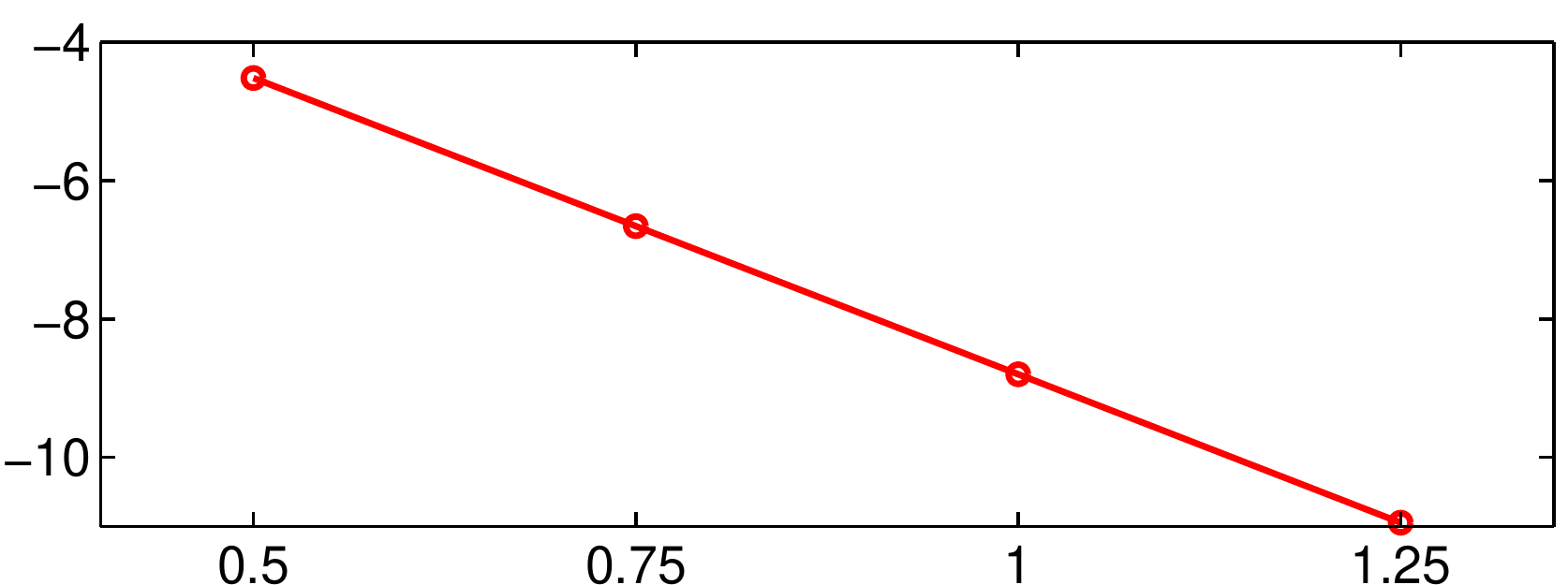}
\put(-12,8){\begin{sideways}Relative error
\end{sideways}}
\put(-6,9.5){\begin{sideways}$\log_{10}\{\mathcal{E}_\text{RoC}\}$
\end{sideways}}
\put(30,-5){Imaginary frequency $-\xi L/2\pi\text{c}$}
\end{overpic}\;\,
\\[2.5mm]
\caption{\label{Fig:barrier_1D_G_at_complex_nR_pi_var_x}Top: Magnitude of the electric field Green function as a function of position for fixed frequencies along the negative imaginary axis, $\omega=-\text{i}\xi$, with $\xi L/2\pi\text{c}=0.5$ (red), $\xi L/2\pi\text{c}=1$ (green) and $\xi L/2\pi\text{c}=1.5$ (blue). Bottom: Relative error $\mathcal{E}_\text{RoC}(\gamma)$ when estimating the boundary of the region of convergence as the intersection of lines as in the top panel calculated using $\xi L/2\pi\text{c}$ and $\xi L/2\pi\text{c}+0.25$.} \end{figure}

To set up a practical calculation method for estimating the region of convergence, we note that the slope of the curves in Fig.~\ref{Fig:abs_G_1D_rel_vs_gamma_nR_pi_a_0p5_x0_0_x_m3_m2_m1} are proportional to the distance of $x$ from the boundary. Therefore, we can estimate the region of convergence by plotting the logarithm of the magnitude of the Green tensor as a function of varying real space position and for different values of $\omega=-\text{i}\xi$. %
This results in a number of straight lines with different slopes, and the crossing points will approximate the boundary of the region of convergence, as illustrated in Fig.~\ref{Fig:barrier_1D_G_at_complex_nR_pi_var_x}. The approximation becomes better when using lines pertaining to larger \change{negative imaginary frequencies $\xi$}. \change{Indeed, c}alculating the intersection between lines calculated using $\xi L/2\pi\text{c}$ and $\xi L/2\pi\text{c}+0.25$, the relative error becomes exponentially smaller with increasing $\xi$, as seen in the bottom panel of Fig.~\ref{Fig:barrier_1D_G_at_complex_nR_pi_var_x}. %
From a practical point of view, the fact that the result of this purely numerical investigation of the Green function agrees with the analytical result gives us confidence that we can apply a similar approach for general structures where no analytical results are available. \change{In particular, we shall apply this method to estimate the region of convergence of the plasmonic dimer in Section~\ref{Sec:convergence_radius_dimer_of_gold_nano_spheres}.}%

\subsubsection*{Extending the region of convergence for the dielectric barrier}
\begin{figure}[htb!]
\centering %
\begin{overpic}[width=9.4cm]{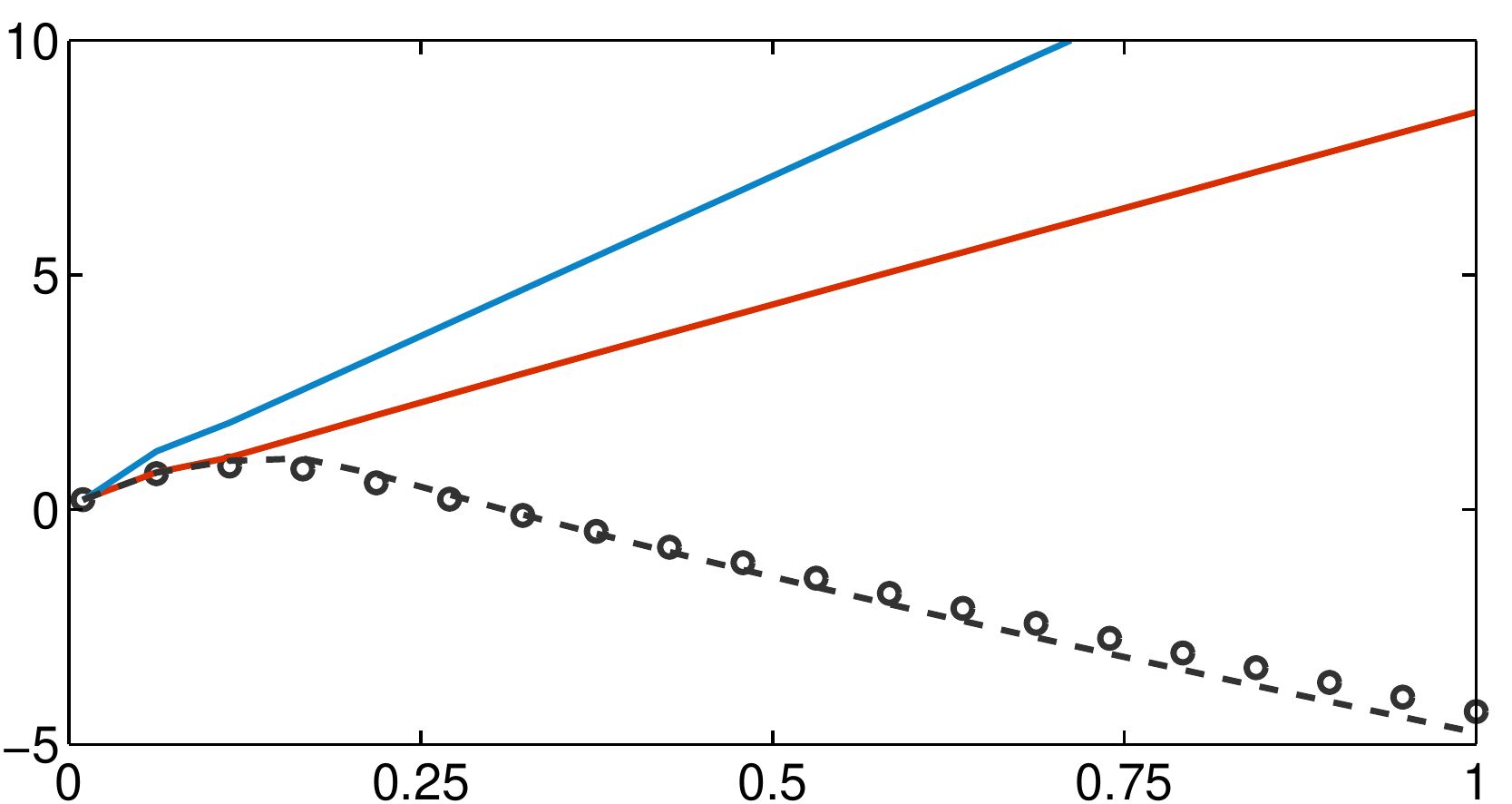}
\put(-12,5){\begin{sideways}Green function magnitude
\end{sideways}}
\put(-6,2){\begin{sideways}$\log_{10}\{|G(x,x',-\text{i}\xi)|/G_0(-\text{i}\xi)\}$
\end{sideways}}
\put(30,45){$G(x_0,x_0,-\text{i}\xi)$}
\put(70,33){$G(x_0,x',-\text{i}\xi)$}
\put(30,-5){Imaginary frequency $-\xi L/2\pi\text{c}$}
\end{overpic}
\\[5mm]
\caption{\label{Fig:abs_G_1D_rel_vs_gamma_nR_pi_w_dEps_at_3_w_approx}Magnitude of different electric field Green functions along the negative imaginary axis $\omega=-\text{i}\xi$ in units of $G_0(-\text{i}\xi) = \text{c}/2\xi$. The positions $x'=L$ and $x_0=3L$ are both outside the original resonator and outside the region of convergence of the single dielectric barrier. Red and blue solid lines show behaviors of the green functions for the single dielectric barrier as indicated, and grey dashed line shows behavior of the approximate total Green function with circles indicating the corresponding reference calculation.} \end{figure}
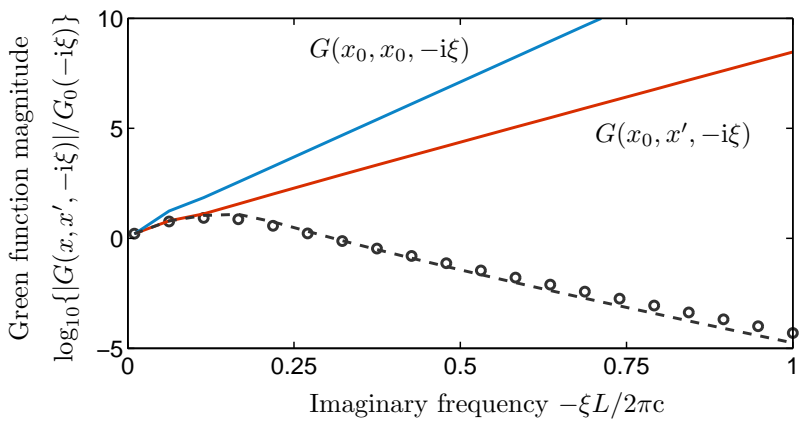

To illustrate the ideas put forward in Section~\ref{Sec:Formally_extending_ROC}, we consider now the addition of a small (constant and real valued) perturbation to the background permittivity of width $\Delta X$ and centered on $x_0=3L$. %
Using the Dyson equation, we can write the total Green function of the dielectric barrier and the additional perturbation as
\begin{align}
G_\text{tot}(x,x',\omega) = G(x,x',\omega) + k_0^2\Delta\epsilon\int_{\Delta X}  G(x,y,\omega)G_\text{tot}(y,x',\omega) \ud y,
\label{Eq:Dyson_eq_1D}
\end{align}
where $k_0 =\omega/\text{c}$. To appreciate that the total Green function may vanish for certain choices of $x$ and $x'$ and increasing values of $\xi$, we consider the case of a sufficiently narrow barrier, so that the integrand is approximately constant. Choosing $x=x_0$, we can then approximate Eq.~(\ref{Eq:Dyson_eq_1D}) as
\begin{align}
G_\text{tot}(x_0,x',\omega) \approx G(x_0,x',\omega) + k_0^2\Delta\epsilon\Delta X G(x_0,x_0,\omega)G_\text{tot}(x_0,x',\omega),
\end{align}
from which we can express the total Green function approximation explicitly as
\begin{align}
G_\text{tot}(x_0,x',\omega) \approx \frac{G(x_0,x',\omega)}{1 - k_0^2\Delta\epsilon\Delta X G(x_0,x_0,\omega)}.
\label{Eq:G_tot_1D_approx}
\end{align}
In this way, we can see how the asymptotic properties of $G_\text{tot}(x_0,x',-\text{i}\xi)$ are related to the asymptotic properties of the Green functions $G(x_0,x',-\text{i}\xi)$ and $G(x_0,x_0,-\text{i}\xi)$. The magnitude of all three Green functions are shown in Fig.~\ref{Fig:abs_G_1D_rel_vs_gamma_nR_pi_w_dEps_at_3_w_approx} along with a direct reference calculation of $G_\text{tot}(x_0,x',-\text{i}\xi)$.

From the above analysis, we conclude that the faster divergence of the denominator in Eq.~(\ref{Eq:G_tot_1D_approx}) results in an overall exponential decay in the asymptotic form of $G_\text{tot}(x_0,x',-\text{i}\xi)$. Notably, the asymptotic form is independent of the actual values chosen for $\Delta\epsilon$ and $\Delta X$. Therefore, we can choose them arbitrarily small so as to not influence a given QNM approximation with fixed number of terms, as argued in Section~\ref{Sec:Formally_extending_ROC}.

\subsubsection{Region of convergence for the plasmonic dimer}
\label{Sec:convergence_radius_dimer_of_gold_nano_spheres}
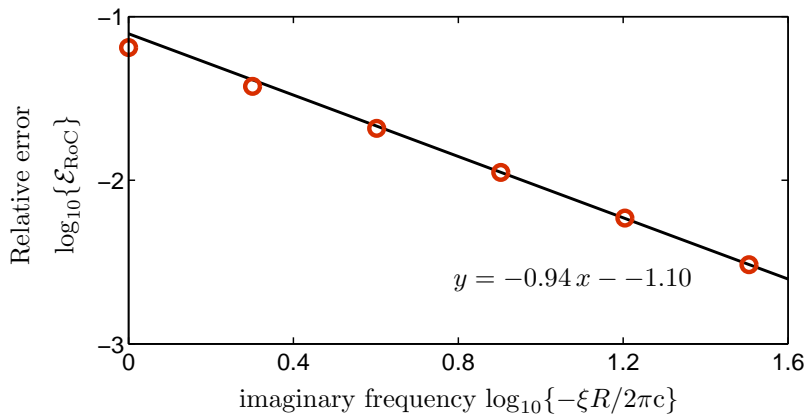
\begin{figure}[htb!]
\centering %
\begin{overpic}[width=9.4cm]{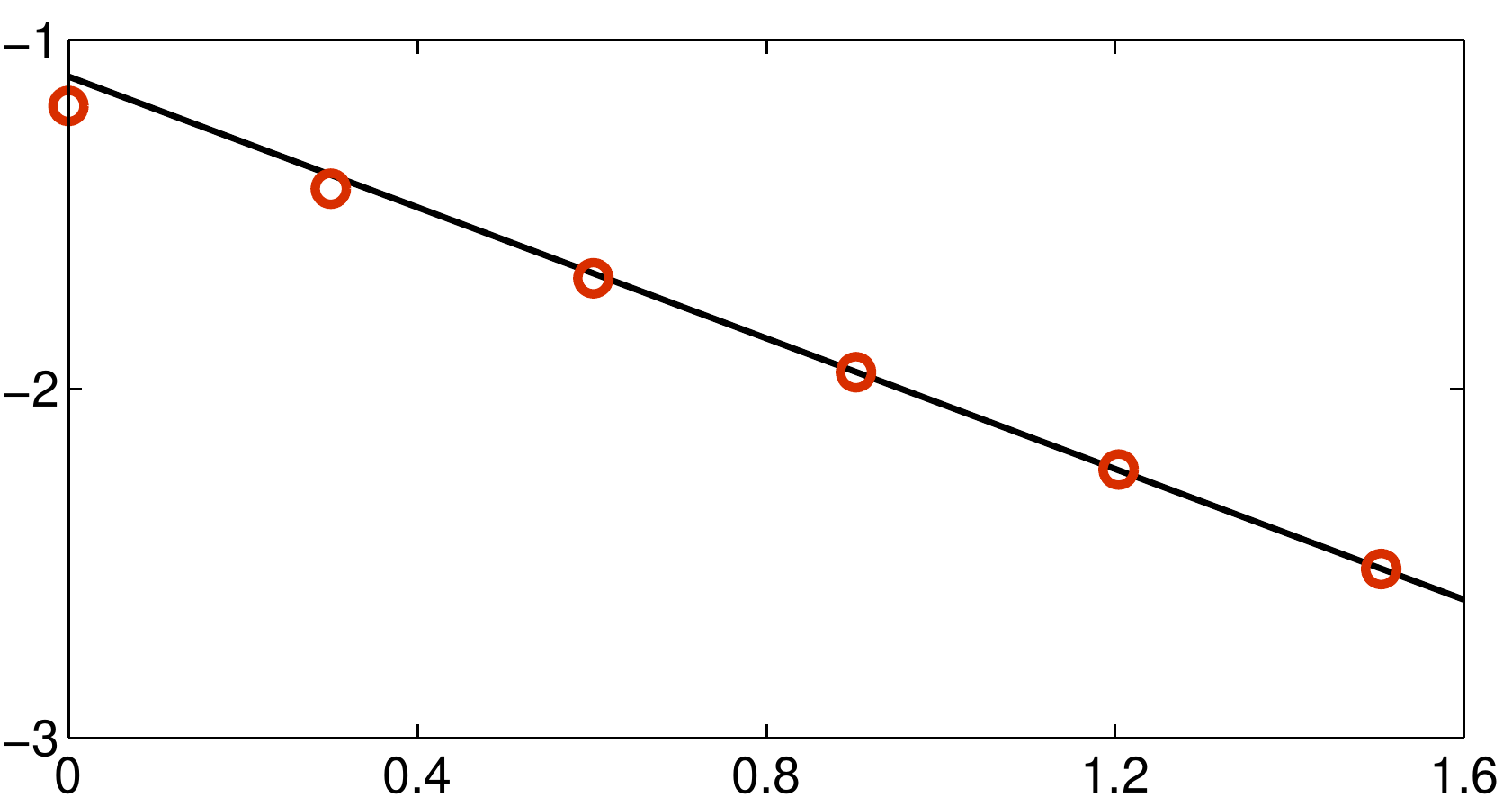}
\put(-12,15){\begin{sideways}Relative error
\end{sideways}}
\put(-6,16.5){\begin{sideways}$\log_{10}\{\mathcal{E}_\text{RoC}\}$
\end{sideways}}
\put(20,-5){imaginary frequency $\log_{10}\{-\xi R/2\pi\text{c}\}$}
\put(50,12){$y=-0.94\,x - -1.10$} %
\end{overpic}\;\,
\\[3mm]
\caption{\label{Fig:singleSphere_R_50nm_G_at_complex_relErr_vs_Gams_xp_25nm}Relative error $\mathcal{E}_\text{RoC}(\xi)$ \change{as a function of imaginary freqency $\xi$} when calculating the radius of the region of convergence for a single gold sphere based on the Green tensor at complex frequencies $\omega-\text{i}\xi$ using $\xi R/2\pi\text{c}$ and $\xi R/2\pi\text{c}+1$.} \end{figure}

Following a similar approach as for the dielectric block in one dimension, we can now investigate the region of convergence for the dimer of gold nano spheres from Fig.~\ref{Fig:Purcell_avs_Lmax_6_rdip_0p5_0p25_0_tloRange_0p05_to_0p25}. To this end, we must rely on numerical calculations of the three-dimensional Green tensor at complex frequencies, which is not as well-behaved as in the one-dimensional case. Therefore, we first verify that the method works by investigating the case of a single gold nano sphere, for which we can calculate the region of convergence analytically, as shown in Appendix~\ref{Sec:Regions_of_completeness_analytical}. From the analysis, we know that the region of convergence for the single sphere is itself a sphere with the same center. Therefore, placing the center of the single golde nono sphere at the origin, we fix the position of $\mr'$ inside the sphere by setting $x'=R/2$ and $y'=z'=0$, and  calculate the radius of the region of convergence for 100 different angles equally distributed on a circle in the plane through the center. %
Figure~\ref{Fig:singleSphere_R_50nm_G_at_complex_relErr_vs_Gams_xp_25nm} shows the average relative error, as a function of the \change{negative imaginary frequency $\xi$}, when estimating the boundary of the region of convergence as the intersection of lines corresponding to various complex frequencies $\omega-\text{i}\xi$ along a line parallel to the imaginary axis with fixed $\omega=\pi\text{c}/R$. %

At sufficiently large values of $\xi$, the matrix problem becomes poorly conditioned, which limits the attainable relative error. This problem becomes particularly severe when using large values of the cut-off parameter $l_\text{max}$ governing the expansion in terms of spherical wave functions, cf. Section~\ref{Sec:Calculations_using_VIE}. From the analysis in Appendix~\ref{Sec:Regions_of_completeness_analytical}, however, we know that the limiting behavior of the Green tensor is governed by the components with $l=0$, so we can carry out the analysis using only these wave functions, which also significantly speeds up the calculations. Despite the limited number of calculation points available, the average relative errors appears to tend to zero as a function of $\xi$ in a polynomial manner, with the lowest attainable relative error on the order of a few parts in a thousand. Although the convergence and the accuracy appears to be markedly different from that of the exponential convergence found in the one-dimensional case, we still consider this a useful approach, since in practice we are mostly interested in the overall shape of the region of convergence.

\begin{figure}[htb!]
\centering %
\begin{overpic}[width=9.4cm]{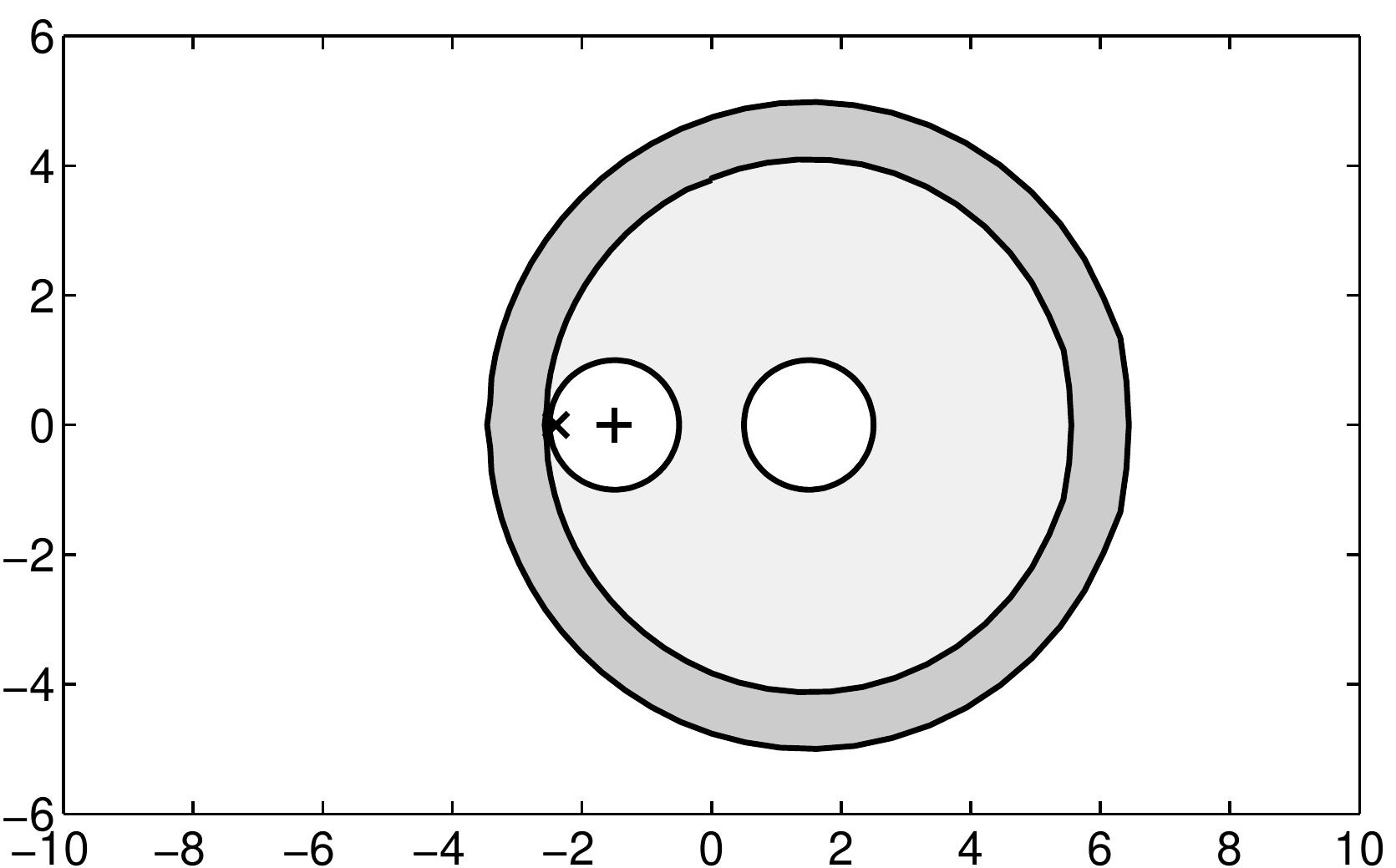}
\put(-7,23){\begin{sideways}Position $y/R$
\end{sideways}}
\put(41,-6){Position $x/R$}
\end{overpic}
\\[5mm]
\caption{\label{Fig:plasmonicDimer_R_50nm_d_50nm_G_at_complex_var_xp}Regions of convergence for the dimer of gold spheres from Fig.~\ref{Fig:Purcell_avs_Lmax_6_rdip_0p5_0p25_0_tloRange_0p05_to_0p25} showing, for different values of $\mr'$ inside one of the spheres, the positions $\mr$ for which the absolute value of the Green tensor tends to zero in the limit $|z|\rightarrow\infty$. Dark gray shading corresponds to $\mr'=\mathbf{R}_1$ in the center of the first sphere (indicated by a $+$) and light gray corresponds to $\mr'=\mathbf{R}_1 - 0.9R\mathbf{x}$ close to the boundary of the first sphere (indicated by a $\times$). Note, the dark shading extends under the light shading.
} \end{figure}

Turning to the case of the plasmonic dimer, we start by fixing $\mr'$ in the center of one sphere and vary the position $\mr$ in the plane through the centers of both spheres. As in the one-dimensional case and the case of a single sphere, we find that the Green tensor behaves in an exponential manner and diverges for positions relatively far away, whereas it tends to zero at positions within a curve in the plane defining the region of convergence, as shown in Fig.~\ref{Fig:plasmonicDimer_R_50nm_d_50nm_G_at_complex_var_xp} for two different positions of $\mr'$ inside the sphere. %
As expected, the region of convergence depends on the choice of $\mr'$, but in both cases, the region of convergence appears to be a sphere centered on the other gold nano sphere. As in the previous cases studied, the boundary of the region of convergence appears to tend to the boundary of the sphere, when $\mr'$ is moved towards the boundary from the inside. For $\mr$ and $\mr'$ both outside the spheres, we have found no numerically reliable examples where the Green tensor does \change{not appear to diverge in the limit $\xi\rightarrow\infty$}.

\section{Applications}
\label{Sec:Applications}
In this Section, we apply the QNM modeling framework to a number of problems of interest in nanophotonics. Section~\ref{Sec:Scattering_calculations} discusses the derivation of the
Coupled Mode Theory (CMT) equations using either the projection operator in Eq.~(\ref{Eq:QNM_innerprod_generalized}) or the so-called Field Equivalence Principle~\cite{Jin_2010}. As an alternative to the CMT --- which provides the QNM expansion of the total field --- we discuss also how one can calculate the scattered field by means of the QNM expansion of the Green function inside the resonator. In Section~\ref{Sec:Coupled_resonators}, we apply the Field Equivalence Principle to \change{investigate QNM hybridization by calculating} the QNMs of coupled systems based on the QNMs of the individual resonators. Section~\ref{Sec:Perturbation_theory} presents the use of QNMs for perturbation theory calculations and lastly, in Section~\ref{Sec:Purcell_factor_calculations}, we discuss the use of QNMs for Purcell factor calculations. In all cases, we show practical applications of the results using one or both of the example material systems from Section~\ref{Sec:Introductory_example}.

\subsection{Scattering calculations and CMT}
\label{Sec:Scattering_calculations}
As we have seen in Section~\ref{Sec:QNMs_as_residues}, the QNMs are intimately related to the poles of the Green tensor. Therefore, in general, it is possible to use the QNMs to calculate the scattered field resulting from a given input field, an exercise generally referred to as the construction of the scattering matrix~\cite{Alpeggiani_PRX_7_021035_2017, Lobanov_PRA_98_033820_2018, Weiss_PRB_98_085433_2018} and directly related to calculations of experimentally relevant quantities such as scattering and extinction \change{cross sections}~\cite{Yan_PRB_97_205422_2018, Unger_PRL_121_246802_2018}. Instead of treating the problem in a scattering framework, one can also take the point of view, that the electromagnetic resonator can act as a temporal energy storage when excited by an incoming pulse. In this case, the problem is conceptually identical to the so-called (temporal) CMT, which represents a physically appealing modeling tool for integrated optical circuitry based on cavities coupled through waveguides~\cite{Haus_1984, Suh_JQE_40_1511_2004, Joannopoulos2008}. Indeed, a QNM framework was used to derive the (temporal) CMT equations of coupled cavity-waveguide systems in Ref.~\cite{Kristensen_JLT_35_4247_2017}, and very similar ideas were recently applied to excitation of plasmonic resonators in Ref.~\cite{Yan_PRB_97_205422_2018}.

In classical scattering calculations, the total electromagnetic field is split in two parts corresponding to the incoming field and the scattered field as $\mF_\text{tot}(\mr,\omega)=\mF_\text{in}(\mr,\omega)+\mF_\text{scat}(\mr,\omega)$. The incoming field is taken to be a solution to Maxwell's equations in the background material without the electromagnetic resonator, and in general it does not obey the Silver-M{\"u}ller radiation condition. The scattered field represents the change in electromagnetic field profile caused by the resonator and does obey the Silver-M{\"u}ller radiation condition. Application of the projection operator in Eq.~(\ref{Eq:QNM_innerprod_generalized}) requires the field $\mF(\mr,\omega)$ to obey the Maxwell curl equations inside the volume $V$, as well as the Silver-M{\"u}ller radiation condition. For a rather general approach to scattering calculations with QNMs, we now artificially change the total field by subtracting off the incoming field at positions outside and on the border of the volume $V$. In essence, this approach is similar to the use of a total field/scattered field technique in numerical electromagnetism~\cite{Mur_IEEE_TEC_23_377_1981, Tavlove_2003, Busch_LaserPhotRev_5_773_2011}. In this way, we define the field $\mF_\text{TFSF}(\mr,\omega)$ as
\begin{align}
\mF_\text{TFSF} = \left\{\begin{array}{ll}
        \mF_\text{tot}(\mr,\omega), & \text{for } \mr\in V\setminus \partial V\\
        \mF_\text{scat}(\mr,\omega), & \text{otherwise.}
        \end{array}\right.
\label{Eq:F_TFSF_def}
\end{align}
To set up equations for the driving of the field in an electromagnetic resonator by an incoming field, we start from defining equation for the total field,
\begin{align}
\mD\,\mF_\text{tot}(\mr,\omega) = -\text{i}\omega\mF_\text{tot}(\mr,\omega),
\end{align}
multiply from the left with $\mFb_m(\mr)\mmW$ and integrate over a volume $V$ containing the electromagnetic resonator. Following the same approach as in Section~\ref{Sec:Formal_expansions}, we can rewrite this in terms of the projection operator in Eq.~(\ref{Eq:QNM_innerprod_generalized}),
\begin{align}
-(\tlo_m-\omega)\langle\langle\mFb_m(\mr)|\mF_\text{tot}(\mr,\omega)\rangle\rangle = 0,
\end{align}
but this expression is not immediately useful, since $\mF_\text{tot}(\mr,\omega)$ does not fulfill the Silver-M{\"u}ller radiation condition. To utilize the projection operator, we now split the surface integral in two terms corresponding to the incoming field and the scattered field,
\begin{align}
\langle\langle\mFb_m(\mr)|\mF_\text{TFSF}(\mr,\omega)\rangle\rangle = -\frac{\text{i}}{\tlo_m-\omega}I_{\partial_V}\big(\mFb_m(\mr),\mF_\text{in}(\mr,\omega)\big).
\end{align}
Last, assuming that $\mF_\text{TFSF}(\mr,\omega)$ can be expanded as in Eq.~(\ref{Eq:mF_pole_expansion_form}), we can use the same arguments as in Section~\ref{Sec:Formal_expansion_of_general_field} to arrive at a CMT type expression for the expansion coefficients of the form
\begin{align}
a_m(\omega) = \frac{\text{i}}{2\epsilon_0}\int_{\partial V} \Big[\mH_\text{in}(\mr,\omega)\times\mft_m(\mr) + \mE_\text{in}(\mr,\omega)\times\mgt_m(\mr)\Big]\cdot \mathbf{\hat n}\,\ud A.
\label{Eq:alpha_m_CMT}
\end{align}

To derive the temporal CMT equations, we can follow the exact same procedure as in Ref.~\cite{Kristensen_JLT_35_4247_2017} by first Fourier transforming Eq.~(\ref{Eq:mF_pole_expansion_form}) and defining
\begin{align}
\mF(\mr,t) = \sum_n F_n(t)\mFt_n(\mr),
\label{Eq:mF_pole_expansion_form_time_dependent}
\end{align}
where
\begin{align}
F_n(t) = \frac{1}{2\pi}\int_{-\infty}^\infty\frac{a_n(\omega)}{\omega-\tlo_n}\text{e}^{-\text{i}\omega t}\ud\omega.
\end{align}
\change{Differentiating} %
with respect to $t$ \change{and rearranging the resulting integrand}, we find the temporal CMT equation %
\begin{align}
\partial_tF_n(t) = -\text{i}\tlo_nF_n(t) - \text{i}a_n(t),
\label{Eq:temporal_CMT_total_field}
\end{align}
where $a_{n}(t)$ is the inverse Fourier transform of $a_{n}(\omega)$.

\subsubsection{Derivation using the Field Equivalence Principle}
The same expression can be derived using the Field Equivalence Principle and the QNM expansion of the Green tensor in Eq.~(\ref{Eq:GE_QNM_expansion}). This approach has previously been applied to derive the CMT equations for coupled cavity-waveguide structures in Ref.~\cite{Kristensen_JLT_35_4247_2017}.  %
The starting point is the equivalent surface currents
\begin{align}
\mathbf{J}_\text{in}(\mr,\omega) &= \mathbf{\hat n}\times\mH_\text{in}(\mr,\omega)\\
\mathbf{M}_\text{in}(\mr,\omega) &= \mE_\text{in}(\mr,\omega)\times\mathbf{\hat n},
\label{Eq:sourceCurrent_app}
\end{align}
where $\mathbf{\hat n}$ is an outward oriented unit vector perpendicular to the surface enclosing the sources of $\mE_\text{in}(\mr,\omega)$ and $\mH_\text{in}(\mr,\omega)$. Using Eq.~(\ref{Eq:F_from_G_and_sources}), we focus on the top equation for the electric field inside the resonator,
\begin{align}
\mE(\mr,\omega) = \text{i}\omega\mu_0\int_{\partial V}\mG_\text{EE}(\mr,\mr',\omega)\cdot\mathbf{J}_\text{in}(\mr',\omega) + \mG_\text{EM}(\mr,\mr',\omega)\cdot\mathbf{M}_\text{in}(\mr',\omega)\,\ud V.
\label{Eq:Field_Equivalence_Principle}
\end{align}
By inserting the relevant components of the QNM expansion for the matrix Green tensor in Eq.~(\ref{Eq:MG_QNM_expansion}) and rearranging the terms in the form of Eq. (\ref{Eq:mF_pole_expansion_form}), we find
\begin{align}
a_m(\omega) &= \frac{\text{i}}{2\epsilon_0}\int_{\partial V} \mft_m(\mr')\cdot\left[\mathbf{\hat n}\times\mH_\text{in}(\mr',\omega)\right] %
- \mgt_m(\mr')\cdot\left[\mE_\text{in}(\mr',\omega)\times\mathbf{\hat n} \right]\ud A',%
\label{Eq:alpha_m_CMT_FEP}
\end{align}
which can be rewritten in the exact form of %
Eq.~(\ref{Eq:alpha_m_CMT}) by use of the vector relation $\mathbf{A}\cdot\left[\mathbf{B}\times\mathbf{C}\right] = \mathbf{B}\cdot\left[\mathbf{C}\times\mathbf{A}\right] = \mathbf{C}\cdot\left[\mathbf{A}\times\mathbf{B}\right]$.

From the derivation, it appears that the integration surface in Eqs.~(\ref{Eq:alpha_m_CMT}) or (\ref{Eq:alpha_m_CMT_FEP}) must be strictly outside the material defining the resonator in order to ensure that the scattered field %
can be written in terms of purely outwards propagating fields, or to ensure that the equivalent surface currents can be written in terms of the incoming field only. In practice, however, the integration can be performed on the inside of the boundary, since changes in the normal components of $\mft_m(\mr')$ and $\mgt_m(\mr')$ across the boundary do not affect the value of the integral. %

\subsubsection{Scattered field calculations}
While Eq.~(\ref{Eq:alpha_m_CMT}) provides the QNM expansion of the total field inside the volume $V$, it may also be interesting to calculate the expansion coefficients of the scattered field explicitly. %
Assuming for simplicity $\mu_\text{r}=1$, we consider the scattered part of the electric field, which \change{can be calculated based on the electric field Green function and the incoming field $\mE_\text{in}(\mr,\omega)$ as}
\begin{align}
\mE_\text{scat}(\mr,\omega) = k_0^2\int_V \mG^\text{EE}(\mr,\mr',\omega)\Delta\epsilon(\mr')\mE_\text{in}(\mr',\omega)\ud V,
\label{Eq:LS_second_form}
\end{align}
where $k_0=\omega/\text{c}$, and $\Delta\epsilon(\mr) = \epsilon_\text{r}(\mr)-\epsilon_\text{B}$ is the local change in permittivity defining the electromagnetic resonator in a homogeneous background with permittivity $\epsilon_\text{B}$. %
Defining
\begin{align}
\mF_\text{scat}(\mr,\omega) = \sum_m\frac{b_m(\omega)}{\omega-\tlo_m}\mFt_m(\mr),
\label{Eq:mF_scat_pole_expansion_form}
\end{align}
we can use Eq.~(\ref{Eq:LS_second_form}) and the QNM expansion of the Green tensor in Eq.~(\ref{Eq:GE_QNM_expansion}) to find that
\begin{align}
b_m(\omega) = -\frac{\omega}{2}\int_V\mft_m(\mr')\fran{\cdot}\Delta\epsilon_\text{r}(\mr')\mE_\text{in}(\mr',\omega)\ud V,
\label{Eq:b_n_def_LS}
\end{align}
which is similar to the expression derived by Yan \emph{et al.} using a slightly different approach~\cite{Yan_PRB_97_205422_2018}. We note, that Eq.~(\ref{Eq:mF_scat_pole_expansion_form}) is valid also in cases where the QNM expansion of the Green tensor does not converge to the proper value at the boundary, since the derivation is fundamentally based on the volume integral in Eq.~(\ref{Eq:b_n_def_LS}).

In cases where the resonator is made from a homogeneous material, we can use the vector Green theorem of the second kind and the wave equation to rewrite the expression as a surface integral,
\begin{align}
b_m(\omega) = \frac{\text{i}}{2\epsilon_0}\frac{\omega\Delta\epsilon}{\tlo_m^2\epsilon_\text{r}-\omega^2\epsilon_\text{B}}\int_{\partial V} \Big[\omega\mH_\text{in}(\mr,\omega)\times\mft_m(\mr) + \tlo_m\mE_\text{in}(\mr,\omega)\times\mgt_m(\mr)\Big]\cdot\mathbf{\hat n}\,\ud A,
\label{Eq:b_n_CMT_LSE}
\end{align}
where the integral is over the resonator surface. Clearly, following %
analogous argumentations as for the QNM expansion of the total field in Eqs.~(\ref{Eq:mF_pole_expansion_form_time_dependent}) to (\ref{Eq:temporal_CMT_total_field}), one can now calculate the temporal CMT equations for the scattered field, as found also in Ref.~\cite{Yan_PRB_97_205422_2018}.

Even though Eqs.~(\ref{Eq:alpha_m_CMT}) and (\ref{Eq:b_n_CMT_LSE}) are very similar in shape, they are not identical. Whereas the former provides the QNM expansion of the total field inside the volume $V$, the latter provides only the scattered field. Consequently, the difference $c_n(\omega) = a_n(\omega) - b_n(\omega)$ gives the QNM expansion of the incoming field inside the volume $V$. %

\subsubsection{Scattering calculations and CMT for the dielectric barrier}
\label{Sec:Scattering_calculations_for_the_dielectric_barrier}
The calculations in Section~\ref{Sec:Scattering_calculations} include a surface integral coupling the incoming field to each of the QNMs at the boundary between the total field and the scattered field regions. For the scheme to work at all positions inside the dielectric barrier, we must choose the boundary of $V$ to coincide with the physical boundaries at $x=\pm L/2$. %
To calculate the transmitted and reflected field at positions outside the resonator, we apply the Field Equivalence Principle in Eq.~(\ref{Eq:Field_Equivalence_Principle}) with equivalent surface currents at $x=L/2$ and $x=-L/2$ given in terms of the QNM expansion of the scattered field in Eq.~(\ref{Eq:mF_scat_pole_expansion_form}). %
\begin{figure}[htb!]
\centering %
\begin{overpic}[width=9.4cm]{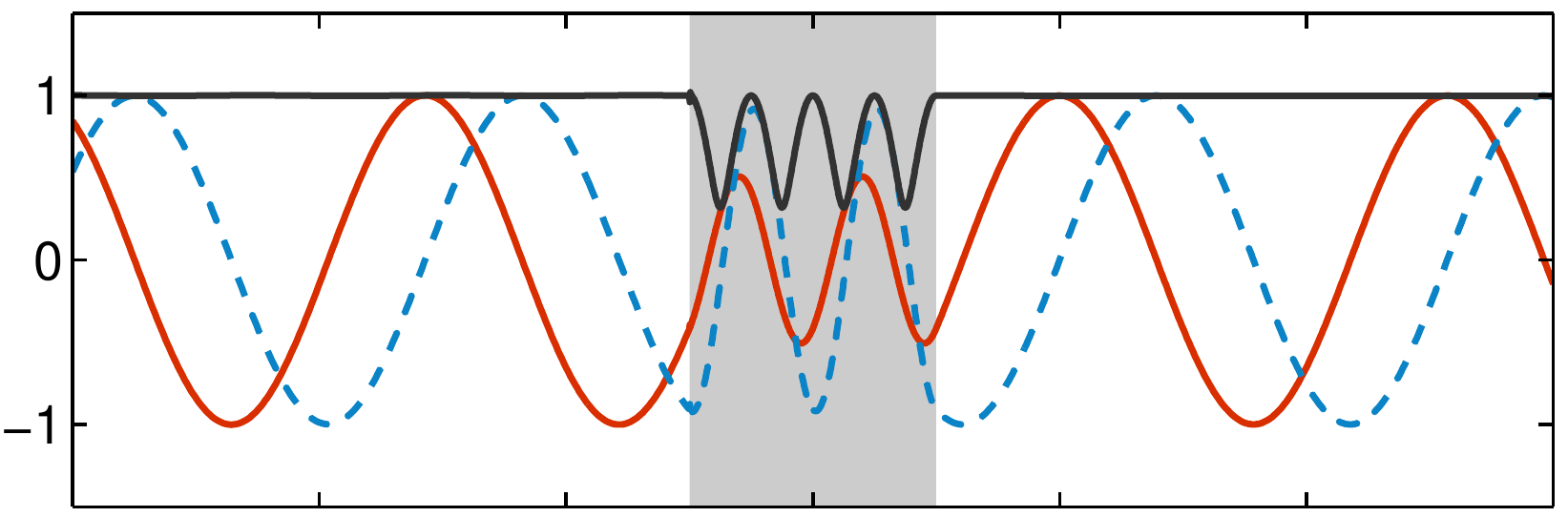}
\put(-12,6){\begin{sideways}Electric field\end{sideways}}
\put(-6,6){\begin{sideways}$E_\text{500}(\omega_0)/E_0$\end{sideways}} %
\put(42,-6){Position, $x/L$}
\end{overpic}\;
\\[2mm]
\begin{overpic}[width=9.45cm]{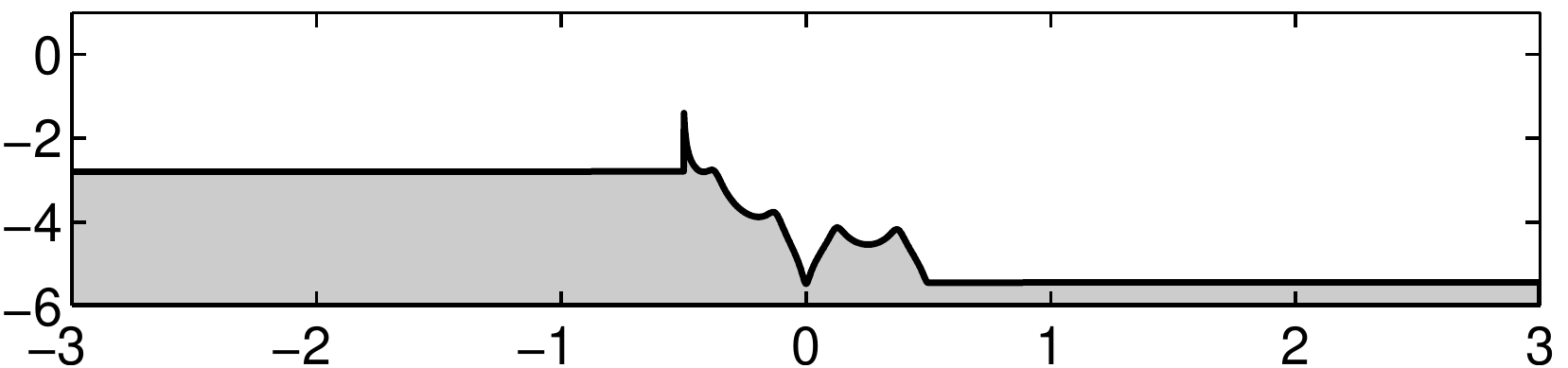}
\put(-11,4.5){\begin{sideways}Rel. Error\end{sideways}}
\put(-6,0){\begin{sideways}$\log_{10}\{\mathcal{E}_{500}(\omega_0)\}$\end{sideways}} %
\put(42,-6){Position, $x/L$}
\end{overpic}
\\[5mm]
\caption{\label{Fig:CMT_k0_4_dielectricBarrier_nR_pi_N_500}\change{Transmission through the dielectric barrier in one dimension from Section~\ref{Sec:1Dresonator}.} Top: QNM approximation to the total electric field in the case of a scatting type calculation with an incoming field from the left of the form $E_\text{in}(x,\omega_0)=E_0\exp\{i\omega_0 x/\text{c}\}$ with $\omega_0=4\text{c}/L$. Red solid and blue dashed curves show the real and imaginary parts, respectively, and the black curve shows the absolute value. Gray shading indicates the extent of the dielectric barrier. Bottom: Relative error as a function of position.}
\end{figure}
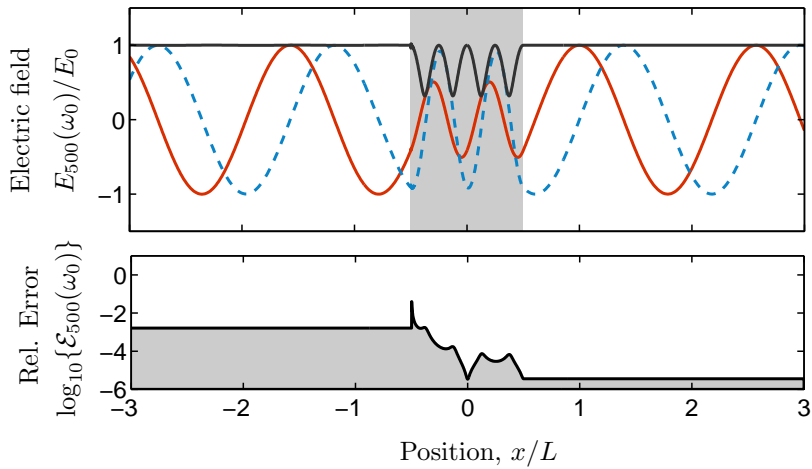
The QNM approximation to the total field inside the volume $V$ can be written as
\begin{align}
E_{\text{tot},N}(x,\omega) = \sum_{n=-N}^N \frac{a_n(\omega)}{\omega-\tlo_n}\mft_n(x),
\label{Eq:E_tot_N}
\end{align}
with $a_n$ given explicitly in Eq.~(\ref{Eq:alpha_m_CMT_FEP}). For any point $-L/2<x\leq L/2$ inside the resonator, $E_N(x,\omega)$ converges to the correct value in the sense that the relative error
\begin{align}
\mathcal{E}_N(x,\omega)=\frac{|E_{\text{tot},N}(x,\omega)-E_\text{tot}(x,\omega)|}{|E_\text{tot}(x,\omega)|}
\label{Eq:E_tot_rel_error}
\end{align}%
can be made arbitrarily small by increasing $N$.

Instead of approximating the total field directly using Eq.~(\ref{Eq:E_tot_N}), we can calculate the QNM approximation to the scattered field as
\begin{align}
E_{\text{scat},N}(x,\omega) = \sum_{n=-N}^N \frac{b_n(\omega)}{\omega-\tlo_n}\mft_n(x),
\label{Eq:E_scat_N}
\end{align}
with $b_n$ given explicitly in Eq.~(\ref{Eq:b_n_def_LS}) or (\ref{Eq:b_n_CMT_LSE}). By adding the incoming field, we find the alternative expression
\begin{align}
E_{\text{tot},N}^\text{s}(x,\omega = E_\text{in}(x,\omega)+E_{\text{scat},N}(x,\omega),
\label{Eq:E_tot_N_from_scat}
\end{align}
and we define the associated relative error as
\begin{align}
\mathcal{E}_N^\text{s}(x,\omega)=\frac{|E_\text{in}(x,\omega)+E_{\text{scat},N}(x,\omega)-E_\text{tot}(x,\omega)|}{|E_\text{tot}(x,\omega)|}.
\label{Eq:E_tot_rel_error_from_scat}
\end{align}%

To calculate the transmitted electric field beyond the barrier at $x>L/2$, one can treat the QNM approximation to the total field at the boundary of the total field region %
as the input field in the Field Equivalence Principle.  %
The corresponding calculation of the total field at positions $x<-L/2$ is much more delicate. In principle, one can calculate the reflected field at the position $x=-L/2$ by subtracting off the incoming field from the total field, but the expansion in Eq.~(\ref{Eq:E_tot_N}) fails at the position $x=-L/2$ because of the manifest outwards propagating nature of the QNMs. One option, then, is to use the approximate field value at a position just inside the resonator where the QNM approximation converges to the correct total field. Another, much more efficient method is to calculate the scattered field at the left boundary directly from the QNM approximation to the scattered field in Eq.~(\ref{Eq:E_scat_N}) with $x=-L/2$.

Considering now the explicit case of an \change{incoming plane wave from the left} of the form $E_\text{in}(x,\omega_0)=E_0\exp\{i\omega_0 x/\text{c}\}$ with $\omega_0=4\text{c}/L$, Fig.~\ref{Fig:CMT_k0_4_dielectricBarrier_nR_pi_N_500} shows the calculated approximation to the total electric field using $N=500$ along with the relative error. %
At positions outside the resonator, the relative error is constant, because the approximate reflected and transmitted fields are calculated using the Field Equivalence Principle. The failure of the QNM approximation to the total field at $x=-L/2$ is evident as a peak in the relative error.

\begin{figure}[htb!]
\centering %
\begin{overpic}[width=9.4cm]{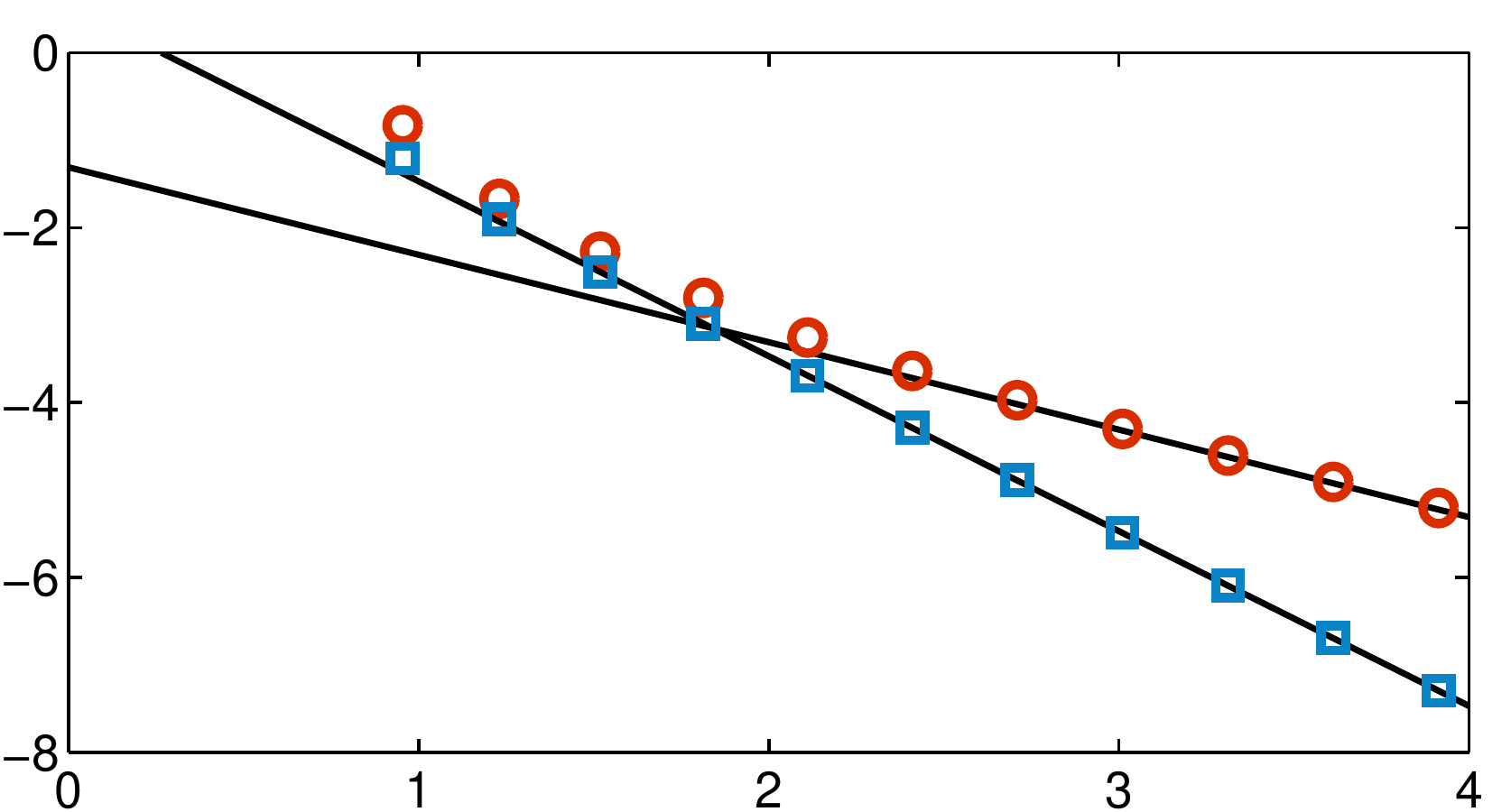}
\put(-13,16){\begin{sideways}Relative error\end{sideways}}
\put(-6,8){\begin{sideways}$\log_{10}\{\mathcal{I}_N\}$ or $\log_{10}\{\mathcal{E}_N\}$\end{sideways}}
\put(24,-6){Number of QNMs, $\log_{10}\{2N+1\}$} %
\put(64,31){$y=-1.00\,x - 1.31$} %
\put(50,8){$y=-2.00\,x +0.53$} %
\end{overpic}
\\[5mm]
\caption{\label{Fig:CMT_k0_4_dielectricBarrier_nR_pi_log10_relErr_vs_log10_N_x_m0p25_0p25}Convergence analysis for the CMT approximation to the total field inside the dielectric barrier. Red circles show the relative integrated error $\mathcal{I}_N(\omega_0)$ in the region $-L/4<x<L/4$ around the center of the barrier, and blue squares show the relative error $\mathcal{E}_N(x,\omega_0)$ at $x=0$ in the center of the barrier. Black lines show linear fits to the last three data points.}
\end{figure}
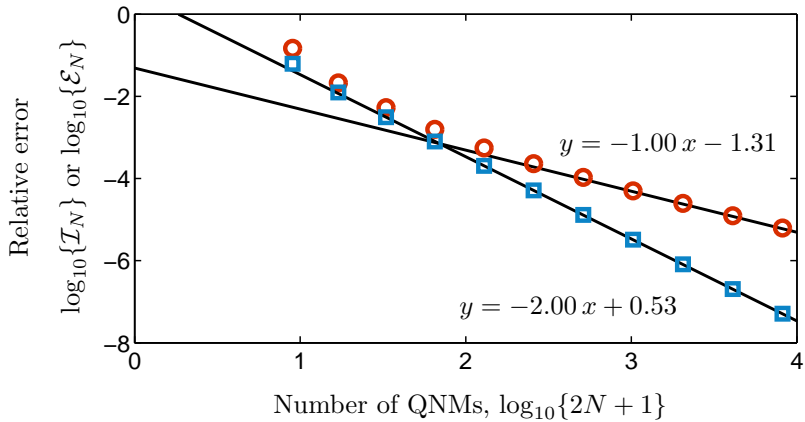

To assess the convergence properties of the QNM approximation to the total field in Eq.~(\ref{Eq:E_tot_N}), Fig.~\ref{Fig:CMT_k0_4_dielectricBarrier_nR_pi_log10_relErr_vs_log10_N_x_m0p25_0p25} shows a double logarithmic plot of the relative integrated error
\begin{align}
\mathcal{I}_N(\omega) = \frac{\int|E_\text{tot}(x,\omega)-E_{\text{tot},N}(x,\omega)| \ud x }{\int|E_\text{tot}(x,\omega)|\ud x},
\end{align}
where the integrals are taken in the region $-L/4<x<L/4$ around the center of the barrier \change{in order to stay clear of the problematic point at the left boundary}. A fit to the last data points indicate a first order polynomial convergence with the number of QNMs. The figure also shows the relative error $\mathcal{E}_N(x,\omega_0)$ at $x=0$ in the center of the barrier, which tends to zero in a second order polynomial fashion. We attribute this increased rate of convergence to the special symmetry of the point in the center. Similar analysis for general off-center points (not shown) show first order polynomial convergence, which evidently limits the rate of convergence of the relative integrated error.

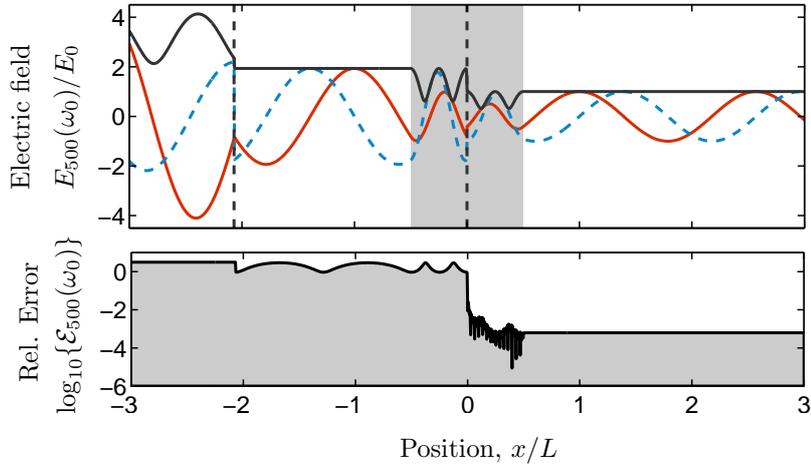
\begin{figure}[htb!]
\centering %
\begin{overpic}[width=9.4cm]{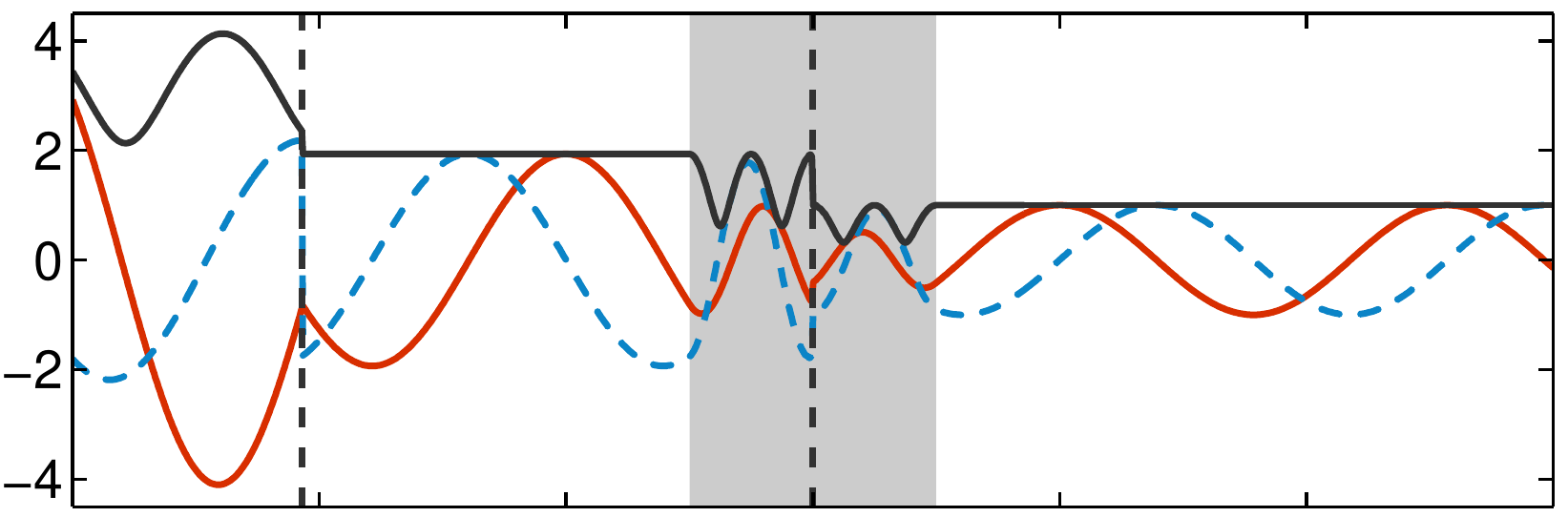}%
\put(-12,6){\begin{sideways}Electric field\end{sideways}}
\put(-6,6){\begin{sideways}$E_\text{500}(\omega_0)/E_0$\end{sideways}} %
\put(42,-6){Position, $x/L$}
\end{overpic}\;
\\[2mm]
\begin{overpic}[width=9.45cm]{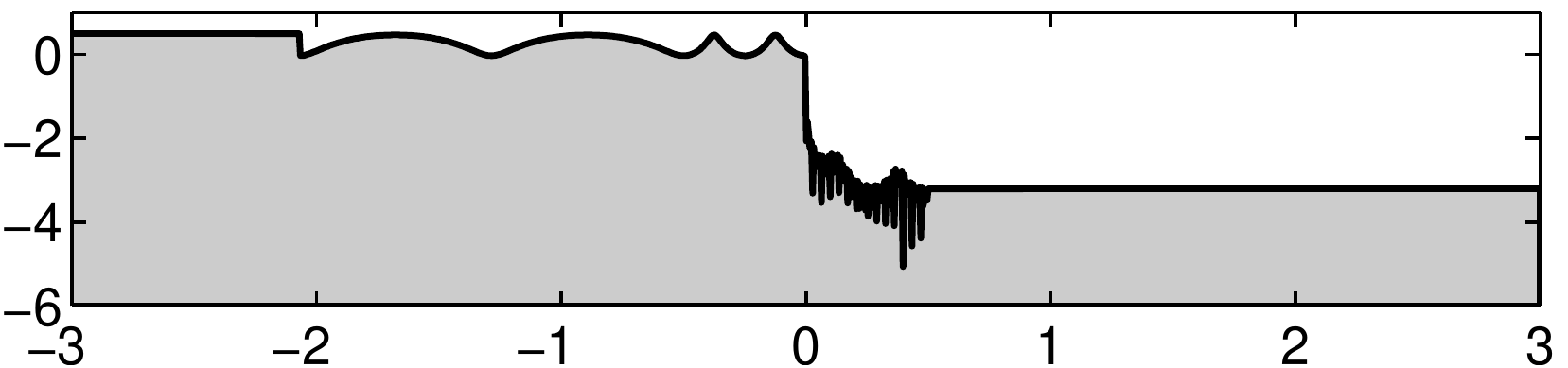}%
\put(-11,4.5){\begin{sideways}Rel. Error\end{sideways}}
\put(-6,0){\begin{sideways}$\log_{10}\{\mathcal{E}_{500}(\omega_0)\}$\end{sideways}} %
\put(42,-6){Position, $x/L$}
\end{overpic}
\\[5mm]
\caption{\label{Fig:CMT_k0_4_dielectricBarrier_nR_pi_N_500_xp_m_1_p_0p5_pi}Top: QNM approximation to the total electric field similar to Fig.~\ref{Fig:CMT_k0_4_dielectricBarrier_nR_pi_N_500}, but calculated for a total field region defined by $-(\pi+1)L/2<x<L/2$. Bottom: Relative error as a function of position.}
\end{figure}

As a curious consequence of the region of convergence, we show in Fig.~\ref{Fig:CMT_k0_4_dielectricBarrier_nR_pi_N_500_xp_m_1_p_0p5_pi} an example of a CMT calculation in which the total field region is defined by $-(\pi+1)L/2<x<L/2$. In this case, it follows from the expression for the region of convergence in Eq.~(\ref{Eq:dielectric_barrier_region_of_convergence_nR_pi}) that the QNM expansion will be convergent only for $x'>0$. This is exactly what we observe as an abrupt jump in the field and in the associated relative errors at $x=0$. At positions left of the total field region, the electric field is the sum of the incident field and the reflected field as calculated at the boundary $x=-(\pi+1)L/2$ where the QNM expansion of the scattered field is not convergent. %
The transmitted field is calculated by use of the Field Equivalence Principle at the right boundary $x=L/2$, where the QNM expansion is convergent. Comparing to Fig.~\ref{Fig:CMT_k0_4_dielectricBarrier_nR_pi_N_500}, the relative error is larger, because the rate of convergence of the total field is not as good as when coupling the field at $x=-L/2$.

\subsubsection*{Transmission spectrum calculations for the dielectric barrier}
\begin{figure}[htb!]
\centering %
\begin{overpic}[width=9.75cm]{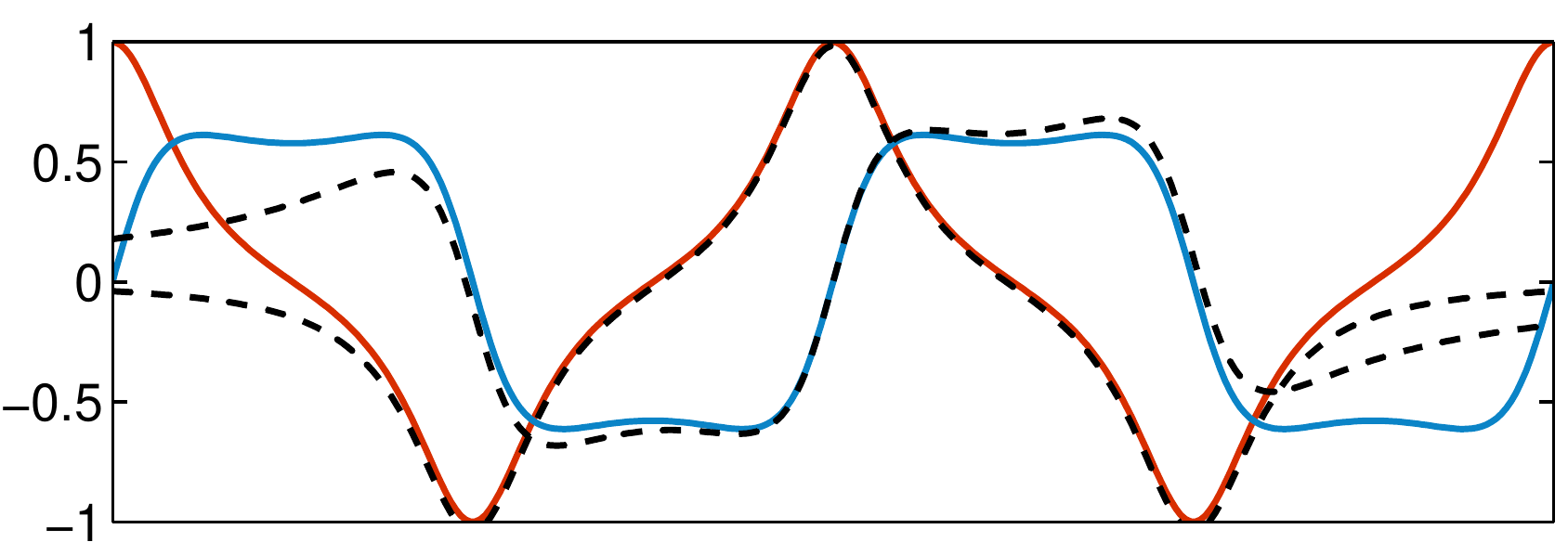}
\put(-5,5){\begin{sideways}Transmission, $T$\end{sideways}} %
\put(22,28.5){Imag}
\put(18,4){Real}
\put(42,-6){Frequency, $\omega L/\text{c}$}
\end{overpic}\;\;\;\;
\\
\begin{overpic}[width=9.55cm]{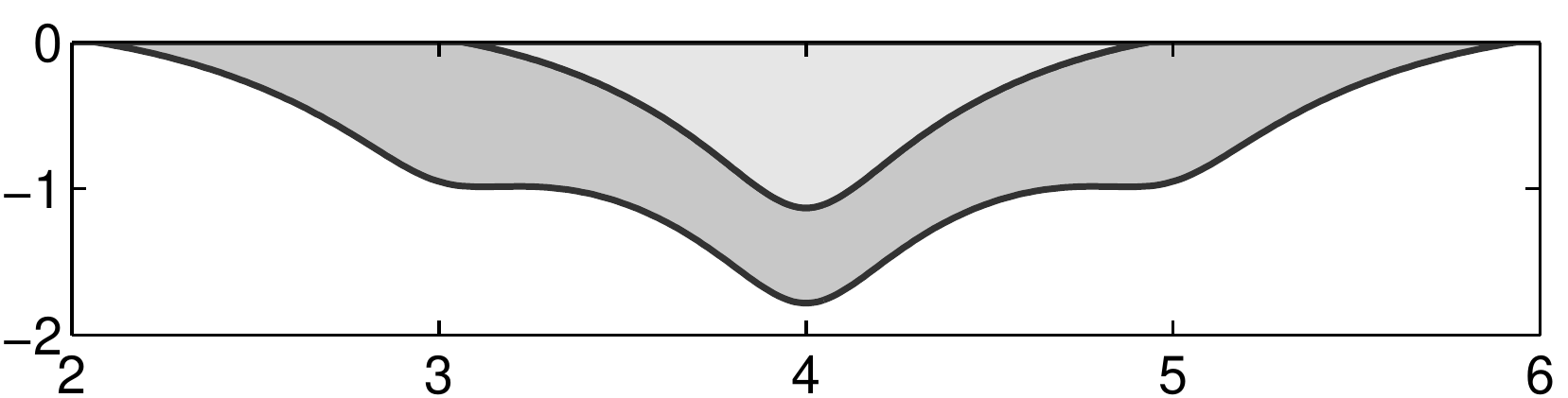}
\put(-11,4.5){\begin{sideways}Rel. Error\end{sideways}}
\put(-5,6){\begin{sideways}$\log_{10}\{\mathcal{E}\}$\end{sideways}} %
\put(36,-6){Frequency, $\omega L/\text{c}$}
\end{overpic}
\\[5mm]
\caption{\label{Fig:transmission_spectrum_dielectricBarrier_nR_pi_center_on_4_real_plus_imag_incl_CMT}Top: Real (red) and imaginary (blue) parts of the transmission through the dielectric barrier as a function of frequency. Dashed curves indicate the CMT approximation using the three QNMs with $n\in\{3,4,5\}$. Bottom: Relative error $\mathcal{E}(\omega)=|T_\text{CMT}(\omega)-T_\text{ref}(\omega)|/T_\text{ref}(\omega)$. Dark shading shows the error in the QNM approximation in the top panel, and light shading shows the error in the single QNM approximation using only $n=4$ (not shown in the top panel).}
\end{figure}
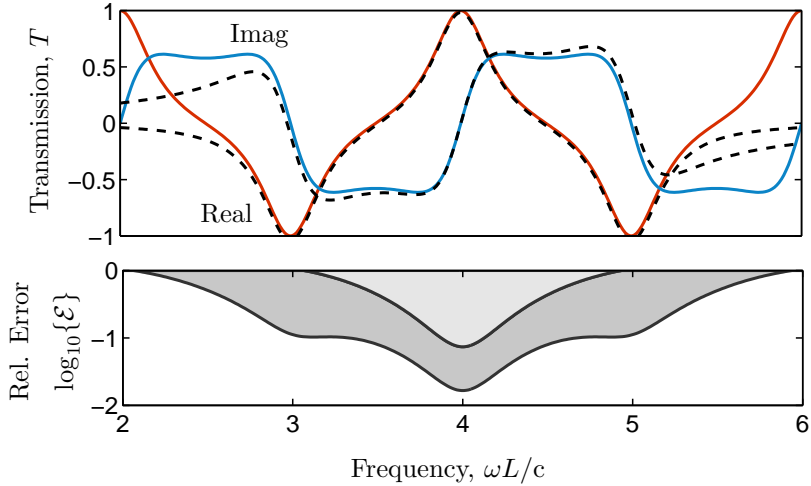

We return now to the transmission calculations in Fig.~\ref{Fig:transmission_spectrum_dielectricBarrier_nR_pi} and the claim, that we can make the error in the QNM approximation to the transmission arbitrarily small. We define the transmission through the dielectric barrier as the ratio between the outgoing electric field at $x=L/2$ to the incoming electric field at $x=-L/2$. There is no scattering back at positions beyond the rightmost boundary, and since the electric field is continuous at the boundary, we can write the transmission as
\begin{align}
T(\omega) = \frac{E_\text{tot}(L/2,\omega)}{E_\text{in}(-L/2,\omega)}.
\end{align}
In Fig.~\ref{Fig:transmission_spectrum_dielectricBarrier_nR_pi_center_on_4_real_plus_imag_incl_CMT}, we show the real and imaginary parts of the transmission spectrum from Fig.~\ref{Fig:transmission_spectrum_dielectricBarrier_nR_pi} along with the CMT approximation as calculated using a sum similar to that in Eq.~(\ref{Eq:E_tot_N}), but using only \change{the three QNMs with indices} $n\in\{3,4,5]\}$. The bottom panel shows the associated relative error as well as the relative error in the single QNM approximation using only $n=4$; the minima in the relative errors are at approximately 2\% and 7\%. As more QNMs are included in the sum, the minimum error decreases and the bandwidth of the approximation increases in a symmetric way around the center point.

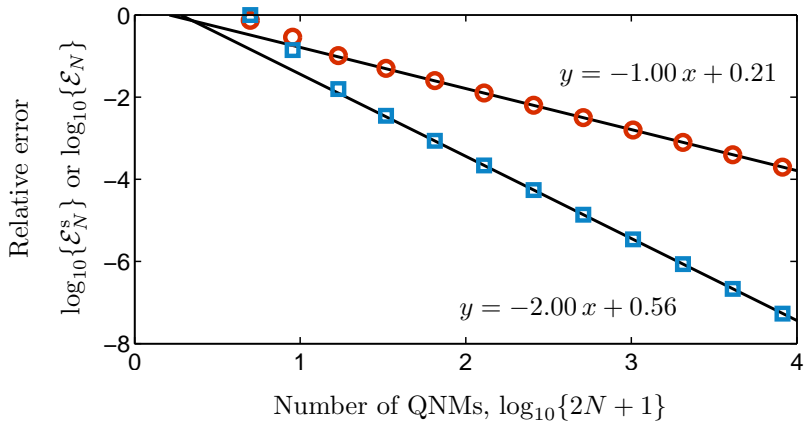
\begin{figure}[htb!]
\centering %
\begin{overpic}[width=9.4cm]{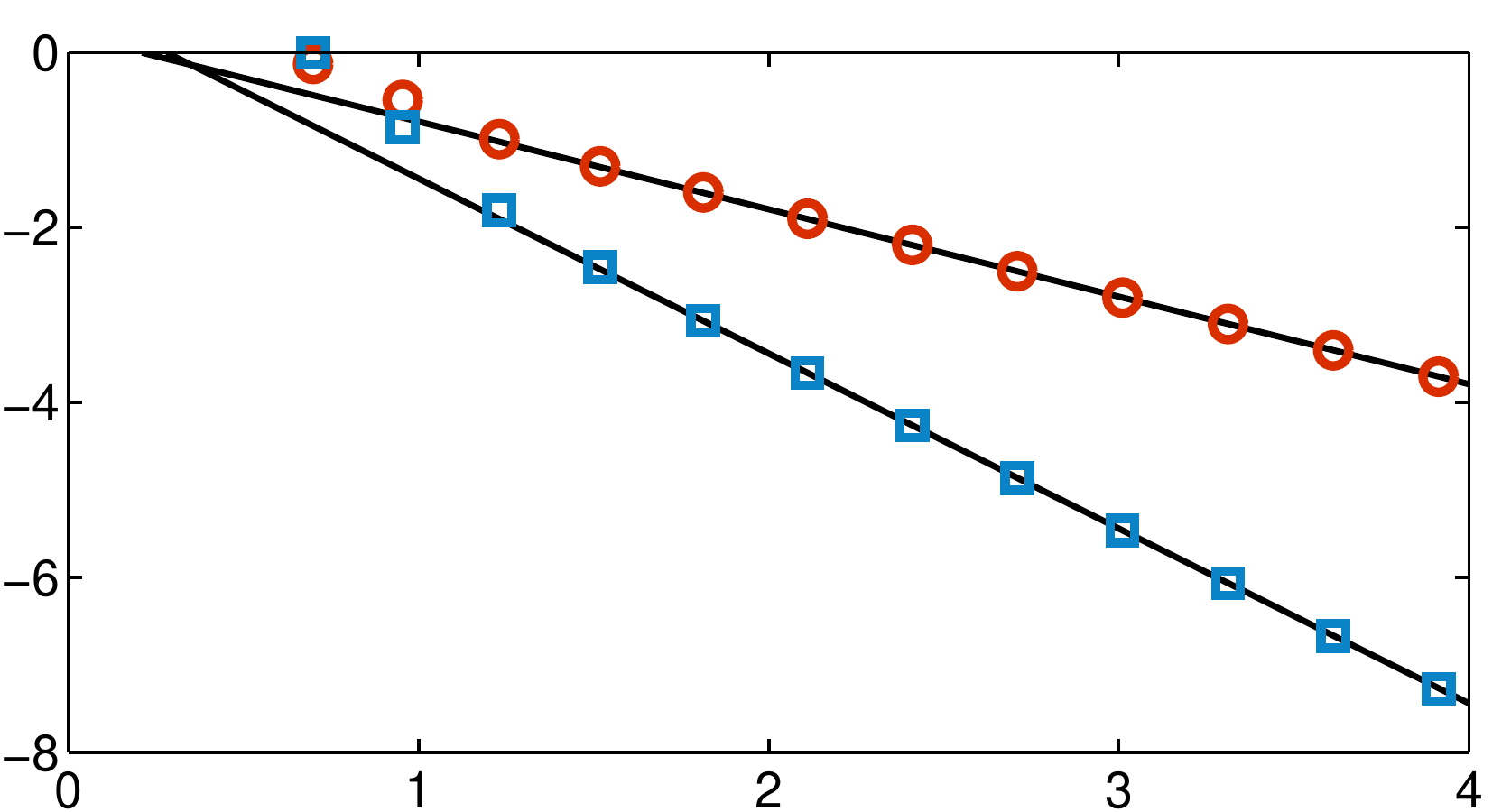}
\put(-13,16){\begin{sideways}Relative error\end{sideways}}
\put(-6,8){\begin{sideways}$\log_{10}\{\mathcal{E}_N^\text{s}\}$ or $\log_{10}\{\mathcal{E}_N\}$\end{sideways}}
\put(24,-6){Number of QNMs, $\log_{10}\{2N+1\}$} %
\put(64,41){$y=-1.00\,x + 0.21$} %
\put(50,8){$y=-2.00\,x +0.56$} %
\end{overpic}
\\[5mm]
\caption{\label{CMT_k0_4_dielectricBarrier_nR_pi_log10_relErrs_at_right_side_vs_log10_2Np1_x_m0p25_0p25}Convergence analysis for the CMT approximation to the transmission through the dielectric barrier at the frequency $\omega_0=4\text{c}/L$. Red circles and blue squares show the relative error $\mathcal{E}_N^\text{s}(L/2,\omega_0)$ in calculations based on the scattered field in Eq. (\ref{Eq:E_tot_N_from_scat}) or the relative error $\mathcal{E}_N(L/2,\omega_0)$ in calculations based on the total field in Eq.~(\ref{Eq:E_tot_N}), respectively. Black lines show linear fits to the last three data points.}
\end{figure}

When using a modest number of QNMs, as in Fig.~\ref{Fig:transmission_spectrum_dielectricBarrier_nR_pi_center_on_4_real_plus_imag_incl_CMT}, it is useful to choose the QNMs symmetrically around the frequency of interest. Indeed, %
while the relative error is as low as a few percent at $\omega L/\text{c}=4$, it is on the order of unity at $\omega L/\text{c}=2$. The rate of convergence, however, is expected to be independent of the center point for the summation, so we can generally %
approximate $E_\text{tot}(x,\omega)$ either directly using Eq.~(\ref{Eq:E_tot_N}) or as the sum of the incoming field and the scattered field as in Eq.~(\ref{Eq:E_tot_N_from_scat}). The associated relative errors in the transmission become simply the relative errors in the total field \change{at $x=L/2$} as calculated using either Eq.~(\ref{Eq:E_tot_rel_error}) or Eq.~(\ref{Eq:E_tot_rel_error_from_scat}), respectively. Figure~\ref{CMT_k0_4_dielectricBarrier_nR_pi_log10_relErrs_at_right_side_vs_log10_2Np1_x_m0p25_0p25} shows the relative errors in the transmission as a function of the number of QNMs in the sum. The error in the direct total field QNM approximation appears to tend to zero in a second order polynomial fashion, which we attribute to the special point at the resonator boundary. The scattered field approach, which is inherently based on integrating across the barrier, appears to inherit the first order polynomial convergence, as was the case also for the relative integrated error, cf. Fig.~\ref{Fig:CMT_k0_4_dielectricBarrier_nR_pi_log10_relErr_vs_log10_N_x_m0p25_0p25}.

\change{
\subsubsection*{Supplementary code} 
With the supplementary code~\cite{dielectricBarrier_arXiv} we provide the files necessary to reproduce the scattering calculation in  Fig.~\ref{Fig:CMT_k0_4_dielectricBarrier_nR_pi_N_500}. We encourage interested readers to vary the number of QNMs in the sum or change the left boundary of the total field region to reproduce the curious example in Fig.~\ref{Fig:CMT_k0_4_dielectricBarrier_nR_pi_N_500_xp_m_1_p_0p5_pi}.}

\subsubsection{Coupled mode theory for the plasmonic dimer}
\label{Sec:CMT_for_plasmonic_dimer}
\begin{figure}[htb!]
\centering %
\quad\;\;
\begin{overpic}[width=9.55cm]{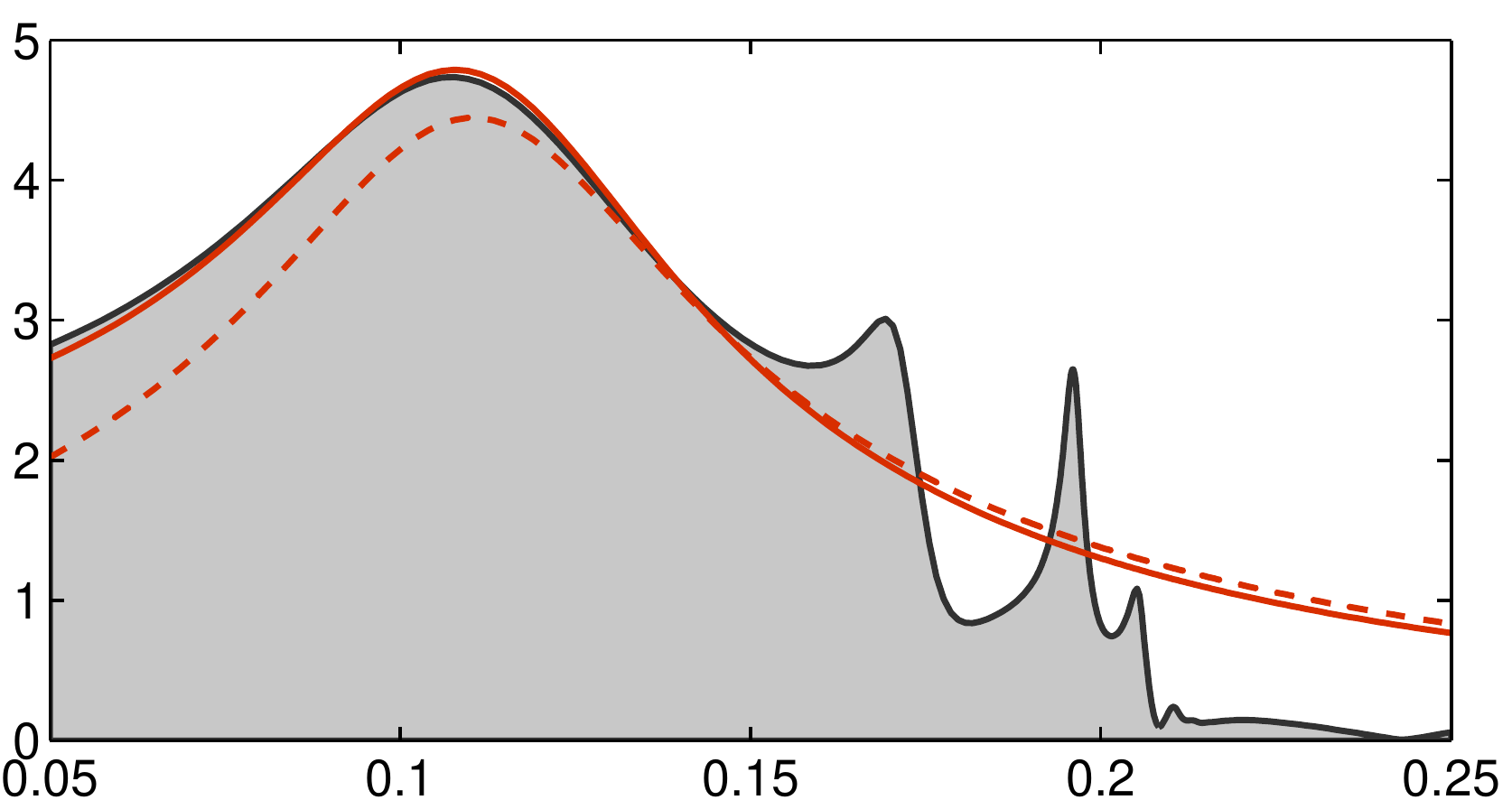}
\put(-13,18){\begin{sideways}Electric field\end{sideways}}
\put(-7,20){\begin{sideways}$E(\mr_0)/E_0$\end{sideways}} %
\put(34,-6){Frequency, $\omega_\text{R}d/2\pi\text{c}$}
\end{overpic}\;\;\;\;\\[5mm]
\caption{\label{Fig:plasmonicDimer_scattering_plus_CMT_abs_wideRange}Magnitude of the local electric field at the point $\mr_0$ in the middle between the two gold spheres of the plasmonic dimer when illuminated by a plane wave of magnitude $E_0$ and polarization along the dimer axis. Red dashed and full curves show the single- and two-QNM CMT approximations when using $\mFt_1(\mr)$ or $\mFt_1(\mr)$ and $\mFt_1^*(\mr)$, respectively, and gray shading shows the reference calculation.}
\end{figure}

We now turn to the plasmonic dimer and consider the electromagnetic response of the dimer when illuminated by a plane wave of magnitude $E_0$ polarized along the dimer axis. Figure \ref{Fig:plasmonicDimer_scattering_plus_CMT_abs_wideRange} shows the broadband electric field at the position $\mr_0$ directly in the middle between the two spheres. %
As in the case of the Purcell factor in Fig.~\ref{Fig:Purcell_avs_Lmax_6_rdip_0p5_0p25_0_tloRange_0p05_to_0p25}, a number of distinct peaks are visible, each of which can be attributed primarily to a single QNM. The peak maximum is found at $\omega d/2\pi\text{c}\approx0.11$, which corresponds roughly to the real part of $\tlo_1$, cf. Eq.~\ref{Eq:plasmonic_dimer_d1_resonance_freq}.

The QNM $\mFt_1(\mr)$ is commonly referred to as the ``dipolar mode'' of the dimer, because of the field pattern, which resembles the field between two particles with opposite charges. From a physical point of view, this mode is therefore expected to couple effectively to the incoming plane wave, wherefore it is also known as a ``bright mode''. Other modes are so-called ``dark modes'', because they cannot be effectively excited from the far field. Typically, the distinction between bright and dark modes are inferred from symmetry considerations of the exciting field and the QNMs as in Ref.~\cite{Moeferdt18}, for example. The expression for the expansion coefficient in Eq.~(\ref{Eq:alpha_m_CMT}) exactly captures the excitation of the different QNMs by an incoming field, and hence provides a precise mathematical distinction between bright and dark modes in general resonators.

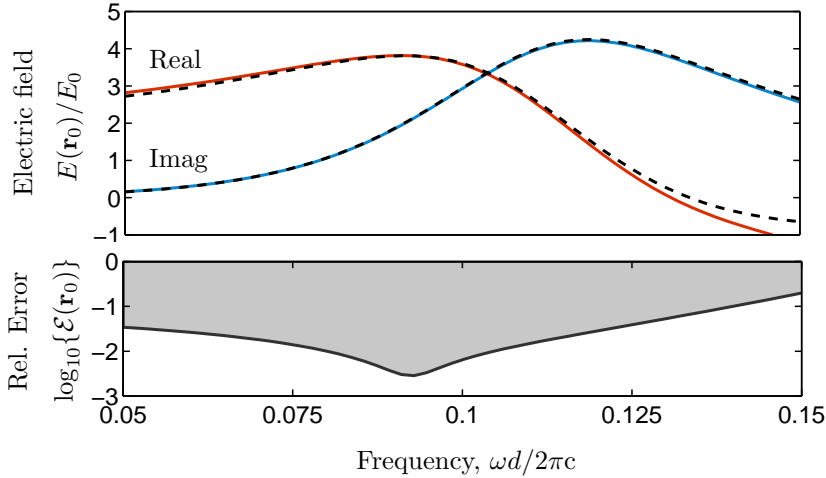
\begin{figure}[htb!]
\centering %
\begin{overpic}[width=9.4cm]{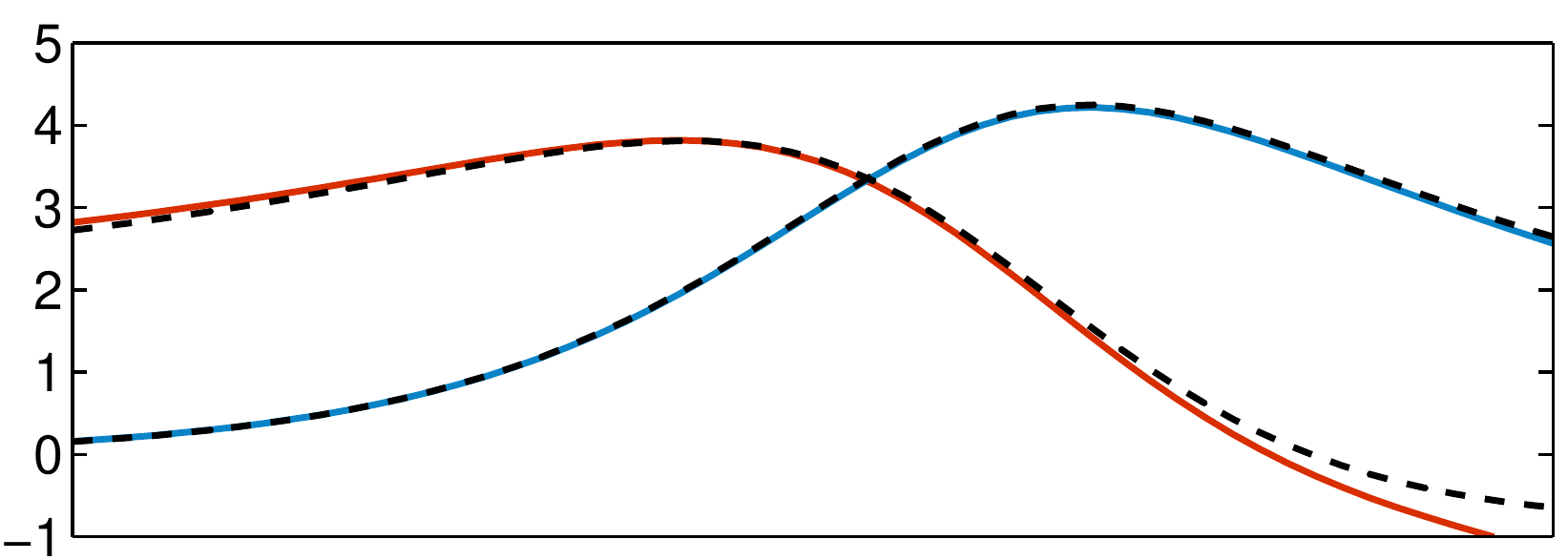}
\put(-11,7){\begin{sideways}Electric field\end{sideways}}
\put(-5,9){\begin{sideways}$E(\mr_0)/E_0$\end{sideways}} %
\put(8,25){Real}
\put(8,11){Imag}
\put(42,-6){Position, $x/L$}
\end{overpic}\;\;\;
\\
\begin{overpic}[width=9.75cm]{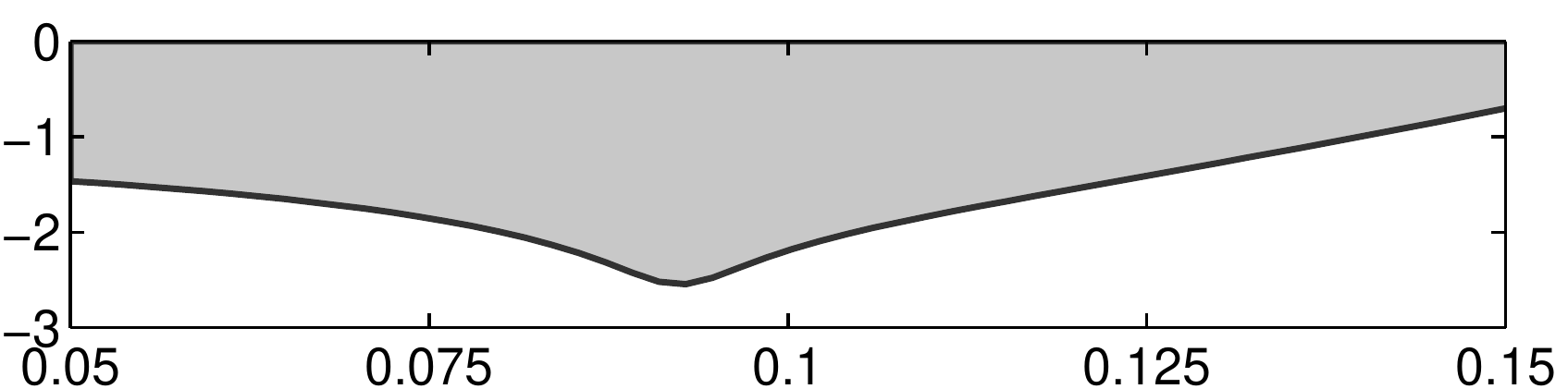}
\put(-11,4.5){\begin{sideways}Rel. Error\end{sideways}}
\put(-5,3){\begin{sideways}$\log_{10}\{\mathcal{E}(\mr_0)\}$\end{sideways}} %
\put(36,-6){Frequency, $\omega d/2\pi\text{c}$}
\end{overpic}
\\[5mm]
\caption{\label{plasmonicDimer_scattering_plus_CMT_real_plus_imag_Lmax_5}Top: Real (red) and imaginary (blue) parts of the electric field at the point $\mr_0$ in the center between the two spheres in the plasmonic dimer, when illuminated by a plane wave of magnitude $E_0$ with polarization along the dimer axis. Dashed curves indicate the CMT approximation using the QNMs $\mFt_1(\mr)$ and $\mFt_1^*(\mr)$. Bottom: Relative error $\mathcal{E}(\mr,\omega)=|E_\text{ref}(\mr,\omega)-E_\text{CMT}(\mr,\omega)|/|E_\text{ref}(\mr,\omega)|$ as a function of frequency.}
\end{figure}

Even if the QNM expansion is not formally convergent at the position of interest between the two spheres, we can extend the region of convergence by the procedure discussed in Section~\ref{Sec:Formally_extending_ROC}. In this case, for example, we can imagine including a sphere of finite but vanishingly small permittivity difference between the gold spheres making up the plasmonic dimer. Alternatively, we can embed the entire dimer in a large sphere of vanishingly small permittivity difference with respect to the background. In both cases, we expect then the QNM expansion to be convergent at the position $\mr_0$. At the same time, the change in the QNMs of interest can be made arbitrarily small, see also Section~\ref{Sec:Perturbation_theory_for_the_plasmonic_dimer}, so the expansion coefficients $a_n(\omega)$ of any truncated series can be calculated simply by use of the QNMs of the original plasmonic dimer. The red dashed and full curves in Fig.~\ref{Fig:plasmonicDimer_scattering_plus_CMT_abs_wideRange} show the magnitude of the CMT approximation to the electric field when using either $\mFt_1(\mr)$ or $\mFt_1(\mr)$ and $\mFt_1^*(\mr)$. As in the case of the dielectric barrier, the CMT approximation captures not only the magnitude, but also the phase of the electromagnetic field. Focusing on the low-frequency region, Fig. \ref{plasmonicDimer_scattering_plus_CMT_real_plus_imag_Lmax_5} shows the real and imaginary parts of the electric field reference calculations, as well as the relative error. The relative error of the two-QNM CMT approximation is below 0.05 over a large bandwidth and as low as a few parts in a thousand at best.

\subsection{Coupled resonators}
\label{Sec:Coupled_resonators}
When two electromagnetic resonators are placed in close proximity, we expect them to couple to each other and produce new, hybridized resonances. Indeed, this is a fundamental phenomenon, which can be observed in all areas of physics. In the so-called tight binding approach, or the linear combination of atomic orbitals (LCAO), the hybridized resonances of coupled, localized wave functions can be calculated through the use of an overlap integral~\cite{Gupta_2016}. In this way, one can immediately appreciate the origin of the coupling, and one can use this methodology to enhance or suppress the coupling, or to calculate approximate band structures, for example. %
The QNMs, however, are not localized, so we cannot immediately apply the LCAO framework. The exponential divergence, in particular, would lead to larger and larger overlap integrals if the distance between the resonators was increased. %
\change{Instead,} we take a different approach and use the QNMs of the individual resonators in combination with the Field Equivalence Principle to set up a self-consistent set of equations for the scattered field expansion coefficients of the compound system.

On the boundary of a given resonator, we assume that the scattered field at frequency $\omega$ can be expanded as in Eq.~(\ref{Eq:mF_scat_pole_expansion_form}). The scattered field is radiated away and may itself lead to scattering off another resonator. In this way, and quite analogous to multiple scattering theory, we can set up a self-consistent set of equations for the scattered field expansion coefficients $b_m^{(i)}$, where $i$ now counts the scatterers. By use of Eq.~(\ref{Eq:Field_Equivalence_Principle}) with equivalent surface currents given in terms of the QNM expansion in Eq.~(\ref{Eq:mF_scat_pole_expansion_form}), the scattered field from resonator $j$ at position $\mr$ is given as
\begin{align}
\mF_\text{scat}(\mr,\omega) = \text{i}\omega\mu_0\sum_n\frac{b_n^{(j)}}{\omega-\tlo_n}\int_{\partial V}\underline{\underline{\mG_\text{B}}}(\mr,\mr',\omega)
\begin{bmatrix}
\mathbf{\hat n}\times\mgt_n(\mr')\\
\mft_n(\mr')\times\mathbf{\hat n}
\end{bmatrix}\ud A,
\end{align}
\change{where $\underline{\underline{\mG_\text{B}}}(\mr,\mr',\omega)$ denotes the matrix Green tensor of  the homogeneous background material}. This field leads to a scattered field in resonators $i\neq j$, for which the expansion coefficients are given by Eq.~(\ref{Eq:b_n_CMT_LSE}) with $\mF_\text{scat}(\mr,\omega)$ in place of $\mF_\text{in}(\mr,\omega)$. Using a fixed number of QNMs in each resonator, one can use this approach to set up a matrix eigenvalue equation for the expansion coefficients of the coupled system of the form %
\begin{align}
\mathbf{b} = \mathbf{B}(\omega)\mathbf{b},
\label{Eq:mmB_coupled_resonators}
\end{align}
from which the QNMs of the coupled system can be found as the points where the eigenvalues of the matrix $\mathbf{B}-1$ vanish. The procedure is thus very similar to the expansion in terms of spherical wave functions discussed in Section~\ref{Sec:Calculations_using_VIE}, but with the important difference that the basis functions are now %
the QNMs of the individual resonators.

\subsubsection{Coupled dielectric barriers}
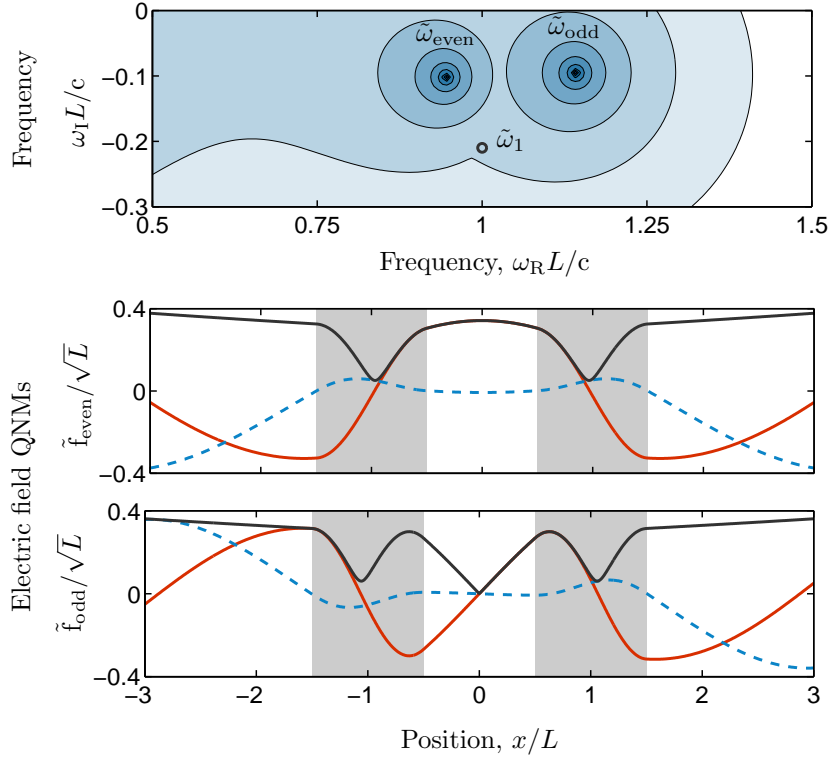
\begin{figure}[htb!]
\centering %
\hspace{1mm}
\begin{overpic}[width=9.65cm]{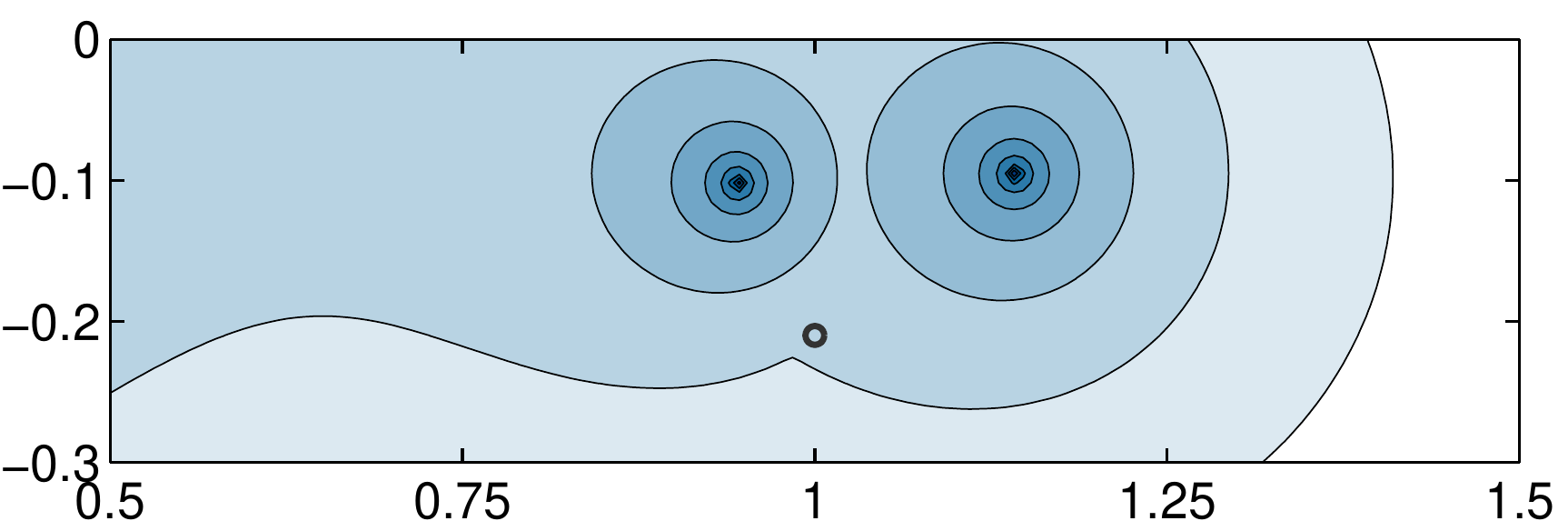}
\put(-12,9){\begin{sideways}Frequency\end{sideways}}
\put(-5,12){\begin{sideways}$\omega_\text{I}L/\text{c}$\end{sideways}}
\put(38,-5){Frequency, $\omega_\text{R}L/\text{c}$}
\put(54,12){$\tlo_1$}
\put(43,26.5){$\tlo_\text{even}$}
\put(61,27.5){$\tlo_\text{odd}$}
\end{overpic}
\\[8mm]
\begin{overpic}[width=9.5cm]{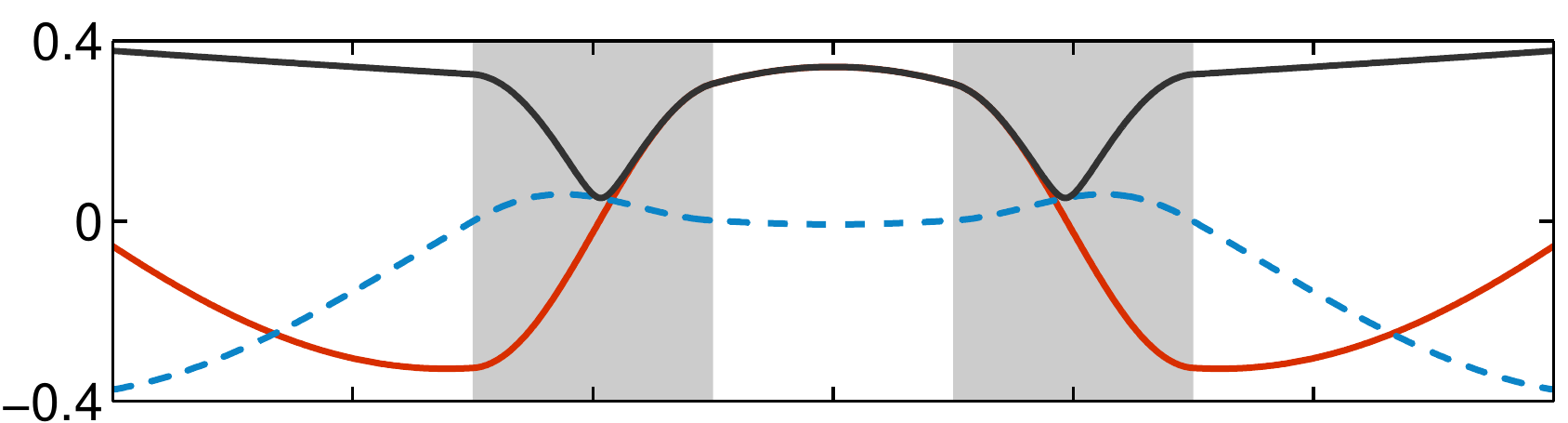}
\put(-12,-18){\begin{sideways}Electric field QNMs\end{sideways}}
\put(-5,5){\begin{sideways}$\ft_\text{even}/\sqrt{L}$\end{sideways}}
\end{overpic}
\\[1mm]
\begin{overpic}[width=9.65cm]{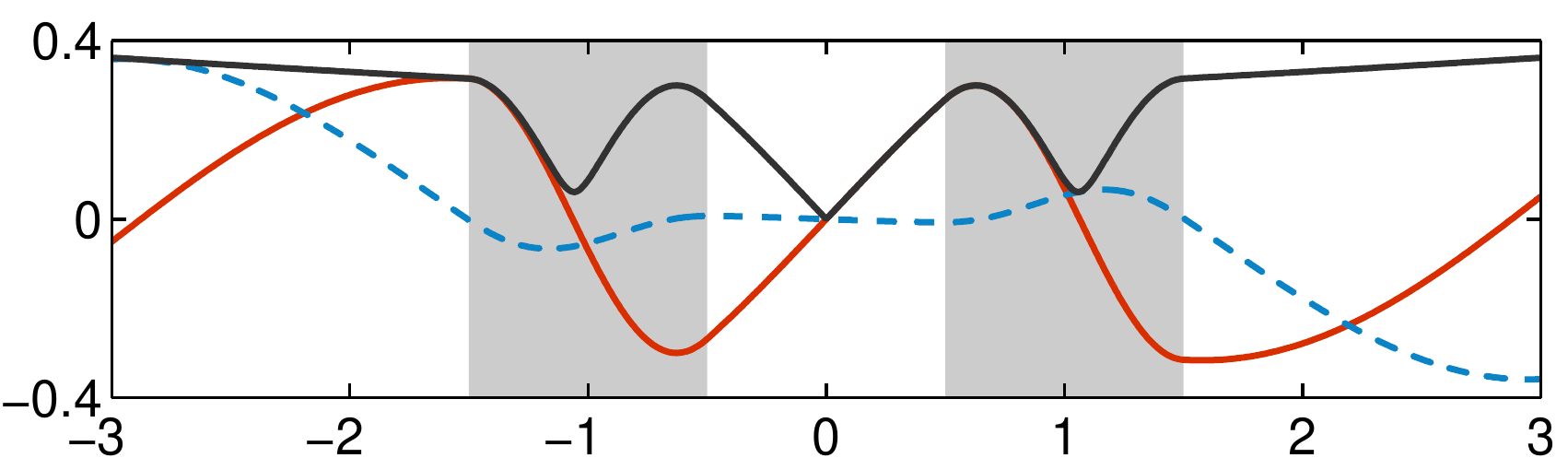}
\put(-5,10){\begin{sideways}$\ft_\text{odd}/\sqrt{L}$\end{sideways}}
\put(42,-6){Position, $x/L$}
\end{overpic}
\\[5mm]
\caption{\label{Fig:double_barrier_1D_errorMap}Top: Complex spectrum of the double barrier system showing the two QNM frequencies near the position of $\tlo_1$ of the single barrier system (indicated by a black circle). Center and bottom: Mode profile showing the absolute values of the electric field QNMs corresponding to $\tlo_\text{even}$ (center) and $\tlo_\text{odd}$ (bottom). Gray shading indicates the extent of the dielectric barriers.}
\end{figure}

We now consider the system made from two copies of the dielectric barrier in Section~\ref{Sec:1Dresonator} that are positioned a distance of $D=L$ apart. This system does not allow a simple, closed form expression for the QNMs or the resonance frequencies, and so we turn to numerical methods. In practice, we use a one-dimensional formulation of the VIE formulation from Section~\ref{Sec:Calculations_using_VIE}, but we note that these calculations can be performed with practically all the methods discussed in Section~\ref{Sec:QNM_calculation_methods}. For the particular choice of separation, we find two QNM resonance frequencies close to $\tlo_1$ of the single barrier system,
\begin{align}
\tlo_\text{even}L/\text{c} &= 0.948576736961 - 0.098814034278\text{i}\label{Eq:double_barrier_tlo_even}\\
\tlo_\text{odd}L/\text{c} &= 1.145319154125 - 0.092211949692\text{i},\label{Eq:double_barrier_tlo_odd}
\end{align}
which, upon inspection of the mode profiles, can be associated with an even and an odd mode, as shown in Fig.~\ref{Fig:double_barrier_1D_errorMap}.

\begin{figure}[htb!]
\centering %
\begin{overpic}[width=9.4cm]{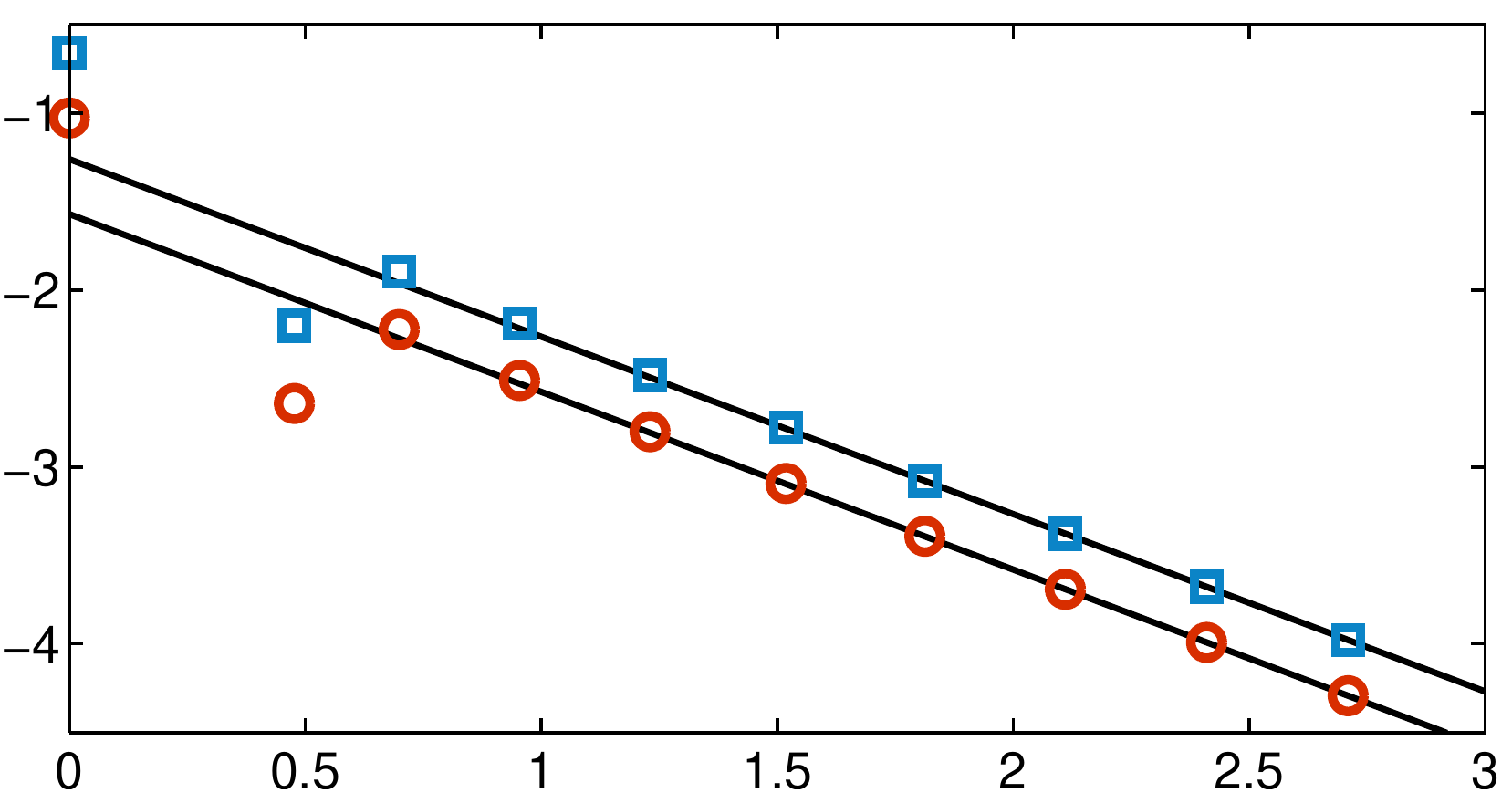}
\put(-6,3){\begin{sideways}Relative error, $\log_{10}\{\mathcal{E}_\text{QNM}\}$\end{sideways}}
\put(14,-6){Number of QNMS per barrier, $\log_{10}\{2N+1\}$} %
\put(64,24){$y=-1.00\,x - 1.26$} %
\put(38,8){$y=-1.00\,x - 1.57$} %
\end{overpic}
\\[5mm]
\caption{\label{Fig:double_barrier_1D_k0_Q_convergence_log10_rel_Error_vs_log10_2Np1}Convergence analysis for the calculation of $\tlo_\text{even}$ (red circles) $\tlo_\text{odd}$ (blue squares) of the double barrier system by means of a QNM expansion showing the relative error $\mathcal{E}_\text{QNM}=|\tlo_\text{QNM}-\tlo_\text{ref}|/|\tlo_\text{ref}|$ as a function of number of QNMs used in the expansions (in each barrier). Black lines show the corresponding fits to the last four data points.}
\end{figure}
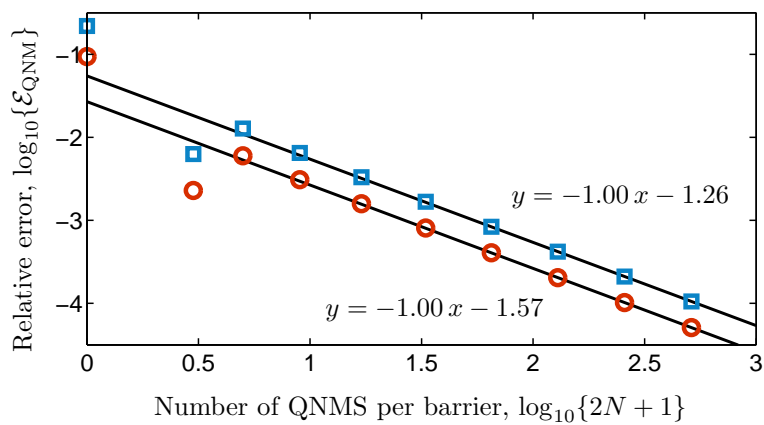

To set up Eq.~(\ref{Eq:mmB_coupled_resonators}) for the double barrier system, we note that the boundaries of the resonators are simply the four points $x_1^-$, $x_1^+$, $x_2^-$ and $x_2^+$, indicating the left ($-$) and right ($+$) boundaries of resonators 1 and 2. Starting from Eq.~(\ref{Eq:b_n_CMT_LSE}), we can use the wave equations to formulate $b_m^{(1)}$ and $b_m^{(2)}$ in terms of the electric field only as
\begin{align}
b_m^{(1)} &= B_m^{(1)}(\omega)\left[ E_\text{inc}^{(1)}(x_1^+,\omega) \partial_x\mft_m(L/2) - \mft_m(L/2) \partial_xE_\text{inc}^{(1)}(x_1^+,\omega) \right. \nonumber\\
& \qquad\qquad\quad \left.  -E_\text{inc}^{(1)}(x_1^-,\omega) \partial_x\mft_m(-L/2) + \mft_m(-L/2) \partial_xE_\text{inc}^{(1)}(x_1^-,\omega) \right] \label{Eq:coupled_barriers_b_m_1} \\[2mm]
b_m^{(2)} &= B_m^{(2)}(\omega)\left[ -E_\text{inc}^{(2)}(x_2^-,\omega) \partial_x\mft_m(-L/2) + \mft_m(-L/2) \partial_xE_\text{inc}^{(2)}(x_2^-,\omega) \right. \nonumber\\
& \qquad\qquad\quad \left.  E_\text{inc}^{(2)}(x_2^+,\omega) \partial_x\mft_m(L/2) - \mft_m(L/2) \partial_xE_\text{inc}^{(2)}(x_2^+,\omega) \right], \label{Eq:coupled_barriers_b_m_2}
\end{align}
where
\begin{align}
B_m^{(i)}(\omega) =  \frac{\text{c}}{2}\frac{\omega\Delta\epsilon}{\tlo_i^2\epsilon_\text{r}-\omega^2\epsilon_\text{B}}	,
\end{align}
and the incoming fields $E_\text{inc}^{(1/2)}(x,\omega)$ are taken to be the scattered fields from resonators $2/1$. The propagation of the scattered fields is particularly simple in one dimension, since it is done by a simple multiplication with the exponential factor $\exp\{\pm\text{i}kx\}$, where $\pm$ denotes the orientation of the normal vector at the boundary. Thus, the scattered fields impinging on resonators 1/2 are given as
\begin{align}
E_\text{inc}^{(1)}(x,\omega) &= E_\text{scat}^{(2)}(x_2^-,\omega)\text{e}^{-\text{i}k(x-x_2^-)}%
\\[2mm]
E_\text{inc}^{(2)}(x,\omega) &= E_\text{scat}^{(1)}(x_1^+,\omega)\text{e}^{\text{i}k(x-x_1^+)}.
\end{align}
Expanding the scattered fields as in Eq.~(\ref{Eq:mF_scat_pole_expansion_form}) symmetrically around $n=1$, %
\begin{align}
E_\text{scat}^{(i)}(x,\omega) = \sum_{n=1-N}^{1+N}\frac{b_n^{(i)}}{\omega-\tlo_n}\mft_n(x),
\end{align}
and inserting in Eqs.~(\ref{Eq:coupled_barriers_b_m_1}) and (\ref{Eq:coupled_barriers_b_m_2}), we can immediately set up Eq.~(\ref{Eq:mmB_coupled_resonators}) for a given choice of $N$ and map out the corresponding frequency landscape (not shown). We solve Eq.~(\ref{Eq:mmB_coupled_resonators}) by an iterative search to find the approximate frequencies $\tlo_\text{QNM}$ and compare to the high accuracy reference calculations in Eqs.~(\ref{Eq:double_barrier_tlo_even}) and (\ref{Eq:double_barrier_tlo_odd}). Figure \ref{Fig:double_barrier_1D_k0_Q_convergence_log10_rel_Error_vs_log10_2Np1} shows the relative error as a function of the number of QNMs used in the expansion.
Using only a single QNM in each resonator (case of $N=0$) gives qualitatively correct results with relative errors of 9\% for $\tlo_\text{even}$ and 22\% for $\tlo_\text{odd}$, and in general we find that the relative error decreases linearly with the number of QNMs used in the expansion, cf. Fig.~\ref{Fig:double_barrier_1D_k0_Q_convergence_log10_rel_Error_vs_log10_2Np1}. The expansion of the double barrier QNMs in terms of the single barrier QNMs thus appears to be convergent, which is in full agreement with the analysis in Section~\ref{Sec:RoC_dielectric_barrier} \change{We remark, that since the proposed calculation scheme relies on expansion of the scattered field in terms of QNMs inside the resonators only, the limited region of convergence of the QNM expansions do not affect the convergence of the coupled resonator calculations}. Last, we note that for this particular, example, the reference calculations were done with two basis functions in each barrier, leading to the solution of a $4\times4$ matrix equation problem. The last points in Fig.~\ref{Fig:double_barrier_1D_k0_Q_convergence_log10_rel_Error_vs_log10_2Np1}, in contrast, were calculated using a $1024\times1024$ matrix equation. The coupled resonator approach, therefore, will likely not be an effective way of calculating QNMs of general coupled resonators, even if it is fundamentally a convergent scheme. Rather, we consider it to be an interesting analytical tool to investigate the coupling mechanism between QNMs in different resonators.

\subsection{First order perturbation theory}
\label{Sec:Perturbation_theory}

If the material defining the electromagnetic resonator is slightly perturbed by a local change $\Delta\epsilon(\mr)$ or $\Delta\mu(\mr)$, we expect to be able to calculate the dominant change to the QNM resonance frequency using first-order perturbation theory~\cite{Lai_PRA_41_5187_1990}. Indeed, we may in principle calculate the change in the QNM resonance frequency --- as well as the changes in the QNM itself --- to arbitrary accuracy by going to perturbation theory of sufficiently high order. In practice, the use of higher order perturbation theory quickly becomes cumbersome, especially in the so-called Rayleigh-Schr{\"o}dinger formulation~\cite{Ballentine_1998}. %
An alternative to the Rayleigh-Schr{\"o}dinger formulation ---  the Brillouin-Wigner formulation --- expresses the energy as an implicit series expansion~\cite{Ballentine_1998}. Alternatively, as was illustrated in Ref.~\cite{Muljarov_EPL_92_50010_2010}, one can use the QNMs of the unperturbed system to formulate a linear eigenvalue problem for the perturbed system.

\subsubsection{Perturbations in the permittivity}
\label{Sec:Perturbations_in_the_permittivity}
To see how the \change{normalization integral} in Eq.~(\ref{Eq:MuljarovNorm_E_H_form}) arises naturally in this process, we %
write the defining equation for the $m$'th QNM of the perturbed system as
\begin{align}
\mD\,\mFt'(\mr,\tlo') + \text{i}\tlo'\mFt'(\mr,\tlo') = -\text{i}\tlo'\mmDelta\,\mFt'(\mr,\tlo'),
\label{Eq:defining_Eq_QNM_perturb}
\end{align}
where $\mmDelta = \text{diag}\{\Delta\epsilon(\mr)/\epsilon_\text{r}(\mr), \Delta\mu(\mr)/\mu_\text{r}(\mr)\}$, %
and
\begin{align}
\mFt'(\mr,\omega') &= \mFt_m(\mr) + \Delta\mFt(\mr,\omega') \\
\tlo' &= \tlo_m + \Delta\tlo.
\end{align}
Inserting in Eq.~(\ref{Eq:defining_Eq_QNM_perturb}) and expanding to first order, we find that the perturbation $\Delta\mFt(\mr,\omega')$ solves the equation
\begin{align}
\left\{\mD + \text{i}\tlo_m\right\}\Delta\mFt(\mr,\omega') + \text{i}\Delta\tlo\mFt_m(\mr) = -\text{i}\tlo_m\mmDelta\,\mFt_m(\mr).
\end{align}
Multiplying from the left with $\mFb_m(\mr)\mmW$ and integrating, we can use %
Eq.~(\ref{Eq:innerProd_adjointDerivation}) to find the expression
\begin{align}
I_{\partial V}\big(\mFb_m(\mr),\Delta\mFt(\mr,\omega')\big) + \text{i}\Delta\tlo\langle\mFb_m(\mr)|\mFt_m(\mr)\rangle = -\text{i}\tlo_n\langle\mFb_m(\mr)|\mmDelta\,\mFt_m(\mr)\rangle,
\end{align}
which can be simplified by noting that
$\Delta\mFt(\mr,\omega') \approx \Delta\tlo\partial_\omega\mFt_m(\mr)$, %
so that
\begin{align}
\Delta\tlo\left[ \langle\mFb_m(\mr)|\mFt_m(\mr)\rangle - \text{i}I_{\partial V}\big(\mFb_m(\mr),\Delta\mFt(\mr,\omega)\big) \right] = -\tlo_m\langle\mFb_m(\mr)|\mmDelta\,\mFt_m(\mr)\rangle.
\end{align}
Last, writing out the expression for the surface integral and making use of %
Eqs.~(\ref{Eq:Muljarov_insight}) and (\ref{Eq:projection_operator}), we can write the first-order correction in the form \begin{align}
\Delta\tlo = -\tlo_m \langle\mFb_m(\mr)|\mmDelta\,\mFt_m(\mr)\rangle. %
\end{align}
In cases with $\Delta\mu(\mr)=0$, %
this reduces to the well-known form%
\begin{align}
\Delta\tlo = -\frac{\tlo_m}{2}\int_V\Delta\epsilon(\mr)\mft_m(\mr)\cdot\mft_m(\mr)\ud V,
\label{Eq:QNM_perturbation_theory}
\end{align}
as also found in Refs.~\cite{Lai_PRA_41_5187_1990,Lee_JOSAB_16_1418_1999}. Higher-order perturbation theory with QNMs has been treated in \change{Refs.~\cite{Leung_JOSAB_13_805_1996, Leung_PRA_49_3068_1994,  Lee_JOSAB_16_1418_1999}, and Ref.~\cite{Ng_JOSAB_19_154_2002} treats the question of degenerate perturbation theory}. First-order perturbation theory of the resonance frequency is easily performed with numerically calculated QNMs for which one does not usually have easy access to the entire spectrum. Because of the coupling between different QNMs, higher-order perturbation theory is best suited to geometries where one has access to analytical expressions for the QNMs.

\subsubsection{Shifting boundaries}
Perturbation theory with shifting boundaries represents a non-trivial extension of the theory, since the local change in material properties close to the boundary of the resonator material will generally change in a non-perturbative way. The problem has been solved by Lai~\emph{et al.}.~\cite{Lai_PRA_41_5187_1990} and by Johnson \emph{et al.}~\cite{Johnson_PRE_65_066611_2002} \change{for Hermitian eigenvalue problems, but for which the central arguments} %
immediately carry over to the case of first-order perturbation theory of QNMs. For the case of a resonator made from a non-magnetic material in a background with permittivity $\epsilon_\text{B}$, we can, therefore, immediately infer that~\cite{Lai_PRA_41_5187_1990,Johnson_PRE_65_066611_2002}
\begin{align}
\Delta\tlo_m &= -\frac{\tlo_m}{2}\int_{\partial V}\left[\left(\epsilon_\text{r}-\epsilon_\text{B}\right)\mft_m^{\|}(\mr)\cdot\mft_m^{\|}(\mr)\right. \nonumber \\ &\left.\qquad\qquad\qquad\qquad- \left(\frac{\epsilon_{\text{r}/\text{B}}^2}{\epsilon_\text{r}}-\frac{\epsilon_{\text{r}/\text{B}}^2}{\epsilon_\text{B}}\right)\mft_m^{\perp}(\mr)\cdot\mft_m^{\perp}(\mr)\right] \Delta h(\mr)\,\ud A,
\label{Eq:pert_w_shifting_boundaries}
\end{align}
\change{in which $\partial V$ is the surface of the original volume,} $\epsilon_{\text{r}/\text{B}}$ denotes the permittivity of the material on the side of the interface where the field is evaluated, and $\Delta h(\mr)$ represents the local shift of the surface in the normal direction. %

\subsubsection{Perturbation theory for the dielectric barrier}
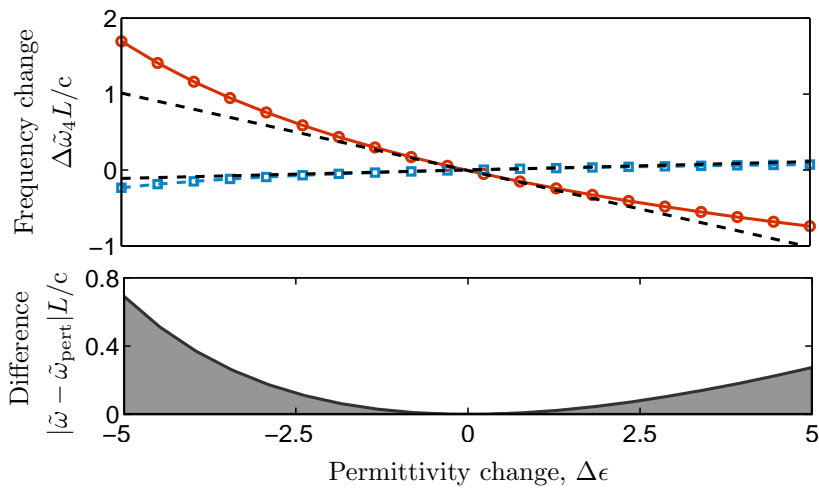
\begin{figure}[htb!]
\centering %
\begin{overpic}[width=9.65cm]{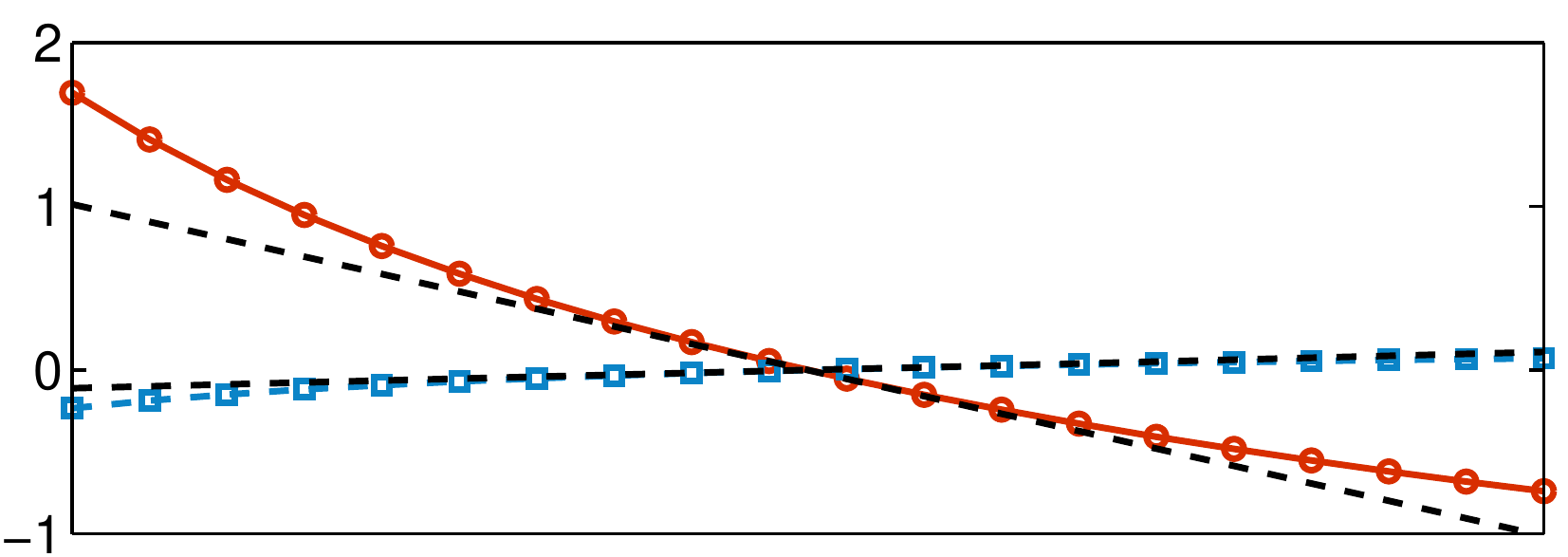}
\put(-10,3){\begin{sideways}Frequency change\end{sideways}}
\put(-5,12){\begin{sideways}$\Delta\tlo_4 L/\text{c}$\end{sideways}}
\end{overpic}\\
\begin{overpic}[width=9.75cm]{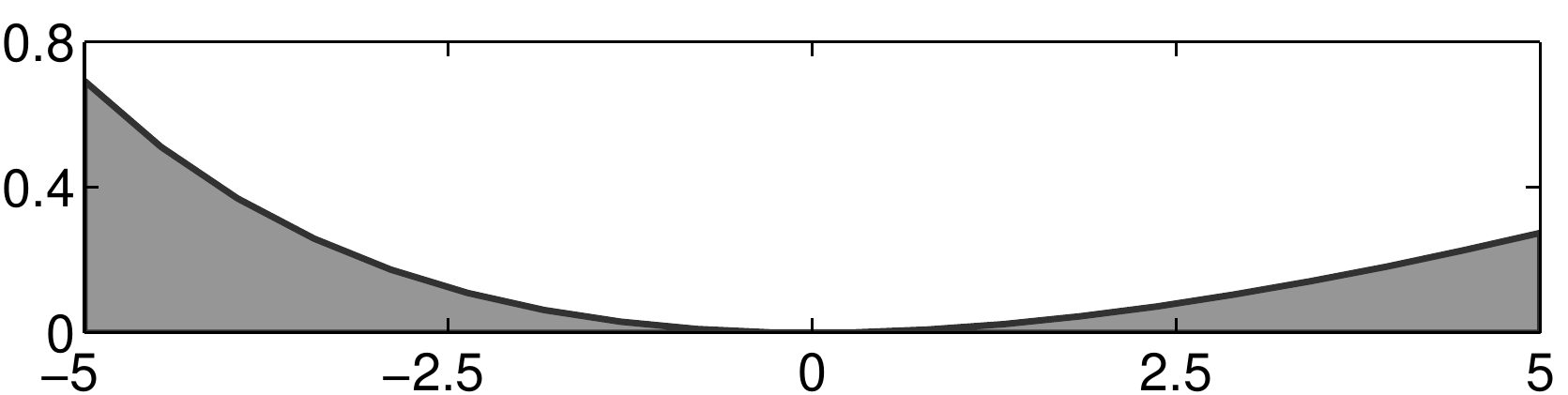}
\put(-10,5){\begin{sideways}Difference\end{sideways}}
\put(-5,1){\begin{sideways}$|\tlo-\tlo_\text{pert}| L/\text{c}$\end{sideways}}
\put(33,-5){Permittivity change, $\Delta\epsilon$} %
\end{overpic}\\[5mm]
\caption{\label{Fig:barrier_1D_complex_freq_and_perturb}Top: Real (red circles) and imaginary (blue squares) parts of the shift in the complex QNM resonance frequency $\tlo_4$ as a function of permittivity change $\Delta\epsilon$ of the dielectric barrier. Dashed curves show the first order perturbation theory result. Bottom: Absolute value of the difference between the exact result and the first-order perturbation theory result.} %
\end{figure}

To illustrate the practical use of Eq.~(\ref{Eq:QNM_perturbation_theory}), we use the one-dimensional example of the dielectric barrier, for which the QNM resonance frequencies are known analytically, cf. Eq.~(\ref{Eq:QNM_freqs_dielectric_barrier}). Assuming the permittivity of the dielectric barrier is changed by a small and constant amount $\Delta\epsilon$, we find from Eq.~(\ref{Eq:QNM_perturbation_theory}) that the first order change in resonance frequency should be
\begin{align}
\Delta\tlo_m = -\frac{\tlo_m}{2}\Delta\epsilon\int_{-L/2}^{L/2}\ft_m(x)^2\ud x,
\end{align}
where the electric field QNMs $\ft_m(x)$ are given by Eq.~(\ref{Eq:QNM_dielectric_barrier_eField}), and the \change{normalization integral} is given by Eq.~(\ref{Eq:dielectric_barrier_normalization_Efield_only}). Depending on the parity of $m$, the electric field QNMs are proportional to either the sine or the cosine functions.
By direct integration, we find that we can write the result as
\begin{align}
\Delta\tlo_mL/\text{c} = -\frac{\Delta\epsilon}{2n_\text{R}^3}\left[n_\text{R}\tlo_mL/\text{c} + (-1)^m\sin\{n_\text{R}\tlo_mL/\text{c}\}\right],
\end{align}
which, by use of the explicit form for the resonance frequencies in Eq.~(\ref{Eq:QNM_freqs_dielectric_barrier}), can be simplified as
\begin{align}
\Delta\tlo_mL/\text{c} = -\frac{\Delta\epsilon}{2n_\text{R}^3}\left[n_\text{R}\tlo_mL/\text{c} -\text{i}\frac{2n_\text{R}n_\text{B}}{n_\text{R}^2-n_\text{B}^2}\right].
\label{Eq:delta_tlo_dielectric_barrier}
\end{align}
To assess the validity of this result, we can compare directly with the explicit expression in Eq.~(\ref{Eq:QNM_freqs_dielectric_barrier}), which is valid for all values of $n_\text{R}$. Substituting $n_\text{R}+\Delta n$ in place of $n_\text{R}$, we can make a series expansion around the point $\Delta n=0$ to find that the first order change in the resonance frequency is indeed given by Eq.~(\ref{Eq:delta_tlo_dielectric_barrier}). Figure~\ref{Fig:barrier_1D_complex_freq_and_perturb} shows the real and imaginary parts of $\tlo_4$ when the permittivity is changed by adding or subtracting as much as $\Delta\epsilon=\pm5$. The first order approximation, as calculated from Eq.~(\ref{Eq:delta_tlo_dielectric_barrier}), provides the tangents to the curves for the exact results in the point $\Delta\epsilon=0$, as expected.

\subsubsection{Perturbation theory for the plasmonic dimer}
\label{Sec:Perturbation_theory_for_the_plasmonic_dimer}
To illustrate the influence of the vector nature of the QNMs, we consider the case where a small sphere with radius $R_\Delta$ and local constant permittivity change $\Delta\epsilon$ is inserted at the point $\mr_0$ directly between the gold spheres of the plasmonic dimer.

For simplicity, %
we approximate the integral by assuming the QNMs are constant throughout the volume $V_\Delta$ of the small sphere, in which case Eq.~(\ref{Eq:QNM_perturbation_theory}) can be written as
\begin{align}
\Delta\tlo_1 -\frac{\tlo_1}{2}\frac{4\pi R_\Delta^3}{3v_1} \Delta\epsilon,
\label{Eq:tlo_1_perturbation_w_sphere}
\end{align}
where %
\begin{align}
v_n = \frac{\langle\langle\mFb_n(\mr)|\mFt_n(\mr)\rangle\rangle}{\epsilon_\text{r}(\mr_0)\mft_n^2(\mr_0)}
\label{Eq:generalized_effective_mode_volume}
\end{align}
is the generalized effective mode volume of QNM $n$,
as introduced for leaky optical cavities in Ref.~\cite{Kristensen_OL_37_1649_2012} and extended to the case of dispersive materials in Ref.~\cite{Sauvan_PRL_110_237401_2013}. Since the location $\mr_0$ is the same location that was chosen for the scaling of the QNMs to unity in Section~\ref{Sec:normalization_of_plasmonic_dimer}, the generalized effective mode volume reduces to the QNM \change{normalization integral} in this case. %

\begin{figure}[htb!]
\centering %
\begin{overpic}[width=9.75cm]{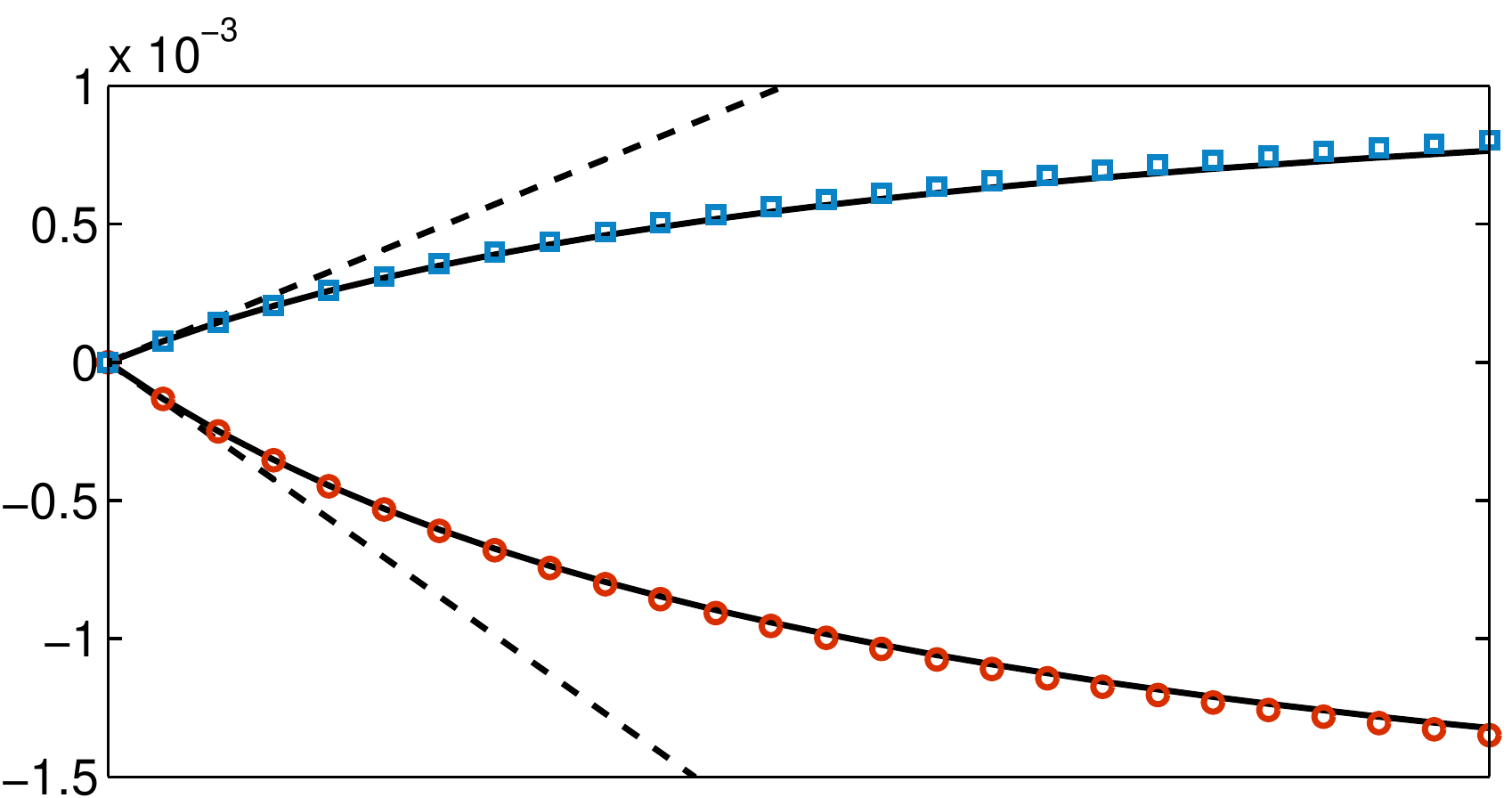}
\put(-11,9){\begin{sideways}Frequency change\end{sideways}}
\put(-6,17){\begin{sideways}$\Delta\tlo_1 L/\text{c}$\end{sideways}}
\end{overpic}\\
\begin{overpic}[width=9.75cm]{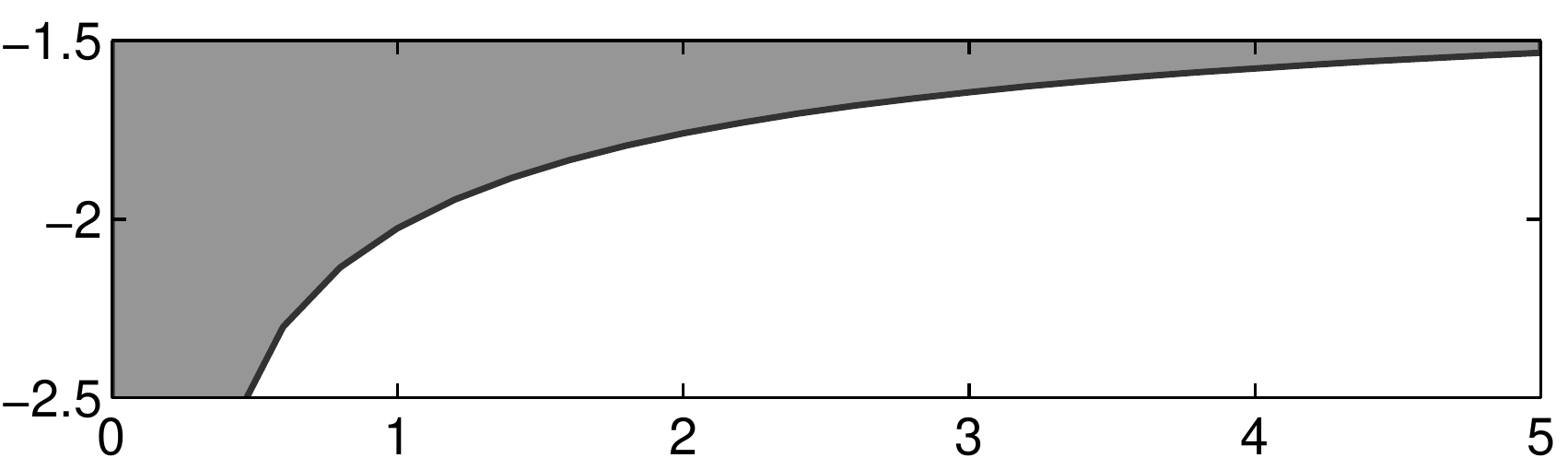}
\put(-11,7){\begin{sideways}Rel. Diff.\end{sideways}}
\put(-6,7.5){\begin{sideways}$\log_{10}\{\mathcal{E}\}$\end{sideways}}
\put(33,-5){Permittivity change, $\Delta\epsilon$} %
\end{overpic}\\[5mm]
\caption{\label{Fig:plasmonicDimer_perturb_in_center_Lmax_4_dEpss_0_to_5}Top: Real (red circles) and imaginary (blue squares) parts of the shift in the complex QNM resonance frequency $\tlo_1$ as a function of permittivity change $\Delta\epsilon$ of a sphere with radius $R_\Delta=R/4$ in the center between the gold spheres of the plasmonic dimer. Dashed curves show the first order perturbation theory result, and full curves show the result of more properly accounting for the vector nature of the QNM by use of the known polarizability of a small sphere. Bottom: Relative difference $\mathcal{E}(\Delta\epsilon)=|\Delta\tlo-\Delta\tlo_1|/|\Delta\tlo|$ between the full curves and the reference calculations in the top panel.} \end{figure}
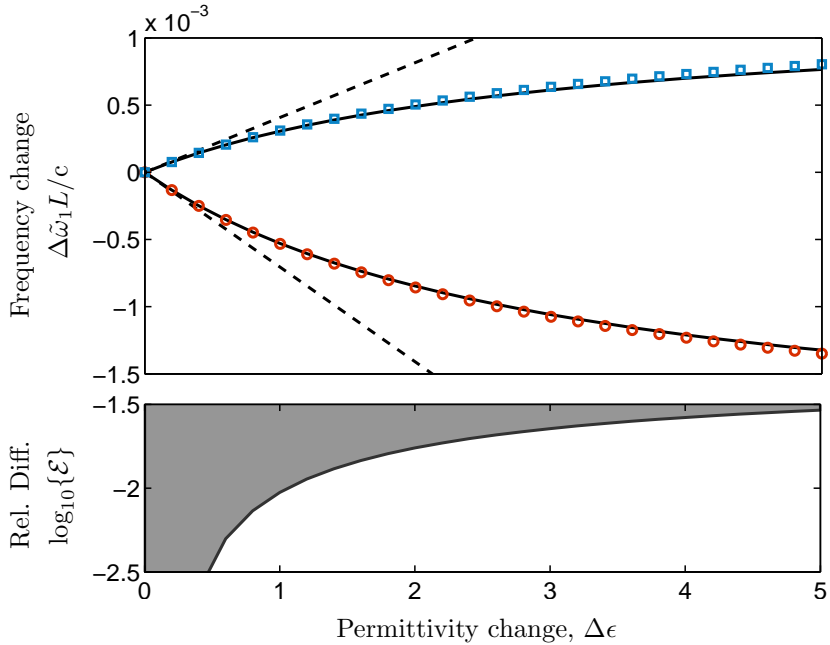

The assumption that the electric field QNM $\mft_1(\mr)$ is constant throughout the volume $\Delta V$ is a remarkably good approximation, even for relatively large spheres, as measured by the difference
\begin{align}
\Delta V = \int_V\frac{\mft_1(\mr)\cdot\mft_1(\mr)}{\mft_1^2(\mr_0)}\ud V - V_\Delta,
\end{align}
where $V_\Delta=\frac{4}{3}\pi R_\Delta^3$ is the volume of the small sphere. For $R_\Delta =0.1R$, the relative difference is $\Delta V/V_\Delta\approx 0.001$, and for $R_\Delta =0.25R$, it is $\Delta V/V_\Delta\approx 0.006$. For the latter case, Fig.~\ref{Fig:barrier_1D_complex_freq_and_perturb} shows the change in resonance frequency as a function of $\Delta\epsilon$ along with the first-order perturbation theory result as calculated from Eq.~(\ref{Eq:tlo_1_perturbation_w_sphere}). The first order perturbation theory correctly captures only the initial change and quickly breaks down as it is incapable of following the nonlinear nature of the complex frequency change at larger values of the permittivity change. Part of the reason for the breakdown can be related to the lack of properly accounting for the vector nature of the QNM. To investigate this effect, %
instead of working with changes in permittivity, we can %
treat the sphere as a constant change in polarization of the form %
\begin{align}
\Delta\mathbf{\tilde{P}} \approx \frac{\alpha_\Delta}{V_\Delta}\mft_1(\mr_0),
\end{align}
where $\alpha_\Delta$ is the polarizability of a sphere with radius $R_\Delta$ and permittivity $\Delta\epsilon=\epsilon_\text{r}-\epsilon_\text{B}$. Using $\epsilon_\text{B}=1$ and the Clausius-Mossotti relation for the polarizability of a sphere with radius $R$ and permittivity $\epsilon_\text{r}$ in the long wavelength limit,
\begin{align}
\alpha_\text{sph} = 4\pi\epsilon_0R^3\frac{\epsilon_\text{r}-1}{\epsilon_\text{r}+2},
\end{align}
we find $\alpha_\Delta = 4\pi\epsilon_0R^3 \Delta\epsilon/(3-\Delta\epsilon)$ and
\begin{align}
\Delta\mathbf{\tilde{P}} = \epsilon_0\frac{\Delta\epsilon}{1+\Delta\epsilon/3}\mft_1(\mr_0).
\end{align}
The factor $\Delta\epsilon/(1+\Delta\epsilon/3)$ evidently plays the role of a linear susceptibility, so we can immediately substitute it in place of $\Delta\epsilon$ in Eq.~(\ref{Eq:tlo_1_perturbation_w_sphere}). In this way, we obtain an approximation which is first order in the constant field, $\mft_1(\mr)\approx\mft_1(\mr_0)$, but which treats the change in polarization to all orders. \change{The solid black curves in Fig.~\ref{Fig:plasmonicDimer_perturb_in_center_Lmax_4_dEpss_0_to_5} shows the results of this approach, which clearly captures the correct resonance frequency change over a much larger range of permittivity changes than the direct application of Eq.~(\ref{Eq:tlo_1_perturbation_w_sphere}). }

\subsubsection*{Shifting boundaries}
\begin{figure}[htb!]
\centering %
\begin{overpic}[width=8cm]{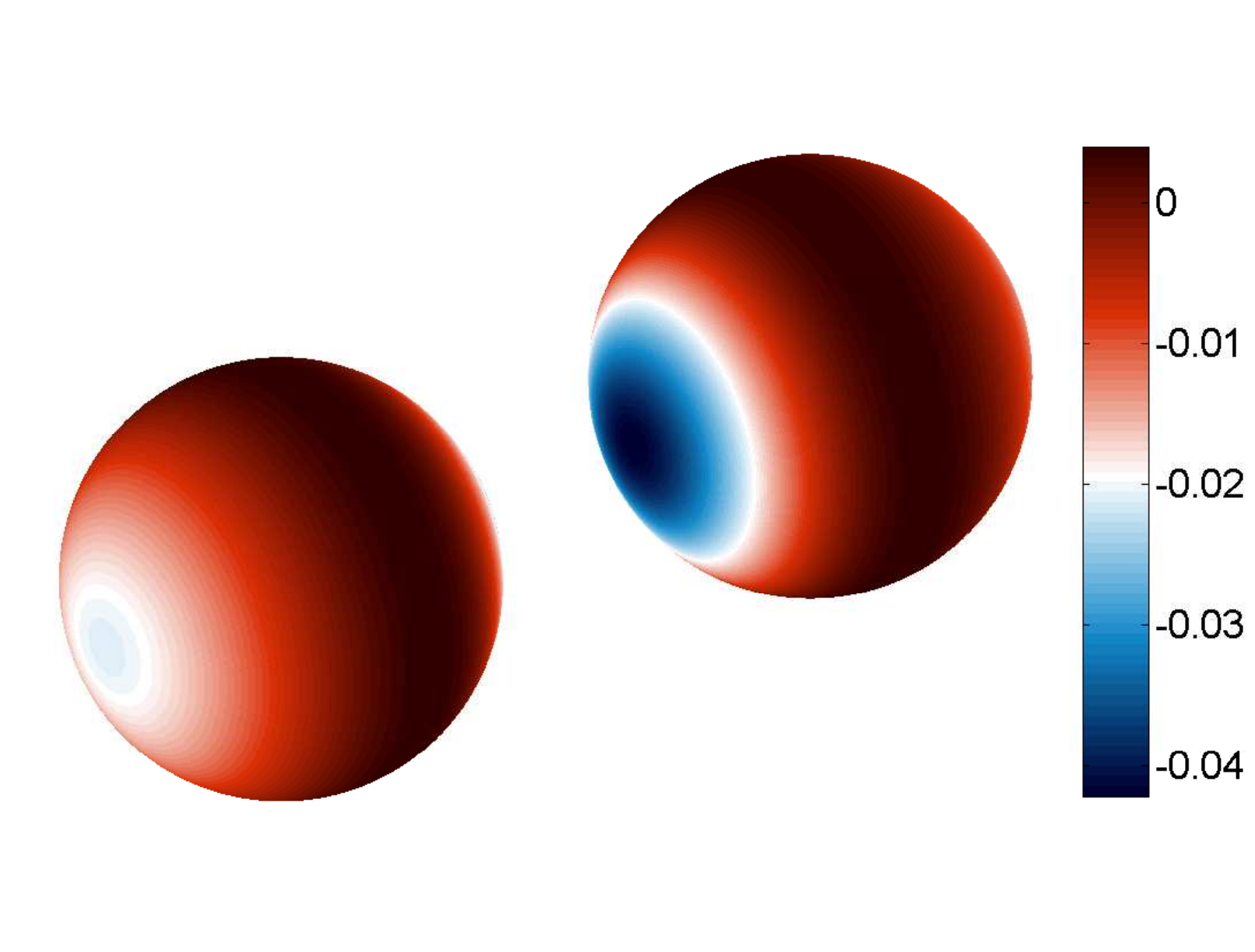}
\end{overpic}\\[-8mm]
\caption{\label{Fig:perturbation_w_shifting_boundaries_integrand}Real part of the differential contribution to the complex frequency shift of $\Delta\tlo_1/\Delta R$ in units of $2\pi\text{c}/d^2$ plotted on the surface of the dimer. } \end{figure}

As an example of the practical use of Eq.~(\ref{Eq:pert_w_shifting_boundaries}), we consider the shift in complex resonance frequency $\tlo_1$ from Fig.~\ref{Fig:Purcell_avs_Lmax_6_rdip_0p5_0p25_0_tloRange_0p05_to_0p25} resulting from increasing or decreasing the size of the gold spheres. At each point on the surface, the (normalized) electric field QNM $\mft_1(\mr)$ is split in a parallel and a normal component to enable the evaluation of the integral in Eq.~(\ref{Eq:pert_w_shifting_boundaries}). Figure~\ref{Fig:perturbation_w_shifting_boundaries_integrand} shows the real part of the differential contribution to the frequency shift plotted on the surface of the dimer. %
Comparing to the field profile in Fig.~\ref{Fig:Purcell_avs_Lmax_6_rdip_0p5_0p25_0_tloRange_0p05_to_0p25}, the shape of the QNM is partly recognized by the relatively large contribution from the parts of the spheres closest to the dimer center. By standard numerical integration, we can now evaluate the surface integrals to find
\begin{align}
\frac{\Delta\tlo_1}{\Delta R}\frac{d^2}{2\pi\text{c}} \approx -0.119440(1) + 0.009168(1)\text{i}. %
\label{Eq:pert_w_shifting_boundaries_tlo_1}
\end{align}

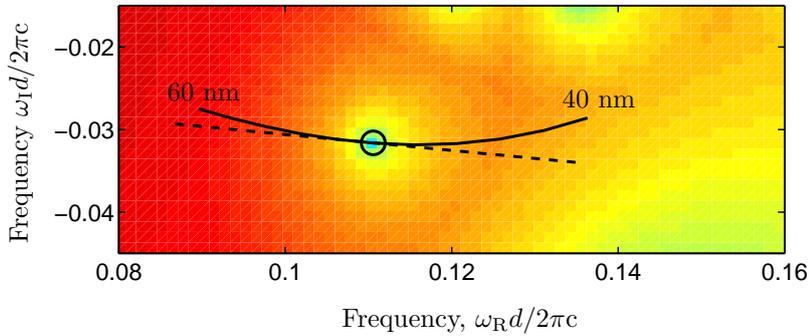
\begin{figure}[htb!]
\centering %
\begin{overpic}[width=10cm]{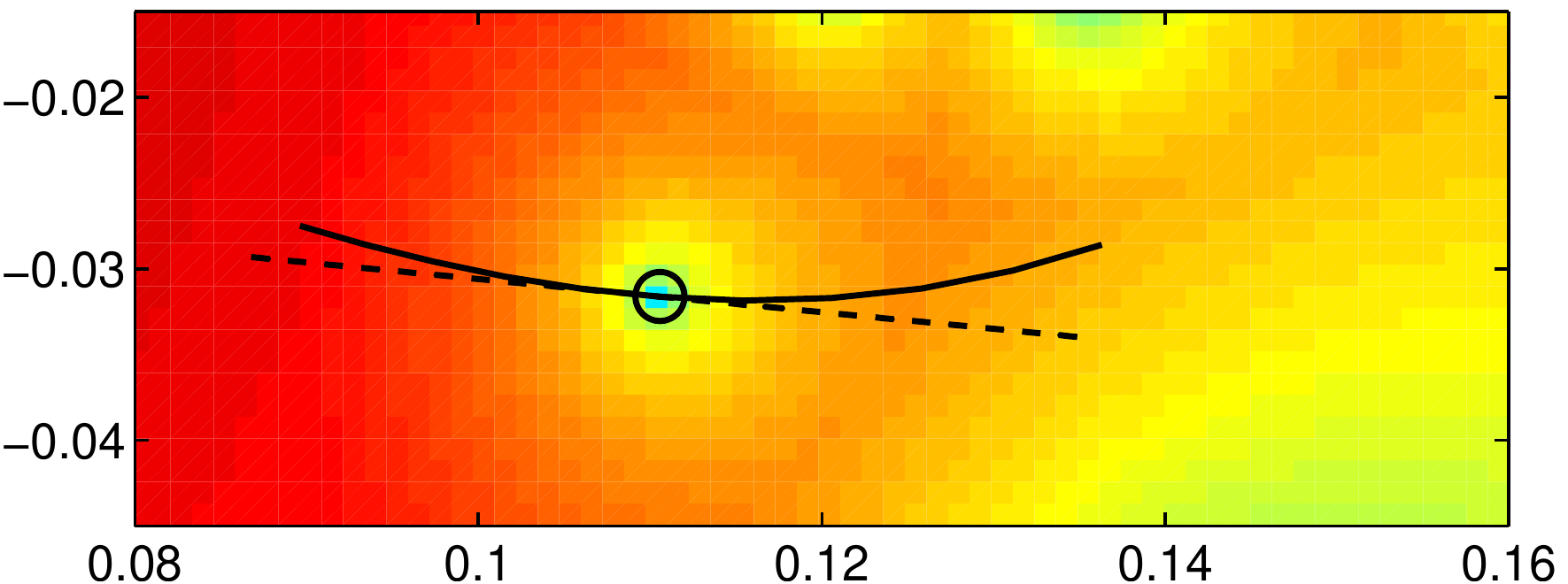}
\put(15,24){60 nm}
\put(67,23){40 nm}
\put(-6,5){\begin{sideways}Frequency $\omega_\text{I}d/2\pi\text{c}$\end{sideways}}
\put(38,-6){Frequency, $\omega_\text{R}d/2\pi\text{c}$}
\end{overpic}\\[5mm]
\caption{\label{Fig:perturbation_w_shifting_boundaries}Zoom of the Complex QNM spectrum in Fig.~\ref{Fig:Purcell_avs_Lmax_6_rdip_0p5_0p25_0_tloRange_0p05_to_0p25} in the vicinity of $\tlo_1$. Full curve shows the movement of the complex resonance frequency when changing the radius of the gold spheres continuously between $40\,$nm and $60\,$nm as indicated. The dashed curve shows the tangent to the curve as given by first order-perturbation theory with shifting boundaries.} \end{figure}

As a reference, we compare to high accuracy calculations of the resonance frequency by use of the VIE method as detailed in Section~\ref{Sec:Calculations_using_VIE}. Figure~\ref{Fig:perturbation_w_shifting_boundaries} shows the movement of $\tlo_1$ in the complex plane as a function of varying radius of the spheres. The first-order perturbation result in Eq.~(\ref{Eq:pert_w_shifting_boundaries_tlo_1}) provides the tangent to the curve in the point $z=\tlo_1$ as expected.

\subsection{Purcell factor calculations}
\label{Sec:Purcell_factor_calculations}
The Purcell factor~\cite{Purcell_PhysRev_69_681_1946} is a common figure of merit for nanophotonic resonators and provides a measure for the enhanced emission rate of an electric dipole source inside the resonator relative to the emission in a homogeneous material. In the so-called weak coupling regime, it is well known that the rate of spontaneous emission of a quantum emitter, with resonance frequency $\Omega$ and dipole moment $\mathbf{d}=d\,\mathbf{e}_\text{d}$, is proportional to the imaginary part of the transverse Green tensor~\cite{Dung_PRA_62_053804_2000,Novotny_2007},
\begin{align}
\Gamma = \frac{2\Omega^2}{\hbar\epsilon_0\text{c}^2}\mathbf{d}\cdot\text{Im}\left\{\mG^\text{EE}_\perp(\mr_0,\mr_0,\Omega)\right\}\cdot\mathbf{d},
\label{Eq:Gamma_general}
\end{align}
where $\hbar$ is the reduced Planck constant, and $\mr_0$ is the location. It follows, that the rate of emission is not an intrinsic property of the emitter, but depends also on its position in 
the environment, as observed also experimentally using a number of different material systems, including fluorescing molecules in front of a dielectric mirror~\cite{Drexhage_JL_1_693_1970}, Rydberg atoms in a superconducting cavity in Ref.~\cite{Goy_PRL_50_1903_1983}, and quantum dots in an optical micro cavity in Ref.~\cite{Gerard_PRL_81_1110_1998}.

Upon dividing the general result in Eq.~(\ref{Eq:Gamma_general}) by the corresponding emission in a homogeneous background medium of refractive index $n_\text{R}$, for which we have the relation
\begin{align}
\mathbf{d}\cdot\text{Im}\left\{\mG^\text{EE}_\text{B}(\mr_0,\mr_0,\Omega)\right\}\cdot\mathbf{d} = \frac{n_\text{R}\Omega}{6\pi\text{c}}d^2,
\end{align}
we obtain the Purcell effect in the form
\begin{align}
F_\text{P} = \frac{6\pi\text{c}}{n_\text{R}\Omega} \mathbf{e}_\text{d}\cdot\text{Im}\left\{\mG^\text{EE}_\perp(\mr_0,\mr_0,\Omega)\right\}\cdot\mathbf{e}_\text{d}.
\label{Eq:Purcell_factor_def}
\end{align}
If the quantum emitter is placed in a resonator, we can %
expand the transverse Green tensor by use of the QNMs. For instance, using the first of the expressions in Eq.~(\ref{Eq:GE_transverse_two_versions}), we find
\begin{align}
F_\text{P} \approx \frac{3\pi\text{c}^3}{n_\text{R}\Omega} \mathbf{e}_\text{d}\cdot\text{Im}\left\{ \sum_m\frac{\mft_m(\mr_0)[\mft_m(\mr_0)]^\text{T}}{\tlo_m(\tlo_m-\Omega)}\right\}\cdot\mathbf{e}_\text{d}.
\label{Eq:F_P_sum_QNMs}
\end{align}
When a single QNM $\mft_m(\mr_0)$ is sufficient to adequately approximate the Green tensor, and when the emitter frequency is tuned to coincide with the real part of the QNM resonance frequency, $\Omega=\omega_m$, we find the approximate relation
\begin{align}
F_\text{P} \approx\frac{3\pi\text{c}^3}{n_\text{R}\omega_m}\text{Im}\left\{\text{i}\frac{\mft_m^2(\mr_0)}{\omega_m\gamma_m}\right\}.
\end{align}
Last, using $Q=\omega_m/2\gamma_m$, as defined in Eq.~(\ref{Eq:Q_n_def}), and $\epsilon_\text{r}=n_\text{R}^2$, we can write the expression in the exact form due to Purcell~\cite{Purcell_PhysRev_69_681_1946},
\begin{align}
F_\text{P} \approx \frac{3}{4\pi^2}\left(\frac{\lambda_\text{res}}{n_\text{R}}\right)^3\frac{Q}{V_\text{eff}},
\label{Eq_Purcell_factor}
\end{align}
where $\lambda_\text{res}=\omega_m/2\pi\text{c}$ is the (real) resonance wavelength, %
and the effective mode volume $V_\text{eff}$ is calculated from the generalized effective mode volume $v_n$ from Eq.~(\ref{Eq:generalized_effective_mode_volume}) as
\begin{align}
\frac{1}{V_\text{eff}} = \text{Re}\left\{\frac{1}{v_m}\right\}.
\end{align}
The expression for the emission rate in Eq.~(\ref{Eq:Gamma_general}) is valid for emitters in the vicinity nano photonic resonators made of general dispersive \change{and absorptive} materials as long as the emitter is placed outside the material, which typically are the positions of interest in physical realizations. For example, in the case of the plasmonic dimer below, the emitter will be placed in vacuum at the position directly in between the two spheres. The above derivations follow the approach of Ref.~\cite{Kristensen_OL_37_1649_2012}. An identical result was presented in the more general case of dispersive \change{and absorptive} materials in Ref.~\cite{Sauvan_PRL_110_237401_2013}, see also Refs.~\cite{Ge_NJP_16_113048_2014, Muljarov_PRB_94_235438_2016, Kristensen_PRA_92_053810_2015, Muljarov_PRA_96_017801_2017, Kristensen_PRA_96_017802_2017}. If the emitter is not tuned into resonance with the QNM, one can write the approximate Purcell factor by multiplying a correction factor onto the expression in Eq.~(\ref{Eq_Purcell_factor}), but it is typically much easier to work directly with the QNM Green tensor expansion in Eq.~(\ref{Eq:F_P_sum_QNMs}), as we do below.

\subsubsection{Purcell factor in gap center of plasmonic dimer }
\label{Sec:Purcell_factor_of_plasmonic_dimer}
The plasmonic dimer supports a large number of QNMs, some of which show up as distinct peaks in scattering spectra or Purcell factor calculations, as noted already in a somewhat qualitative manner in connection with Figs.~\ref{Fig:Purcell_avs_Lmax_6_rdip_0p5_0p25_0_tloRange_0p05_to_0p25} and \ref{Fig:errorMap_wideRange}. In this Section, we discuss the practical application of the QNM approximations in detail. %

The red dashed curve in Fig.~\ref{Fig:Purcell_avs_Lmax_6_rdip_0p5_0p25_0_tloRange_0p05_to_0p25} shows the result of approximating the electric field Green tensor by a single term as
\begin{align}
\mG^\text{EE}_\perp(\mr,\mr',\omega) \approx \frac{\text{c}^2}{2}\frac{\mft_1(\mr)\mft_1(\mr')}{\tlo_1(\tlo_1-\omega)}.
\label{Eq:G_GoldDimer_0_0_0_oneMode}
\end{align}
The approximate expression arguably captures the main qualitative features of the spectrum around the resonance, and the relative error at the peak is as low as a few percent. The agreement can be dramatically improved, however, by using the symmetry property of the QNMs to ensure the correct crossing behavior of the Green tensor. %
By adding to Eq.~(\ref{Eq:G_GoldDimer_0_0_0_oneMode}) the corresponding term from the mode $\mft_1^*(\mr)$ at the complex resonance frequency $-\tlo_1^*$, we find
\begin{align}
\mG^\text{EE}_\perp(\mr,\mr',\omega) \approx \frac{\text{c}^2}{2}\frac{\mft_1(\mr)\mft_1(\mr')}{\tlo_1(\tlo_1-\omega)} + \frac{\text{c}^2}{2}\frac{\mft_1^*(\mr)\mft_1^*(\mr')}{\tlo_1^*(\tlo_1^*+\omega)},
\label{Eq:G_GoldDimer_0_0_0_oneMode_star}
\end{align}
the result of which is shown with the red solid curve in in Fig.~\ref{Fig:Purcell_avs_Lmax_6_rdip_0p5_0p25_0_tloRange_0p05_to_0p25}. Moreover, in this case, the peak relative error is of the order of just one part in a thousand, and the range of validity of the approximation is greatly improved as well. %
It is also interesting to note that using a single QNM approximation with the alternative Green tensor expansion %
\begin{align}
\mG^\text{EE}_\perp(\mr,\mr',\omega) \approx \frac{\text{c}^2}{2}\frac{\mft_1(\mr)\mft_1(\mr')}{\omega(\tlo_1-\omega)}
\label{Eq:G_GoldDimer_0_0_0_oneMode_alt}
\end{align}
results in an approximation that is worse (not shown in Fig.~\ref{Fig:Purcell_avs_Lmax_6_rdip_0p5_0p25_0_tloRange_0p05_to_0p25}) because of the built-in divergence of the approximate Green tensor at $\omega=0$, which is not present in the full (transverse) Green tensor. %

As this example shows, even at the position $\mr_0$ \change{in the middle} between the spheres, %
where we do not expect the QNM expansion to converge to the correct value of the Green tensor, an expansion in terms of a few QNMs can be a very good approximation. For Purcell factor calculations as these, there is also the interesting fact, that most emitters would themselves be treated one way or another as having an intrinsic passive electromagnetic response. The passive response of a quantum dot, for example, could be modeled as a tiny dielectric sphere. Because of the small size, the inclusion of a tiny sphere will only cause a perturbative shift in the lower lying QNMs, but the position $\mr=\mr'$ will be inside the tiny sphere, wherefore the QNM expansion is expected to be convergent, as discussed in Section~\ref{Sec:convergence_radius_dimer_of_gold_nano_spheres}.

\subsubsection*{Large bandwidth approximation}
\begin{figure}[htb]
\centering %
\begin{overpic}[width=9.4cm]{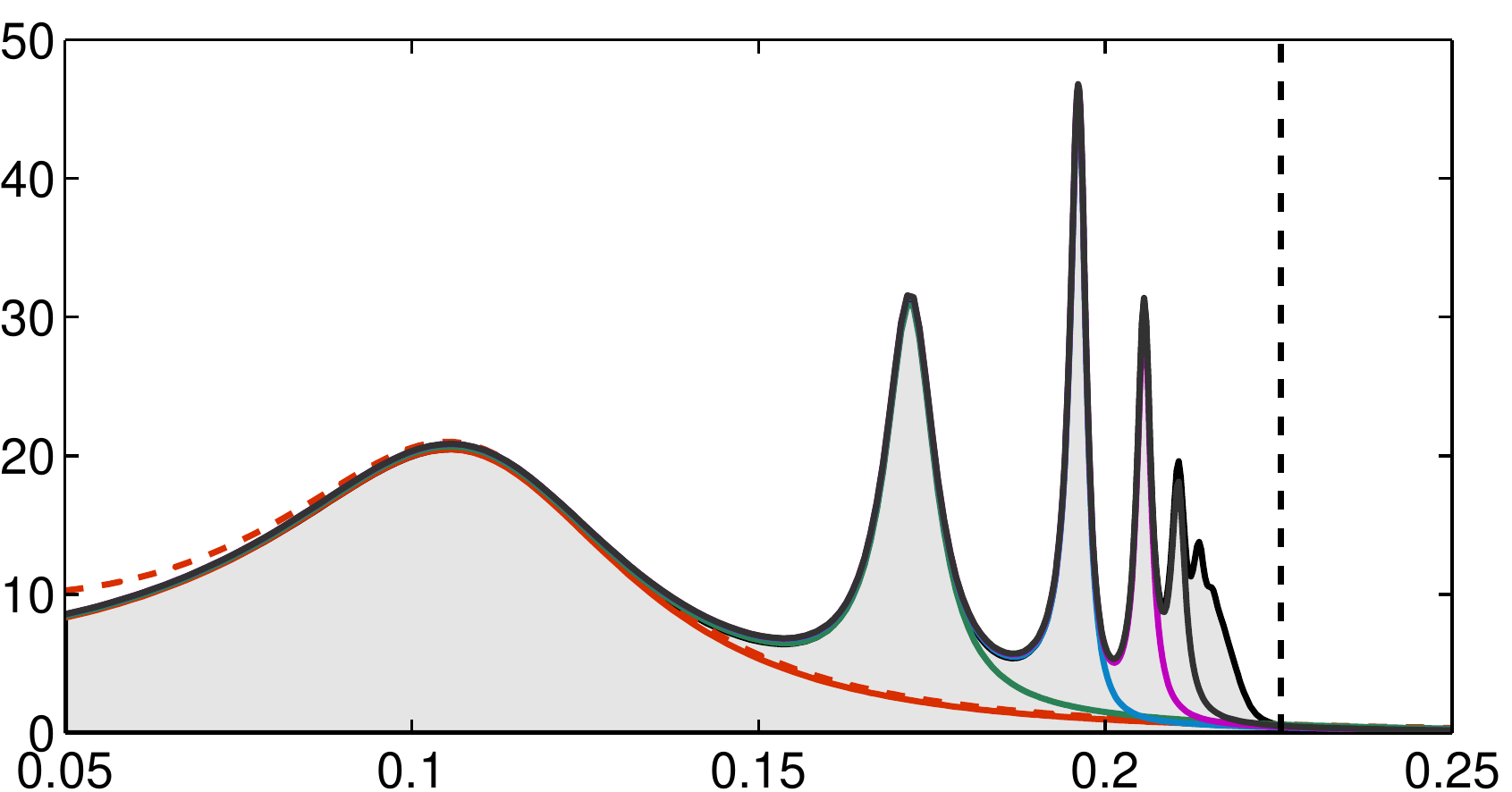}
\put(-14,8){\begin{sideways}Purcell factor, $F_\text{P}$\end{sideways}}
\end{overpic}\\[-2.4mm]
\begin{overpic}[width=9.4cm]{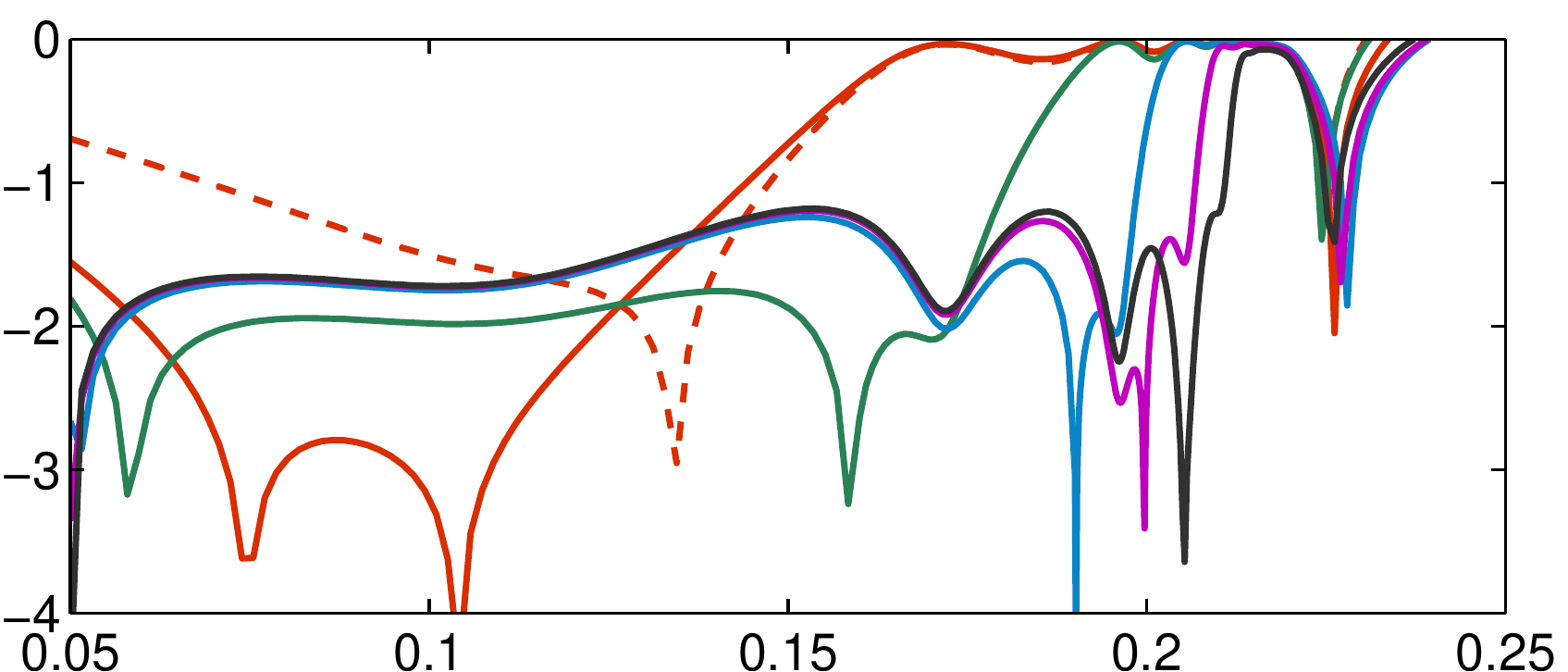}
\put(-14,8){\begin{sideways}Relative Error\end{sideways}}
\put(-7,5){\begin{sideways}$\text{log}_{10}|\Delta F_\text{P}|/|F_\text{P}^\text{ref}|$\end{sideways}}
\put(30,-6){Resonance frequency $\omega d/2\pi\text{c}$}
\end{overpic}\\[5mm]
\caption{\label{Fig:Purcell_Y_pos_0_0_0_Lmax_12_w_QNM_1_to_5_spectrum}Top: Purcell factor, Eq.~(\ref{Eq:Purcell_factor_def}), for a dipole oriented along the $y$-axis at the position $\mr_0$ in the center between the two gold spheres as in Fig.~\ref{Fig:Purcell_avs_Lmax_6_rdip_0p5_0p25_0_tloRange_0p05_to_0p25} but showing a larger bandwidth. Grey shading %
shows the reference calculation $F_\text{P}^\text{ref}$. \change{The red dashed curve shows the approximate Purcell factor resulting from Eq.~(\ref{Eq:G_GoldDimer_0_0_0_oneMode}), and the solid colored lines show the approximations $F_{\text{P},N}$ resulting from Eq.~(\ref{Eq:Purcell_factor_def}) by substituting the QNM approximations to the Green tensor in Eq.~(\ref{Eq:G_QNM_1_to_N_modes_star}) for increasing values of $N$. Red, green, blue, purple and dark grey curves show the result $F_{\text{P},N}$ with $N=1,2,3,4,5$, respectively. The dashed vertical line indicates the position of $\omega_{\text{sp}}$.} Bottom: Relative error \change{$|\Delta F_\text{P}(\omega)|/|F_\text{P}^\text{ref}(\omega)|$ corresponding to the calculations in the top panel.}} %
\end{figure}

To obtain a deeper understanding of the convergence properties of the QNM Green tensor expansion, it is illustrative to calculate it in a wider frequency range that covers several resonances. The high-symmetry point between the two nano spheres is well suited for this purpose, because a large fraction of the QNMs vanish at this point. Tables~\ref{Tab:k0_Qs_d} and \ref{Tab:v_ns} in Appendix~\ref{Sec:Calcualtion_details_for_plasmonic_dimer} list the complex resonance frequencies as well as the generalized effective mode volumes of the five first QNMs of interest in the region $0<\omega_n<\omega_\text{sp}$ and $\gamma_1\leq\gamma_n$, \change{where $\omega_\text{sp}=\omega_{\text{p}}/\sqrt{2}$ is the frequency %
at which the in-plane wave vector $k_{\|}(\omega)$ %
of a surface plasmon polariton on a planar Drude metal surface in air tends to infinity~\cite{Novotny_2007}. In this limit, the associated in-plane wavelength $\lambda_{\|}(\omega)=2\pi/k_{\|}(\omega)$, which describes propagation along the interface, tends to zero, wherefore the finite curvature of the spheres becomes irrelevant and the spheres look locally like plane surfaces. Nevertheless, the spherical shapes lead to a periodicity requirement on the QNMs, which can only be met at certain in-plane wavelengths. As the spacing between the in-plane wavelengths becomes smaller, one can fit an ever increasing number of oscillations around the sphere, and this effect shows up as an accumulation point in the spectrum at $\omega\approx\omega_\text{sp}$.} %
Figure~\ref{Fig:Purcell_Y_pos_0_0_0_Lmax_12_w_QNM_1_to_5_spectrum} shows the Purcell factor from Fig.~\ref{Fig:Purcell_avs_Lmax_6_rdip_0p5_0p25_0_tloRange_0p05_to_0p25} %
along with Purcell factors calculations based on Green tensor approximations of the form
\begin{align}
\mG^\text{EE}_N(\mr,\mr',\omega) = \sum_{n=1}^N\frac{\text{c}^2}{2}\frac{\mft_n(\mr)\mft_n(\mr')}{\tlo_n(\tlo_n-\omega)} + \frac{\text{c}^2}{2}\frac{\mft_n^*(\mr)\mft_n^*(\mr')}{\tlo_n^*(\tlo_n^*+\omega)},
\label{Eq:G_QNM_1_to_N_modes_star}
\end{align}
for increasing values of $N$. \change{To simplify the notation, we dropped the ``$\perp$'' in Eq.~(\ref{Eq:G_QNM_1_to_N_modes_star}), since the QNM approximations are inherently transverse, as discussed in Section~\ref{Sec:Electric_field_Green_tensor_in_three_dimensions}.} %
Each additional term in the sum results in the sampling of an extra peak in the spectrum. \change{To quantify the error, we consider the Purcell factor difference $\Delta F_\text{P}=F_{\text{P},N}-F_\text{P}^\text{ref}$, where $F_{\text{P},N}$ is calculated with Eq.~(\ref{Eq:Purcell_factor_def}) by substituting the QNM approximation to the Green tensor in Eq.~(\ref{Eq:G_QNM_1_to_N_modes_star}), and $F_\text{P}^\text{ref}$ is the reference calculation.} From the plot of the relative error, the convergence appears to be non-trivial (at least initially), in that, for a given peak in the spectrum, the error is seen to initially drop as the corresponding QNM is included in the sum, but then subsequently increase as more terms are added and the width of the overall agreement increases. When including the first five QNMs in the region with $\omega_n<\omega_\text{sp}$ and $\gamma_n\leq\gamma_1$ , there is a residual relative error on the order of several percent. Obviously, if the Green tensor expansion is \change{consistent}, we expect that this error can be lowered by sufficiently many additional QNMs. \change{We note, that the estimated relative error in the numerical calculation of both the QNMs and the reference spectrum are several orders of magnitude lower than the disagreement %
Therefore, we trust that the observed residual error in Fig.~\ref{Fig:Purcell_Y_pos_0_0_0_Lmax_12_w_QNM_1_to_5_spectrum} is entirely due to the approximation in Eq.~(\ref{Eq:G_QNM_1_to_N_modes_star}) and is not the result of limited numerical accuracy.} 
In practice, however, there will be an infinite number of QNMs in the region of interest at smaller and smaller distances to an accumulation point around $\omega_{\text{sp}}$. The ever decreasing distance between the QNMs makes the numerical determination and normalization of the modes increasingly difficult. For practical purposes, therefore, a controlled approximation of the full spectrum by use of an expression such as Eq.~(\ref{Eq:G_QNM_1_to_N_modes_star}) with $N\rightarrow\infty$ seems to be prohibitively difficult for plasmonic resonators, %
unless the QNMs can be calculated analytically. %
For dispersionless materials, \change{or possible other materials which show no accumulation point in the spectrum,} these shortcomings may not be as severe. Nevertheless, the example arguably serves to illustrate some general points about the power and shortcomings of QNM Purcell factor calculations, namely that the QNM approximations in general are remarkably good when only the dominant QNMs are included, but the accuracy or band width provided by more QNMs may not be worth the additional computational costs and complexity of the model.

\section{Conclusions}
\label{Sec:Conclusions}
In this Tutorial, we have presented a bi-orthogonal approach to modeling electromagnetic resonators using QNMs, and used this approach, in combination with examples of a dielectric barrier and a plasmonic dimer in one and three dimensions, respectively, to analyze and illustrate the usefulness and limitations of resonator models based on QNMs.

Starting from the QNMs as solutions to the Maxwell curl equation with the Silver-M{\"u}ller radiation condition, we have illustrated how a bi-orthogonal approach leads to a useful definition of adjoint QNMs and an associated projection operator. When the operator is applied to project a QNM onto itself, it provides a well-known formulation of the QNM norm. By applying the operator to a certain class of electromagnetic field problems, including Green tensor calculations, one can derive formal expansions of the fields in terms of QNMs. For the case of the Green tensor, we have discussed how this approach compares to an alternative approach in which the QNMs are directly associated with the residues of the analytical continuation of the Green tensor.

The question if and when the formal solutions are convergent and consistent was discussed in some detail, building on ideas from the literature. %
Much of the convergence analysis is based on the central enabling fact, that QNM expansions of the electromagnetic Green tensor $\mmG(\mr,\mr',\omega)$ is possible for positions within the resonators and possibly beyond, depending on the parameters of the resonator and position $\mr'$, as discussed in Section~\ref{Sec:Regions_of_completeness}. In general, for a given choice of $\mr'$, one can numerically trace out a region of convergence for which the Green tensor expansion is expected to converge to the correct result in the limit when infinitely many QNMs are included in the expansion. Even then, in certain cases there might be additional contributions stemming from branch cuts in the analytical continuation of the Green tensor. %
At positions inside the region of convergence, %
the actual convergence is \change{generally a relatively slow function of the number of QNMs $N$; for} %
the dielectric barrier, where we can investigate the convergence explicitly, we find it to be of the order $1/N$ in general, cf. Figs.~\ref{Fig:G_k0_4_dielectricBarrier_nR_pi_log10_relErr_vs_log10_2Np1_xp_0} and \ref{Fig:double_barrier_1D_k0_Q_convergence_log10_rel_Error_vs_log10_2Np1}.  For certain special points in the Coupled Mode Theory (CMT) calculations, such as the center of the barrier and the rightmost boundary, the QNM approximation was found to sometimes converge as $1/N^2$ as seen in Figs.~\ref{Fig:CMT_k0_4_dielectricBarrier_nR_pi_log10_relErr_vs_log10_N_x_m0p25_0p25} and \ref{CMT_k0_4_dielectricBarrier_nR_pi_log10_relErrs_at_right_side_vs_log10_2Np1_x_m0p25_0p25}. %

Depending on the calculation method and level of accuracy, the computational costs for individual QNMs are relatively large. In combination with the modest convergence properties, this means that the appeal of a QNM decomposition is largest when the response of the system is governed mainly by a few QNMs. In such cases, a QNM approach provides a physically appealing and analytically tractable mode decomposition which is often very accurate. In the transmission spectrum for the dielectric barrier in Fig.~\ref{Fig:transmission_spectrum_dielectricBarrier_nR_pi}, the relative error of the one and three QNM approximations are as low as seven percent and two percent, respectively, as analyzed in Section~\ref{Sec:Scattering_calculations_for_the_dielectric_barrier}. Similarly, in the Purcell factor approximation in Fig.~\ref{Fig:Purcell_avs_Lmax_6_rdip_0p5_0p25_0_tloRange_0p05_to_0p25}, the relative errors of the one and two QNM approximations are as low as a few percent and one part in a thousand, as discussed in Section~\ref{Sec:Purcell_factor_of_plasmonic_dimer}.

The final part of the Tutorial was devoted to four examples of QNM modeling: Scattering calculations and CMT, coupled resonators, perturbation theory, and Purcell factor calculations. The derivation of the scattering calculations in Section~\ref{Sec:Scattering_calculations} provided an interesting display of the importance of the radiation conditions. Indeed, the total field in a scattering calculation does not fulfill the radiation condition, and therefore cannot be \change{expanded in terms of the QNMs by use of the projection operator derived in Section~\ref{Sec:adjoint_QNMs_and_normalization}.} %
The scattered field, however, does fulfill the radiation condition, and by splitting the total field in an incoming and a scattered field part, the projection onto the QNMs lead to a physically appealing expression in which the incoming field acts as a source for the total field inside the resonator. To derive expressions for the scattered field, one can instead work with an integral equation and the well-known QNM expansion of the Green tensor. The same integral equation can be used to set up a linear equation system for calculating the hybridized QNMs in coupled resonances, as illustrated  in Section~\ref{Sec:Coupled_resonators}. Conceptually, such an approach is well known from tight binding type theories, but the exponential divergence of the QNMs outside the individual resonators complicates the formulation in practice. Sec.~\ref{Sec:Perturbation_theory} presented a number of applications of first-order perturbation theory in order to illustrate how the QNM \change{normalization integral} arises naturally in this process, and to show practical limitations and extensions, such as the case of shifting boundaries and the use of the polarizability of small scatterers to improve the accuracy. Lastly, in sec.~\ref{Sec:Purcell_factor_calculations} we discussed the use of QNMs in the calculation of Purcell factors, and showed how the original formula due to Purcell arises naturally from a single QNM approximation to the Green tensor.

We hope that this Tutorial can serve to illustrate how the use of QNMs provides a mathematically rigorous framework for the modeling of a multitude of different phenomena associated with electromagnetic resonators. As argued in the introduction, many models of resonator phenomena --- such as laser models or models for light propagation through coupled cavity-waveguide structures --- have been implicitly relying on a mode decomposition, although almost exclusively by treating the cavity as a closed system and rarely by explicitly defining the cavity modes as QNMs obeying a radiation condition. By fully exploiting a QNM approach, one can usually obtain models of the same complexity, but with explicit and precise definitions of the various parameters in the models such as mode volumes and coupling constants.\newline

\section*{Acknowledgments}
We would like to express our sincere gratitude to Dirk Hundertmark (KIT), Herbert Koch (Bonn University), \change{ Stephen Hughes (Queen's University), Felix Binkowski and Lin Zschiedrich (Zuse Institute Berlin), and Christian Wolff (SDU) for numerous inspiring discussions. A special thank to Jakob Rosenkrantz de Lasson, who developed and implemented the code for the VIE formulation calculations.}

\subsubsection*{Funding}
Deutsche Forschungsgemeinschaft (DFG), SFB (951 HIOS B10, Project 182087777); German Federal Ministry of Education and Research,  Photonics Research Germany (Project~13N14149);. Deutsche Forschungsgemeinschaft (DFG), DIP (grants FO 703/2-1 and SCHM 1049/7-1).

\appendix

\section{Practical convergence studies}
\label{App:Practical_convergence_studies}
Contemporary modeling of electromagnetic scattering relies heavily on numerical solutions of partial differential equations. These numerical solutions naturally come with associated numerical errors, and it is therefore of considerable interest to have a systematic way of estimating the accuracy. In this appendix, we discuss how one can use the mathematical definitions of convergence and consistency to assess the convergence and ultimately assign an estimated error to a numerical calculation in a systematic way.

\subsection{Convergence and consistency}
\label{Sec:Convergence_and_consistency}
Any discretization will introduce some kind of parameter $h$ controlling how fine the discretization is. The implicit assumption is that, as the size of $h$ is varied to make the discretization finer and finer, the resulting calculated values become closer and closer, so that the sequence of results is convergent:
\subsubsection*{Convergence (Cauchy criterion)}
A sequence of numbers $s_1, s_2, s_3, ... , s_n$ is convergent if for any  $\epsilon>0$, there exists a number $N$, so that $|s_m-s_n|<\epsilon$ for all $n,m>N$.
\newline

In practice, however, we require not only that a sequence is convergent, but also that it tends to the correct result --- a requirement known as consistency:
\subsubsection*{Consistency}
A sequence of numbers $s_1, s_2, s_3, ... , s_n$ is convergent with limit $S$ if for any $\epsilon>0$, there exists a number $N$, so that $|S-s_n|<\epsilon$ for all $n>N$. %
\newline

The ideas behind the two definitions are reflected in different approaches for assessing the convergence of a given sequence, which we shall refer to as convergence and consistency studies. %
Convergence studies rely on the Cauchy criterion and therefore can be performed on any set of data. Consistency studies, on the other hand, can only be applied to cases in which the result is known analytically. In such (rare) cases, one can directly plot the absolute or relative error as a function of the parameter(s) limiting the accuracy. In both cases, if the norm $|S-s_n|$ or $|s_m-s_n|$ decrease in a systematic way as we increase the parameter(s) limiting the accuracy, we shall assume that the sequence is convergent. As noted in the introduction, however, we should also be interested in somehow estimating the error in the calculations. While this is straightforward in cases where the result is known, it requires a bit of work (and additional assumptions) for convergence studies based on the Cauchy criterion. Ideally, one should use both approaches to assess the accuracy. Indeed, there can easily be systematic errors that only appear in direct comparison to analytical solutions --- this is the case, for example, with reflections from PML boundaries. One would then first check for consistency by use of an auxiliary problem with comparable physical dimensions and materials and comparing to a high-accuracy reference calculation to identify the parameter(s) limiting the accuracy. Only thereafter does it make sense to worry about the convergence properties and the accuracy of the actual problem at hand.

In practical calculations, one can vary a (sometimes rather large) number of parameters, and the error will depend stronger on some of these parameters than on others. A prominent example of a parameter liming the accuracy is the size of the discrete triangles making up the calculation mesh in a BEM calculation, in which case the error is expected to vanish only in the limit of vanishing discretization size. For a generic parameter $h$ and a convergent series of function values $f(h)$ with limit $f_0$, we can write the calculated value at any $h$ as the correct value plus an error term as
\begin{align}
f(h) = f_0 + \mathcal{E}(h),
\label{Eq:f_w_error}
\end{align}
where $\mathcal{E}(h)\rightarrow0$ for $h\rightarrow0$. Assuming that one can identify the parameters limiting the accuracy, one can typically also identify the functional form of the error term - even if one does not know the correct limit $f_0$. To this end, the observations in Sections~\ref{Sec:Eq:Diff_vs_h} and \ref{Sec:Eq:Diff_vs_p_exp} below may be of help. 

In cases where we are able to obtain a model for the error term, we can estimate the limiting value $f_0$ by rewriting Eq.~(\ref{Eq:f_w_error}) as

\begin{align}
f_0 \approx  f(h_\text{min}) - \mathcal{E}(h_\text{min}),
\end{align}
where $h_\text{min}$ is the smallest value of $h$ used in the calculations. In addition to the estimated value of $f_0$, we shall generally use the absolute value of $\mathcal{E}(h_\text{min})$ as a conservative estimate of the numerical error on the estimated value. We write the estimated numerical error in parenthesis immediately behind the digit(s) to which it pertains. A value of $\pi\approx3.15(2)$ thus signifies that there is an estimated error of $\Delta\pi=\pm0.02$, so that we expect the true value of $\pi$ to lie in the interval $3.13<\pi<3.17$.

\subsubsection{Polynomial convergence}
\label{Sec:Eq:Diff_vs_h}
In many numerical calculations, the error tends to zero in a polynomial fashion for which we can write the dominant term as
\begin{align}
\mathcal{E}(h) = \mathcal{E}_0 h^\alpha,
\label{Eq:polynomialError}
\end{align}
\change{where $\mathcal{E}_0$ and $\alpha$ are the initially unknown parameters characterizing the dominant polynomial behavior.}
\label{Sec:Diff_x}
Assuming Eq.~(\ref{Eq:f_w_error}) to hold, we can eliminate the unknown correct limit $f_0$ by considering the difference
\begin{align}
f(h) - f(xh) &= \mathcal{E}(h)-\mathcal{E}(xh) \\
&=\mathcal{E}_0h^\alpha[1-x^\alpha] \label{Eq:Diff_vs_h}
\end{align}
for $x<1$. Taking the logarithm on both sides of the equation we then find that
\begin{align}
\log\left(f(h) - f(xh)\right) = \alpha\log(h) + \log\left(\mathcal{E}_0[1-x^\alpha]\right),
\label{Eq:logDiff_vs_log_h}
\end{align}
so that one can conveniently extract the exponent in the (dominant) polynomial error term by a simple fit of the logarithm of the differences as a function of the logarithm of $h$. One can conveniently limit the numerical work, by only calculating the function at exponentially spaced values as $f(h), f(xh), f(x^2h), f(x^3h), ...\;$. Traditionally, $x=1/2$ is often used, corresponding to successively halving the parameter $h$. %

\subsubsection{Exponential convergence}
\label{Sec:Eq:Diff_vs_p_exp}
In some cases, the numerical error tends to zero for increasing $p$ in an exponential fashion as
\begin{align}
\mathcal{E}(p) = \mathcal{E}_0 k^p,
\label{Eq:exponentialError}
\end{align}
for $k<1$. To make this model fit with Eq. (\ref{Eq:f_w_error}), we can set $h=k^p$. As in the case of a polynomial convergence, we can eliminate the unknown limit $f_0$ by forming the difference
\begin{align}
f(p) - f(p+q) &= \mathcal{E}_0\big[k^p-k^{p+q}\big]\\
&=\mathcal{E}_0k^p\big[1-k^{q}\big], \label{Eq:Diff2_vs_p}
\end{align}
for $q>0$. Taking the logarithm on both sides we find
\begin{align}
\log(f(p) - f(p+q)) &= p\log(k)+\log(\mathcal{E}_0[1-k^q]),
\label{Eq:logDiff_vs_p}
\end{align}
so that one can convenient extract the base $k$ and subsequently the parameter $\mathcal{E}_0$ from a simple fit of the logarithm of the differences as function of $p$.

\section{Calculation details for the plasmonic dimer}
In the VIE calculations for the plasmonic dimer, the spherical wave functions constitute a complete basis within the individual spheres. Therefore, we can assume the expansion to converge to the correct solution in the limit $l_\text{max}\rightarrow\infty$, provided the numerical solution of the matrix problem in Eq.~(\ref{Eq:LippmannSchwinger_discretized}) does not itself introduce errors. In practice, there will be truncation errors associated with the representation of the data in finite precision, but these errors are expected to be at least thousand times smaller than the estimated errors below, so we shall ignore them in this analysis. In practice, the calculations were performed by assuming that the material parameters of the Drude permittivity is given exactly by $\hbar\omega_\text{p}=7.9$\,eV and $\hbar\gamma=0.06$\,eV\change{, which are similar to values that have been found to provide a reasonable description for gold, see for example Refs.\cite{Kreiter_PRB_65_125415_2002,Grady_CPL_399_167_2004}}.  %

For the calculation of $\tlo_1$, Fig.~\ref{Fig:Dp1_vs_Lmax_real_and_imag} shows, as a function of the cut-off parameter $l_\text{max}$, the logarithm of $D_{+1}(l_\text{max}) = \tlo_\text{num}(l_\text{max})-\tlo_\text{num}(l_\text{max}+1)$. %
To a good approximation, the data points for both the real and the imaginary parts in Fig.~\ref{Fig:Dp1_vs_Lmax_real_and_imag} fall on a straight line, indicating an exponential convergence in both cases, cf. the discussion in Section~\ref{Sec:Eq:Diff_vs_p_exp}. For the other QNM resonance frequencies of interest, we find a similar behavior. Assuming that the finite value of $l_\text{max}$ is the dominating source of error in the calculation, we can use fitted values of $k$ and $\mathcal{E}_0$ in Eq.~(\ref{Eq:exponentialError}) to get a model for $\mathcal{E}(l_\text{max})$. Rewriting Eq.~(\ref{Eq:f_w_error}) by setting $f_0=\tlo_n$ and $f(p)=\tlo_\text{num}(p)$, we can then use the largest value of $l_\text{max}=8$ to estimate the true value as
\begin{align}
\tlo_n \approx \tlo_\text{num}(8) - \mathcal{E}_0k^{8}.
\end{align}

The calculated values are listed in Tab.~\ref{Tab:k0_Qs_d} along with the estimated errors $\Delta\omega_n=|\Delta\omega_n-\text{i}\Delta\gamma_n|$, which grow as a function of $n$, because all calculations were performed with a fixed number of spherical wave functions set by $l_\text{max}=8$. The QNM wave functions of interest grow in complexity with increasing values of $n$, and this translates into a slower rate of convergence and hence an increase in numerical error for a given value of $l_\text{max}$. The associated generalized effective mode volumes of the five QNMs are listed in Table~\ref{Tab:v_ns}. They were all calculated using integration around the complex resonance frequency as described in Section~\ref{Sec:normalization_of_plasmonic_dimer} with post processing of the data to estimate the numerical error as described above. %

\label{Sec:Calcualtion_details_for_plasmonic_dimer}
\begin{table}[htb]
\centering
\begin{tabular}{ l | c | c }
$n$ & $\tlo_n d/2\pi\text{c}$ & $|\Delta\tlo_n| d/2\pi\text{c}$ \\
\hline
  1 & $0.11057832294(5) -0.03161631327(9)$ & $1\times10^{-10}$  \\
  2 & $0.171958419(6) - 0.004371575(6)\text{i}$ & $7\times10^{-10}$  \\
  3 & $0.19622876(2) - 0.00140568(9)\text{i}$ & $3\times10^{-9}$ \\
  4 & $0.20561015(0) - 0.00117600(5)\text{i}$ & $8\times10^{-9}$ \\
  5 & $0.2105409(5) - 0.0011639(6)\text{i}$ & $1\times10^{-8}$ \\
\end{tabular}
\caption{\label{Tab:k0_Qs_d}Complex resonance frequencies of the QNMs of interest with non-vanishing electric field components along the dimer axis at the point in the center of the gap between the two spheres.}
\end{table}

\begin{table}[htb]
\centering
\begin{tabular}{ l | c | c }
$n$ & $v_n^{-1}d^3 $ & $|\Delta v_n^{-1}| d^3$\\
\hline
  1 & $0.2101658(4) - 0.0527264(3)\text{i}$ & $5\times10^{-7}$ \\
  2 & $0.0977(7) - 0.0013(3)\text{i}$ & $1\times10^{-5}$ \\
  3 & $0.06278(0) + 0.00919(2)\text{i}$ & $6\times10^{-6}$\\
  4 & $0.0383(30) + 0.00091(2)\text{i}$ & $4\times10^{-6}$\\
  5 & $0.02185(7) + 0.00008(8)\text{i}$ & $2\times10^{-6}$\\
\end{tabular}
\caption{\label{Tab:v_ns}Generalized (inverse) effective mode volumes $v_n^{-1}=\mft_{yn}^2(\mr_0)\epsilon_\text{r}(\mr_0)/\langle\langle\mft_n(\mr)|\mft_n(\mr)\rangle\rangle$ of the QNMs of interest with non-vanishing electric field components along the dimer axis at the point in the center of the gap between the two spheres.}
\end{table}

\section{Dispersive materials}
\label{App:dispersiveMaterials}
Upon adding Eq.~(\ref{Eq:DrudeCurrent_governingEq}), the governing matrix equation for the QNMs takes the form

\begin{align}
\begin{bmatrix}
0& [\epsilon_0]^{-1}\nabla\times & -[\epsilon_0]^{-1} \\
-[\mu_0]^{-1}\nabla\times &0 &0 \\
\epsilon_0\omega_\text{p}^2(\mr) & 0 & -\gamma(\mr)
\end{bmatrix}
\begin{bmatrix}
\mE(\mr,\omega)\\
\mH(\mr,\omega)\\
\mJ(\mr,\omega)
\end{bmatrix}
=-\text{i}\omega
\begin{bmatrix}
\mE(\mr,\omega)\\
\mH(\mr,\omega)\\
\mJ(\mr,\omega)
\end{bmatrix},
\label{Eq:MaxwellEquations_matrixForm_QNM_w_J}
\end{align}
with $\mJ(\mr,\omega) = 0$ at positions outside the material. In the Drude model, the current density is directly proportional to the electric field, so we do not introduce additional boundary conditions for $\mJ(\mr,\omega)$ at the boundary of the material. The weight function in the inner product is extended to $\mathbf{W}=\text{diag}\{\epsilon_0, \mu_0,[\epsilon_0\omega_\text{p}^2]^{-1}\}$, and the arguments of Sec. \ref{Sec:adjoint_QNMs_and_normalization} can be repeated to find, that if $\mFt_m(\mr) = [\mft_m(\mr),\mgt_m(\mr),\mjt_m(\mr)]^\text{T}$ is a solution to Eq.~(\ref{Eq:Maxwell_QNM_spinorForm}), where $\mathbf{D}$ is the matrix in  Eq.~(\ref{Eq:MaxwellEquations_matrixForm_QNM_w_J}) and the electromagnetic fields obey the Silver-M{\"u}ller radiation condition, then $\mFb_m(\mr) = [\mft_m(\mr),-\mgt_m(\mr),-\mjt_m(\mr)]^\text{T}$ is a solution to Eq.~(\ref{Eq:adjointQNMsDiffEq}), in which
\begin{align}
\mDb = \begin{bmatrix}
0& -[\epsilon_0]^{-1}\nabla\times & [\epsilon_0]^{-1} \\
[\mu_0]^{-1}\nabla\times &0 &0 \\
-\epsilon_0\omega_\text{p}^2(\mr) & 0 & -\gamma(\mr)
\end{bmatrix}
,	
\label{Eq:adjoint_D_with_J}
\end{align}
and the electromagnetic fields obey the adjoint radiation condition. The projection operator in Eq.~(\ref{Eq:QNM_innerprod_generalized}) can now be generalized as
\begin{align}
\langle\langle\mFb_m(\mr)|\mF(\mr,\omega)\rangle\rangle &= \frac{1}{2\epsilon_0}\int_V\epsilon_0\mft_m(\mr)\cdot\mE(\mr,\omega) - \mu_0 \mgt_m(\mr)\cdot\mH(\mr,\omega)\nonumber\\&\qquad\qquad\qquad\qquad\qquad\qquad-\mjt_m(\mr)\mJ(\mr,\omega)/\epsilon_0\omega_\text{p}^2\,\ud V \nonumber \\
&\quad+\frac{\text{i}}{2\epsilon_0(\tlo_m-\omega)}\int_{\partial V} [\mE(\mr,\omega)\times\mgt_m(\mr)-%
\mft_m(\mr)\times\mH(\mr,\omega)]\cdot\mathbf{n}\, \ud A,
\label{Eq:QNM_innerprod_generalized_Drude_with_J}
\end{align}
and the rest of the arguments in Sec.~\ref{Sec:adjoint_QNMs_and_normalization} remains valid, as do the arguments surrounding the projection properties in Section~\ref{Sec:Formal_expansions}. Note, that %
the surface integral is unaffected, because the current density vanishes outside the material. 

It is illustrative to rewrite the projection operator in terms of the electromagnetic fields only. From Eq.~(\ref{Eq:MaxwellEquations_matrixForm_QNM_w_J}), it is clear that the current density is directly proportional to the electric field,
\begin{align}
\mJ(\mr,\omega) = \epsilon_0\frac{\omega_\text{p}^2(\mr)}{\gamma(\mr)-\text{i}\omega}\mE(\mr,\omega).
\end{align}
Inserting this expression in Eq.~(\ref{Eq:QNM_innerprod_generalized_Drude_with_J}), we can write the projection operator in the compact form of Eq.~(\ref{Eq:QNM_innerprod_generalized_Drude}). %

\section{Calculation details for the dielectric barrier}
The one dimensional example of a dielectric barrier represents the special case of a dielectric slab in three dimensions, when considering only fields of perpendicular incidence. For definiteness, we consider electromagnetic waves moving along the $x$ axis and choose the polarizations of the electric and magnetic fields to be along the $y$ and $z$ axes, respectively. Writing $\mE(\mr,\omega)=E(x,\omega)\hy$ and $\mH(\mr,\omega)=H(x,\omega)\hz$, we find that the Silver-M{\"u}ller radiation condition takes the form
\begin{align}
E(x,\omega) \rightarrow \sqrt{\frac{\mu_0}{\epsilon_\text{B}\epsilon_0}}H(x,\omega),\quad x\rightarrow\infty,
\end{align}
The general solutions to the wave equation is known to be a sum of forwards and backwards propagating plane waves of the form $E(x,\omega)\propto\exp\{\pm\text{i}\omega x/\text{c}\}$. Therefore, if the radiation condition is satisfied in the limit $x\rightarrow\infty$, it must also be satisfied at all finite values of $x$ outside the scattering region. Considering also positions left of the barrier, we can therefore rewrite the radiation condition as a boundary condition in the exact form in Eq.~(\ref{Eq:QNM_boundary_conditoion_1D}), which is satisfied by both the QNMs and the Green function for the dielectric barrier.

\subsection{QNMs of dielectric barrier}
\label{App:QNMs_of_dielectric_slab}
To calculate the QNMs of the dielectric barrier, we %
start with the Ansatz for the electric field QNM %
\begin{align}
\ft_m(x)=&\begin{cases}
    A_m \text{e}^{-\text{i}n_\text{B}\tlk_m x}& x<-L/2\\
    B_m\text{e}^{\text{i}n_\text{R}\tlk_mx}+C_m\text{e}^{-\text{i}n_\text{R}\tlk_mx} & -L/2<x<L/2\\
    D_m \text{e}^{\text{i}n_\text{B}\tlk_mx} & L/2<x
   \end{cases}
\label{Eq:QNM_dielectric_barrier_eField_app}
\end{align}
where $\tlk_m=\tlo_m/\text{c}$. Because of the symmetry of the problem, the solutions will be either even or odd with respect to the point $x=0$. Therefore, we can set $C_m=(-1)^mB_m$ and $D_m=(-1)^mA_m$. Moreover, since the QNMs are eigenfunctions, we can introduce a scaling of choice by setting $B_m=1$. Finally, for a given $m$ we can then calculate $A_m$ using the continuity at $x=-L/2$ to find the general form for the electric field QNM in Eq.~(\ref{Eq:QNM_dielectric_barrier_eField}). From the Maxwell curl equation it now follows that the corresponding magnetic field QNM must be of the form
\begin{align}
\gt_m(x)=&\sqrt{\frac{\epsilon_0}{\mu_0}}\begin{cases}
    -n_\text{B}A_m  \text{e}^{-\text{i}n_\text{B}\tlk_m x}& x<-L/2\\
    n_\text{R}\left(\text{e}^{\text{i}n_\text{R}\tlk_mx} -(-1)^m\text{e}^{-\text{i}n_\text{R}\tlk_mx}\right) & -L/2<x<L/2\\
    (-1)^m n_\text{B}A_m \text{e}^{\text{i}n_\text{B}\tlk_mx} & L/2<x
   \end{cases}.
\label{Eq:QNM_dielectric_barrier_hField_app}
\end{align}

To calculate the complex QNM frequencies, we use the requirement of continuity of both $\ft_m(x)$ and $\gt_m(x)$. At $x=-L/2$ we find the conditions
\begin{align}
A_m\text{e}^{\text{i}n_\text{B}\tlk L/2} &= \text{e}^{-\text{i}n_\text{R}\tlk_m L/2} +(-1)^m\text{e}^{\text{i}n_\text{R}\tlk_m L/2}
\end{align}
and
\begin{align}
-n_\text{B}A_m\text{e}^{\text{i}n_\text{B}\tlk L/2} &= n_\text{R}\left(\text{e}^{-\text{i}n_\text{R}\tlk_m L/2} -(-1)^m\text{e}^{\text{i}n_\text{R}\tlk_m L/2}\right),
\end{align}
and combining the two, we can eliminate $A_m$ to find the condition
\begin{align}
(n_\text{R}+n_\text{B})^2\text{e}^{-\text{i}n_\text{R}\tlk L} = (n_\text{R}-n_\text{B})^2\text{e}^{\text{i}n_\text{R}\tlk L}.
\label{Eq:1D_QNM_freq_condition_app}
\end{align}
Rearranging the terms by expanding the squares, this relation can be rewritten in the form of Eq.~(\ref{Eq:1D_errorMap}) for which the solutions are given in Eq.~(\ref{Eq:QNM_freqs_dielectric_barrier}). Up to a normalization factor, this completely specifies the QNMs of the dielectric barrier.

\subsection{Electric field Green function for the dielectric barrier}
\label{App:Green_function_for_the_dielectric_slab}
Combining the equations for $\mG^\text{EE}(\mr,\mr',\omega)$ and $\mG^\text{HE}(\mr,\mr',\omega)$ in Eq.~(\ref{Eq:Green_def}), it follows that the electric field Green tensor solves the equation
\begin{align}
\nabla\times\nabla\times\mG^\text{EE}(\mr,\mr',\omega) - \epsilon_\text{R}(\mr,\omega)\frac{\omega^2}{\text{c}^2}\mG^\text{EE}(\mr,\mr',\omega) = \delta(\mr-\mr').
\end{align}
In the one dimensional case, we limit the discussion to the $\hy\hy$ component of the dyadic, and we %
define $G(x,x',\omega)=\hy\cdot \mG^\text{EE}(x,x',\omega)\cdot\hy$. Writing out the curl operator, we find that the one dimensional Green function solves the equation
\begin{align}
\partial^{2}_{xx}G(x,x',\omega) + \epsilon_\text{R}(x,\omega)k^2G(x,x',\omega) = -\delta(x-x'),
\label{Eq:G_defining_equation_1D}
\end{align}
where $k=\omega/\text{c}$, and we further impose the condition, that the Green tensor should also satisfy the Silver-M{\"u}ller radiation condition. To calculate $G(x,x',\omega)$, we choose a fixed $x'$ within the barrier and expand the function in forward and backwards traveling waves in the various regions defined by $x'$ and the boundaries at $x=\pm L/2$. For the case of $x'$ inside the resonator, this gives the ansatz
\begin{align}
G(x,x',\omega)=&\begin{cases}
    A(x') \text{e}^{-\text{i}n_\text{B} k x}& x<-L/2\\
    B(x')\text{e}^{\text{i}n_\text{R} k x}+C(x')\text{e}^{-\text{i}n_\text{R} k x} & -L/2<x<x'\\
    D(x')\text{e}^{\text{i}n_\text{R} k x}+E(x')\text{e}^{-\text{i}n_\text{R} k x} & x'<x<L/2\\
    F(x') \text{e}^{\text{i}n_\text{B} k x} & L/2<x
   \end{cases},
\label{Eq:electric_field_G_dielectric_barrier_app}
\end{align}
for which we must determine the six expansion coefficients. Demanding continuity of the Green function at $x=\pm L/2$, we can express $A(x')$ and $F(x')$ as
\begin{align}
A(x') &= \text{e}^{\text{i}(n_\text{R}-n_\text{B})k L/2}C(x') + \text{e}^{-\text{i}(n_\text{R}+n_\text{B})k L/2}B(x')\label{Eq:G_1D_A} \\
F(x') &= \text{e}^{\text{i}(n_\text{R}-n_\text{B})k L/2}D(x') + \text{e}^{-\text{i}(n_\text{R}+n_\text{B})k L/2}E(x')\label{Eq:G_1D_F},
\end{align}
and by combining with the requirement of differentiability, we can eliminate $A(x')$ and $F(x')$ to find the relations
\begin{align}
C(x')&=\alpha\text{e}^{-\text{i}n_\text{R}L}B(x')\label{Eq:G_1D_C}\\
D(x')&=\alpha\text{e}^{-\text{i}n_\text{R}L}E(x')\label{Eq:G_1D_D},
\end{align}
in which $\alpha=(n_\text{R}+n_\text{B})/(n_\text{R}-n_\text{B})$. To find expressions for $B(x')$ and $E(x')$, we combine the requirement of continuity at $x=x'$ and the requirement that the Green function should be a solution to Eq.~(\ref{Eq:G_defining_equation_1D}). To this end, we integrate Eq.~(\ref{Eq:G_defining_equation_1D}) from $x=x'-\epsilon$ to $x=x'+\epsilon$ to find the condition
\begin{align}
\lim_{\epsilon\rightarrow0}\big\{G'(x'+\epsilon,x',\omega) - G'(x'-\epsilon,x',\omega) \big\} = -1,
\end{align}
where $G'(x,x',\omega)$ denotes the derivative of $G(x,x',\omega)$ with respect to $x$. In this way, we find that $B(x')$ and $E(x')$ solve the equation
\begin{align}
\begin{bmatrix}
\text{e}^{\text{i}n_\text{R}k x'}& -\alpha\text{e}^{-\text{i}n_\text{R}k L}\text{e}^{\text{i}n_\text{R}k x'} \\
\alpha\text{e}^{-\text{i}n_\text{R}k L}\text{e}^{-\text{i}n_\text{R}k x'} &-\text{e}^{-\text{i}n_\text{R}k x'}
\end{bmatrix}
\begin{bmatrix}
B(x')\\
E(x')
\end{bmatrix}
=-\frac{\text{i}}{2n_\text{R}k}
\begin{bmatrix}
1\\
-1
\end{bmatrix},
\label{Eq:Green_function_1D_B_and_E_coeffs}
\end{align}
for which we can write the solutions as
\begin{align}
B(x') &= -\frac{\text{i}}{2n_\text{R}k}\frac{ \text{e}^{\text{i}n_\text{R}k(L-x')} + \alpha\,\text{e}^{\text{i}n_\text{R}k x'} }{\text{e}^{\text{i}n_\text{R}k L} - \alpha^2\,\text{e}^{-\text{i}n_\text{R}k L}} \label{Eq:G_1D_B}\\
E(x') &= -\frac{\text{i}}{2n_\text{R}k}\frac{ \text{e}^{\text{i}n_\text{R}k(L+x')} + \alpha\,\text{e}^{-\text{i}n_\text{R}k x'} }{\text{e}^{\text{i}n_\text{R}k L} - \alpha^2\,\text{e}^{-\text{i}n_\text{R}k L}}\label{Eq:G_1D_E}.
\end{align}
Inserting in Eq.~(\ref{Eq:electric_field_G_dielectric_barrier_app}), this completely specifies the electric field Green function for the dielectric barrier in the case where at least one of the two point $x$ or $x'$ are inside the material. 

From the expansion coefficients in Eqs.~(\ref{Eq:G_1D_B}) and (\ref{Eq:G_1D_E}), we can find the poles of the Green function as the solutions to the equation
\begin{align}
\text{e}^{\text{i}n_\text{R}k L} = \alpha^2\,\text{e}^{-\text{i}n_\text{R}k L},
\end{align}
which is identical to the condition for the QNM resonance frequencies in Eq.~(\ref{Eq:1D_QNM_freq_condition_app}). In addition, the Green function has a pole at $k=0$, which has consequences for the QNM expansions, as pointed out in Section~\ref{Sec:Alternative_expression_for_the_Green_tensor}

\subsection{Region of convergence for the dielectric barrier}
From the analytical form of the Green function, we can calculate the boundaries of the region of convergence for the dielectric barrier, by analyzing the analytical continuation of the Green tensor as discussed in Section~\ref{Sec:Regions_of_completeness}. For $x'$ inside the resonator, and $x>L/2$, we focus initially on the functional form of $E(x')$ as given in Eq.~(\ref{Eq:G_1D_E}) and investigate the behavior for complex values of $k$ in the lower half of the complex plane. 

Simplifying the expression for $E(x')$ by the factor $\exp\{\text{i}n_\text{R}kL\}$, we can rewrite it as
\begin{align}
E(x') &= R(k)\frac{ \text{e}^{\text{i}n_\text{R}k x'} + \alpha\,\text{e}^{\text{i}n_\text{R}k (x'-L)} }{1 - \alpha^2\,\text{e}^{-2\text{i}n_\text{R}k L}},
\label{Eq:G_1D_E_modified}
\end{align}
where $R(k)=-\text{i}/2n_\text{R}k$ is unimportant in determining the region of convergence. The second term in the denominator tends to zero in an exponential fashion as $k$ is varied along any straight line downwards in the lower half of the complex plane, and the first term in the numerator will be decisive in determining the region of convergence. %
Combining Eqs.~(\ref{Eq:G_1D_D}) and (\ref{Eq:G_1D_F}), we can write 
\begin{align}
F(x')\text{e}^{in_\text{B}kx} = (\alpha+1)\text{e}^{-\text{i}(n_\text{R}+n_\text{B})kL/2}E(x'),
\end{align}
and inserting Eq.~(\ref{Eq:G_1D_E_modified}), we find, that the numerator tends to zero when $k$ is varied along any straight line downwards in the complex plane, provided 
\begin{align}
n_\text{R}(x'-L/2)+n_\text{B}(x-L/2) < 0. 
\end{align}
As the position of $x'$ is varies towards the boundary from the inside, the region of convergence closes in on the boundary from the outside. For $x'=L/2$, a detailed analysis shows that $x=L/2$ is not in the region of convergence. Due to the symmetry of the problem, we can immediately infer a similar behavior for $x<L/2$, and rewriting slightly, we find the general condition
\begin{align}
-\frac{L}{2} - \frac{n_\text{R}}{n_\text{B}}\left(\frac{L}{2}+x'\right) < x < \frac{L}{2} + \frac{n_\text{R}}{n_\text{B}}\left(\frac{L}{2}-x'\right).
\end{align}

\section{Independence of integration volume in Eq.~(\ref{Eq:QNM_innerprod_generalized})}
\label{App:integrationAppendix}
Starting from Eq.~(\ref{Eq:QNM_innerprod_generalized}) and considering any part of the volume with $\epsilon_\text{r}(\mr)=\epsilon_\text{B}$, denoted by $V'_\text{B}$, we show below that the contribution to the entire integral from this part vanishes.
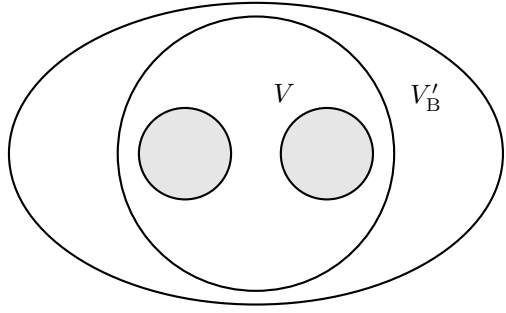
\begin{figure}[htb!]
\centering %
\begin{tikzpicture}[>=latex]
\begin{axis}[
xmin = -4, xmax = 4,
ymin = -3,  ymax = 3,
width=10.2cm,
axis on top,
axis equal,%
height=0.5\textwidth,
hide axis,
xtick=\empty,
ytick=\empty,
]
\fill[color=light-gray] (axis cs: -1.25,0) circle (0.05\textwidth);
\draw[thick] (axis cs: -1.25,0) circle (0.05\textwidth);
\fill[color=light-gray] (axis cs: 1.25,0) circle (0.05\textwidth);
\draw[thick] (axis cs:  1.25,0) circle (0.05\textwidth);
\draw[thick] (axis cs:  0,0) circle (0.15\textwidth);
\draw[thick] (axis cs: 0,0) ellipse (3.25cm and 2cm);
\node at (axis cs: .5, .75) [above]{$V$};
\node at (axis cs: 3, .6) [above]{$V'_\text{B}$};
\end{axis}
\end{tikzpicture}\quad\quad\;%
\caption{\label{Fig:integration_V_and_Vp}Schematic showing an electromagnetic resonator in the form of two spheres within an integration volume $V$. For calculations of the projection of a field onto the QNMs, the additional integration through the volume $V'_\text{B}$ does not contribute.}
\end{figure}

Considering two volumes $V$ and $V+V'_\text{B}$, as illustrated in Fig.~\ref{Fig:integration_V_and_Vp}, the contribution to the projection operation from integration throughout $V'_\text{B}$ can be written as
\begin{align}
I_{V'_\text{B}} &= \frac{1}{2\epsilon_0}\int_{V'_\text{B}} \epsilon_0\epsilon_\text{B}\mft_m(\mr)\cdot\mE(\mr,\omega) - \mu_0\mgt_m(\mr)\cdot\mH(\mr,\omega)\ud V'_\text{B}\nonumber \\
&\qquad +\frac{\text{i}}{2\epsilon_0(\tlo_m-\omega)}\int_{\partial V'_\text{B}}\left[\mH(\mr,\omega)\times\mft_m(\mr) + \mE(\mr,\omega)\times\mgt_m(\mr)\right]\cdot \hat{\mathbf{n}}\,\ud A.
\label{Eq:Projection_Vp_integral}
\end{align}
Considering the first term in the surface integral, we can express the magnetic field in terms of the electric field using the curl equation and apply Green's vector theorem of the first kind,
\begin{align}
\int_V \big[\nabla\times\mathbf{P}\big]\cdot\big[\nabla\times\mathbf{Q}\big]&- \mathbf{P}\cdot\nabla\times\nabla\times\mathbf{Q}\,\ud V =\int_{\partial V} \hat{\mathbf{n}}\cdot\left[\mathbf{P}\times\nabla\times\mathbf{Q}\right]\,\ud A,
\end{align}
to find
\begin{align}
\int_{\partial V'_\text{B}}\left[\mH(\mr,\omega)\times\mft_m(\mr) \right]\cdot\mathbf{\hat n}\, \ud A&= \frac{\text{i}}{\mu_0\omega}\int_{V'_\text{B}}\big[\nabla\times\mft_m(\mr)\big]\cdot\big[\nabla\times\mE(\mr,\omega)\big]\nonumber \\
&\qquad\qquad\qquad - \mft_m(\mr)\cdot\nabla\times\nabla\times\mE(\mr,\omega)\,\ud V.
\end{align}
Rewriting the first term using the curl equation, and the second term using the electric field wave equation, we can further simplify the expression as
\begin{align}
\int_{\partial V'_\text{B}}\left[\mH(\mr,\omega)\times\mft_m(\mr) \right]\cdot\mathbf{\hat n}\, \ud A &= -\text{i}\int_{V'_\text{B}}\tlo_m\mu_0\mgt_m(\mr)\cdot\mH(\mr,\omega) + \omega\epsilon_0\epsilon_\text{B}\mft_m(\mr)\cdot\mE(\mr,\omega)\,\ud V.
\end{align}
In a similar fashion, we can rewrite the second term in the surface integral as
\begin{align}
\int_{\partial V'_\text{B}} \left[\mE(\mr,\omega)\times\mgt_m(\mr)\right]\cdot \hat{\mathbf{n}}\, \ud A &= \text{i}\int_{V'_\text{B}}\omega\mu_0\mH(\mr,\omega)\cdot\mgt_m(\mr) + \tlo_m\epsilon_0\epsilon_\text{B}\mE(\mr,\omega)\cdot\mft_m(\mr)\,\ud V.
\end{align}
Finally, inserting the previous expressions in Eq.~(\ref{Eq:Projection_Vp_integral}), we find that the contributions from the surface integrals exactly cancel the volume integral, so that
\begin{align}
I_{\partial V'_\text{B}}=0.
\end{align}
We conclude that the regions outside the scatterers do not contribute to the value of the integral in Eq.~(\ref{Eq:QNM_innerprod_generalized}), so one is free to choose any integration volume of convenience as long as it contains the scatterers for which $\epsilon_\text{R}(\mr)\neq \epsilon_\text{B}$.

\section{Region of completeness for the sphere}
\label{Sec:Regions_of_completeness_analytical}
To investigate the region of completeness for the sphere, we %
shall focus initially on the scalar electric field Green function $g(\mr,\mr',\omega)$ \change{of} the Helmholtz equation
\begin{equation}
\nabla^{2}g(\mr,\mr',\omega)+\epsilon(\mr,\omega)\frac{\omega^{2}}{c^{2}}g(\mr,\mr',\omega)=\delta(\mr-\mr').
\end{equation}
As we shall see, the arguments leading to the region of convergence in the scalar case are identical to those in the vectorial case. The scalar Green function fulfills the Dyson equation
\begin{align}
g(\mr',\mr,\omega) = g_\text{B}(\mr',\mr,\omega) + k_0^2\int g_\text{B}(\mr',\mathbf{s},\omega)\Delta\epsilon(\mathbf{s},\omega) g(\mathbf{s},\mr,\omega) \ud V,
\label{Eq:scalarDyson_eq_3D}
\end{align}
\change{in which $k_0=\omega/\text{c}$, $g_\text{B}(\mr,\mr',\omega)$ is the Green function of the homogeneous background material with permittivity $\epsilon_\text{B}=n_\text{B}^2$, and $\Delta\epsilon(\mr,\omega)=\epsilon(\mr,\omega)-\epsilon_\text{B}$. For the present calculations, we shall assume that the material of the sphere is homogeneous, so that $\Delta\epsilon(\mr,\omega)=\Delta\epsilon(\omega)$ inside the sphere, and zero outside.} The Green function in the homogeneous background material can be written in terms of the spherical Hankel function of the first kind as
\begin{align}
g_\text{B}(\mr,\mr',\omega) = \frac{\text{i}k_\text{B}}{4\pi}h_0(k_\text{B}|\mr-\mr'|),
\label{Eq:g_scalar_hom} %
\end{align}
where $k_\text{B}=n_\text{B}\omega/\text{c} = n_\text{B}k_0$. 

Since we are ultimately interested in the limiting behavior of the Green function $g(\mr,\mr',-\text{i}\xi)$ in the limit $\xi\rightarrow\infty$, we shall make use of the fact that the spherical Hankel functions of the first kind, in general, can be written in the form~\cite{Abromowitz_1964}
\begin{align}
h_n(z) = R_n(z)\text{e}^{\text{i}z},
\end{align}
where~\cite{Abromowitz_1964}
\begin{align}
R_n(z) = \frac{\text{i}^{-(n+1)}}{z}\sum_{k=0}^n \frac{(n+k)!}{k!\Gamma(n-k+1)(-2\text{i}z)^k}
\end{align}
is a rational function of polynomials; $R_1(z) = -(z+\text{i})/z^2$, for example. Similarly, since the spherical bessel function $j_n(z)$ can be written in terms of the spherical Hankel functions of the first and second kind, we can write
\begin{align}
j_n(z) =  P_n(z)\text{e}^{\text{i}z} + Q_n(z)\text{e}^{-\text{i}z},
\label{Eq:sphBessel_from_sphHankel}
\end{align}
where $P_n(z)$ and $Q_n(z)$ are rational functions of polynomials. Clearly, when $z$ is varied in the direction of the negative imaginary axis, terms with $\exp\{\text{i}z\}$ will diverge (exponentially). As we shall see, the exponentially divergent terms will be decisive in determining the region of convergence, whereas the terms with $\exp\{-\text{i}z\}$ will be unimportant rest terms in the calculations below.

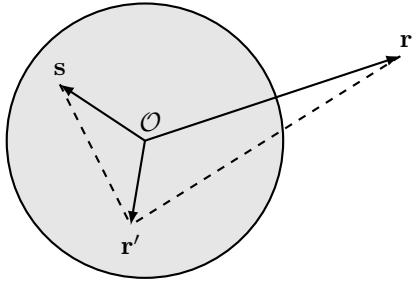
\begin{figure}[htb!]
\centering %
\begin{tikzpicture}[>=latex]
\begin{axis}[
xmin = -4, xmax = 4,
ymin = -3,  ymax = 3,
width=10.2cm,
axis on top,
axis equal,%
height=0.5\textwidth,
hide axis,
xtick=\empty,
ytick=\empty,
]
\fill[color=light-gray] (axis cs: 0,0) circle (0.15\textwidth);
\draw[thick] (axis cs: 0,0) circle (0.15\textwidth);
\draw[thick,->] (axis cs: 0,0) -- (axis cs: 4.5,1.5); %
\draw[thick,->] (axis cs: 0,0) -- (axis cs: -.25,-1.5); %
\draw[thick,->] (axis cs: 0,0) -- (axis cs: -1.5,1); %

\draw[thick, dashed] (axis cs: -1.5,1) -- (axis cs: -.25,-1.5); %
\draw[thick, dashed] (axis cs: 4.5,1.5) -- (axis cs: -.25,-1.5); %

\node at (axis cs: 0.1, 0) [above]{$\mathcal{O}$};
\node at (axis cs: 4.6, 1.5) [above]{$\mr$};
\node at (axis cs: -.25, -1.5) [below]{$\mr'$};
\node at (axis cs: -1.5, 1) [above]{$\mathbf{s}$};
\end{axis}
\end{tikzpicture}\quad\quad\;%
\caption{\label{Fig:regionOfConvergence_singleSphere}Schematic of a single homogeneous sphere with coordinates used for the derivation of the region of convergence. In the calculations using the Dyson equation, the background Green functions, which depend on the distances $|\mr-\mr'|$ and $|\mr'-\mathbf{s}|$ (indicated by dashed lines), are reformulated in terms of vectors defined with respect to the center of the sphere $\mathcal{O}$.}
\end{figure}

We consider the case of a single spherical and homogeneous resonator with radius $R$. For our analysis $\mr'$ is inside the sphere and $\mr$ is outside, as illustrated in Fig.~\ref{Fig:regionOfConvergence_singleSphere}.
The integral is over the volume of the sphere, and the general approach shall therefore be to express the Green functions in terms of spherical Bessel and Hankel functions defined with respect to the center of the sphere. For $s>r'$, for example, we can write the spherical Hankel function as~\cite{Martin_MultipleScattering}
\begin{align}
h_0(k_\text{B}|\mr'-\mathbf{s}|) = 4\pi\sum_{l,m} h_l(k_\text{B}s)\left\{Y_l^m(\mathbf{\hat{s}})\right\}^* j_l(k_\text{B}r')Y_l^m(\mathbf{\hat{r}'}),
\label{Eq:h_0_expansion}
\end{align}
where $Y_l^m(\mathbf{\hat{v}})$ denotes the spherical harmonic of order $l,m$ evaluated at the direction of the vector $\mathbf{v}$. Even if we do not know the explicit expression for $g(\mr',\mr,\omega)$ at this point, we know that we can expand it in terms of spherical Bessel functions defined with respect to the center of the sphere as~\cite{deLasson_JOSAB_30_1996_2013}
\begin{align}
g(\mr',\mr,\omega) = \sum_{\nu,\mu} j_n(n_\text{R}k_0r')Y_\nu^\mu(\mathbf{\hat{r}'})\alpha_{\nu,\mu}(\mr,\omega),
\label{Eq:g_sph_expansion}
\end{align}
where the unknown expansion coefficients $\alpha_{\nu,\mu}(\mr,\omega)$ depend on both $\mr$ and $\omega$ as indicated.

Inserting now Eqs.~(\ref{Eq:h_0_expansion}) and (\ref{Eq:g_sph_expansion}), we can express the integral in Eq.~(\ref{Eq:scalarDyson_eq_3D}) in terms of spherical wave functions %
defined with respect to the center of the sphere. For the angular part of the integration, the orthogonality relation~\cite{Martin_MultipleScattering}
\begin{align}
\int_\Omega Y_n^m(\mathbf{\hat{s}})\left\{Y_\nu^\mu(\mathbf{\hat{s}})\right\}^*\ud\Omega = \delta_{n\nu}\delta_{m\mu},
\end{align}
where $\Omega$ denotes the surface of the sphere, leads to considerable simplifications. %
For the integration over $s$ %
we rewrite the spherical Bessel function as in Eq.~(\ref{Eq:sphBessel_from_sphHankel}).
Looking at the integrand of this expression, the limiting behavior is given by the terms in the integrand with $s>r'$ and divergent exponential factors of the form $\exp\{\text{i}k_\text{B} s\}\exp\{\text{i}n_\text{R} k_\text{B}s\}$. In particular, by treating the integral as a Riemann sum, we can appreciate that there will be contributions to the integral with exponential factors of the form $\exp\{\text{i}k_\text{B} R\}\exp\{\text{i}n_\text{R}k_\text{B}R\}$, and since $s\leq R$, terms with these factors will be the fastest growing terms. With this argumentation, we %
write the resulting expression in the form
\begin{align}
g(\mr',\mr,\omega) = g_\text{B}(\mr',\mr,\omega) &+ \sum_{l,m} R_{l,m}^\text{max} \text{e}^{\text{i}k_\text{B} R }\text{e}^{\text{i}n_\text{R}k_0R} \text{e}^{\text{i}k_\text{B} r' } \alpha_{lm}(\mr,\omega)\nonumber \\
&+ \text{rest terms},
\label{Eq:scalarDyson_eq_3D_w_rest}
\end{align}
where $R_{l,m}^\text{max}$ contains factors from $Y_l^m(\mathbf{\hat{r}'})$ and the rational functions multiplying onto the spherical Hankel and Bessel functions; they depend at most in a non-exponential manner on $k_\text{B}$.
As suggested by the superscript, the sum contains the fastest growing exponential terms that will be decisive in determining the limiting behavior of the Green function; the rest terms contain terms that grow slower or decay (exponentially) to zero when $k_\text{B}$ is varied in the direction of the negative imaginary axis.

Next, we rewrite the expression as %
\begin{align}
g(\mr',\mr,\omega) \approx \frac{g_\text{B}(\mr',\mr,\omega)}{1 - \Xi},
\label{Eq:scalarDyson_eq_3D_fraction}
\end{align}
where
\begin{align}
\Xi = \sum_{l,m} R_{l,m}^\text{max} \alpha_{lm}(\mr,\omega) \frac{ \text{e}^{\text{i}k_\text{B} R }\text{e}^{\text{i}n_\text{R}k_0R} \text{e}^{\text{i}k_\text{B} r' } }{g(\mr',\mr,\omega)}.
\label{Eq:Xidef}
\end{align}
The use of ``$\approx$'' in Eq.~(\ref{Eq:scalarDyson_eq_3D_fraction}) signifies that we dropped the rest terms in Eq.~(\ref{Eq:scalarDyson_eq_3D_w_rest}), which are unimportant in the limit of interest. From Eqs.~(\ref{Eq:sphBessel_from_sphHankel}) and (\ref{Eq:g_sph_expansion}) we can appreciate, that if $g(\mr',\mr,\omega)$ tends to zero as $k_\text{B}$ is varied in the direction of the negative imaginary axis, then $\alpha_{lm}(\mr,\omega)$ must also tend to zero in this limit --- and it must do this faster than $\exp\{-\text{i}n_\text{R}k_\text{B}r'\}$.  One term in the expansion for $g(\mr',\mr,\omega)$ will tend to zero slower than the other terms. Denoting this term $l=L$ and $m=M$, we can simplify the fraction in Eq.~(\ref{Eq:Xidef}) by $\exp\{\text{i}n_\text{R}k_\text{B}r'\}\alpha_{LM}$ to find
\begin{align}
\Xi &= \frac{ R_{L,M}^\text{max} \text{e}^{\text{i}k_0( n_\text{B}R + n_\text{R} R + n_\text{B}r' - n_\text{R} r') }}{1 + \text{rest terms}} \nonumber \\[2mm]
&\quad+ \sum_{\substack{ l\neq L \\ m\neq M}}R_{l,m}^\text{max} \frac{\alpha_{lm}(\mr,\omega)}{\alpha_{LM}(\mr,\omega)} \frac{  \text{e}^{\text{i}k_0( n_\text{B}R + n_\text{R} R+ n_\text{B}r' - n_\text{R} r') } }{1 + \text{rest terms}},
\label{Eq:Xi_rewriting}
\end{align}
where the rest terms in the numerators all tend to zero as $k_\text{B}$ is varied in the direction of the negative imaginary axis. By construction, the limiting behavior of $\Xi$ is now governed by the first term in Eq.~(\ref{Eq:Xi_rewriting}).

In the last step, we expand the background Green function $g_\text{B}(\mr',\mr,\omega)$ in spherical Hankel and Bessel functions defined with respect to the center of the sphere using Eq.~(\ref{Eq:h_0_expansion}) and simplify the fraction in Eq.~(\ref{Eq:scalarDyson_eq_3D_fraction}) by $\exp\{\text{i}k_0( n_\text{B}R + n_\text{R} R+ n_\text{B}r' - n_\text{R} r') \}$. In this way, we can see, that the Green function $g (\mr',\mr,-\text{i}\xi)$ tends to zero in the limit $\xi\rightarrow\infty$ if
\begin{align}
r < R + \frac{n_\text{R}}{n_\text{B}}(R - r').
\label{Eq:region_of_convergence_sphere}
\end{align}
Thus, the region of convergence for the spherical resonator is itself a sphere with the same center and with a radius which depends on $r'$. As $r\rightarrow R$, the radius of the region of convergence tends to the radius of the spherical resonator.

\subsubsection*{Generalization to the tensor case}
The electric field Green tensor in general satisfies the Dyson equation
\begin{align}
\mG^\text{EE}(\mr,\mr',\omega)= \mG_\text{B}^\text{EE}(\mr',\mr,\omega) + k_\text{B}^2\Delta\epsilon\int_{\Delta V}  \mG_\text{B}^\text{EE}(\mr',\mathbf{s},\omega)\mG^\text{EE}(\mathbf{s},\mr,\omega) \ud V,
\label{Eq:Dyson_eq_3D}
\end{align}
where the Green tensor in the homogeneous background material can be written in terms of the scalar background Green function in Eq.~(\ref{Eq:g_scalar_hom}) as
\begin{align}
\mG^\text{EE}_\text{B}(\mr,\mr',\omega) = \left(1 + \frac{\nabla\nabla}{k_\text{B}^2}\right)g_\text{B}(\mr,\mr',\omega).
\label{Eq:G_EE_B_from_scalar}
\end{align}
As in the scalar case, even if we do not know the explicit expression, we know that we can expand all components of $\mG^\text{EE}(\mr,\mr',\omega)$ in terms of spherical wave functions as in Eq.~(\ref{Eq:g_sph_expansion}). %
In this way, we can expand all terms in the integrand to end up with 9 separate equations of the form in Eq.~(\ref{Eq:scalarDyson_eq_3D_w_rest}). The rest of the argumentation remains valid, so we conclude that the region of convergence is the same as in the scalar case in Eq.~(\ref{Eq:region_of_convergence_sphere}).

\reminder{

\subsection{Region of convergence for a dimer of spheres}
In the case of two spheres, we shall follow the same general approach as for a single sphere and focus on the case of $\mr'$ in sphere 1, as illustrated in Fig.~\ref{Fig:regionOfConvergence_twoSpheres}.
\begin{figure}[htb!]
\centering %
\begin{tikzpicture}[>=latex]
\begin{axis}[
xmin = -4, xmax = 10,
ymin = -3,  ymax = 3,
width=\textwidth,
axis on top,
axis equal,%
height=0.5\textwidth,
hide axis,
xtick=\empty,
ytick=\empty,
]
\fill[color=light-gray] (axis cs: 0,0) circle (0.15\textwidth);
\draw[thick] (axis cs: 0,0) circle (0.15\textwidth);
\draw[thick,->] (axis cs: 0,0) -- (axis cs: 4.5,1.5); %
\draw[thick,->] (axis cs: 0,0) -- (axis cs: -.25,-1.5); %
\draw[thick,->] (axis cs: 0,0) -- (axis cs: -1.5,1); %
\draw[thick, dashed] (axis cs: -1.5,1) -- (axis cs: -.25,-1.5);
\draw[thick, dashed] (axis cs: 4.5,1.5) -- (axis cs: -.25,-1.5);

\node at (axis cs: 0.05, 0) [above]{$\mathcal{O}_1$};
\node at (axis cs: 4.6, 1.5) [above]{$\mr$};
\node at (axis cs: -.25, -1.5) [below]{$\mr'$};
\node at (axis cs: -1.5, 1) [above]{$\mathbf{s}_1$};

\fill[color=light-gray] (axis cs: 7,0) circle (0.125\textwidth);
\draw[thick] (axis cs: 7,0) circle (0.125\textwidth);
\draw[thick,->] (axis cs: 7,0) -- (axis cs: 4.5,1.5); %
\draw[thick,->] (axis cs: 7,0) -- (axis cs: 6.5,-1.25); %
\draw[thick, dashed] (axis cs: -.25,-1.5) -- (axis cs: 6.5,-1.25); %

\node at (axis cs: 7.05, 0) [above]{$\mathcal{O}_2$};
\node at (axis cs: 6.5, -1.25) [below]{$\mathbf{s}_2$};

\end{axis}
\end{tikzpicture}\quad\quad\;%
\caption{\label{Fig:regionOfConvergence_twoSpheres}Schematic of two homogeneous spheres with coordinates %
for the derivation of the region of convergence. In the calculations using the Dyson equation, the background Green functions, which depend on the distances $|\mr-\mr'|$, $|\mr'-\mathbf{s}_1|$, or $|\mr'-\mathbf{s}_2|$ (indicated by dashed lines), are reformulated in terms of vectors defined with respect to either of the centers of the spheres $\mathcal{O}_1$ and $\mathcal{O}_2$.}
\end{figure}
For positions $\mathbf{s}_1$ in sphere 1, we express the background Green function using Eq.~(\ref{Eq:h_0_expansion}). For positions $\mathbf{s}_2$ in sphere 2, if $\mathbf{b}$ denotes the vector connecting the centers of the two spheres, we can use a so-called two-center expansion to write the spherical Hankel function in the general form~\cite{Martin_MultipleScattering}
\begin{align}
h_0(k_\text{B}|\mr'-\mathbf{s}_2|) = \sum_{n,m,\nu,\mu}S_{n\nu}^{m\mu}(k_\text{B}\mathbf{b}) j_n(k_\text{B}s_2)\left\{Y_n^m(\mathbf{\hat{s}_2})\right\}^* j_\nu(k_\text{B}r')Y_\nu^\mu(\mathbf{\hat{r}'}),
\label{Eq:h_0_two_center_expansion}
\end{align}
where the separation matrices $S_{n\nu}^{m\mu}(k_\text{B}\mathbf{b})$ have the functional form
\begin{align}
S_{n\nu}^{m\mu}(k_\text{B}\mathbf{b}) = \sum_q K(n,m,\nu,\mu,q) h_q(k_\text{B}b)\left\{Y_q^{\mu-m}(\mathbf{\hat{s}})\right\}^*,
\end{align}
and the konstants $K(n,m,\nu,\mu,q)$, which can be expressed in terms of the so-called Gaunt coefficients~\cite{Martin_MultipleScattering}, do not depend on $k_\text{B}$. %

\begin{align}
g(\mr',\mr,\omega) = g_\text{B}(\mr',\mr,\omega) &+ \sum_{l,m} R_{lm}^\text{max} \text{e}^{\text{i}k_\text{B} R_1 }\text{e}^{\text{i}n_1k_\text{B}R_1} \text{e}^{\text{i}k_\text{B} r' } \alpha_{lm}^{(1)}(\mr,\omega)\nonumber \\
&+ \sum_{n,m,\nu,\mu} R_{nm\nu\mu}^\text{max}\text{e}^{\text{i} k_\text{B} b} \text{e}^{\text{i}k_\text{B} R_2 }\text{e}^{\text{i}n_2 k_\text{B}R_2} \text{e}^{\text{i}k_\text{B} r' } \alpha_{nm\nu\mu}^{(2)}(\mr,\omega)\nonumber \\
&+ \text{rest terms},
\end{align}

... \color{red}[Plus additional argumentation]\color{black}

The background Green function can be written in terms of the spherical Hankel function of the first kind as
\begin{align}
sdfsf
\end{align}

$g_\text{B}(x,x',\omega)$ denotes the background Green tensor~\cite{Tai_1994}
\begin{align}
\mG_\text{B}(\mr,\mr',\omega) = \left( 1 + \frac{1}{k_\text{B}^2}\nabla\nabla^\text{T}\right)g_\text{B}(\mr,\mr',\omega),
\end{align}

To investigate the region of completeness for the dimer of spheres, we start by looking at the case of a single spherical and homogeneous resonator and

write the total electric field Green tensor in terms of the Dyson equation as
\begin{align}
\mG(\mr,\mr',\omega) = \mG_\text{B}(\mr,\mr',\omega) + k_\text{B}^2\Delta\epsilon\int_{\Delta V}  \mG_\text{B}(\mr,\mathbf{s},\omega)\mG(\mathbf{s},\mr',\omega) \ud V,
\label{Eq:Dyson_eq_3D}
\end{align}
in which $\mG_\text{B}(x,x',\omega)$ denotes the background Green tensor~\cite{Tai_1994}
\begin{align}
\mG_\text{B}(\mr,\mr',\omega) = \left( 1 + \frac{1}{k_\text{B}^2}\nabla\nabla^\text{T}\right)g_\text{B}(\mr,\mr',\omega),
\label{Eq:backGroundGreenTensor}
\end{align} %
where
\begin{align}
g_\text{B}(\mr,\mr',\omega) = \frac{1}{4\pi}\text{e}^{\text{i}k_\text{B}|\mr-\mr'|}.
\end{align}
is the scalar background Green function in the homogeneous background material. The matrix charactor of Eq.~(\ref{Eq:Dyson_eq_3D}) leads to a

Although immensely complicated in general, each element of the $3\times3$ matrix on the right hand side of Eq.~(\ref{Eq:Dyson_eq_3D}) can be written as a sum of terms of the form
\begin{align}
k_\text{B}^2\Delta\epsilon\int_{\Delta V}
\end{align}

, wherefore we can pull the entire differential operator
is related to the electric field Green tensor $\mG^\text{EE}(\mr,\mr',\omega)$ via the relation

\section{Regions of completeness analytical approach}
\label{Sec:Regions_of_completeness_analytical}
One option for estimating the region of completeness for a certain class of resonators made from individual point scatterers is the Dyson equation,
\begin{align}
\mG(\mr,\mr',\omega) = \mG_\text{B}(\mr,\mr',\omega) + \frac{\omega^2}{\text{c}^2}\int_V\mG_\text{B}(\mr,\mr'',\omega)\Delta\epsilon_\text{r}(\mr'')\mG(\mr'',\mr',\omega) \ud V''
\label{Eq:Dyson}
\end{align}
in which $\Delta\epsilon_\text{R}(\mr) = \epsilon_\text{r}(\mr)-\epsilon_\text{B}$, and $\mG^\text{B}(\mr,\mr',\omega)$ is the electric field Green tensor in a homogeneous background of permittivity $\epsilon_\text{B}$. In higher dimensions, one must in principle include a de-polarization term~\cite{Yaghjian_proc_IEEE_68_248_1980, Martin_PRE_58_3009_1998} on the right hand side of Eq.~(\ref{Eq:Dyson}), but this does not affect the analysis below.

If the spatial extent of $\Delta\epsilon(\mr)$ is sufficiently small, we can approximate it as  $\Delta\epsilon_n(\mr)\approx\Delta\epsilon\delta(\mr-\mr_n)$. By exploiting Eq.~(\ref{Eq:Dyson}), we can in principle iteratively calculate the Green tensor for a large collection of point scatterers as in Ref.~\cite{Martin_JOSAA_11_1073_1994}. %
Denoting by $\mG^{(n)}(\mr,\mr')$ the Green tensor in the presence of $n$ scatterers, the $n$'th iteration of the program gives
\begin{align}
\mG^{(n)}(\mr,\mr') = \mG^{(n-1)}(\mr,\mr') + k_0^2\Delta\epsilon \frac{\mG^{(n-1)}(\mr,\mr_n)\mG^{(n-1)}(\mr_n,\mr')}{1-k_0^2\Delta\epsilon\mG^{(n-1)}(\mr_n,\mr_n)}.
\end{align} %
The strategy now, is to find bounds on $\mr$ and $\mr'$ for which the magnitude of the Green tensor vanishes in the limit $|\omega|\rightarrow\infty$.

\subsection{Scattering by two points in one dimension}
\label{Sec:Scattering_by_two_points}
By introducing the notation
\begin{align}
R_0 &= \text{i}k_0|x-x'| \\
R_n &= \text{i}k_0|x-x_n|\\
R'_n &= \text{i}k_0|x'-x_n|\\
L =&=\text{i}k_0|x_2-x_1|,
\end{align}
and substituting the explicit expression for the Green function from Eq.~(\ref{Eq:G_B_1D}), we can simplify the analysis substantially. For the one point Green function, we find
\begin{align}
G^{(1)}(x,x') = \frac{\text{i}}{2k_0}\text{e}^{R_0} -  \frac{\Delta\epsilon}{4} \frac{\text{e}^{(R_1+R'_1)}}{1-\text{i}k_0\Delta\epsilon/2}= \frac{\text{i}}{2k_0}\Big(\text{e}^{R_0}+X(k_0)\text{e}^{(R_1+R'_1)}\Big),
\end{align}
and
\begin{align}
G^{(1)}(x_2,x_2) = \frac{\text{i}}{2k_0} -  \frac{\Delta\epsilon}{4} \frac{\text{e}^{2L}}{1-\text{i}k_0\Delta\epsilon/2} = \frac{\text{i}}{2k_0}\Big(1+X(k_0)\text{e}^{2L}\Big),
\label{Eq:G_1_x2_x2}
\end{align}
where
\begin{align}
X(k_0) = \frac{\text{1}}{2/\text{i}k_0\Delta\epsilon - 1} \rightarrow -1,\; \text{for}\;|k_0|\rightarrow\infty.
\end{align}
Using these expressions in the two point Green function, we find
\begin{align}
G^{(2)}(x,x') = \frac{\text{i}}{2k_0}\left( \text{e}^{R_0}+X(k_0)\text{e}^{(R_1+R'_1)} + \frac{\left(\text{e}^{R_2}+X(k_0)\text{e}^{(R_1+L)}\right)\left(\text{e}^{R'_2}+X(k_0)\text{e}^{(L+R'_1)}\right)}{2/\text{i}k_0\Delta\epsilon - \left(1+X(k_0)\text{e}^{2L}\right)}   \right).
\label{Eq_G_2_x_xp}
\end{align}
This expression reveals the tricky interference dictating the limiting behavior of the Green functions for this family of geometries made from point scatterers. Whereas the first term in Eq.~(\ref{Eq_G_2_x_xp}) tends to zero as $1/k_0$, the second and third terms diverge exponentially, unless $x=x'=x_1$ (in which case $R_1=R'_1=0)$. At large values of $|k_0|$, however, the two divergent terms differ in sign. To evaluate this limit, we set $2/\text{i}k_0\Delta\epsilon=0$ and $x(k_0)=-1$ and write the expression as a single fraction as
\begin{align}
G^{(2)}(x,x') \approx \frac{\text{i}}{2k_0}\left( \frac{ \left(\text{e}^{R_0}-\text{e}^{(R_1+R'_1)}\right)\left(1-\text{e}^{2L}\right) - \left(\text{e}^{R_2}-\text{e}^{(R_1+L)}\right)\left(\text{e}^{R'_2}-\text{e}^{(L+R'_1)}\right)}{ 1-\text{e}^{2L}}   \right).
\label{Eq_G_2_x_xp_single_fraction}
\end{align}
Expanding the numerator, it follows that the large argument behavior of the Green function is of the form
\begin{align}
G^{(2)}(x,x') \approx \frac{\text{i}}{2k_0}\left( \text{e}^{R_2+R'_1-L} + \text{e}^{R_1+R'_2-L} - \text{e}^{R_0} \right),
\label{Eq_G_2_x_xp_three_terms_left}
\end{align}
which is still problematic, since $R_0>0$. For positions $x$ and $x'$ directly between the two scatterers, however, it will always be the case, that either $R_2+R'_1-L=R_0$ or $R_1+R'_2-L=R_0$, in which case two of the terms in Eq.~(\ref{Eq_G_2_x_xp_three_terms_left}) exactly cancel. Assuming that the first condition is true, the second condition gives the region of convergence as
\begin{align}
|x-x_1| + |x'-x_2| < |x_2-x_1| .
\end{align}
A very similar condition is found for the one-dimensional dielectric resonator in section~\ref{Sec:1Dresonator}.

\question{On would think, that this would generalize directly to higher dimensions, %
since the Hankel (2D) and spherical Hankel (3D) functions behave as complex exponentials at large arguments. Numerical tests, however, seems to suggest that the QNM expansion for the dimer is only convergent close to (or on) the line between the centers of the spheres, and only if $\mr=\mr'$, in which case the Green tensor seems to be falling off as $1/|\omega|$, which might, in fact, not be enough because of the factor of $|\omega|$ from the Jacobian of the integral} %

} %

\bibliography{qnmbib}

\end{document}